\documentclass[a4paper,11pt]{article}

\newenvironment{dedication}
  {\clearpage           
   \thispagestyle{empty}
   \vspace*{\stretch{1}}
   \itshape             
   \raggedleft          
  }
  {\par 
   \vspace{\stretch{1.5}} 
   \clearpage           
  }

\usepackage[a4paper,left=2.73cm,right=2.7cm,top=3cm,bottom=3.5cm]{geometry}

\usepackage{graphicx}
\usepackage{amsmath,epsfig}
\usepackage{amssymb,amsfonts}
\usepackage{latexsym}
\usepackage{epstopdf}  
\usepackage{slashed}
\usepackage{subfig}
\usepackage[colorlinks=true,linktocpage=true,linkcolor=blue,citecolor=blue]{hyperref}

\numberwithin{equation}{section}

\def\be{\begin{equation}}
\def\ee{\end{equation}}

\newcommand{\morder}[1]{\mathcal{O}\left(#1\right)}

\allowdisplaybreaks[2]

\newcommand{\bear}{\begin{align}}
\newcommand{\eear}{\end{align}}
\newcommand{\bea}{\begin{align}}
\newcommand{\eea}{\end{align}}
\newcommand{\nn}{\nonumber}

\newcommand{\LUV}{\Lambda_\mathrm{UV}}
\newcommand{\LIR}{\Lambda_\mathrm{IR}}
\newcommand{\vs}{u}

\def\hri#1#2{\href{http://arxiv.org/abs/#1}{[ArXiv:#1]#2}}
\def\hre#1#2{\href{http://arxiv.org/abs/#1/#2}{[ArXiv:#1/#2]}}

\newbox\pippobox

\def\II{\relax{\rm I\kern-.18em I}}

\def\l{\lambda}
\def\m{\mu}
\def\n{\nu}

\def\s{\sigma}
\def\pa{\partial}
\def\trc{\mathbb Tr}

\def\t{\tau}

\def\L{\Lambda}

\def\h{\kappa}
\def\gf{w}

\def\ag{\mathfrak{a}}
\def\ie{{\em i.e.},\ }

\def\Awf{{A}} 
\def\G{G} 
\def\tG{\tilde G}

\def\ax{\mathfrak{a}}

\def\Ca{C_a}
\def\thf{\theta_T}
\def\tt{{\tilde\theta}}
\def\mh{{\hat M}}

\begin{document}

\begin{titlepage}



\hfill{CCQCN-2016-1511}

\hfill{CCTP-2016-8}

\hfill{MPP-2016-167}

\vskip 40pt

\begin{center}
 {\LARGE \bf \sc The CP-odd sector and $\theta$ dynamics in holographic QCD}
\\[1.52cm]
{\large Daniel Are\'an$^{a,}$, Ioannis Iatrakis$^{b}$,
Matti J\"arvinen$^{c}$, and Elias Kiritsis$^{d,e,f}$}

\bigskip

{}$^{a}${\small \it
\href{https://www.mpp.mpg.de/english/index.html}{Max-Planck-Institut f\"ur Physik (Werner-Heisenberg-Institut)},\\
F\"ohringer Ring 6, D-80805, Munich, Germany.}\\

\vskip 5pt

{}$^{b}${\small \it \href{http://web.science.uu.nl/itf/}{Institute for Theoretical Physics},
Utrecht University Leuvenlaan 4,\\ 3584 CE Utrecht, The Netherlands.}\\

\vskip 5pt

{}$^{c}${\small \it
\href{http://www.lpt.ens.fr/}{Laboratoire de Physique Th\'eorique de l' \'Ecole Normale Sup\'erieure}
\& \\ Institut de Physique Th\'eorique Philippe Meyer,
PSL Research University, \\ CNRS,  Sorbonne Universit\'es, UPMC
Univ.\,Paris 06, \\ 24 rue Lhomond, 75231 Paris Cedex 05, France.}\\

\vskip 5pt

{}$^{d}${\small \it \href{http://hep.physics.uoc.gr}{Crete Center for Theoretical Physics,
Institute for Theoretical and Computational Physics},
Department of Physics, University of Crete, 71003 Heraklion, Greece.}\\

\vskip 5pt

{}$^{e}${\small \it \href{http://qcn.physics.uoc.gr/}{Crete Center for Quantum Complexity and Nanotechnology},\\
Department of Physics, University of Crete, 71003 Heraklion, Greece.}

\vskip 5pt

{}$^{f}${\small \it \href{http://www.apc.univ-paris7.fr}{APC, Universit\'e Paris 7, Diderot},
CNRS/IN2P3, CEA/IRFU,\\ Obs. de Paris, Sorbonne Paris Cit\'e,
B\^atiment Condorcet, \\F-75205, Paris Cedex 13, France (UMR du CNRS 7164).}

\end{center}

\vskip 17pt

\abstract{
\large
The holographic model of V-QCD is used to analyze the physics of QCD in the Veneziano large-N limit.
An unprecedented analysis of the CP-odd physics is performed going beyond the level of effective field theories. The structure of holographic saddle-points at finite $\theta$ is determined, as well as its interplay with chiral symmetry breaking. Many observables (vacuum energy and higher-order susceptibilities, singlet and non-singlet masses and mixings) are computed as functions of $\theta$ and the quark mass $m$.
Wherever applicable the results are compared to those of chiral Lagrangians, finding agreement.
In particular, we recover the Witten-Veneziano formula in the small $x\to 0$ limit, we compute the $\theta$-dependence of the pion mass and we derive the hyperscaling relation for the topological susceptibility in the conformal window in terms of the quark mass.}

\vskip 80pt


\end{titlepage}

\begin{dedication}
{\Large Dedicated to the memory of Ioannis Bakas:\\ a fine physicist, a kind man and a gentleman.}
\end{dedication}

\newpage


\tableofcontents

\section{Introduction and Outlook}

The axial anomaly plays an important role in the physics of strong interactions and is inherently related to the $U(1)_A$ problem of QCD. The massless QCD Lagrangian enjoys a flavor symmetry, $SU(N_f)_V \times SU(N_f)_A \times U(1)_V \times U(1)_A$. The $U(1)_V$ part is conserved and results in the baryon number conservation.
$SU(N_f)_V \times SU(N_f)_A$ is spontaneously broken down to $SU(N_f)_V$. The spontaneous breaking is signaled by the existence of Goldstone bosons in the low energy spectrum of the theory. However, there is neither any trace of $U(1)_A$ symmetry in the spectrum of the theory nor any light Goldstone boson which would signal its spontaneous breaking. Instead, a large mass of $\eta'$, compared to standard expectations from current algebra \cite{Weinberg:1975ui}, was observed experimentally. Historically this is known as the  $U(1)_A$ problem in QCD, \cite{Crewther:1978zz}.

't Hooft proposed, \cite{'tHooft:1976up} that the nontrivial topological gauge field configurations,
instantons,  violate the $U(1)_A$ symmetry. Classical instanton solutions lead to nonzero $\int d^4x\, {\mathbb Tr}\, G \wedge G$, where $G$ is the gluon field strength. This leads to tunneling among different  vacua with different topological charge. Most importantly, a nontrivial $\int d^4x\, {\mathbb Tr}\, G \wedge G$  in QCD results in the nonconservation of the axial current, due to the axial anomaly. This implies that a nonzero CP-odd term  in the QCD Lagrangian, known as the $\theta$-term, ${\theta \over 32 \pi^2} \,{\mathbb Tr}\, G \wedge G$, can affect non-perturbatively the dynamics of the theory.  In \cite{Crewther:1977ce}, it was pointed out that in the context of the instanton picture, certain anomalous Ward identities are not satisfied and the expectation values of certain operators do not have the correct $\theta$-dependence.

The $U(1)_A$ problem was further studied in the large $N_c$ limit where an additional puzzle appeared: if the $U(1)_A$ anomaly is responsible for the would be Goldstone boson (the $\eta'$) having a mass, then the mass must be due to the instantons.
Therefore it
should be proportional to the standard instanton factor that is exponentially small at large $N_c$.
On the other hand, Ward identities seemed to indicate an inverse power law dependence of the $\eta'$ mass on $N_c$.
Witten, \cite{Witten1}, by studying a similar model in two dimensions argued that instantons do not behave as a gas (as is usually assumed in instanton calculations), but rather the instanton number becomes continuous,
and this is responsible for the power law dependence of the $\eta'$ mass.

Along the same line of thought, Veneziano, \cite{vu1}, introduced the limit where quark loops contribute to leading order to the $U(1)_A$ anomaly:
\be
N_c \to \infty \,,\qquad N_f \to \infty \,,\qquad {N_f \over N_c}=x=\mathrm{fixed} \,,\qquad \lambda= g_{YM}^2 N_c =\mathrm{fixed} \,.
\label{vl}\ee
In this limit he was able to resolve the inconsistencies pointed out in  \cite{Crewther:1977ce},
to rederive the $\eta'-$mass formula, Eq. (\ref{etapmassfinal}), which was earlier advocated by
Witten, \cite{Witten2}, and to show that the anomalous Ward identities are satisfied.

The dependence of low energy QCD physics on the $\theta$-angle was further studied in the context of the low energy effective
Lagrangians, \cite{divechia,Rosenzweig:1979ay, DiVecchia:1980yfw, Witten:1980sp,Kawarabayashi:1980dp,Ohta:1981ai}.
More recently, lattice field theory methods have also been employed
to study the topological
dynamics of QCD, \cite{Vicari:2008jw, Bonati:2013dza, Bonati:2015sqt, Bonati:2016tvi}.

The inclusion of a $\theta$-term in QCD obviously leads to $CP$ violation effects in strong interactions.
As it was shown in \cite{Shifman:1979if}, independently of the confining gluon dynamics,
the presence of instantons leads to  $P$- and $T$-odd effects. However, no such experimental signal
has been observed until now: the experimental bound for the value of $\theta$ is
$|\theta| \le 3\cdot 10^{-10}$. Several attempts have been made
to solve the problem,
\cite{Shifman:1979if,Peccei:1977hh,WeinbergAxion,Wilczek:1977pj,Kim:1979if,
Zhitnitsky:1980tq,Dine:1981rt,Vafa:1984xg,diCortona:2015ldu}.
The  straightforward solution proposes the existence of a (fundamental) pseudoscalar axion field which
couples to the topological operator ${\mathbb Tr}\, G\wedge G$.
This coupling is suppressed by a large scale, that makes the axion interactions weak.
In this way the $\theta$-angle becomes now a dynamical variable (the expectation value of the axion field)
and the QCD dynamics forces this expectation value to relax to zero, \cite{Witten2}.

The topological effects in QCD, have recently attracted much attention due to the exciting discovery of the
Chiral Magnetic Effect, which takes place when the Quark Gluon Plasma (QGP) moves in a background magnetic
field, as soon as there is chiral charge imbalance in the medium.
It has been claimed that such an imbalance is created due to topological fluctuations of the medium and
their connection to the axial anomaly, \cite{Kharzeev:2007jp}. Even though it has been argued that the
instanton contributions at finite temperature are exponentially suppressed, \cite{Shuryak:1978yk},
topological fluctuations of the medium, due to sphalerons, at finite temperature, \cite{Kuzmin:1985mm},
lead to a net axial charge, \cite{Iatrakis:2014dka}.  Anomalous conductivities were also studied in holography in \cite{Jimenez-Alba:2014iia} and  their renormalization for non conformal theories in \cite{Gursoy:2014ela}.

The effects of the axial anomaly and the $\theta$-term in low energy QCD have been also studied in the
context of holography.
In \cite{Witten:1998uka}, Witten studied the $\theta$-dependence of the $D_4$ brane holographic model,
dual to a certain  pure four-dimensional Yang Mills theory, \cite{Witten:1998zw}.
The $\theta$-angle was
introduced as the source of a Ramond-Ramond (RR) bulk field. Then, the energy density of the vacuum was
computed as a function of $\theta$ and it was shown that for every $\theta$ there are infinite distinct
vacua.
Similar conclusions hold in bottom-up holographic models of pure Yang Mills such as  the Improved
Holographic QCD (IHQCD) model, \cite{ihqcd,ihqcd2}. In \cite{Bigazzi:2015bna},  similar observables were computed, building on the background solutions with backreacting $\theta$ of \cite{Barbon:1999zp,Dubovsky:2011tu} in case of $\theta \sim {\mathcal O}(N_c^2)$.

The
Witten-Veneziano formula for the mass of $\eta'$ in the black $D_4$ brane theory was
derived holographically in \cite{Barbon:2004dq}, by including probe $D_6$ flavor branes in the $D_4$
background, \cite{Kruczenski:2003uq}. A similar result was drawn in a different holographic model,
where $D_3$ branes were embedded on a ${\mathbb C_3 / (Z_3 \otimes Z_3)}$ orbifold
singularity, \cite{Armoni:2004dc}. Later, the Witten-Veneziano formula was verified in more realistic
holographic models such as the Witten-Sakai-Sugimoto model, \cite{Sakai:2004cn}  and the
tachyon AdS/QCD model, \cite{ckp, ikp}, where a $\theta$ angle of order ${\mathcal O}{(1)}$ was considered. The backreaction of the flavor to the geometry in the  Witten-Sakai-Sugimoto model and the effect of finite $\theta$ angle were considered in \cite{Bigazzi:2014qsa}.  In models with flavors, the coupling of the flavor branes to the
RR fields is found by anomaly inflow arguments, which were presented in \cite{Green:1996dd}.

The 't Hooft large-$N_c$ limit is a excellent technical tool to study non-perturbative dynamics.
Concerning the physics of the axial $U(1)_A$ anomaly, however, the Veneziano large-$N_c$ limit in (\ref{vl})  is more appropriate as in this limit the $U(1)_A$ anomaly is a leading effect.
This limit, was advocated already in order to describe holographic models similar to QCD, that
exhibit a conformal window in some part of their phase diagram, \cite{jk}. This led to a class
of holographic theories under the name of V-QCD, whose properties were analyzed in several
contexts with interesting and sometimes unexpected results, \cite{jk}-\cite{Jarvinen:2015ofa}.

The purpose of the present paper is to fully analyze the CP-odd dynamics of V-QCD associated to
the $\theta$-dynamics as well as to the dynamics of the phases of the quark mass matrix.

 \subsection{Summary of results}

 We will first describe the complete V-QCD action with CP-odd terms which contains the physics of the
 axial anomaly and the $\theta$-angle.
We will use this theory to analyze, among other things, the phase diagram (as a function of
the {\em complex} quark mass and the $\theta$-angle) and the meson spectrum. In more detail,
the main results are as follows.

In Sec.~\ref{sec:VQCDdef} we indicate the general structure of CP-odd terms that are added in the V-QCD
models. Our method is based on earlier work~\cite{ckp,Bigazzi:2005md,Casero:2006pt}, and its adaptation to
the fully backreacted models was initially studied in~\cite{Arean:2013tja}. The CP-odd sector arises from
the Wess-Zumino-Witten term for the $N_f$ space-filling pairs of $D4-\overline{D4}$ branes. However,
following the reasoning in the glue~\cite{ihqcd} and flavor~\cite{jk} sectors, we introduce potential
functions depending on the bulk scalars, the dilaton and the tachyon, in the CP-odd action, therefore
switching from top-down to bottom-up approach.  We restrict to flavor independent backgrounds
(respecting the $SU(N_f)_V$ symmetry), and in particular, to flavor independent quark mass, writing
the complex tachyon field as $T = \tau e^{i\xi} \mathbb{I}$, where $\mathbb{I}$ is the unit matrix in
flavor space. The final CP-odd action is then given in Eq.~\eqref{samain}. We stress that full backreaction
between all terms in the action (glue, flavor, and CP-odd) is included.

The CP-odd fields are
\begin{itemize}
 \item The axion $\ag$ which is dual to the operator $\mathbb{T}r\, G \wedge G$ and sources
 the $\theta$-angle on the boundary.
 \item The phase of the tachyon $\xi$ which is (roughly) dual to the operator $\bar \psi\gamma_5 \psi$ and
 sources the phase of the quark mass.
 \item These are related by the anomaly to the divergence of the axial current
 (the longitudinal component of the $U(1)_A$ vector of the bulk theory).
\end{itemize}
The precise dictionary is specified through the boundary coupling to field theory in~\eqref{dict}.
We  demonstrate that V-QCD  models are consistent with the periodicity of the $\theta$-angle in QCD.
As expected for QCD in the Veneziano limit~\cite{Witten:1998uka}, the vacua
related by $\theta \mapsto \theta +2\pi$ are not linked by continuous deformation of
the $\theta$-angle, but there is branch structure instead.  Moreover, we argue that the axial
anomaly, given in Eq.~\eqref{axcuranom1}, is correctly reproduced. The gauge-independent CP-odd
source is identified with the gauge invariant $\bar\theta$-angle in QCD:
$\bar\theta = \theta +\arg\det M_q$, where $M_q$ is the (complex) quark mass matrix.

In Sec.~\ref{sec:thetaback} we derive the equations of motion for V-QCD at finite $\theta$-angle
and analyze their asymptotic solutions. The axial $U(1)_A$ symmetry implies that the CP-odd degrees
of freedom, the axion and the tachyon phase, can be integrated out. After taking into account symmetry
and regularity in the IR, their effect reduces to an additional integration constant $C_a$,
which is seen to be proportional to the VEV of the   $\mathbb{T}r\, G \wedge G$ operator. After
solving the background equations, this integration constant can be mapped to the (UV value of the)
$\bar\theta$-angle.

Perturbative analysis of the solutions near the boundary shows that solutions at finite $C_a$
(and therefore nontrivial $\bar\theta$) must always have nonzero quark mass and that $\bar\theta$
becomes ambiguous as  $m_q \to 0$. This reflects properties of QCD: the $\theta$-angle can be
gauged away if any of the quarks is massless. We also note that the VEVs of the quark bilinears
$\bar\psi\psi$,  $\bar\psi\gamma_5\psi$, and the VEV of  $\mathbb{T}r\, G \wedge G$ respect axial
symmetry. The IR regularity of the (fully backreacted) solutions is seen to give constraints
to the dependence of the flavor and CP-odd terms of the V-QCD action on the dilaton and the tachyon.
In particular, we point out that the string theory prediction for the dilaton dependence of the flavor action in the IR falls in the narrow range of acceptable behaviors which produce fully regular solutions, complementing earlier, similar results~\cite{ihqcd,Arean:2013tja}.

In Sec.~\ref{sec:potentials} we review how the various potentials in the glue and flavor
sectors of the V-QCD action are constrained due to regularity, asymptotic behavior near the
boundary and in the UV, and by agreement with QCD at the qualitative level. In particular we combine previous results with the additional constraints from the asymptotic analysis of the CP-odd solutions in Sec.~\ref{sec:thetaback}.
We determine explicit choices for
potentials which satisfy the constraints, therefore finalizing the construction of V-QCD  at
finite $\bar\theta$.

In order to compare our results to the
chiral Lagrangians, they  are derived in the Veneziano limit in Sec.~\ref{sec:EFT}.
We note that
\begin{itemize}
 \item There is a delicate issue in the ordering of the chiral ($m_q \to 0$) and 't Hooft or
 probe ($x \to 0$) limits: the chiral limit needs to be taken before the probe limit or simultaneously
 with it for the chiral Lagrangians to be applicable.  This issue is not relevant  in the Veneziano limit,
 where $x$ is finite.
 \item Unlike in the 't Hooft limit, the glueballs and mesons mix at leading order in the Veneziano limit.
 We argue that this mixing does not affect the chiral Lagrangian~\eqref{cc2} for the Goldstone modes nor the GOR relation.
 \item The chiral Lagrangian in the Veneziano limit, \eqref{cc2}, has two important terms which are
 suppressed in the 't Hooft limit but are leading in the Veneziano limit: one term is responsible for the
 chiral anomaly and another  allows the decay constants of the pions and the $\eta'$ meson to be
 different.
\end{itemize}

In Sec.~\ref{sec:thetavac} V-QCD is used to analyze (analytically and
numerically) the vacuum structure as a function of $m_q$, $x=N_f/N_c$, and $\bar\theta$.
As a  function of $x$ QCD has two phases of interest: the QCD-like phase for $0<x<x_c$, and the
conformal window for $x_c\le 0 < 11/2 \equiv x_\mathrm{BZ}$ where the model has an IR fixed point
(see~\cite{jk} and the review in Sec.~\ref{sec:bg}). In addition inside the low-$x$ phase,
for $x_c-x \ll 1$, there is a region where the RG flow includes walking, or quasiconformal behavior:
the coupling constant varies very slowly for a large range of energies.

A rich and interesting structure is found in the QCD-like phase. We solve numerically the vacua in V-QCD and when applicable we  compare the results to those derived from chiral Lagrangians.

The main results from the analysis of V-QCD in the QCD-like phase $x<x_c$ are:
\begin{itemize}

 \item In the limit $m_q \to 0$, where chiral Lagrangians for QCD are reliable, they agree with V-QCD.
 In particular, the leading terms of the free energy as a function of $\bar\theta$ and topological
 susceptibility,
 which can be derived analytically in V-QCD, match exactly with the predictions of the effective (chiral)
 theory.

 \item The analytic agreement with the effective chiral field theory is present  both when (only) the pions are light
 ($m_q \to 0$ with $x=\morder{1}$), and also when both the pions and the $\eta'$ mesons are light ($m_q \to 0$ and
 $x \to 0$ with $m_q \sim x$).

 \item We carry out a detailed numerical analysis of the vacua at finite $\bar\theta$ in V-QCD in
 regimes where chiral Lagrangians are not expected do be valid
and analytic approximations are not known ({\it e.g.},
 intermediate quark masses), determining key observables such as the topological
 susceptibility and free energy.

 \item As $m_q \to \infty$, the quarks decouple and the dynamics becomes that of Yang-Mills
 theory.\footnote{Decoupling here means that observables as functions of the gauge
 invariant $\bar\theta$-angle, such as the topological susceptibility and the free energy, approach
 their Yang-Mills form. The phase of the quark mass does not decouple, but appears only through the
 gauge invariant variable $\bar\theta$.}

 \item The final result for the free energy for the energetically favored phase in Eq.~\eqref{freeenfinal}
 is similar in form to the result in the 't~Hooft limit~\cite{Witten:1998uka}.

 \item We demonstrate that the dependence between the (complex) source and VEV of the tachyon has a
 complicated structure, implied by IR regularity, which naturally appears in holographic models
 but is hard to describe by using field theory techniques. The results in this article generalize
 the spiral dependence to complex variables and is linked to a tower of perturbatively unstable
 Efimov vacua.
\end{itemize}

The phase structure as a function of $x$, in the conformal window, and near the conformal transition
at $x= x_c$ is studied as well.
In the conformal window, the vacuum structure is simpler than in the QCD-like phase as the structure
related to the Efimov vacua is absent. In this phase and near the conformal phase transition,
the dependence on the quark mass of observables (such as the topological susceptibility) is
understood in terms of the separation of UV and IR scales, 
in agreement with the behavior at $\bar\theta=0$~\cite{Jarvinen:2015ofa}.

Specifically, we find that:
\begin{itemize}
 \item
The scale separation gives rise to the hyperscaling relation for the topological susceptibility in the
conformal window:
\be
 \chi \propto m_q^\frac{4}{1+\gamma_*} \quad \mathrm{as} \quad m_q \to 0\,,\qquad \left(x_c < x < x_\mathrm{BZ} \right) \ ,
\ee
where $\gamma_*$ is the anomalous dimension of the quark mass at the IR fixed point.
\item
 In the walking regime, $x_c-x \ll 1$, the topological susceptibility obeys an intermediate scaling
 law, $\chi \propto m_q^2$, which holds for longer and longer range of masses as $x \to x_c$ from below.
\item The agreement with chiral Lagrangians as $m_q \to 0$ in the QCD-like phase, is found for all $x$ within the range $0<x<x_c$, even in the walking regime.
We check this explicitly for the topological susceptibility.
\end{itemize}

We perform two separate calculations of meson masses. First, in Sec.~\ref{sec:singletPSzeroth} we complete
the analysis of~\cite{Arean:2013tja} by computing at vanishing $\bar\theta$-angle the spectra of the
flavor singlet pseudoscalar modes, which involve the fields of the CP-odd action $S_a$. The results
are as follows:

\begin{itemize}
 \item The pseudoscalar glueball modes mixes with the $\bar \psi \gamma_5 \psi$ states at generic values
 of $x$, and the mixing is suppressed for $x \to 0$.
 \item The $\eta'$ meson is identified as the lightest state in this tower as $x \to 0$. It is shown
 analytically (in Appendix~\ref{app:WV}) and verified numerically that its mass satisfies the
 Witten-Veneziano relation
\be
 m_{\eta'}^2 = m_\pi^2  + x\, \frac{N_f\,N_c\,\chi_\mathrm{YM}}{f_\pi^2} \,,
\ee
where $\chi_\mathrm{YM}$ is the Yang-Mills topological susceptibility, when both $x$ and the quark
mass are small.
\item Apart from the $\eta'$ meson, a numerical study shows that the dependence on the masses
 on $x$ and $m_q$ is similar to other sectors, discussed in~\cite{Arean:2013tja}. In particular,
 when $m_q=0$  the dependence on $x$ is mild for $x= \morder{1}$, and as $x \to x_c{}^-$ all masses
 follow the Miransky or Berezinskii-Kosterlitz-Thouless (BKT) scaling law of Eq.~\eqref{scalescal}.
\end{itemize}

Finally, in Sec.~\ref{sec:nonsingletfl}, we analyze the spectrum of flavor non-singlet fluctuations at
finite $\bar\theta$-angle. Specifically, we use the branch of vacua that are continuous deformations
of the ``standard'' background at $\bar\theta = 0$.
\begin{itemize}
 \item At finite $\bar\theta$, the scalar and pseudoscalar mesons mix. The mixing vanishes as expected when
 $\bar\theta \to 0$. There is no mixing in the spin-one sector because vectors and axial vectors
 transform with opposite signs under charge conjugation, which remains as a good quantum number even
 at finite $\bar\theta$ (whereas parity and CP are broken).
 \item The pion mass
 is shown to satisfy the generalized Gell-Mann-Oakes-Renner relation
 \be
  f_{\pi}^2\, m_\pi^2 =  -\langle \bar \psi\psi\rangle\big|_{m_q=0}\, m_q\, \cos \frac{\bar\theta}{N_f} +\morder{m_q^2}
 \ee
 analytically and numerically.
 \item Apart from the pion mode, the dependence of the meson masses on $\bar\theta$ is weak.
\end{itemize}

Briefly the structure of this paper is as follows. In Sec.~\ref{sec:VQCDdef} we review the V-QCD model and its background solutions for $\bar\theta=0$. In Sec.~\ref{sec:thetaback} we derive the equations of motion with the CP-odd sector included, and analyze their asymptotic solutions. In Sec.~\ref{sec:potentials} we present choices of potentials for the V-QCD action which satisfy all known constraints.  Sec.~\ref{sec:EFT} contains a detailed analysis of the chiral Lagrangians for QCD in the Veneziano limit. In Sec.~\ref{sec:thetavac} we carry out a detailed analysis of the vacuum structure of the V-QCD models at finite $\bar\theta$. In Sec.~\ref{sec:singletPSzeroth} we compute the spectra of flavor singlet CP-odd fluctuations at $\bar\theta = 0$. Finally, in Sec.~\ref{sec:finitethspc} we analyze the flavor nonsinglet meson spectra at finite $\bar\theta$.

Technical details are presented in the Appendix. In Appendix~\ref{app:Asymptotics}, we carry out a detailed analysis of the UV and IR asymptotics of the backgrounds at finite $\bar\theta$. In Appendix~\ref{app:thetabackgrounds} we discuss technical details of the vacuum solutions at finite $\bar\theta$. Appendix~\ref{app:quadfluctdet} contains the fluctuation equations for the flavor singlet CP-odd modes. In Appendix~\ref{app:thbfluc} we derive the fluctuation equations for the flavor nonsinglet modes at finite $\bar\theta$. Appendix~\ref{app:WV} has details of the proof of the Witten-Veneziano formula for the mass of the $\eta'$ meson in V-QCD. In Appendix~\ref{app:gmor} we prove the Gell-Mann-Oakes-Renner relation at finite $\bar\theta$ in V-QCD.

\section{V-QCD} \label{sec:VQCDdef}

We shall start by writing down the action for V-QCD

\be
 S = S_g + S_f + S_a\,,
\ee
where $S_g$ and $S_f$ are the pieces corresponding to the glue and flavor sectors, while $S_a$, which will be the
central piece of our analysis, describes the CP-odd sector.
The first two contributions have been carefully analyzed in \cite{jk}, since they are the only ones contributing to the vacuum structure
at zero $\theta$-angle. We will briefly discuss them before focusing on the CP-odd piece $S_a$.

The glue action, introduced in {\cite{ihqcd}, takes the form
\be
S_g= M^3 N_c^2 \int d^5x \ \sqrt{-g}\left(R-{4\over3}{
(\partial\lambda)^2\over\lambda^2}+V_g(\lambda)\right)
\label{vg}
\ee
with $\l=e^\phi$ the exponential of the dilaton, dual to the operator ${\mathbb Tr} F^2$.
Hence we identify the background value of $\l$ with the 't Hooft coupling.
As for the metric, the Ansatz for the background solution reads
\be
ds^2=e^{2 \Awf(r)} (dx_{1,3}^2+dr^2)\,,
\label{bame}
\ee
where the warp factor $A$ is identified  with the logarithm of the energy scale in the field theory.
In our conventions, the UV boundary is at $r=0$ (and $A\to\infty$), and the radial coordinate is then
in the range $r\in[0,\infty)$. Moreover, the metric will be close to that of AdS near the UV boundary.
Therefore  $A \sim -\log(r/\ell)$ with $\ell$ being the (UV) AdS radius, and in the UV $r$ is roughly dual
to the inverse of the energy scale of the field theory.

As shown in \cite{Arean:2013tja} the fluctuations of the action for the flavor sector $S_f$  mix with those of the CP-odd term $S_a$. Therefore we first write the action for the flavor sector in general~\cite{ckp} (see also~\cite{sstachyon}),
\be
S_f= - \frac{1}{2} M^3 N_c  {\mathbb Tr} \int d^4x\, dr\,
\left(V_f(\l,T^\dagger T)\sqrt{-\det {\bf A}_L}+V_f(\l, TT^\dagger)\sqrt{-\det {\bf A}_R}\right)\,,
\label{generalact}
\ee
with the radicands defined as
\begin{align}
{\bf A}_{L\,MN} &=g_{MN} + \gf(\l,T) F^{(L)}_{MN}
+ {\h(\l,T) \over 2 } \left[(D_M T)^\dagger (D_N T)+
(D_N T)^\dagger (D_M T)\right] \,,\nonumber\\
{\bf A}_{R\,MN} &=g_{MN} + \gf(\l,T) F^{(R)}_{MN}
+ {\h(\l,T) \over 2 } \left[(D_M T) (D_N T)^\dagger+
(D_N T) (D_M T)^\dagger\right] \,,
\label{Senaction}
\end{align}
and the covariant derivative given by
\be
D_M T = \partial_M T + i  T A_M^L- i A_M^R T\,.
\ee
We notate the 5-dimensional indices with capital Latin letters $M,N,\ldots$ and the 4-dimensional Lorentz indices with Greek letters $\mu,\nu,\ldots$ throughout the article.
The fields  $A_{L}$, $A_{R}$ and $T$ are $N_f \times N_f$ matrices in the flavor space, and under the left and right
$U(N_f)$ gauge transformations they transform as
\begin{align}
&A_L\to V_L\,A_L\,V_L^\dagger-idV_L\,V_L^\dagger\,,\qquad
A_R\to V_R\,A_R\,V_R^\dagger-idV_R\,V_R^\dagger\,,\nonumber\\
&T\to V_R\,T\,V_L^\dagger\,,\hspace{3.1cm}
T^\dagger\to V_L\,T^\dagger\,V_R^\dagger\,,
\label{gaugetransf}
\end{align}
with $V_L\,V_L^\dagger=\mathbb{I} =V_R\,V_R^\dagger$.
$A_{L}$, $A_{R}$ are dual to the left and right flavor currents of the theory while $T$ is dual to the quark mass operator.

It is also useful to define
 \be
 x\equiv {N_f\over N_c}\;.
 \ee

In general, it is not known how to perform the trace in
(\ref{generalact}) when the arguments of the determinants
are non-Abelian
matrices in flavor space. However, since we will be considering cases where the quarks are either massless or
have all the same mass, the background solution will be proportional to the unit matrix. Additionally,
for this kind of backgrounds, the fluctuations of the Lagrangian are unambiguous up to quadratic order.

As in \cite{Arean:2013tja} we will consider the following form of the tachyon potential
\be
V_f(\l,TT^\dagger)=V_{f0}(\l) e^{- a(\l) T T^\dagger} \,,
\label{tachpot}
\ee
and restrict the functions $\h(\l,T)$ and $\gf(\l,T)$ to be independent of $T$.
Moreover, the functions $V_{f0}(\l)$, $a(\l)$, $\h(\l)$, and $\gf(\l)$ are constrained
by requiring agreement with the dynamics of QCD \cite{jk,alho,Arean:2013tja}.
We will review the suitable choices for these potentials in Section~\ref{sec:potentials}.

\subsection{The CP-odd sector and the U(1)$_A$ anomaly} \label{sec:CP-oddaction}
The action of the CP-odd sector results from the WZ term coupling the closed string axion to the
phase of the tachyon and the U(1)$_A$ gauge boson. This term was discussed in \cite{ckp}, and further adapted
to our model of holographic QCD in \cite{Arean:2013tja}.
Since we will consider only the case where the quarks are massless, or have all the same mass, we can write
the tachyon as
\be
T=\tau(r)\,e^{i\,\xi(r)}\,\mathbb{I}_{N_f}\,,
\label{comptach}
\ee
where $\mathbb{I}_{N_f}$ denotes the $N_f \times N_f$ unit matrix in flavor space.

Next, following \cite{ckp,Arean:2013tja}, we write the action of the CP-odd sector as
\be
S_a=S_\mathrm{closed}+S_\mathrm{open}\, ,\qquad
S_\mathrm{closed}=-{M^3\over2}\int d^5x\, \sqrt{-g}\,{|H_4|^2\over Z(\l)}\, ,\qquad H_4=dC_3\,,
\ee
where $C_3$ is the RR three-form axion,
and
\be
S_\mathrm{open}=i\int C_3\wedge\, \Omega_2=i\int C_3\wedge \,d\Omega_1\, ,\qquad
\Omega_1=i\,N_f\left[2V_a(\l,\tau)\,A-\xi\, dV_a(\l,\tau)\right]\, ,\nn\\
\ee
where $A$ is the flavor singlet term of the axial gauge boson
\be \label{U1Afielddef}
A_{M}={A_M^{L}-A_{M}^{R}\over 2}\,.
\ee

The potential $V_a(\l,\tau)$ is known in flat-space tachyon condensation, in which case is
the same as that appearing in the tachyonic DBI (\ie $V_f$) and is independent of the dilaton
\cite{sen,Kraus:2000nj,ckp}.
Although in our model $V_a$ might be different from $V_f$, it must satisfy the same basic properties; it becomes a constant (related to the anomaly) at $T=0$, and it vanishes exponentially at $T=\infty$.
Hence we will initially take $V_a$ to be of the form
\be
V_a(\tau)=e^{-b\,\tau^2}\,,
\label{vadef}
\ee
and discuss possible alternatives in Sec.~\ref{sec:potentials}.

After dualizing  the three-form $C_3$ to a pseudoscalar axion field $\tilde a$ via
\be
{H_4 \over  {Z(\l)}}={^*}\left(d\tilde a+i\,\Omega_1\right)\,,
 \ee
the CP-odd action becomes
\be
S_a=-{M^3\,N_c^2\over2}\int d^5x\, \sqrt{- g}\,Z(\l)\left[\partial_M\ag-x\left(2V_a(\l,\tau)\,A_M-\xi\, \partial_MV_a(\l,\tau)
\right)\right]^2\,,
\label{samain}
\ee
in terms of the QCD axion
\be
\ag={\tilde a\over N_c}\,,
\label{axdef}
\ee
which, as we will discuss below, is normalized so that $\ag$ is dual to $\theta/N_c$.

In order to establish the holographic dictionary for the CP-odd part of the boundary theory,
we start by writing the Lagrangian of QCD as
\be
{\cal S}_\mathrm{QCD} = \int d^4x\,\bigg [
-{1\over2g^2}\,{\mathbb Tr}\, G_{\mu\nu}\,G^{\mu\nu}+i\bar \psi \slashed D \psi
-\bar \psi_R\,M_q\,\psi_L-\bar \psi_L\,M_q^\dagger\,\psi_R
+{\theta \over 32\pi^2}\,\epsilon^{\m\n\rho\sigma}\,{\mathbb Tr}\, G_{\mu\nu}\,G_{\rho\sigma}
\bigg]
\label{QCDL}
\ee
where $\psi_L = (1 +\gamma^5)\,\psi/2$, $\psi_R = (1 -\gamma^5)\,\psi/2$, and $M_q$ is the (potentially complex) quark mass matrix.

The bulk action (\ref{samain}) is invariant under the gauge transformation
\be
A_M\to A_M+\pa_M\varepsilon\,,\qquad
 \xi\to \xi-2\varepsilon\,,\qquad
 \ag\to \ag+2x\,V_a\,\varepsilon\,,
\label{u1transf}
\ee
where the first two transformations follow
from \eqref{gaugetransf} for a gauge transformation of the form
$V_L=V_R^\dagger=e^{i\varepsilon\,\mathbb{I}_{N_f}}$.
Notice that on the boundary this transformation realizes the QCD axial anomaly
upon assuming that the boundary values of the fields $\ag$, $A$ and $\xi$ source the
operators $\epsilon^{\m\n\rho\sigma}\,{\mathbb Tr} (G_{\m\n} G_{\rho\sigma}), J_{\mu}^5$ and
$m_q\, \bar\psi\,\gamma^5\,\psi$ respectively,
according to the following boundary action
\begin{align}
\label{dict}
S_{\delta} &= {N_c \over 32 \pi^2} \int_{r=\delta} d^4 x \, \ag\,
\epsilon^{\m\n\rho\sigma}{\mathbb Tr} (G_{\m\n}\, G_{\rho\sigma})
+ \int_{r=\delta} d^4 x\, J^{(L)ij}_{\mu}{A^{(L)\mu\, ij}}
+ \int_{r=\delta} d^4 x\, J^{(R)ij}_{\mu}A^{(R)\mu\, ij} \nonumber \\
&\phantom{=}- K_T \int_{r=\delta} d^4 x\,\frac{1}{\ell\, \delta}\,\bar \psi_R\,T\,\psi_L
-  K_T \int_{r=\delta} d^4 x\,\frac{1}{\ell\, \delta}\,\bar \psi_L\,T^\dagger\,\psi_R
\end{align}
where $\delta$ is a UV cutoff, and $J^{(L/R)ij}_{\mu} = \bar \psi^i\, \gamma_\mu(1\pm\gamma_5)\, \psi^j/2\,$,
(with $i,j=1\dots N_f$).

The proportionality constants between boundary values of the bulk fields and the sources of their
dual operators on the field theory are not fixed in~\eqref{dict}. They nevertheless disappear from any Renormalization Group (RG)
-invariant quantity (like the product of a source times its VEV). Indeed, we have included the parameter $K_T$ which will
appear in the relation between the quark mass and the boundary value of the tachyon.

We could include a second free parameter
in front of the term $\sim \ag\,\epsilon^{\m\n\rho\sigma}{\mathbb Tr} (G_{\m\n} G_{\rho\sigma})$, see~\cite{cs}.
 However, we have chosen
to fix the normalization of the axion $\ag$ such that its boundary value is precisely $\theta/N_c$, as seen by comparing~\eqref{dict} and~\eqref{QCDL}.
Further requiring that the potential $V_a$ approaches unity at the boundary, $V_a(\l,T=0)=1$,
the $U(1)_A$ gauge transformation~\eqref{u1transf} implies that the axial anomaly is correctly reproduced. We will see this explicitly below in Eq.~\eqref{axcuranom1}.

The normalization of the couplings of the gauge fields $A^{(L/R)}$ was chosen to be consistent with the gauge transformations~\eqref{gaugetransf}.

Further,
recall that the CP transformation of the fermion bilinears is given by
\be
\bar \psi^i \psi^j(t,\mathbf{x}) \mapsto \bar \psi^j \psi^i(t,-\mathbf{x})\,,\qquad
\bar \psi^i \gamma_5 \psi^j(t,\mathbf{x}) \mapsto -\bar \psi^j \gamma_5 \psi^i(t,-\mathbf{x}) \,.
\ee
We require the proportionality coefficient $K_T$ to be real so that the corresponding terms in~\eqref{dict}
are CP invariant if the tachyon transforms as $T(r,t,\mathbf{x}) \mapsto T(r,t,-\mathbf{x})^*$,
which is indeed the transformation found in~\cite{ckp}.

Notice also that for a diagonal tachyon as in \eqref{comptach}, the last two terms of \eqref{dict} take the form
\be
-{K_T\over2} \int_{r=\delta} d^4 x\,\frac{1}{\ell\, \delta}\,\tau\left(e^{i\xi}+e^{-i\xi}\right)\bar \psi\,\psi
-{K_T\over2} \int_{r=\delta} d^4 x\,\frac{1}{\ell\, \delta}\,\tau\left(e^{i\xi}-e^{-i\xi}\right)\bar\psi\,
\gamma^5\psi\,,
\label{tachcoupl}
\ee
which in the limit of small phase $\xi\ll1$ reduce to
\be
- K_T \int_{r=\delta} d^4 x\,\frac{1}{\ell\, \delta}\,\tau\,\bar \psi\,\psi
- K_T \int_{r=\delta} d^4 x\,\frac{1}{\ell\, \delta}\,\xi\,\tau\,i\,\bar\psi\,
\gamma^5\psi\,.
\label{tachcouplsxi}
\ee
Finally, we point out that $\ag(r)$ and $\xi(r)$ both transform under \eqref{u1transf}
reflecting the transformation of $\theta$ and the quark mass phase $\xi$ under the anomalous $U(1)_{\rm A}$.
It is the gauge invariant combination
\be
\bar \ax = \ax + x\, \xi\, V_a\,,
\label{baraxdef}
\ee
which is dual to the $U(1)_A$
invariant combination $ \bar\theta/N_c = \theta/N_c +\arg(\det M_q)/N_c$,
upon taking into account that $V_a(\l,T=0)=1$.

Notice that because $\xi$ is a phase, any solution is unchanged under the shift $\xi \mapsto \xi + 2 \pi$. By using the dictionary, this shift implies $\bar \theta \mapsto \bar\theta +2 \pi N_f$, so the results in our model  will be $2\pi N_f$-periodic in $\bar\theta$. But it is known that QCD has a much shorter $2\pi$-periodicity in this angle. This periodicity will be less obvious from our analysis, because it is linked to non-Abelian $SU(N_f)$ transformations. We have already restricted our study to backgrounds where the tachyon is proportional to the unit matrix, which effectively excludes such transformations.

In order to see how the $2\pi$-periodicity arises, notice that we have made a branch choice when defining the CP-odd action in~\eqref{samain}. Here the phase of the tachyon could be written for general $T$ as
\be \label{xiTdef}
 \xi = \frac{1}{2iN_f}\left(\log\det T-\log\det T^\dagger\right)=\frac{1}{N_f}\arg\det T\,.
\ee
We observe  that the branch ambiguity of $\arg$ in~\eqref{xiTdef} corresponds to $\xi \mapsto \xi +2\pi/N_f$, which gives the desired $\bar\theta\mapsto \bar\theta +2 \pi$ in the boundary theory. The branches are connected via non-Abelian transformations. To make this explicit, we may start from a background with a diagonal tachyon, choose $V_R = V_L^\dagger = \mathrm{diag}\left(e^{i\varphi},\ldots,e^{i\varphi},e^{-i(N_f-1)\varphi}\right)$ in~\eqref{gaugetransf}, and apply the transformation as $\varphi$ varies from zero to $\pi/N_f$. Since the transformation matrices belong to $SU(N_f)$ the CP-odd action~\eqref{samain} transforms trivially. In particular,~\eqref{xiTdef} remains constant under the transformation. The end result is, however, that the tachyon changes by $T \mapsto e^{i2\pi/N_f} T$, corresponding to a shift of the tachyon phase by $2\pi/N_f$. Therefore, the transformation connects two ``adjacent'' branch choices in~\eqref{xiTdef}.

According to AdS/CFT,  the boundary field theory generating functional is given by
$W_{\rm QFT}[\ag(x,\delta),\,A^{\m}(x,\delta),\, \xi(x,\delta)]\equiv
\langle e^{i\,S_\delta}\rangle=e^{i\,S_a}$,
where the bulk action $S_a$
is taken to be on-shell.
Applying the transformation (\ref{u1transf}) one obtains
$\delta_{\varepsilon} W_{\rm QFT}\propto  \delta_{\varepsilon} S_a$, and since
$S_a$ is invariant under (\ref{u1transf}),
one obtains $\langle e^{iS_\delta}\,\delta_\varepsilon S_\delta\rangle=0$. Because $\langle e^{i\,S_\delta}\rangle$ defines the generating functional, taking functional derivatives with respect to the sources we see that all correlators of the form
$\langle \dots\,\cdot\,\delta_\varepsilon S_\delta\rangle$ vanish.
Therefore, the following equation holds  for all correlators accessible to our
holographic model (and thus corresponds to an operator identity in the dual QFT):
\begin{align}
\partial_{\mu} J^{(5)\,\m} = &{N_f \over 16 \pi^2}\, \epsilon^{\m\n\rho\sigma}\,
{\mathbb Tr}\, (G_{\m\n}\, G_{\rho\sigma})\nonumber\\
&+2i\,K_T\,\bar\psi^i\gamma^5\psi^i\,
{1\over\delta}\left[\tau \cos(\xi)\right]_{r=\delta}
-2K_T\,\bar\psi^i\psi^i\,{1\over\delta}\left[\tau \sin(\xi)\right]_{r=\delta}\,.
\label{axcuranom1}
\end{align}
Here
$J^{5}_{\mu} = \bar \psi^i\, \gamma_\mu\,\gamma^5\, \psi^i$, and
we have used the fact that $V_a(\l,T=0)=1$, and that $T$ vanishes in the UV.
Next, upon identifying the energy scale as the metric factor $e^A$, which behaves as $e^A\sim 1/r$
in the UV, we define the running quark mass (evaluated at the energy scale $\mu=e^{A}|_{r=\delta}$) as
\be
\bar m_q|_{\mu}=K_T\,{\tau(\delta)\over\ell\,\delta}\,,
\label{eq:bmqdef}
\ee
while $\xi(\delta)=\xi_0$ denotes the phase of the quarks, and $\ell$ is the AdS radius introduced below
\eqref{bame}.
Hence we can write
\begin{align}
\partial_{\mu} J^{(5)\,\m} = &{N_f \over 16 \pi^2}\, \epsilon^{\m\n\rho\sigma}\,
{\mathbb Tr}\, (G_{\m\n}\, G_{\rho\sigma})\nonumber\\
&+2i\,\bar m_q\cos(\xi_0)\,\bar\psi^i\gamma^5\psi^i\,
-2 \bar m_q \sin(\xi_0)\,\bar\psi^i\psi^i\,,
\label{axcuranom2}
\end{align}
where  $\bar\psi\psi$ and $\bar\psi\gamma^5\psi$ stand for the corresponding renormalized operators at
the energy scale $\mu$.
Notice that in the following sections we will instead consider the operators
sourced by the renormalized mass
$m_q$ which is defined
through the  UV asymptotics of the tachyon (see Appendix \ref{subapp:UVback} for more details)
\begin{align}
\label{TUVresmt}
 \frac{1}{\ell}\tau(r) \ =\ &m_q\, r\,
(-\log(r\Lambda))^{-\rho} \left[1+ {\cal
O}\left(\frac{1}{\log(r\Lambda)}\right)\right]
\\ \nn
&
+\sigma\, r^3\,
(-\log(r\Lambda))^{\rho} \left[1+ {\cal
O}\left(\frac{1}{\log(r\Lambda)}\right)\right] \,,
\end{align}
where $m_q$ is a constant and equals the running quark mass $\bar m_q$ at some fixed renormalization scale
(while $\sigma$ corresponds to the renormalized chiral condensate $\sim \langle\bar\psi\,\psi\rangle$,
and $\rho$ is defined in~\eqref{rhodef} in terms of the parameters of the model).
To be precise,
in view of \eqref{eq:bmqdef}, $\bar m_q$ and $m_q$ are related via
\be
\bar m_q = K_T\,m_q\,(-\log(\delta\,\Lambda))^{-\rho}
\left[1+ {\cal O}\left(\frac{1}{\log(\delta\Lambda)}\right)\right]\,,
\label{eq:mqbarmq}
\ee
for small $\delta$ and this same relation will hold between the renormalized operators sourced by $\bar m_q$
(see Eq.~\eqref{axcuranom2} above), and those sourced by $m_q$.

Initially, we will be interested in background solutions with both $A_{\m}$ and $\xi$ vanishing. Therefore, we need solve the equations stemming from the ${\cal O}(N_c^2)$ action $S_g+S_f$ to determine $g_{\m\n},\l,$ and $\tau$.
For the background solutions with a nonzero $\theta$-angle analyzed in
Sec.~\ref{sec:thetavac}
we will have to consider also a contribution from $S_a$.

\subsection{The background solutions at $\theta = 0$}
\label{sec:bg}

In this subsection we review some general features of the background solutions of V-QCD  at zero $\theta$-angle.
We will only consider the standard case that displays a phase diagram similar to what is expected in QCD
(see \cite{Arean:2013tja} for a thorough analysis of the constraints this requirement imposes on the different potentials entering the theory).

The background solutions follow from an Ansatz where $\l$, $\Awf$, and $T$ are functions of the radial coordinate $r$, while
the rest of the fields in the model are consistently set to zero. The Ansatz for the tachyon is further restricted to
$T=\t(r)\mathbb{I}_{N_f}$, corresponding to all quarks having the same real-valued mass (hence setting $\xi=0$ in
\eqref{comptach}).
Two  types of (zero temperature) vacuum solutions were found in  \cite{jk}:
\begin{enumerate}
\item Backgrounds with identically zero tachyon and nontrivial $\l(r)$, $\Awf(r)$. These solutions correspond to chirally symmetric vacua with zero quark mass. In this case,  analytical integration of the equations of motion leaves us with a single first order
differential equation that can be easily solved numerically.
\item Solutions with nonzero $\l(r)$, $\Awf(r)$ and $\t(r)$. These describe vacua with broken chiral symmetry,
with the quark mass and the chiral condensate corresponding respectively to the non-normalizable and normalizable
modes of the tachyon \cite{Arean:2013tja}.
These backgrounds follow from the numerical integration of a set of coupled differential equations.
\end{enumerate}
As shown in Appendix \ref{app:Asymptotics},  we can obtain analytic expansions of the solutions in the UV and IR regions of the geometry
(see \cite{jk,Arean:2013tja} for more detailed analyses).

The standard phase diagram of the theory at zero quark mass is parametrized in terms of the ratio $x=N_f/N_c$, which is
constrained to the range $0 \le x <11/2 =x_\mathrm{BZ}$ since with our normalization the upper bound corresponds to the  Banks-Zaks (BZ)
value in QCD for which the leading coefficient of the  $\beta$-function becomes positive.

The phase diagram within this range consists of two phases, corresponding to the two types of backgrounds above, separated by a phase transition at a critical value $x=x_c$.
\begin{itemize}
 \item For the range $x_c\le x<x_\mathrm{BZ}$, the dominant vacuum solution (at zero quark mass) is of the first type above, with an identically zero tachyon,
 and therefore chiral symmetry is preserved~\cite{Jarvinen:2009fe}. The IR geometry is asymptotically AdS$_5$.
\item For $0<x<x_c$ the dominant background corresponds to solutions of the second kind, hence the tachyon presents a
nontrivial profile even if the quark mass is zero. Chiral symmetry is broken in this phase and the geometry ends in a singularity in the IR.
\end{itemize}

The phase transition at $x=x_c$ (only present at zero quark mass~\cite{Jarvinen:2015ofa}) displays
BKT \cite{bkt} or Miransky \cite{miransky} scaling, in accordance with predictions from the Schwinger-Dyson approach (see, {\it e.g.},~\cite{walk2}). The chiral condensate $\sigma \sim \langle \bar \psi \psi \rangle$,
which is the order parameter of the phase transition, vanishes exponentially as $x\to x_c$ from below. As shown in \cite{jk},
\be
\label{condscaling}
\sigma \sim \exp\left(-\frac{2 \hat K}{\sqrt{x_c-x}}\right)\,,
\ee
with $\hat K$ being a positive constant, while $\sigma$ vanishes identically in the region $x>x_c$ where chiral
symmetry is unbroken. Linked to this scaling is the ``walking'' behavior of the coupling constant for
$x\lesssim x_c$. The field $\l(r)$ dual to the coupling constant becomes approximately constant,
$\l\sim\l_*\,$, for a large range of $r$, and the size of this scaling region enjoys the same scaling as
(the square root of) the condensate \eqref{condscaling}. The physics near the transition has also been studied in other models: a top-down setup~\cite{kutasov}, using a tachyonic DBI action without backreaction~\cite{kutasovdbi}, models with Einstein-dilaton gravity tuned to produce walking~\cite{kajantieIRFP}, and in dynamic AdS/QCD models which are simple bottom-up models where the holographic RG flow is tuned to match with QCD~\cite{Alvares:2012kr}.

The appearance of a region displaying walking behavior and the mechanism for the phase transition at $x=x_c$,
are related to the existence of an IR fixed point for $x\geq x_c$.
First, notice that for the first type of backgrounds in the classification above, $\l(r)\to\l_*$ as $r\to\infty$,
and the solution becomes AdS also in the IR. The region $x\geq x_c$ is therefore called the ``conformal window''.
Second, the violation of the Breitenlohner-Freedman (BF) bound by the tachyon in the IR
fixed point gives rise to an instability that
is responsible for the phase transition at the end of the conformal window ($x=x_c$).

The BF bound is given in terms of the effective IR mass of the tachyon $m_{\tau*}$ as
\be
-m_{\tau*}^2\,\ell_*^2<4\,,
\label{irbfbound}
\ee
where $\ell_*$ is the radius of the IR AdS geometry.
When the bound is violated, solutions where the tachyon has been turned on are favored,
implying spontaneous breaking of chiral symmetry (remember that
we are setting the source of the tachyon -- corresponding to the quark mass -- to zero). In \cite{jk} it was indeed
found that the bound \eqref{irbfbound} is saturated exactly at $x=x_c$, where the BKT transition described above occurs.
This is in agreement with general arguments showing that the violation of the BF bound at an IR fixed point
leads to a BKT transition~\cite{son}.
Additionally, in \cite{jk} the constant $\hat K$ of~\eqref{condscaling} was expressed in terms of $m_{\tau*}$ and $\ell_*$, which
are functions of $x$, as
\be \label{hatKres}
 \hat K = \frac{\pi}{\sqrt{\frac{d}{dx}\left(m_{\tau*}^2 \, \ell_*^2\right)\big |_{x=x_c}}}\,.
\ee

In \cite{jk} it was also shown how the Miransky scaling manifests itself in the ratio of the scales
of the model as $x\to x_c$. One can define the UV and IR scales $\Lambda_{\rm UV}= \Lambda$, $\Lambda_{\rm IR}=1/R$,
in terms of the constants appearing respectively in the UV and IR solutions (see Appendix \ref{app:Asymptotics}).
For the solutions with $x<x_c$ and $x_c -x$ large enough, $\Lambda_\mathrm{IR}/\Lambda_\mathrm{UV} = \morder{1}$,
reflecting the fact that there is only one scale in the model, as happens normally in QCD (where the single scale is denoted by $\Lambda_\mathrm{QCD}$). Instead, when $x\to x_c$,
there is a clear separation of scales, and their ratio behaves as
\be \label{scalescal}
 \frac{\Lambda_\mathrm{UV}}{\Lambda_\mathrm{IR}} \sim  \exp\left(\frac{\hat K}{\sqrt{x_c-x}}\right)\,,
\ee
hence featuring Miransky scaling.

 It is worth pointing out that even as $x\to x_c\,$, $\Lambda_{\rm UV}$ is still the scale at
 which the coupling constant
becomes small. In that limit, the range where the coupling ``walks'' is characterized by the two scales as
$\Lambda_\mathrm{UV}^{-1}\ll r \ll \Lambda_\mathrm{IR}^{-1}$, and the coupling diverges for
$r\gtrsim\Lambda_\mathrm{IR}^{-1}$.
Moreover, in terms of the two scales, the chiral condensate can be expressed as
$\sigma \sim \Lambda_\mathrm{UV} (\Lambda_\mathrm{IR})^2\,$.
Therefore, the Miransky scaling featured in Eq.~\eqref{condscaling}
follows from \eqref{scalescal}
when the condensate is expressed in units of $\Lambda_\mathrm{UV}\,$,
\ie for $\sigma/(\Lambda_\mathrm{UV})^3$.

When the quark mass is nonzero, the phase transition at $x=x_c$ becomes a crossover: chiral symmetry is broken and the dominant solution changes smoothly as $x$ is varied. Even though there are no transitions, one can identify various regions where the dependence of the background on the quark mass is different~\cite{Jarvinen:2015ofa}:
\begin{enumerate}
 \item In the QCD-like regime, with $0<x<x_c$, the background at finite $m_q$ approaches the solution at $m_q=0$ uniformly as $m_q \to 0$. For small enough $m_q$ the mass dependence is therefore perturbative. A characteristic feature in this regime is the light pion mode.
\item Adding a finite quark mass in the conformal window drives the model away from the IR fixed point. For $m_q \ll \LUV$ the background walks, and the amount of walking is controlled by the value of the quark mass. This leads to the hyperscaling relations~\cite{Evans:2014nfa,Jarvinen:2015ofa} for the meson masses.
\item At large quark mass, $m_q \gg \LUV$, the background solution of the tachyon field is large (except for very close to the UV boundary) which leads to the decoupling of quarks from the gluons and a large mass gap for the meson states.
\end{enumerate}

The results at finite quark mass agree with other approaches, in regimes of parameter space where such approaches can be trusted. In particular, the behavior within the conformal region and close to the critical value $x=x_c$ is in agreement with the analysis of RG flows~\cite{Dietrich:2009ns,DelDebbio:2010ze}.

Finally, there also exist solutions corresponding to subdominant vacua of the model. Allowing for a finite
quark mass, one finds the following generic structure \cite{jk}:
\begin{itemize}
 \item For $x_c\le x<x_\mathrm{BZ}$, only one vacuum solution exists, even at finite quark mass.
 \item When $0<x<x_c$ and the quark mass is zero, there is an infinite tower of (unstable) Efimov saddle-point solutions in addition to the
 standard, dominant solution.\footnote{For simplicity it is assumed that there exists an IR fixed point for any positive value
 of $x$, and that the BF bound is violated at the fixed point for any $x$ down to $x=0$.
 }
 \item When $0<x<x_c$ and the quark mass is nonzero, there is an even number (possibly zero) of Efimov vacua.
 The number of vacua increases with decreasing quark mass for fixed $x$.
\end{itemize}
The subdominant Efimov vacua at finite (real) quark mass were carefully studied in \cite{Jarvinen:2015ofa}.
In Sec. \ref{sec:spirals} we will generalize that analysis to the case of a complex quark mass.

\section{Equations of motion and asymptotic solutions at finite $\theta$-angle} \label{sec:thetaback}

\subsection{Equations of motion for the background}\label{subsec:eoms}

In order to study the physics of the model at finite $\theta$-angle, we will solve  the equations of motion
when the QCD axion $\ag$ is finite (and ${\cal O}(1)$ in the large $N_c$ expansion).
As implied by the axial anomaly (\ref{u1transf}) we must also allow for a nonzero overall phase of the tachyon,
and then all three sectors, glue, flavor and CP-odd, contribute to the action. We also need to consider a $U(1)_A$ flavor singlet gauge field given in~\eqref{U1Afielddef}, while the other components of the gauge field are set to zero.
Notice that we consider a tachyon field of the form $T= \t e^{i\xi} \mathbb{I}$ corresponding to all the quarks having equal complex
mass. This allows us to write the action as
\begin{align} \label{radialaction}
 S&=S_g+S_a+S_f=M^3\,N_c^2\int d^5x \nonumber
 \,\Bigg\{\sqrt{-g}\bigg[R-{4\over3}{(\partial_M\lambda)^2\over\lambda^2}+V_g(\lambda) \\\nonumber
 &\phantom{=S_g+S_a+S_f=}
 -{Z(\l)\over2}\left(\partial_M\ax+x\,\xi\,\partial_MV_a(\l,\tau)-2 x\,A_M\,V_a(\l,\tau)\right)^2
 \bigg]\nonumber\\
 &\phantom{=S_g+S_a+S_f=}
 -\frac{x\,V_f(\l,\tau)}{2}\,\left[\sqrt{-\det{\bf A}_{(+)}} + \sqrt{-\det{\bf A}_{(-)}}\right]
 \Bigg\}\,,
 \end{align}
where
\begin{align}
\label{radpmdef}
 {\bf A}_{(\pm)\,MN} &= g_{MN}+\kappa(\l)\left[
 (\partial_M \tau)\, (\partial_N \tau)
 +\tau^2(\partial_M \xi+2 A_M)\,(\partial_N \xi+2A_N)
 \right] \pm w(\l) F_{MN} \,,
\end{align}
with $F_{MN} = \partial_M A_N - \partial_N A_M$.

We list the full equations of motion in Appendix~\ref{app:eoms}, and restrict here to the case (relevant for the background) where all fields only depend on the radial coordinate $r$. We set the sources for the four-vector $A_\mu$ to zero, and as argued in Appendix~\ref{app:eoms}, the solution for $A_\mu$ then vanishes in the bulk also. Moreover, we choose the gauge $A_r=0$.

Taking the Ansatz (\ref{bame}) for the metric, we are then
left with the five fields $A(r)$, $\l(r)$, $\ax(r)$, $\tau(r)$, and $\xi(r)$.
The equation of motion for $\ax$ (Eq.~\eqref{axeom} in Appendix~\ref{app:eoms}) allows us to solve for $\ax'$ as
\be
\ax'={\Ca\,e^{-3A}\over Z}-x\,\xi\,V_a'\,,\label{axsol}
\ee
where $\Ca$ is a constant. Moreover, substituting this solution into the equation of motion for $A_r$~\eqref{Areomfinal}, we obtain
\be
{e^{3A}\over \tG}\,\h\,V_f\,\tau^2\,\xi'-\Ca\,V_a=0\,,
\label{xieqint}
\ee
where
\begin{equation}
\tG(r)=\sqrt{1+\kappa\,e^{-2A}\,(\tau'^2+\tau^2\,\xi'^2)}\,.
\label{gdef}
\end{equation}
Solving for $\xi'$ we arrive at
\be
\xi'={\Ca\, V_a\,\sqrt{e^{2A}+\h\,\tau'^2}
\over
\h^{1/2}\,\tau\sqrt{e^{8A}\,V_f^2\,\h\,\tau^2-\Ca^2\, V_a^2}}\,.
\label{xieqs2}
\ee
As we will see below when analyzing the asymptotic behavior,
the requirement of finding regular solutions with nonvanishing $\xi'$ restricts the form of the
potentials in our model.

For the EoMs of the other fields we obtain
\begin{align}
  &6A''+6A'^2=-{4\over3}{\l'^2\over\l^2}+e^{2A}\,V_g-x\,e^{2A}\,V_f\,\tG
 -\Ca^2\,{e^{-6A}\over 2Z}\,,\label{einseq1s}\\
&12A'^2={4\over3}{\l'^2\over\l^2}+e^{2A}\,V_g-x\,V_f\,{e^{2A}\over \tG}
+\Ca^2\,{e^{-6A}\over 2Z}\,,\label{einseq2s}\\
&\l''-{\l'^2\over\l}+3A'\,\l'={3\over8}\,e^{2A}\,\l^2\bigg\{-{\partial V_g\over\partial \l}+x\,
{\partial V_f\over\partial \l}\,\tG+{x\over2}\,e^{-2A}\,{\partial \h\over\partial \l}{V_f\over \tG}\,
(\tau'^2+\tau^2\,\xi'^2)\nonumber\\
&\qquad\qquad\qquad\qquad\;+\Ca^2\,{e^{-8A}\over2Z^2}\,{\partial Z\over\partial \l}
-x\,\Ca\,e^{-5A}\,\xi'\,{\partial V_a\over\partial \l}
\bigg\}
\,,\label{lambdaeqs}\\
&\partial_r\left({e^{3A}\over \tG}\,\h\,V_f\,\tau'\right)-e^{5A}\,\tG\,{\partial V_f\over\partial \tau}
-{e^{3A}\over \tG}\,V_f\,\h\,\tau\,\xi'^2-e^{3A}{V_f\over2\tG}\,(\tau'^2+\tau^2\,\xi'^2)\,
{\partial\h\over\partial \tau}
+\Ca\,\xi'\,{\partial V_a\over\partial \tau}
=0\,,\label{taueqs}
\end{align}
where we eliminated $\ax$ by using~\eqref{axsol}.

Notice that as the potentials of the action~\eqref{radialaction} are independent of $\ax$ and $\xi$,
it is invariant under the reflection $\ax\mapsto -\ax$, $\xi \mapsto -\xi$, which corresponds to the
CP transformation on the field theory side. This is reflected in the invariance of the equations of
motion~\eqref{xieqs2} and \eqref{einseq1s}--\eqref{taueqs} under $\Ca \mapsto -\Ca$, $\xi \mapsto -\xi$.

\subsection{Asymptotics}
We will now analyze the asymptotic solutions corresponding to backgrounds
at finite $\theta$-angle.
We concentrate on the asymptotics which are affected nontrivially by finite $\bar\theta$. Other results are listed in Appendix~\ref{app:Asymptotics}.

\subsubsection{UV}
\label{ssec:uvasymp}
We begin by considering the effect of a nonzero $\xi$ on the UV solutions of
the equations (\ref{einseq1s} - \ref{lambdaeqs}) for $A(r)$ and $\l(r)$.
It is easy to check that the standard ($\theta=0$) UV asymptotics for $A(r)$ and $\l(r)$, Eqs.
(\ref{UVexpsapp}, \ref{UVexpsapp2}),
solve those equations in the UV upon assuming that $\tau$ vanishes at least as $\tau\sim r$ and $\xi'$
is regular there (the new terms sourced by $\xi'$ are suppressed at least as $r^2$).

Next, assuming that the UV metric is close to AdS, namely $e^A\simeq \ell/r$ as implied by the standard
UV asymptotics
of $A(r)$, and that the tachyon is at most $\tau\sim r$, Eqs. \eqref{xieqs2} and \eqref{taueqs} for
the modulus and phase of the tachyon become
\be \label{smalltaueqs}
 \t'' + \pa_r\log\left(  e^{3A}\,\h\, V_{f0}\right)\, \t' - e^{2A}\,m_{\t}^2\,\t - \t (\xi')^2 \simeq 0 \ ,
 \qquad
 \xi' \simeq \frac{\Ca}{\h\,V_{f0}\,e^{3A}\,\t^2}\,,
\ee
where
\be
 m_{\t}^2 = -\frac{2 a}{\h} \ .
\ee
These two real equations are equivalent to the single linear complex valued equation
\be \label{complextauUV}
 \left(\t e^{i\xi}\right)'' +  \pa_r\log\left(  e^{3A} \h V_{f0}\right)\, \left(\t e^{i\xi}\right)'
 - e^{2A} m_{\t}^2 \t e^{i\xi} \simeq 0 \,.
\ee
Therefore, towards the UV boundary the complex tachyon satisfies this linearized equation of motion where
$C_a$ does not appear explicitly.
In particular, the equation is the same as that for the (real) tachyon at zero $\theta$-angle or equivalently
at zero $\Ca$.

Inserting the UV asymptotics of $A(r)$ and $\l(r)$ in~\eqref{complextauUV}, we obtain the UV asymptotic solution for the complex tachyon,
\be
\begin{split}\label{ctachuv}
\frac{1}{\ell}\tau\,e^{i\xi} &=  e^{i\xi_0}
m_q\,r\,(-\log(\Lambda r))^{-\rho}\,\left(1+{\mathcal O}\left( \log(\Lambda\,r)^{-1}\right)\right)\\
&\phantom{=}+e^{i\xi_0} \hat\sigma\,r^3\,(-\log(\Lambda r))^\rho\,\left(1+{\mathcal O}\left( \log(\Lambda\,r)^{-1}\right)\right)\,,
\end{split}
\ee
in terms of two real valued constants $m_q$ and $\xi_0$, and one complex constant $\hat\sigma$
while $\rho$ is defined in Eq. (\ref{rhodef}).
Notice that $m_q$ and $\xi_0$ are the modulus and phase of the source dual to the complex tachyon,
and thus correspond to the absolute value of the mass of the quarks and its phase phase. This solution satisfies the assumptions we made above and is therefore valid, as can also be verified by inserting it together with the asymptotics for $A(r)$ and $\l(r)$ in the full system
(\ref{einseq1s} - \ref{taueqs}).

To obtain the relation between the integration constant $C_a$ and the coefficients of the expansion, we insert it in the second equation in~\eqref{smalltaueqs}, which gives
\be \label{Carel}
 C_a =  2\ell^5\,\h_0\,W_0\,m_q\,\mathrm{Im}\, \hat \sigma \,,
\ee
where $\h_0 = \kappa(\l=0)$ and $W_0 = V_{f0}(\l=0)$.
For positive quark mass, the UV expansion of $\xi'$ can be read
from~\eqref{ctachuv}:
\be
\xi'=r\,(-\log(\Lambda r))^{2\rho}\left[{\Ca
\over \ell^5\,\h_0\,W_0\,m_q^2}\right]\left(1+{\mathcal O}\left( \log(\Lambda\,r)^{-1}\right)\right)\,.
\label{xiuv}
\ee
Similarly, the expansion of the absolute value is
\be\label{tauuv}
\frac{1}{\ell}\tau =
m_q\,r\,(-\log(\Lambda r))^{-\rho}\,\left(1+{\mathcal O}\left( \log(\Lambda\,r)^{-1}\right)\right)\\
+ \sigma\,r^3\,(-\log(\Lambda r))^\rho\,\left(1+{\mathcal O}\left( \log(\Lambda\,r)^{-1}\right)\right)\,,
\ee
where $\sigma = \mathrm{Re}\, \hat \sigma$.
At nonzero quark mass, the relations between the VEVs can therefore be written as
\be
\hat\sigma = \sigma+{i\Ca
\over 2\ell^5\,\h_0\,W_0\,m_q}\,.
\label{cvev}
\ee

In the case of massless quarks ($m_q=0$), as the form of the complex tachyon solution already makes clear, the
physical solution corresponds to a constant $\xi = \xi_0$, for which $C_a = 0$, in agreement with~\eqref{Carel}.
Notice finally that a constant $\xi$ can be gauged away via (\ref{u1transf}) as expected for QCD with massless
quarks.

We shall finish this subsection by discussing the relation between the subleading terms in the UV asymptotics
of the complex tachyon and the corresponding VEVs on the field theory side. To establish that relation we
compare
the variation of the free energies of the field theory and its holographic dual. We allow the quark mass $m_q$, the phase $\xi_0$,
and the boundary value of the axion $a_0$ to vary keeping $\Lambda $ fixed.

Then, for an IR regular variation, the free energy density satisfies the standard formula
\be \label{dEform}
 \delta \mathcal{E} = -\delta \bar \ax(r) \left.\frac{\pa \mathcal{L}}{\pa \bar\ax'(r)}\right|_{r=0}^\infty - \delta \xi(r) \left.\frac{\pa \mathcal{L}}{\pa\xi'(r)}\right|_{r=0}^\infty
 - \delta \tau(r) \left.\frac{\pa \mathcal{L}}{\pa\tau'(r)}\right|_{r=0}^\infty\,,
\ee
where $\mathcal{L}$ is the (complete) V-QCD Lagrangian, and $\bar\ax$ was defined in \eqref{baraxdef}.
Here the second term vanishes due to the $A_r$ equation of motion\footnote{After switching to the gauge invariant axion $\bar\ax$, the action only depends on $\xi$ through its derivative and invariance under~\eqref{u1transf} implies that $\frac{\pa \mathcal{L}}{\pa\xi'(r)} = \frac{1}{2}\frac{\pa \mathcal{L}}{\pa A_r}$.}.
The third term is UV divergent and needs to be regulated. We should also make sure that the variation of the metric
does not enter the formula when $\Lambda = \LUV$ is kept fixed. These issues were analyzed in detail for the case
of zero $\theta$-angle in~\cite{Jarvinen:2015ofa}. By using the UV expansion of the phase in~\eqref{xiuv} we observe that the nontrivial phase or the CP-odd action do not add any nonzero contributions to this
analysis, and the result for the regularized third term is unchanged. That is,~\eqref{dEform} becomes
\be \label{dEres}
 \delta \mathcal{E} = - M^3 N_c^2 \Ca\, (\delta \ax_0 + x \delta \xi_0)
 - 2 M^3 N_c\, N_f\, W_0\, \kappa_0\, \ell^5\, \sigma\,\delta m_q \,.
\ee
where the first term arises from the first term of~\eqref{dEform} and the second term can be computed as in~\cite{Jarvinen:2015ofa}.
As a check, this result may also be written as
\be
 \delta \mathcal{E} = - M^3 N_c^2 \Ca\, \delta \ax_0
 - 2 M^3 N_c\, N_f\, W_0\, \kappa_0\, \ell^5\, \mathrm{Re}\left[\hat\sigma^*e^{-i\xi_0}\,\delta\! \left(m_q e^{i\xi_0}\right) \right] \,,
\ee
\ie the terms involving the tachyon admit a simple expression in terms of the complex source and VEV
in~\eqref{ctachuv}, which is consistent with the complex tachyon being the natural field to consider in the UV as seen from Eq.~\eqref{complextauUV}.

By using the QCD Lagrangian~\eqref{QCDL}, and the Feynman-Hellmann theorem, or equivalently by using the dictionary implied by~\eqref{dict}, we obtain
\be
\begin{split}
 \langle \bar \psi_R \,\psi_L \rangle
 = \frac{e^{i\xi_0}}{2}\left(\frac{\pa \mathcal{E}}{\pa m_q}
 + \frac{i}{m_q} \frac{\pa \mathcal{E}}{\pa \xi_0}\right) &=-M^3 N_c\, N_f\, W_0\,\kappa_0\, \ell^5\, e^{i\xi_0}\, \hat \sigma\\
 &=
 -M^3 N_c\, N_f\, W_0\,\kappa_0\, \ell^5\, e^{i\xi_0}\left(\sigma
 + \frac{i\Ca}{2\ell^5\,\h_0\,W_0\,m_q}\right)\,,
 \end{split}
\ee
so that the expectation value is given by the coefficient of the subdominant term of the complex tachyon
in~\eqref{ctachuv}. The other condensates are found similarly, {\em e.g.},
\begin{align} \label{condeqs}
 \langle \bar \psi\, \psi \rangle &= -2 M^3 N_c\,N_f\,W_0\,\kappa_0\,\ell^5\,
 \mathrm{Re}\left[e^{i\xi_0}\left(\sigma + \frac{i\Ca}{2\ell^5\,\h_0\,W_0\,m_q}\right)\right]\,,&
\\ i \langle \bar \psi\, \gamma_5\, \psi \rangle &= -2 M^3 N_c\,N_f\,W_0\,\kappa_0\,\ell^5\,
\mathrm{Im}\left[e^{i\xi_0}\left(\sigma + \frac{i\Ca}{2\ell^5\,\h_0\,W_0\,m_q}\right)\right]\,,&
\\
\langle\epsilon^{\m\n\rho\sigma}\,
{\mathbb Tr} (G_{\m\n} G_{\rho\sigma})\rangle&=
32\pi^2\,M^3\,N_c\,C_a\,.&
\end{align}
The relation between the VEVs
\be
 \frac{N_f}{32 \pi^2}\langle\epsilon^{\m\n\rho\sigma}\,{\mathbb Tr} (G_{\m\n} G_{\rho\sigma})\rangle = m_q\, \sin \xi_0\, \langle \bar \psi\, \psi \rangle - im_q\, \cos \xi_0\,\langle \bar \psi\, \gamma_5\, \psi \rangle
\ee
is
consistent with~\eqref{axcuranom2}.

The results above show that the phase of the tachyon is a perturbative correction in the UV when
$m_q\,\sigma  \gg  \Ca $, under the natural assumption that the factor $\ell^5\, W_0\, \kappa_0$
is of order one.
It can be fixed by requiring the UV behavior of the scalar two-point correlator to match with
perturbative QCD~\cite{Arean:2013tja}, which results in
\be \label{scalar2pt}
M^3\, W_0\,\kappa_0\,\ell^5 = \frac{1}{4\pi^2}\,.
\ee
The value of $M^3$ can also be constrained independently by comparing the pressure of the model to that of
high temperature QCD~\cite{alho,data}.

\subsubsection{IR}\label{ssec:thbckir}

We then consider the asymptotic solutions in the IR ($r\to\infty$) at finite $\theta$-angle. For physically relevant solutions we expect that the tachyon diverges in the IR as it does for $\theta=0$~\cite{Arean:2013tja}\footnote{The stable minimum of the tachyon potential is at $T\to\infty$.} so the tachyon potentials $V_f$ and $V_a$ vanish in the IR (and this is indeed what we will find for all regular solutions in the IR). This together with the regularity of $\xi'$ implies that the glue degrees of freedom $A(r)$ and $\l(r)$ satisfy the same asymptotics as at $\theta = 0$, given in Eqs.~(\ref{IRresA}, \ref{IRresl}) in Appendix~\ref{app:Asymptotics}. The asymptotics of the tachyon field is, however, modified as one turns on a finite $\theta$-angle.

The most relevant IR constraint arises as the requirement
of having a regular $\xi'$ from Eq.~(\ref{xieqs2}). Indeed,
by demanding that the denominator of (\ref{xieqs2}) does not become complex one obtains the inequality
\be
e^{4A}V_f\sqrt{\kappa} \t - |\Ca| V_a >0\,,\qquad \left(r\gg \LIR^{-1}\right)\,.
\label{irvineq}
\ee
We have identified two choices for the potentials that satisfy this inequality and lead to sensible IR solutions.
First, if one considers an Ansatz
where the exponential factors in $V_f$ and $V_a$ can be different functions of $\tau$,
the inequality will be satisfied by a $V_a$ that vanishes faster than $V_f$ for an IR diverging
tachyon. A simple choice would be to take $V_f \propto \exp(-a\tau^2)$ and $V_a = \exp(-b\tau^2)$ where $a$ and $b$ are constants satisfying $b>a$. Another choice, which we will use below, is to modify the tachyon dependence of $V_a$, to, {\it e.g.}, $V_a = \exp(-a_q\tau^2 - a_l |\tau|)$.

Second, if one insists on keeping the same exponential factors (see \eqref{tachpot}, \eqref{vadef}), and then $V_a\sim e^{-a\tau^2}$
as in flat space String Theory~\cite{Kraus:2000nj}, then the inequality above will constrain $V_{f0}$.
Namely, the inequality \eqref{irvineq} reduces to
\be \label{IRreq}
 e^{4A}\,V_{f0}\sqrt{\kappa}\, \t > |\Ca| \,,
\ee
and this condition results in a constraint on the potential $V_{f0}$ as we now explain.
In \cite{Arean:2013tja} it was shown that in order to have  linear meson  trajectories, $\h$ should
behave in the IR as $\sim\l^{-4/3}$, while for the flavor potential we had $V_{f0}\sim\l^{v_p}$.
Although $v_p$ was not fixed by the spectrum analysis of \cite{Arean:2013tja}, it was already
shown in \cite{jk} that for $v_p> 10/3$ no acceptable solutions existed.
Potentials with $\h\sim\l^{-4/3}$ and $v_p<10/3$ were indeed analyzed in \cite{Arean:2013tja},
and shown to display linear meson trajectories.
However, after inserting the IR asymptotic expansions of $A(r)$, $\l(r)$ and $\tau(r)$ for that case,
the inequality (\ref{IRreq}) would only be satisfied if we had $v_p\geq 10/3$.
This leads one to consider
potentials having $v_p = 10/3$:
\be
\h\sim \h_c\, \l^{-4/3}\,(\log\l)^{-\h_\ell}\,,\quad
V_{f0} \sim v_c\,\l^{10/3}\, (\log \l)^{-v_\ell}\,; \qquad (\l\to\infty)\,.
\label{vfmir}
\ee
As shown in Appendix \ref{subapp:IRback}, this choice results in regular IR asymptotics 
if certain constraints for $\h_\ell$ and $v_\ell$ are fulfilled. In particular, the asymptotics for the phase is
\be
\xi'\sim \left({2\over3}\right)^{v_l+{\kappa_\ell\over2}}{\Ca\,e^{-4A_c-{8\over3}\l_c}\over v_c\,\sqrt{\kappa_c}}\,
{\tau'\over r^{2-2v_l-\kappa_\ell}\,\tau^2}\,,\qquad (r\to\infty)\,.
\label{xipir}
\ee

However, we have not been able to find potentials with $v_p=10/3$ for which the (numerically constructed) solutions would
have been regular both in the IR and in the UV. Regular solutions are found for $4/3\le v_p \lesssim 3$. The
prediction from string theory, $v_p=7/3$ (see Sec. 2.4 in~\cite{Arean:2013tja}), falls in the middle of the acceptable range.
Similar observations have been made earlier for the other potentials $V_g$, $a$, and $\kappa$ in the model: the power laws
given by string theory arguments,
possibly with multiplicative logarithmic corrections,
give the best match with QCD physics. The result for $V_{f0}$ is, however, different from those for the other potentials,
because the power law is nontrivial (i.e., having power different from zero) even after the transformation from Einstein frame to
string frame. That is, choosing all potentials to have exactly ``critical'' asymptotics in the string frame does not lead to regular
solutions, but following the string theory prediction does, and even results in physics which is close to QCD.

In the numerical analysis below, we will consequently use the first option discussed above and choose the tachyon
dependence in the exponential factor of $V_a$ so that (\ref{irvineq}) is satisfied independently of the asymptotic
form of $V_{f0}$.

\section{Choice of potentials} \label{sec:potentials}

In this section we present concrete choices for the potentials appearing in the action of our
holographic model. These choices will be used in the numerical analysis of subsequent sections, and
largely agree with those introduced in \cite{jk}, and further constrained in \cite{alho,Arean:2013tja},
as we now review.

Two classes of potentials $V_g$, $V_{f0}$, $\h$, and $a$ were considered in
\cite{jk,alho,Arean:2013tja}; they are called potentials I and potentials II.

\begin{itemize}
\item \textbf{Potentials I} were chosen such that the IR power behavior of $\h(\l)\sim\l^{-4/3}$
and $a(\l)\sim\l^0$ is the critical one; as shown in \cite{Arean:2013tja} these values correspond to
the critical point at the edge of the region of acceptable IR asymptotics. These potentials
admit a regular IR solution with exponential tachyon, $\t \sim \t_0 e^{C_\mathrm{I} r}$,
where $C_\mathrm{I}$ can be computed in terms of the potentials, and $\t_0$ is an integration
constant (see Appendix D of \cite{Arean:2013tja} for details).
For the resulting mesonic spectra, the asymptotic trajectories of masses in all towers are linear,
but have logarithmic corrections.
Exactly linear asymptotic trajectories for the mesons can be obtained by considering a slight
modification of potentials I such that $\h(\l)\sim\l^{-4/3}(\log\l)^{1/2}$ in the IR.

 \item \textbf{Potentials II} behave instead as $\h(\l)\sim\l^{-4/3}$ and $a(\l)\sim\l^{2/3} $ in the IR.
 These potentials admit a regular IR solution with $\t \sim \sqrt{C_\mathrm{II}\, r +\t_0}$, and the asymptotic trajectories
 of masses in all towers are quadratic.
\end{itemize}

In order to fully fix the action for this article we will also need to specify the potentials appearing in the
CP-odd action~\eqref{samain}. Notice that he function $Z(\l)$ there, contributes even when $x=0$,
and therefore has been considered in the context of IHQCD~\cite{ihqcd,data,cs}.
In \cite{data} it was shown that the asymptotic behavior of $Z$ at $\l\to 0$ and $\l\to \infty$ was fixed from general principles.
We adopt a similar Ansatz compatible with the asymptotic behaviors as was used there:
\be
 Z(\l)= Z_0 \left[1 + c_a \left(\frac{\l}{\l_0}\right)^4\right] \, .
 \label{zdef}
\ee

For the physics at finite $x$ the choice of $V_a$ is even more relevant. Based on earlier studies and
observations made above, we can immediately set some constraints.
As pointed out in Sec.~\ref{sec:CP-oddaction}, for the anomaly structure to be reproduced correctly we need that
$V_a(\l,\t=0)=1$. That is, in the UV we must have $V_a = 1$ up to terms suppressed by powers of $\t$, and in
particular the leading term is independent of $\l$. Dependence on $\l$ would have introduced perturbative
corrections to  triangle diagrams giving rise to the axial anomaly, which would have conflicted with our
knowledge of QCD. Here we will impose the stricter but natural constraint that $V_a$ only depends on $\t$.
The independence of $V_a$ on $\l$ is consistent with the analysis of boundary string field theory~\cite{Kraus:2000nj}
where the flavor and CP-odd potentials were found to be $\propto \exp(-a\,\t^2)$ with the same $\l$-independent
factor $a$ in both potentials. The $\l$-independence of the flavor potential in V-QCD is also supported by
the analysis of the asymptotic radial trajectories of the meson spectrum~\cite{Arean:2013tja} and the
behavior of the meson masses at high quark mass~\cite{Jarvinen:2015ofa}.

As we have seen in Sec.~\ref{ssec:thbckir}, extra constraints 
result from demanding a regular
solution for $\xi'$ in the IR; in particular, the inequality (\ref{irvineq}) must be satisfied.
The options which lead to regular IR asymptotics can be summarized as
\begin{enumerate}
 \item Choose $V_{f0}(\l)$ with the asymptotics \eqref{vfmir},
 take $V_a = \exp(-a\t^2)$, and use potentials~I for the other functions.
 \item Add a linear term in the exponential factor of the CP-odd potential: $V_a = \exp(- a \t^2 - a_l|\t| )$ (or modify the tachyon dependence in some other way such that $V_a$ is suppressed with respect to $V_f$). Then the IR behavior $V_{f0}(\l)$ can be chosen more freely, {\it e.g.}, as in previous work~\cite{Arean:2013tja}.
\end{enumerate}
We show in Appendix \ref{subapp:IRback} that the first choice still gives regular IR asymptotics for vanishing
$\theta$-angle, and with a good choice of $v_\ell$ the tachyon phase is also regular when
$\theta\ne 0$. The second option above is non-analytic at $\t=0$, but the extra linear term in the
exponential guarantees that the term involving $V_a$ in \eqref{irvineq} is suppressed, and the inequality is
fulfilled. As we mentioned above, we have not found potentials satisfying the first option for which the solutions would have been regular both in the IR and in the UV. Therefore we selected the latter option when performing the numerical analysis of
vacua at finite $\theta$-angle.

Next we summarize the explicit choices of potentials that were used to carry out the numerical analysis of Secs.~\ref{sec:thetavac},~\ref{sec:singletPSzeroth} and \ref{sec:finitethspc}.

\subsection{Potentials I}

The motivation for this choice is to
 mimic (at qualitative level, without fitting any of the numerical results to QCD data) the physics of
 real QCD in the Veneziano limit~\cite{Arean:2013tja}.

With potentials~I we have used in this article the choice $V_a = \exp(- a_q\t^2 - a_l|\t| )$, suggested above,
which also makes the backgrounds at finite $\theta$-angle well defined.  As pointed out above,
we require that the leading dependence on the tachyon in $V_a$ is the same as in $V_f$, \ie
$a_q=\mathrm{const.}=a$. In addition, we choose $a_l$ to be a constant having a sufficiently large value so
that   $e^{4A}V_f\,\sqrt{\kappa}\, \t$ in~\eqref{irvineq} dominates over $V_a$ for all $r \gtrsim 1/\LIR$.
A convenient choice is $a_l = 10/C_\mathrm{I}$, where $C_\mathrm{I}$ is the coefficient in the IR asymptotics of
the tachyon: $\t \sim e^{C_\mathrm{I}r}$. For the numerical coefficients in the function $Z(\l)$ of~\eqref{zdef}
we chose $Z_0 = 1$ and $c_a = 0.1$. This choice was seen to lead to reasonable behavior for all observables
depending on this function.

In summary, the potentials~I are given by
  \begin{align} \label{potIandIIcommon}
    V_{g}(\l)  & = V_0\left[1+V_1 \l + V_2 \l^2 \frac{\sqrt{1+\log(1+\frac{\lambda}{\l_0})}}{\left(1+\frac{\lambda}{\l_0}\right)^{2/3}}\right]\,, \nonumber\\
    V_{f0}(\l) & = W_0\left[1+W_1 \l + W_2 \l^2\right]\,,\\
a(\l)   &= a_0\,,\qquad   \h(\l)  = \frac{1}{\left(1+\frac{3a_1}{4}\l\right)^{4/3}} = w(\l)\,,\\
 Z(\l) &= Z_0 \left[1 + c_a \left(\frac{\l}{\l_0}\right)^4\right] \,,\qquad V_a(\t) = \exp(- a_q\t^2 - a_l|\t| )\,,
\label{Valindef}
 \end{align}
where the coefficients satisfy
\begin{align} \label{potcoeffs}
 V_0 &= 12\, , \qquad V_1 = \frac{11}{27 \pi^2}\,,\qquad  V_2= \frac{4619}{46656 \pi ^4}\, ; \nonumber \\
 W_1 &= \frac{24+(11-2 x) W_0}{27 \pi ^2 W_0}\,,\qquad W_2 = \frac{24 (857-46 x)+\left(4619-1714 x+92 x^2\right) W_0}{46656 \pi ^4 W_0}\,; \nonumber\\
 a_0 &= \frac{12-x W_0}{8} = a_q\, , \qquad a_1 =  \frac{115-16 x}{216 \pi ^2} \, , \qquad \l_0 = {8 \pi^2} \,;\\
 Z_0 &= 1\,,\qquad c_a = 0.1 \,,
\end{align}
and with $a_l$ chosen as explained above. For potentials~I we have used\footnote{As was shown in \cite{alho},
the finite temperature phase diagram is not of the standard type for potentials~I
if $W_0$ is close to its upper limit $24/11$; a chirally symmetric phase is present at small $x$. Therefore we pick a value near the lower end of the possible range.} $W_0=3/11$.

All computations at finite $\theta$-angle in this article were done by using
this choice of potentials.

 \subsection{Potentials II}

 This choice might not model QCD as well as
 potentials~I, but the motivation is to pick a background with different IR structure in order to see how much
 this affects our results for the backgrounds at zero $\theta$-angle. The numerics for potentials~II in this
 article were done with the choice $V_a=\exp(-a(\l)\t^2)$, where $a(\l)$ is the same function which appears
 in $V_f$ and is given explicitly in~\eqref{potIIdefs}.

Explicitly we used
\begin{align} \label{potIIdefs}
a(\l)   &= a_0\,\frac{1+a_1 \l + \frac{\l^2}{\l_0^2}}{\left(1+\frac{\l}{\l_0}\right)^{4/3}}\,, \qquad    \h(\l)
= \frac{1}{\left(1+\frac{\l}{\l_0}\right)^{4/3}}\,,\qquad w(\l) =1 \,;\\
V_a &=\exp(-a(\l)\t^2)\,,\qquad
\end{align}
and all the other functions as for potentials~I above, except that we chose
\be
 W_0 = \frac{12}{x}\left[1-\frac{1}{(1+\frac{7}{4}x)^{2/3}}\right]
\ee
instead of $W_0=3/11$.
With this choice, the pressure agrees with the Stefan-Boltzmann (SB) result at high temperatures \cite{alho} (without the need to introduce an $x$ dependence in the normalization
 of the action). In this article, we used potentials~II in the numerical analysis of Sec.~\ref{sec:singletPSzeroth} (at zero $\theta$-angle).

\section{The chiral Lagrangian analysis in the Veneziano limit}\label{sec:EFT}

In this section we will take a detour and consider the problem from the effective chiral theory point of view. To do this, we must
assume that the bare quark masses $m\ll \Lambda_{UV}$ so that the pions are very light compared to other particles, and
therefore it makes sense to write down an effective action for them integrating out all other particles. This analysis is useful as
we can see what can be determined from low energy symmetries alone and what needs a full non-perturbative computation
(using holography in this paper). Moreover the chiral Lagrangian results provide consistency checks for our holographic
analysis.

We will start by writing the effective action for the expectation values
of the QCD order parameters  $W_{ij}$ which is an $N_f\times N_f$ complex matrix
(the expectation value of $\bar \psi^i_{R}\,\psi^j_L$), as well as $G$ and $\Theta$ the
pseudoscalar and scalar glueball
related expectation values following \cite{Schechter} and
references therein.
\def\mb{{\bar M}}
\def\mbb{{\bar M^{\dagger}}}

We start from the $U(1)_A$ anomaly equation that can be written as
\begin{align}
\langle\partial_{\m}J_5^{\mu}\rangle&={g^2N_f\over 16\pi^2}\,\langle \trc[F\tilde F]\rangle+i(M_{ij}\langle \bar\psi^i_R\psi^j_L
\rangle-M_{ij}^{\dagger} \langle \bar\psi^i_L\psi^j_R\rangle )=\nonumber\\
&=N_fN_c\,G +i\trc[\mb \,W-\mbb \,W^{\dagger}]\,,
\label{c1}
\end{align}
\be
{\rm with}\qquad W_{ij}\equiv \langle \bar\psi^j_R\psi^i_L\rangle\,,\qquad
W^{\dagger}_{ij}\equiv \langle \bar\psi^j_L\psi^i_R\rangle\,, \label{c2}\ee
as well as the conformal anomaly equation (in flat space)
\begin{align}
\langle {T^{\m}}_{\m}\rangle &=-{\beta(g)\over 2g}\langle \trc[F^2]\rangle +(1+\gamma(g))(M_{ij}\langle \bar \psi^i_{R}\,\psi^j_R\rangle +{\rm c.c.})\nonumber\\
&=N_c\,\Theta+(1+\gamma(g))\trc[\mb\,W+\mbb\,W^{\dagger}]\,,
\label{c3}
\end{align}
where $\Theta,G$ were defined so that they are ${\cal O}(1)$ in the Veneziano limit and $\bar M$ is the quark mass matrix.
Here $\beta(g)$ is the QCD $\beta$-function and $\gamma(g)$ is the fermion anomalous dimension.
Because of energy and charge conservation $G$ and $H$ have canonical dimension 4.
Note that although the product $\trc[\mb\,W]$ is RG invariant, W is RG-dependent. For the purposes of an effective theory $W$
will be defined at low energies and therefore $\mb$ will be the renormalized quark matrix at low energies. It will be linearly
related to the UV quark mass matrix for small enough quark masses.

Flavor $U(N_f)_L\times U(N_f)_R$ transformations act as
\be
W\to V_L\,W\,V_R^{\dagger}\,,
\label{c4}\ee
where $V_L,V_R$ are $U(N_f)$ matrices. $G,\Theta$ are flavor invariants.
We can also construct flavor invariants from the matrix $W$
\be
I_n\equiv {1\over N_f (g^2\,N_c)^{6n}}{\rm \trc}[(WW^{{\dagger}})^n]
\label{c5}\ee
which are also ${\cal O}(1)$ in the Veneziano limit.
In the absence of masses and the anomaly, the effective potential takes the following form
\be
V_\mathrm{eff}(G,\Theta,I_n)=N_c^2 \,V_0(G,\Theta,I_n)\,,
\label{c6}\ee
where $V_0$ is an arbitrary function that due to parity invariance must satisfy
\be
V_0(G,\Theta,I_n)=V_0(-G,\Theta,I_n)\,.
\label{c7}\ee

To accommodate the $U(1)_A$ anomaly we must consider that the $U(1)_A$ transformation acts as
\be
W\to e^{i\epsilon}W\,,\qquad G\to G+\epsilon\,.
\label{c8}\ee
Therefore the full effective potential that includes the anomaly is
\be
V_\mathrm{eff}= N_c^2 \,V_0\left(\Theta,I_n,G+{i\over 2N_f}\log\det{W\over W^{\dagger}}\right)\,.
\label{c9}\ee
In the presence of a (complex) mass matrix $\bar M$ at linear order, we have in
addition the associated term from the QCD Lagrangian
\be
V_\mathrm{eff}= N_c^2 \,V_0\left(\Theta,I_n,G+{i\over 2N_f}\log\det{W\over W^{\dagger}}\right)+\trc[\mb \,W+\mbb\,W^{\dagger}]
\label{c10}\ee
By a chiral rotation we can also introduce the $\theta$-angle:
\be
V_\mathrm{eff}= N_c^2 \,V_0\left(\Theta,I_n,G+{i\over 2N_f}\log\det{W\over W^{\dagger}}-{\theta\over N_f}\right)+\trc[\mb \,W+\mbb \,W^{\dagger}]
\label{c11}\ee
and by a phase redefinition of $W$ it can be moved to the masses
\be
V_\mathrm{eff}= N_c^2 \,V_0\left(\Theta,I_n,G+{i\over 2N_f}\log\det{W\over W^{\dagger}}\right)
+\trc[e^{i{\theta\over N_f}}\mb \,W+e^{-i{\theta\over N_f}}\mbb \,W^{\dagger}]\,.
\label{c12}\ee
When quark masses are small (compared to $\Lambda_{\rm QCD}$) the $G$ and $\Theta$ glueballs are
much
heavier than the mesons. We will therefore neglect their kinetic terms and their equations of
motion amount to minimizing their potential.
It is interesting that there seems to be more things that can be said about the dependence of
this action concerning its dependence on the $G$ and $\Theta$ condensates.
A glimpse of this was indicated first in \cite{rg1} and in a more targeted way in \cite{rg2}
where the effective potential for the trace of the stress tensor was calculated holographically
in a single scalar gravitational theory.

We next move to the two derivative level.  We will concentrate on the kinetic terms of the quark condensate $W$ which is the only light remaining field.
To write such kinetic terms, we must introduce flavor invariants with up to two derivatives
\be
I_{\m}^n\equiv {1\over N_f}\trc[(WW^{\dagger})^n\,(\pa_{\m}W)W^{\dagger}]\,,\qquad
\bar I_{\m}^n\equiv  {1\over N_f} \trc[(WW^{\dagger})^n\,W(\pa_{\m}W^{\dagger})]\,,
\label{c13}\ee
and
\be
J_{mn}\equiv
{1\over N_f} \trc[(WW^{\dagger})^m(\pa_{\m}W)(WW^{\dagger})^n (\pa^{\mu}W^{\dagger})]\,,
\label{c14}\ee
and then
\be
S_2=\int d^4x\,\sum_{m,n=0}^{\infty}\left[C_{nm}J_{mn}+\hat C_{mn}I_{\m}\bar I^{\m}\right]\,.
\label{c15}\ee

We now decompose $W$ as a product of an Hermitian ($H$) and a unitary matrix ($U$)
\be
W=H\,U\,,\qquad W^{\dagger}=U^{\dagger}H\,,
\label{c16}
\ee
and thus,
\be
WW^{\dagger}=HUU^{\dagger}H=H^2\,,\qquad \det{W\over W^{\dagger}}=(\det U)^2\,.
\label{c16p2}
\ee
Moreover, we shall write $U$ as
\be
U=\exp\left[{i\over \sqrt{N_f}}\eta'+i\pi^aT^a\right]\,,\qquad
\trc[T^aT^b]=\delta^{ab}\,,
\label{c17}\ee
where $T^a$ are the (traceless) generators of $SU(N_f)$.

Chiral symmetry breaking will give an expectation value to $H$, while $U$ remains free and
parametrizes the Goldstone bosons, namely, $\eta'$ and the generalized pions.

Minimizing now the potential in the massless case (by setting first $\bar M=0$) with respect to
the ``heavy" fields $\Theta, G, H_{ij}$ we will obtain nontrivial VEVs for $G$,
$\Theta$ and $H$ that are function only of $\Lambda_{\rm QCD}$\footnote{Once quark
masses are turned on, there will be ${\cal O}(m^2)$ corrections to such VEVs.}. In particular,
\be
\langle H_{ij}\rangle =\sigma \,\delta_{ij}\,.
\label{c18}\ee

The VEV of $G$ can be absorbed in the $\theta$ phase changing it to $\bar\theta$. The only part of $U$ that appears in the potential is $\log\det U$ due to the anomaly. As argued in the end of Sec. \ref{free} the dependence in the Veneziano limit is quadratic like in the 't Hooft limit,
\be
V_\mathrm{eff}=\kappa N_cN_f \,(-i\log\det U)^2+\cdots\,.
\label{c19}\ee

\subsection{The effective action for the Goldstone modes}

We now set $\Theta$, $G$ and $H$ equal to their VEVs and we rewrite the effective Lagrangian
(up to two derivatives) for the Goldstone modes described by  an $N_f\times N_f$ unitary matrix
$U$ as\footnote{The pion decay constant $\hat f_\pi$ which is normalized as is usual for chiral
Lagrangians differs from the constant $f_\pi$ used elsewhere in this article
by $N_f\,\hat f_\pi^2 = f_\pi^2$.},
\cite{divechia,Rosenzweig:1979ay,DiVecchia:1980yfw,Witten:1980sp,Nath:1979ik}
\begin{align}
{\cal L}_{\rm chiral}=&{ \hat f_{\pi}^2\over 2}\left[\trc[\pa_{\m}U\pa^{\m}U^{-1}]
+\trc[\hat MU+\hat M^{\dagger}U^{\dagger}]\right]-{a\hat f_{\pi}^2\over 2N_c}\,(-i\log\det U)^2+
\nonumber\\
+&{ \hat f_{\eta'}^2-
\hat f_{\pi}^2\over 2N_f}\,\trc[U^{\dagger}\pa_{\m}U]\,\trc[U\pa^{\m}U^{\dagger}]\,.
\label{cc2}
\end{align}
The last term in (\ref{cc2}) originates in the factorized terms in (\ref{c15}),
and although it is subleading in the 't Hooft limit, it is ${\cal O}(1)$ in the Veneziano limit.
It is responsible for the fact that the decay constant of the $\eta'$ is different from the rest
of the Goldstone modes (pions).

The term $\propto(\log\det U)^2$ in (\ref{cc2}) is the anomaly term and is of order
${\cal O}(\eta'^2)$ giving a mass to the $\eta'$. As discussed in the end of Sec.~\ref{free} this is a good estimate also in the Veneziano limit.
The matrix $\mh$ is a renormalized quark mass matrix.
We lump the $\theta$ parameter inside $\mh$
via a chiral rotation.
We will consider the $SU(N_f)$ invariant case where all quark masses are equal to $m_q$.

Finally, the coefficients are written in terms of the physical pion and $\eta'$ decay constants,
$\hat f_{\pi,\eta'}$, as well as the parameter $a$ that as we will soon see is related to the
topological susceptibility.
It should be noted that in the Veneziano limit
\be
\hat f^2_{\pi}\sim \hat f^2_{\eta'}\sim N_c\,,\qquad m_{\eta'}\sim {\cal O}(1)\,.
\label{c20}\ee

We now consider the case explored in this paper: quark masses that are $SU(N_f)$ invariant.
In this case the mass term in the effective chiral Lagrangian can be parametrized in terms of
the pion mass $m_{\pi}$ as
\be
\hat M_{ij}=e^{i{\theta\over N_f}}~m^2_{\pi}\delta_{ij}\,.
\label{cc3}\ee
We now transform $U\to U\,e^{-i{\theta\over N_f}}$ to obtain
\begin{align}
{\cal L}_{\rm chiral}=&{ \hat f_{\pi}^2\over 2}
\left[\trc[\pa_{\m}U\pa^{\m}U^{-1}]+m^2_{\pi}\,\trc[U+U^{\dagger}]-{a\over N_c}(-i\log\det U-\theta)^2\right]+
\nonumber\\
+&{ \hat f_{\eta'}^2
-\hat f_{\pi}^2\over 2N_f}\,\trc[U^{\dagger}\pa_{\m}U]\,\trc[U\pa^{\m}U^{\dagger}]\,,
\label{cc4}
\end{align}
from which the pion potential reads
\be
V(U)={ \hat f_{\pi}^2\over 2}\left[m^2_{\pi}\,\trc[U+U^{\dagger}]
-{a\over N_c}\,(-i\log\det U-\theta)^2\right]\,.
\label{cc5}\ee
By symmetry we should look for a minimum of $V$ of the form  $U_{ij}=e^{i\phi}\delta_{ij}$. However,
we should remember that $2\pi$ rotations of individual masses give the same theory.
Therefore, a better parametrization of the vacua is
\be
U_{ij}=e^{i\phi}e^{2\pi i n_i}~\delta_{ij}\,.
\label{eq:mmatrix}
\ee
Then the potential becomes
\be
V(\phi)= \hat f_{\pi}^2\left[-N_f\,m_{\pi}^2\cos\phi
+{a\over 2N_c}(\theta-N_f\phi-2\pi N)^2\right]\,.
\label{cc6}
\ee
with $N=\sum_{i=1}^{N_f}n_i$. In the sequel, we will set $N=0$,
but we will consider all branches with $\theta$ shifted by multiples of $2\pi$.

We define
\be
\tilde \theta={\theta\over N_f}
\,,\qquad
x={N_f\over N_c}\,,
\label{c21}\ee
and  obtain
\be
V(\phi)=N_f~ \hat f_{\pi}^2\left[-m_{\pi}^2\cos\phi+{x\,a\over 2}(\tilde\theta-\phi)^2\right]
=f_{\pi}^2\left[-m_{\pi}^2\cos\phi+{x\,a\over 2}(\tilde\theta-\phi)^2\right]\,.
\label{cc7}
\ee
Note that $V\sim {\cal O}(N_cN_f)$.

Consider first the case with zero quark mass. In this case $m_{\pi}=0$ and the extremum is
at $\phi=\tilde \theta$. Hence the vacuum energy is independent of $\theta$ as expected.

When  $m_\pi\not= 0$, the extrema satisfy the equation
\be
m_{\pi}^2\sin\phi=a\,x\,(\tilde\theta-\phi)\,,
\label{cc8}\ee
which can be rewritten as
\be
\sin\phi-{\zeta}(\tilde\theta-\phi)=0\,,\qquad {\rm with}\quad \zeta={a\,x\over m_{\pi}^2}\,,
\label{cc1}
\ee
$\zeta$ being a dimensionless parameter which is ${\cal O}(1) $ in the Veneziano limit.
There are three different parameters that enter in $\zeta$: the QCD scale as
$a\sim \Lambda_{\rm QCD}^2$, the bare quark masses $m_q$, and the flavor parameter $x$.
For small $m_q$, $m_{\pi}^2\sim m_q$. The assumption for the validity of the effective chiral
theory implies that $a\gg m_{\pi}^2$.
Therefore, for generic values of $x$, $\zeta\gg 1$.
Only in the 't Hooft limit, $x\to 0$,
can $\zeta$ become much smaller than one.

\begin{figure}[!tb]
\begin{center}
\includegraphics[width=0.49\textwidth]{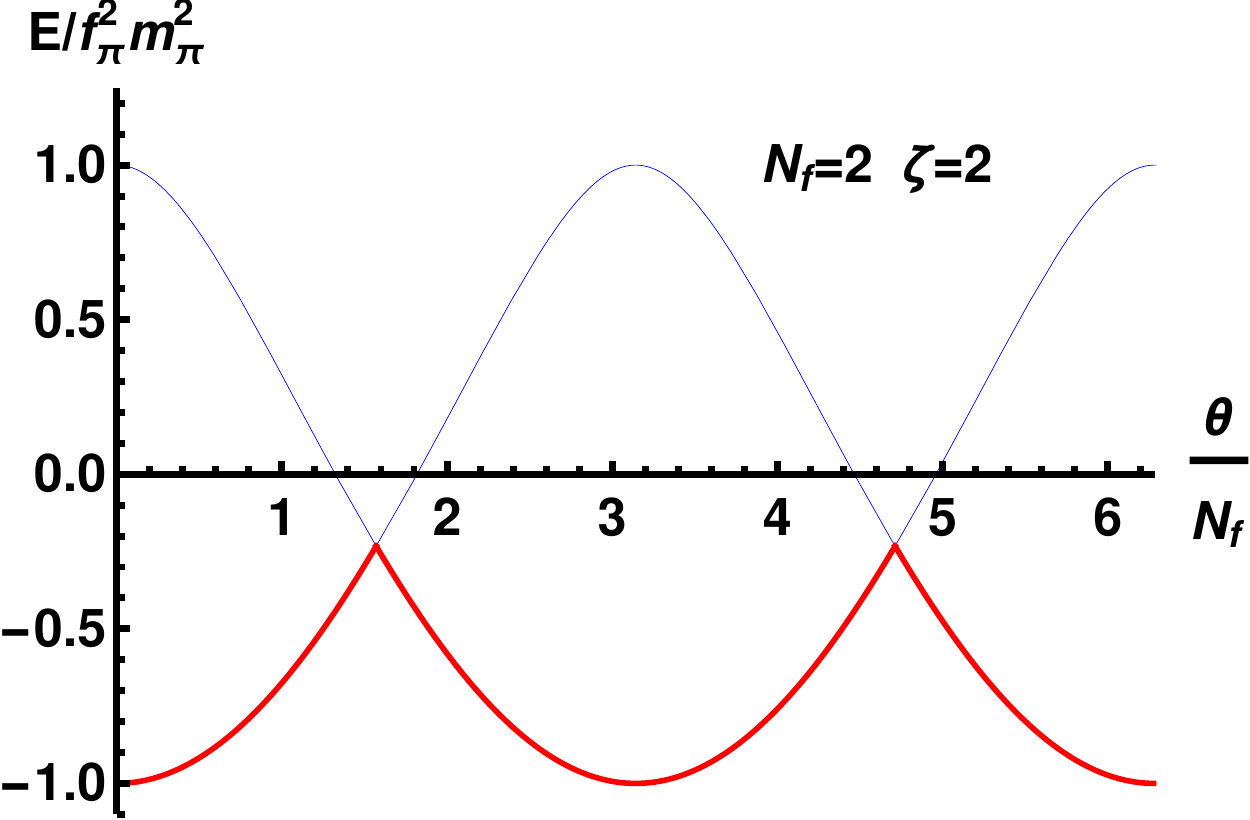}
\includegraphics[width=0.49\textwidth]{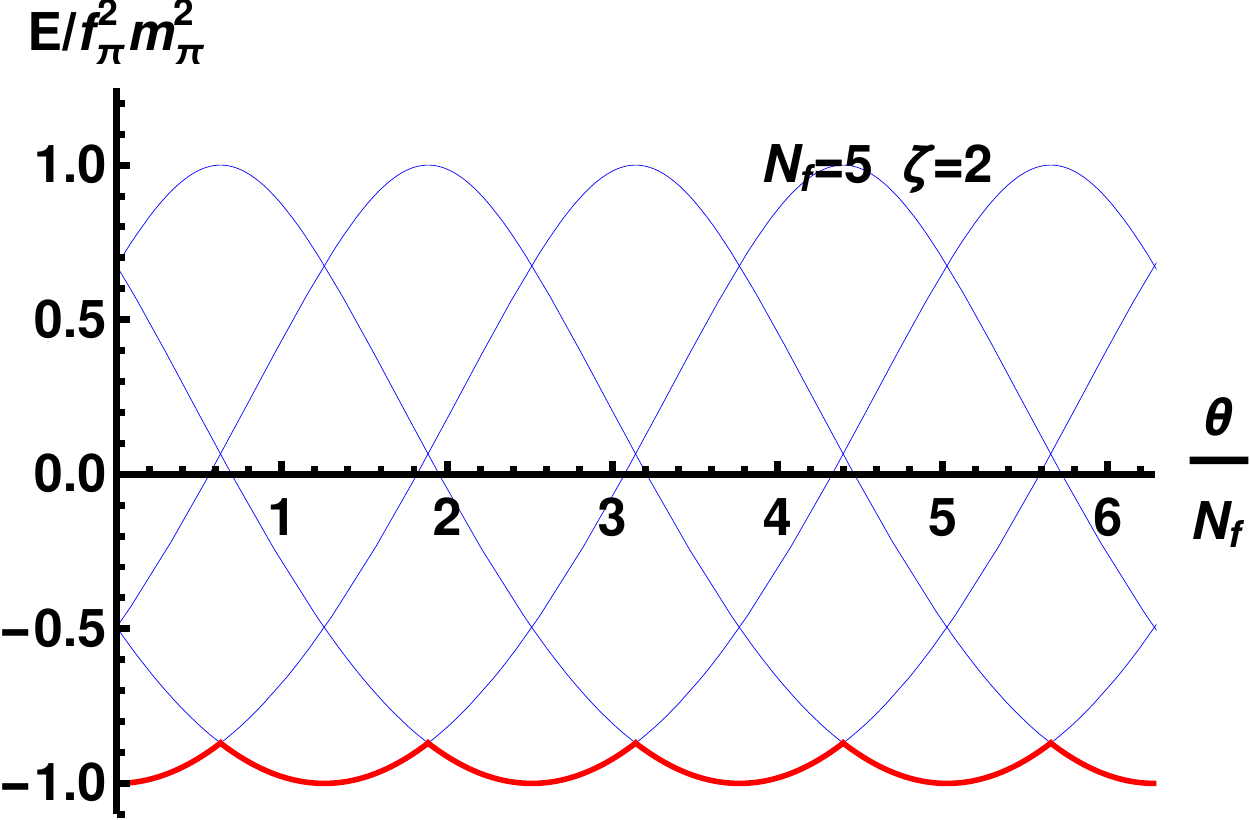}
\includegraphics[width=0.49\textwidth]{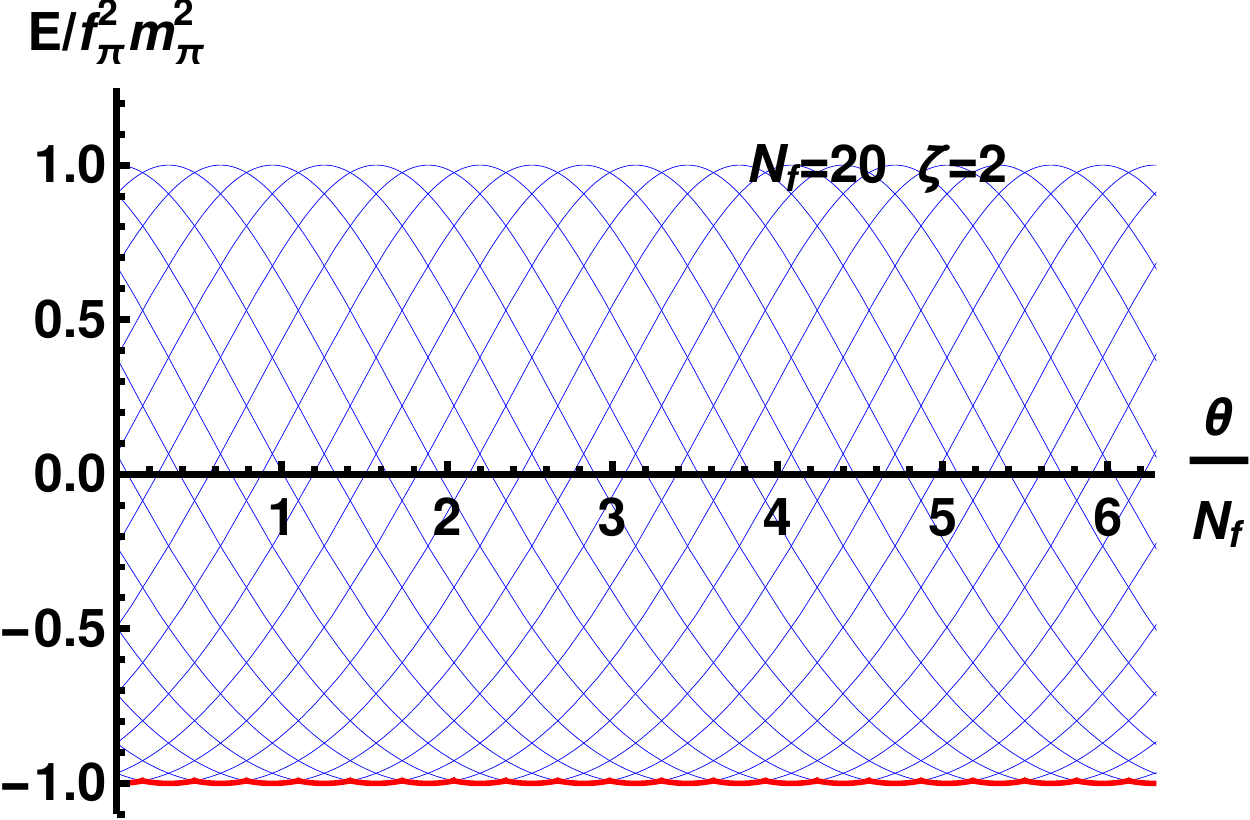}
\end{center}
\caption{The branches of vacua according to the chiral Lagrangian analysis. The thin blue lines show (normalized) energy for all branches, and the thick red line is the final energy for the dominant vacuum.  We chose $\zeta=2$ in all plots with $N_f =2$, $5$, and $20$, in the top-left, top-right, and bottom plots, respectively. }
\label{fig:energybranches}\end{figure}

When $\zeta\geq 1$ there is a unique solution to (\ref{cc8}) as the left hand side of (\ref{cc1})
is monotonic.
But for $\zeta<1$ there is a range of values of $\theta$ where there are two or more minima.
We denote these extrema, which are functions of $\tilde\theta$, as $\phi_i(\tilde\theta)$.
They are minima of the potential if
\be
V''(\phi_i)= f_{\pi}^2~m_{\pi}^2\left[\cos\phi_i+\zeta\right]\geq 0\,.
\label{cc9}\ee
The deepest minimum is the one that minimizes
\be
V(\phi)=f_{\pi}^2\,m_{\pi}^2\left[-\cos\phi+{\zeta\over 2}(\tilde\theta-\phi)^2\right]\,.
\label{cc10}\ee
Therefore, the $\theta$-dependent vacuum energy is
\be
E(\tt)=\min_\phi\,f_{\pi}^2\,m_{\pi}^2\left[-\cos\phi+{\zeta\over 2}(\tilde\theta-\phi)^2\right]\,.
\label{cc100}\ee

In Fig.~\ref{fig:energybranches} we present a calculation for the energy as a function of
$\tt$ for $\zeta=2$, where we now have added the $2\pi\over N_f$ shifts as argued above,
and we have denoted by red lines the minimum values that  indicate the true $\tt$-dependence
of the ground state energy.

To compute the derivatives of the vacuum energy we need the derivatives of $\phi$ from (\ref{cc1})
\be
{\pa\phi\over \pa \tt}={\zeta\over \zeta+\cos\phi}\,,\qquad
{\pa^2\phi\over \pa\tt^2}={\zeta^2\sin\phi\over\left(\zeta + \cos\phi\right)^3}\,.
\label{c22}\ee
We obtain
\be
{\pa E\over \pa\tt}=f_{\pi}^2\, a\,x\,(\tt-\phi)\,,\qquad
\hat \chi_\mathrm{top}(\tt)\equiv{\pa^2 E\over \pa\tt^2}
=f_{\pi}^2\,m_{\pi}^2{\zeta\cos\phi\over \zeta+\cos\phi}\,.
\label{c23}\ee
At $\tt=0$ we have $\phi=0$ and
\be
{\pa E\over \pa\tt}\Big|_{\tt=0}=0\,,\qquad
\hat \chi_\mathrm{top}(\tt=0)\equiv {\pa^2 E\over \pa\tt^2}\Big|_{\tt=0}
=f_{\pi}^2\,m_{\pi}^2{\zeta\over 1+\zeta}\geq 0\,.
\label{c24}\ee
The topological susceptibility in the normalization used in the latter sections of
the article is then
\be
 \chi(\tt) \equiv \chi_\mathrm{top}(\tt) \equiv {\pa^2 E\over \pa\theta^2}
 = \frac{\hat \chi_\mathrm{top}(\tt)}{N_f^2}\,.
\ee
Note that $\chi_\mathrm{top} = \morder{1}$ and $\hat \chi_\mathrm{top} = \morder{N_f^2}$.

We can now compute the meson masses in the nontrivial vacuum by expanding $U$ around it
\be
U=e^{i\phi}\,\cdot\,\exp\left[{i\over \sqrt{N_f}}\eta'+i\pi^aT^a\right]\,.
\label{c25}\ee
We obtain
\be
V=E(\tt)-{1\over 2}\,\hat f_{\pi}^2\left[m_{\pi}^2\cos\phi~\pi^a\pi^a
+(m_{\pi}^2\cos\phi+a\,x)\,\eta'^2\right]+\cdots\,.
\label{c26}\ee
From (\ref{cc4}) we observe that $f_{\pi}$ and $f_{\eta'}$ are $\theta$-independent.

We obtain for the $\tt$-dependent meson masses
\be
m_{\pi}^2(\tt)=m_{\pi}^2\,\cos\phi\,,\qquad m_{\eta'}^2(\tt)
={f^2_{\pi}\over f_{\eta'}^2}\left[ m_{\pi}^2(\tt)+a\,x\right]\,,
\label{c27}\ee
where $f_{\eta'}^2 = N_f \hat f_{\eta'}^2$.
The last relation can be written in terms of the topological susceptibility by solving (\ref{c23})
\be
a\,x={m_{\pi}^2(\tt)~\hat \chi_\mathrm{top}(\tt)
\over f_{\pi}^2 m_{\pi}^2(\tt)-\hat \chi_\mathrm{top}(\tt)}\,,
\label{c32}\ee
as
\be
m_{\eta'}^2(\tt)={f^2_{\pi}\over f_{\eta'}^2}{f_{\pi}^2 m_{\pi}^2(\tt)
\over f_{\pi}^2 m_{\pi}^2(\tt)-\hat \chi_\mathrm{top}(\tt)}\ m_{\pi}^2(\tt)\,.
\label{c31}\ee
Equation (\ref{c31}) is the analogue of the Witten-Veneziano formula in the Veneziano limit.
In the 't Hooft limit, $x\to 0$, we find that
\be
\hat \chi_\mathrm{top}= f_{\pi}^2\,a\,x\left[1+{\cal O}(x)\right]\,,\qquad
\chi_\mathrm{top} = \frac{f_\pi^2}{N_f N_c} a + \morder{x} \,,\qquad
m_{\pi}^2(\tt)=m_{\pi}^2+{\cal O}(x^2)\,,
\ee
where in the last estimate we used (\ref{cc15}).
Since the kinetic and mass terms of mesons are group theoretically similar we also have
\be
f_{\pi}=f_{\eta'}\left[1+{\cal O}(x^2)\right]\,,
\ee
and then (\ref{c31}) can be written as
\be
m_{\eta'}^2(\tt)=m_{\pi}^2(0)+{\hat \chi_\mathrm{top}(0)\over  f_{\pi}^2}+{\cal O}(x^2)
= m_{\pi}^2(0)+x\ {N_f N_c\ \chi_\mathrm{top}(0)\over  f_{\pi}^2}+{\cal O}(x^2)\,,
\ee
which is the standard Witten-Veneziano relation and is also in agreement
with (\ref{etapmassfinalapp}).

\subsection{The Large $\zeta$ limit}

We will investigate now the limit where $\zeta\gg1$.
This is reached when the masses of the quarks are much smaller than the characteristic QCD scale.
For $\zeta\gg1$ the unique solution
of \eqref{cc1} is
\be
\phi_*=\tilde \theta-{\sin\tilde\theta\over \zeta}
+{1\over 2}{\sin(2\tilde\theta)\over \zeta^2}
+{\sin\tilde\theta-3\sin(3\tilde\theta)\over 8\zeta^3}+\cdots\,,
\label{cc11}\ee
and
\be
V''(\phi_*)= f_{\pi}^2~m_{\pi}^2\left[\zeta+\cos\tilde\theta
+{\sin^2\tilde\theta\over \zeta}+\cdots\right]>0\,,
\label{cc12}\ee
while
\be
E(\tilde\theta)=V(\phi_*)=
- f_{\pi}^2~m_{\pi}^2\left[\cos\tilde\theta+{\sin^2\tilde\theta\over 2\zeta}
-{\cos\tilde\theta\sin^2\tilde\theta\over 2\zeta^2}
+{1\over 6\zeta^3}\sin^2\tilde\theta(1+2\cos(2\tilde\theta))+\cdots\right]\,.
\label{cc13}\ee
From (\ref{c23}) and (\ref{c27}) we obtain
\begin{align}
\chi_\mathrm{top}&= f_{\pi}^2m_{\pi}^2\left[\cos\tt
-{\cos(2\tt)\over \zeta}+{\cal O}(\zeta^{-2})\right]\,,\\
m_{\pi}^2(\tt)&=m_{\pi}^2\left[\cos\tt+{\sin^2\tt\over \zeta}+{\cal O}(\zeta^{-2})\right]\,.
\label{c30}
\end{align}

In the large $N_f$ limit $\tilde\theta\to 0$, and (\ref{cc13}) becomes
\be
E(\theta)=- f_{\pi}^2~m_{\pi}^2\left[1
+\left(-1+{1\over \zeta}-{1\over \zeta^2}+\cdots\right){\theta^2\over 2N_f^2}
+{\cal O}(N_f^{-4})\right]\,.
\label{cc14}\ee

In this case, if we are interested in the limit $\tilde\theta\to 0$ we can solve (\ref{cc1})
to all orders in $1/\zeta$ as follows
\be
\phi={\zeta\over \zeta+1}\tilde\theta+{\zeta^3\over 6(\zeta+1)^4}\tilde\theta^3
+{\cal O}(\tilde\theta^5)\,,
\label{c28}\ee
obtaining the following formula for the energy density
\begin{align}
E(\tilde\theta)&=- f_{\pi}^2~m_{\pi}^2\left[1
-{\zeta\over 2(1+\zeta)}\tt^2+{\zeta^4\over 4(1+\zeta)^4}\tt^4+{\cal O}(\tt^6)\right]
\nonumber\\
&=- f_{\pi}^2~m_{\pi}^2\left[1-{ax\over 2(m_{\pi}^2+ax)}\tt^2
+{(ax)^4\over 4(m_{\pi}^2+ax)^4}\tt^4+{\cal O}(\tt^6)\right]\,,
\label{c29}
\end{align}
where in the second line we substituted the value of $\zeta$ from (\ref{cc1}).

\subsection{The $\zeta\to 0$ limit}

This limit can be reached as $x\to 0$ and coincides with the 't Hooft large-N$_c$ limit.
In this limit,  $\zeta\to 0$,  there is an infinite number of extrema that can be found perturbatively in $\zeta$
\be
\phi_n=n\,\pi+(-1)^n\,(\tilde\theta-n\pi)\,\zeta-(\tilde\theta-n\pi)\,\zeta^2+\cdots\,,
\label{cc15}\ee
and thus
\be
V''\sim (-1)^n+\zeta+\cdots
\label{cc16}\ee
which implies that only $n={\rm even}$ are minima.
Evaluating the vacuum energy at the 2n-th minimum we obtain
\be
V_n=- f_{\pi}^2m_{\pi}^2\left[1-{\zeta-\zeta^2\over 2}(\tilde\theta-2n\pi)^2+\cdots\right]\,,
\label{cc17}\ee
and
\be
E(\tt)=\min_{n}\left\{- f_{\pi}^2m_{\pi}^2 \left[1
-{\zeta-\zeta^2\over 2}(\tilde\theta-2n\pi)^2+\cdots\right]\right\}\,.
\ee

\section{Vacua of V-QCD at finite $\theta$-angle}
\label{sec:thetavac}

In this section we analyze vacuum solutions of V-QCD at finite $\theta$-angle.
First we write down explicitly the solutions for the axion $\ax$ and the phase $\xi$. We denote
\be
 f_a(r) \equiv {V_a\sqrt{e^{2A}+\h\,\tau'^2}
\over
\h^{1/2}\,\tau\sqrt{e^{8A}\,V_f^2\,\h\,\tau^2-\Ca^2\, V_a^2}}
\ee
so that~\eqref{xieqs2} implies
\be
\xi'(r) = \Ca\, f_a(r)\,,
\qquad
\xi (r) = \xi_0 + \Ca \int_0^r dr' f_a(r') \, .
\ee
Here $\xi_0$ is identified as the phase of the quark mass on the field theory side. The solution for
$\ax$ can be obtained from~\eqref{axsol}:
\be \label{baraflow}
\ax=\bar \ax_0-x\,\xi\,V_a + \Ca \int_0^r  {dr' \over e^{3A} Z}+x\, \Ca \int_0^r dr' f_a\, V_a\,,\qquad
\bar \ax_0\equiv\ax_0+x\, \xi_0\,,
\ee
where the integration constant $\ax_0$ is related to the standard $\theta$-angle by
$\ax_0 = \ax(r=0) = \theta/N_c$,
and as explained in  Sec.~\ref{sec:CP-oddaction}, the gauge invariant combination $\bar \ax_0$ is related to the
gauge invariant $\bar\theta$-angle through $\bar \ax_0 = \bar\theta/N_c$. Recall that
$\bar\theta = \theta +\arg\det M_q$, where $M_q$ is the quark
mass matrix.  We could use the transformation \eqref{u1transf} to set either $\xi_0$ or $\ax_0$ to zero,
but equivalently we can postpone the gauge fixing and continue working with
$\bar \ax_0$.
The value of $\ax$ at the tip (which will be determined below) is then given by
\be \label{atip}
\ax(\infty) = \bar \ax_0  +\Ca \int_0^\infty dr \left({1\over e^{3A} Z}+x\, f_a\, V_a\right)\,.
\ee
Notice that it is also gauge invariant.

In order to demonstrate the dependence of the free energy on $\bar\theta$, we first analyze the contribution solely from $S_a$ which is obtained from the on-shell
value of the Euclidean action (the overall sign of which is opposite to that of~\eqref{radialaction}):
\begin{align}
 \mathcal{E}_a &= \frac{1}{2}M^3 N_c^2 \int_0^\infty dr\, e^{3A} Z \left(\ax'+x\,\xi\, V_a'\right)^2 &\nonumber \\
  &= \frac{1}{2}M^3 N_c^2\, \Ca \int_0^\infty dr \left(\ax'+x\,\xi\, V_a'\right) = \frac{1}{2}M^3 N_c^2 \Ca \left(\ax(\infty)-\bar \ax_0
 - x\, \Ca \int_0^\infty dr f_a\, V_a\right) \,,&
\end{align}
where we used~\eqref{axsol} to obtain the expression on the second line. We stress that this expression is not the complete free energy, which will be analyzed below, but it is the most important contribution
for the $\bar\theta$-dependence. By using~\eqref{atip} the result may be written as
\be \label{Eafinal}
 \mathcal{E}_a = \frac{1}{2} M^3 N_c^2  \frac{\int_0^\infty  dr  e^{-3A} Z^{-1}}{\left[ \int_0^\infty dr \left(e^{-3A} Z^{-1}+x f_a V_a\right)\right]^2} \left(\bar \ax_0 - \ax(\infty)\right)^2 \,.
\ee
This result is analogous to what was found for the $\theta$ dependence in the Yang-Mills case in~\cite{ihqcd}.
Similarly as in that case, we expect that
the contribution to the energy from the IR singularity, \ie $\ax(\infty)$, vanishes. Otherwise, the IR
singularity would play the role of a second boundary. Vanishing of $\ax(\infty)$ also leads to
$\mathcal{E}_a \propto \bar \ax_0^2 \propto \bar \theta^2$ (for small $\bar \theta$ so that the implicit dependence of the integrals in~\eqref{Eafinal} on $\bar\theta$ can be neglected) which agrees with the large $N_c$ analysis of
QCD~\cite{Witten:1998uka}. We will argue below that after setting $\ax(\infty)$ to zero, also the full free energy, not only $\mathcal{E}_a$, has quadratic behavior for small  $\bar \ax_0$.

The issue described above applies to all string theory ``axions", namely scalars without a potential. As argued above, in all such cases, an explicit boundary condition must be imposed in the IR that is not dictated by regularity. In many cases such axions are internal components of gauge fields or higher forms (even the ten-dimensional IIB axion can be T-dualized to such
a form). A concrete example of this is the case of the black $D_4$ soliton where the $\theta$ angle is generated by a six-dimensional vector field, \cite{Witten:1998uka}. In all such cases, the usual regularity condition for the form field indicates that it should vanish on the extremal horizon, not unlike the boundary condition we chose above.

Setting $\ax(\infty)=0$, the relation between the source $\bar \ax_0$ and the VEV $\Ca$ follows from~\eqref{atip}:
\be \label{axrelation}
 \frac{\bar\theta}{N_c} = \bar \ax_0 = -\Ca \int_0^\infty dr \left({1\over e^{3A} Z}+x\, f_a\, V_a\right)\,,
\ee
where one should recall that the integral also depends implicitly on $\Ca$ so that the relation is not exactly linear.

\begin{figure}[!tb]
\begin{center}
\includegraphics[width=0.49\textwidth]{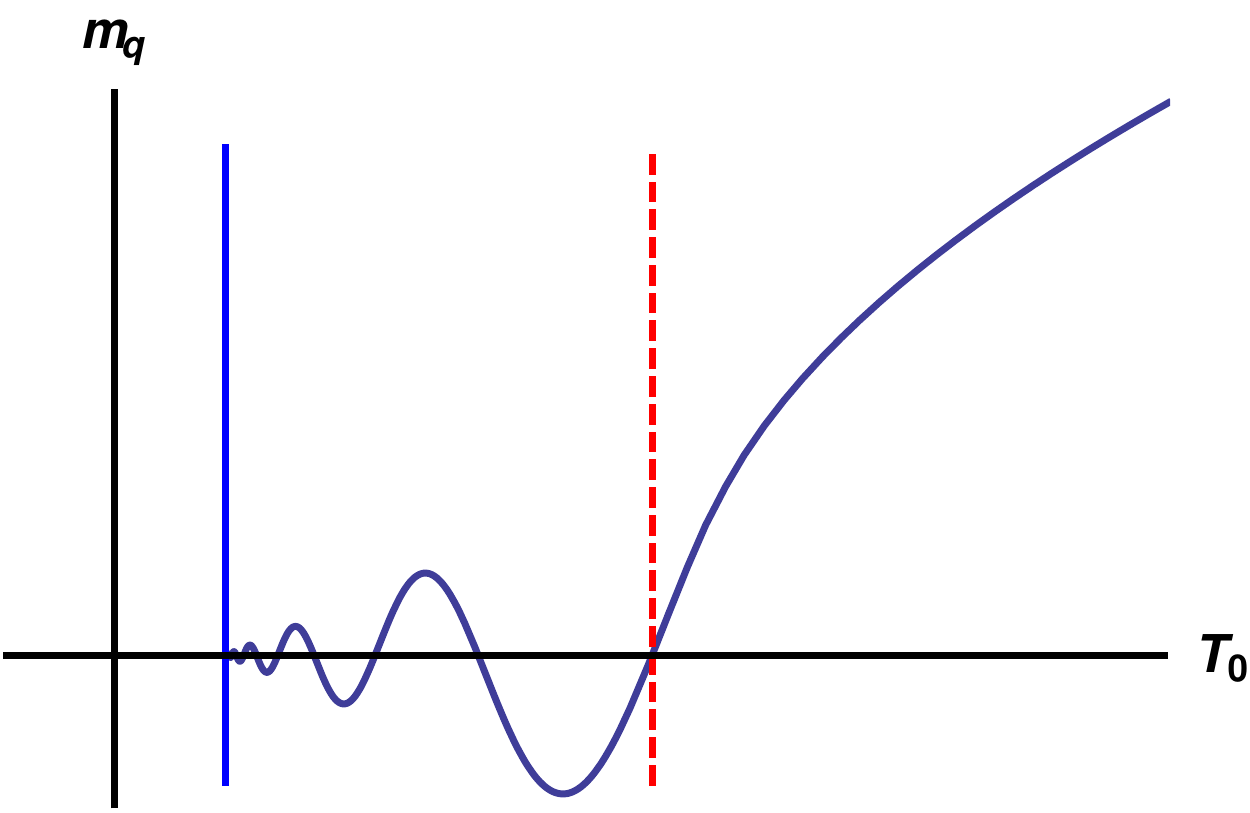}
\end{center}
\caption{Sketch of the dependence of the quark mass  on $T_0$ for $\Ca=0=\bar\theta$, \ie on the horizontal
axis of Fig.~\protect\ref{fig:masscontours} (left). Solutions exist right of the vertical blue line. The dashed
red vertical line denotes the location of the standard vacuum with zero quark mass.}
\label{fig:massdepzerotheta}\end{figure}

\subsection{Construction of backgrounds and their generic properties}

Recall that at zero $\bar \theta$,
the chirally broken backgrounds could be parametrized in terms of a single variable defined through the IR asymptotics of the solution~\cite{jk}. For potentials~I, this variable was denoted by $T_0$ and controlled the normalization of the tachyon in the IR. The value of $T_0$ could be mapped to the physical parameter in QCD, the quark mass, after constructing the background solution. At zero $\bar\theta$ it was natural to choose the tachyon to be real, and to define $\tau$ as the real part of the complex field, so that it could become negative. The source of $\tau$ in the earlier work, \ie the quark mass, consequently maps to the real part of the source of the complex tachyon. An example of the dependence of the quark mass on $T_0$ (in the QCD regime $0<x<x_c$) is given in Fig.~\ref{fig:massdepzerotheta}.  Notice also that negative values of $T_0$ were allowed, but the solutions with opposite values of $T_0$ were related by the reflection $\tau \mapsto -\tau$ which left the action invariant.

\begin{figure}[!tb]
\begin{center}
\includegraphics[width=0.49\textwidth]{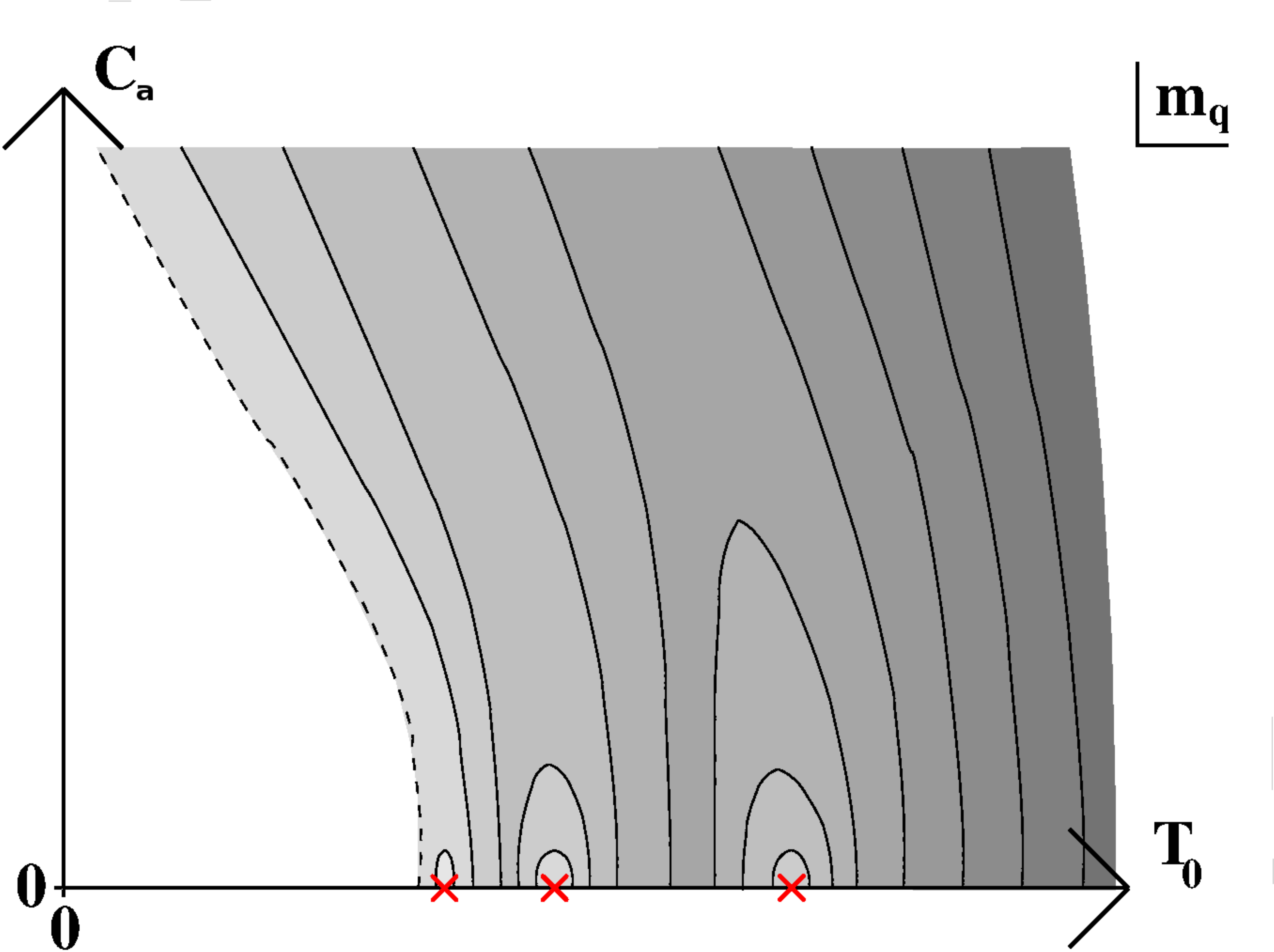}%
\hspace{2mm}\includegraphics[width=0.49\textwidth]{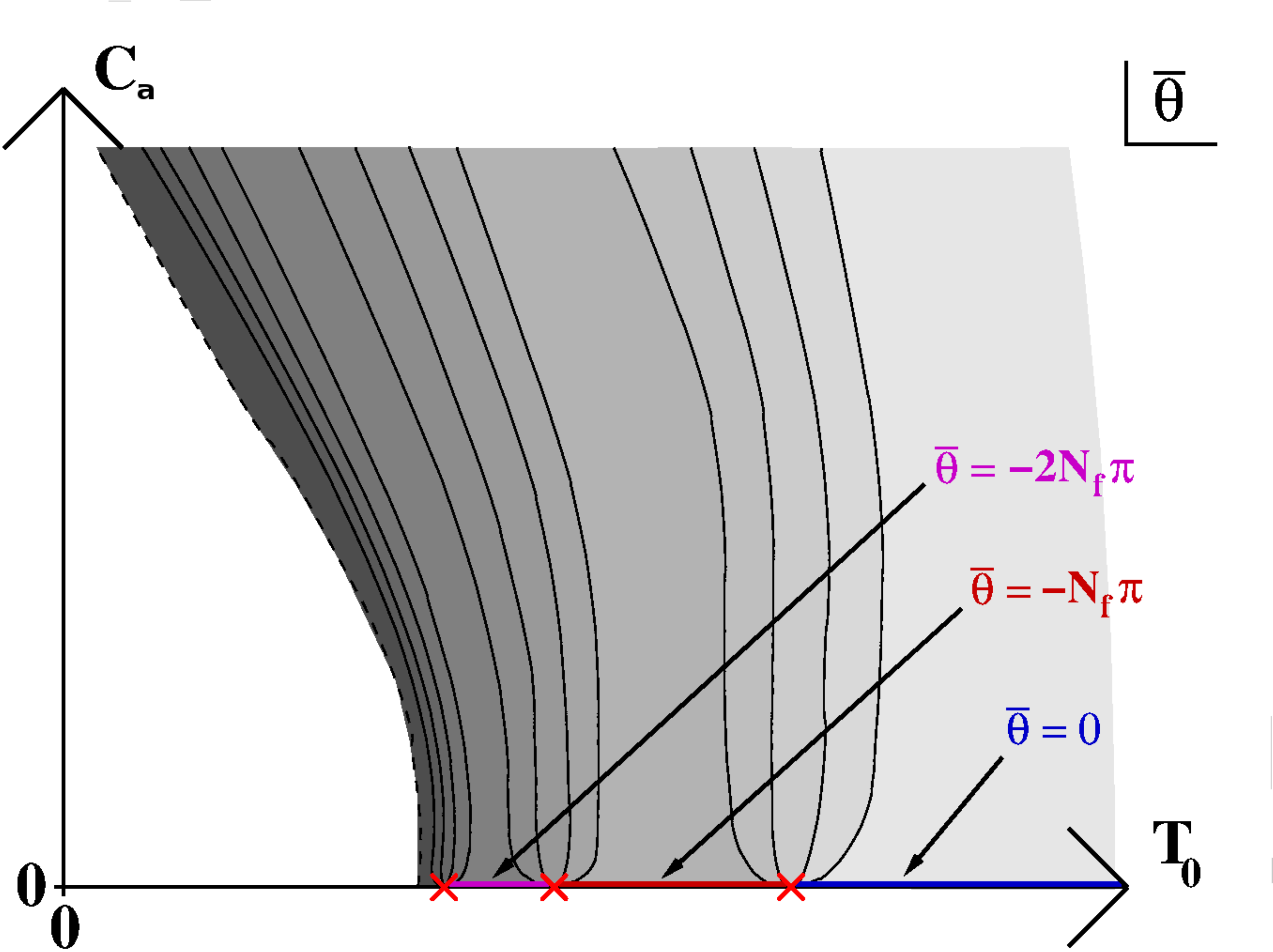}
\end{center}
\caption{Sketch of the dependence of the quark mass (left hand plot) and the $\bar \theta$-angle
(right hand plot) on the parameters $T_0$ and $\Ca$ (with $\Ca$ given in IR units). The contours
lie at fixed quark mass or $\bar \theta$-angle, and the red crosses denote points where the quark mass vanishes.
Solutions exist in the shaded region. The $\bar \theta$-angle takes piecewise constant values on the intervals of the horizontal axis between the crosses as indicated in the right hand plot.}
\label{fig:masscontours}\end{figure}

\begin{figure}[!tb]
\begin{center}
\includegraphics[width=0.49\textwidth]{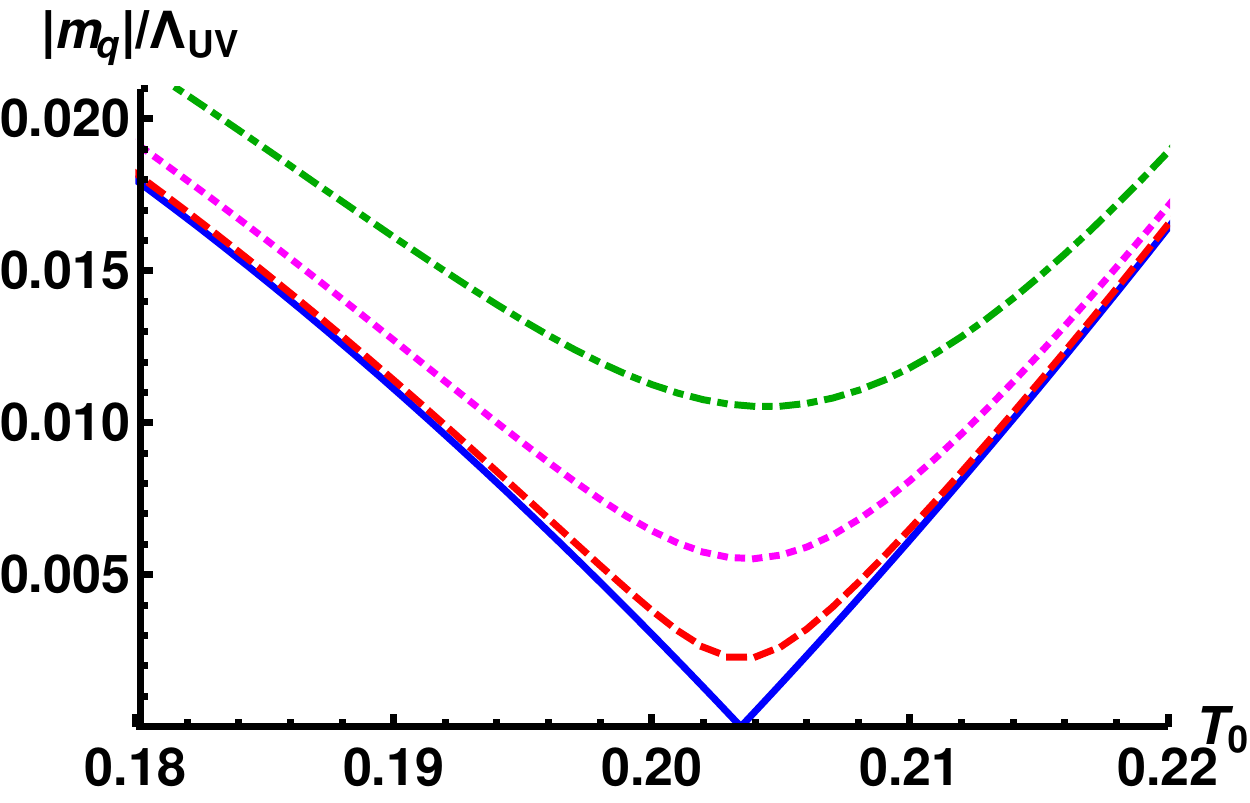}%
\hspace{2mm}\includegraphics[width=0.49\textwidth]{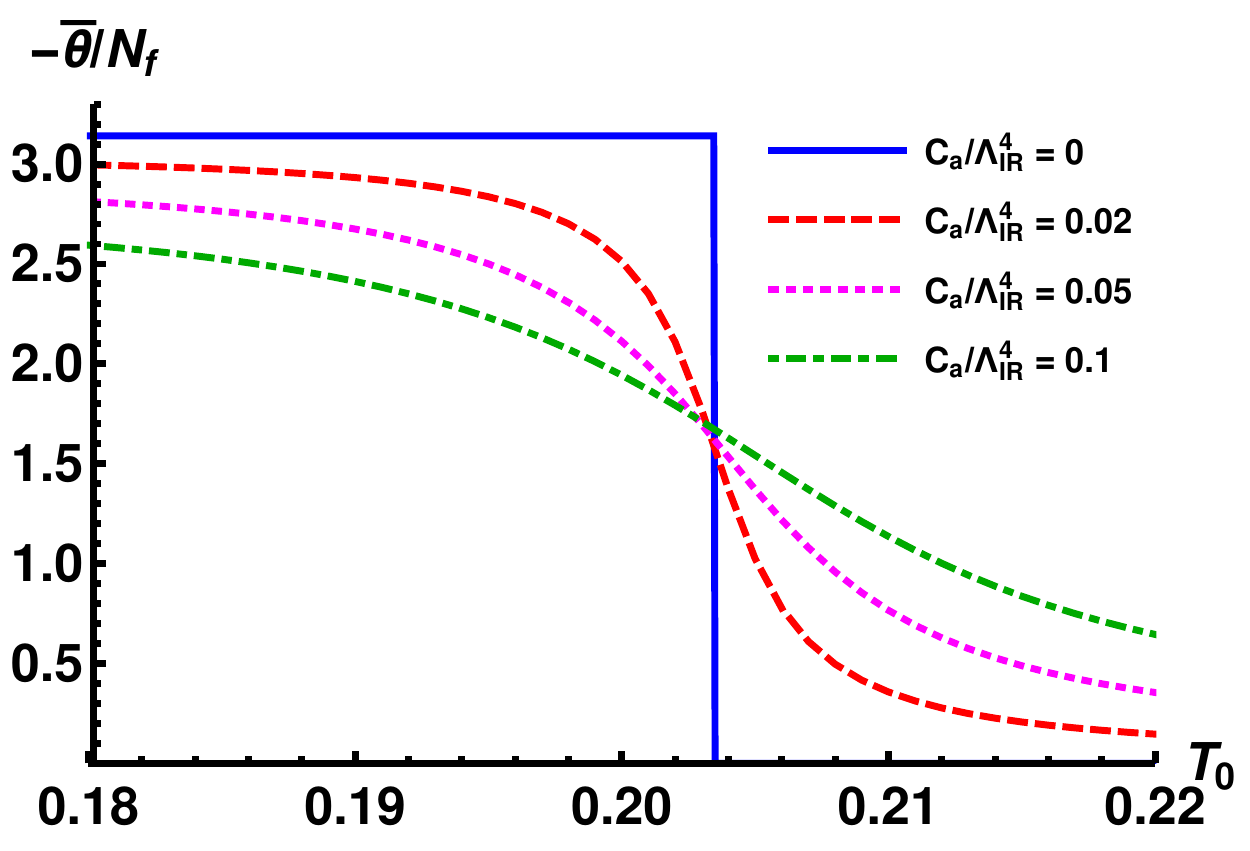}
\end{center}
\caption{The behavior of the quark mass (left hand plot) and the $\bar \theta$-angle (right hand plot) for the ``QCD-like'' potentials I with $x=2/3$ near the rightmost node in Fig.~\protect\ref{fig:masscontours} (left), \ie the standard zero mass vacuum. The solid blue, dashed red, dotted magenta, and dot-dashed green curves have $\Ca/\LIR^4 = 0$, $0.02$, $0.05$, and $0.1$, respectively.}
\label{fig:massthetaT0}\end{figure}

\begin{figure}[!tb]
\begin{center}
\includegraphics[width=0.49\textwidth]{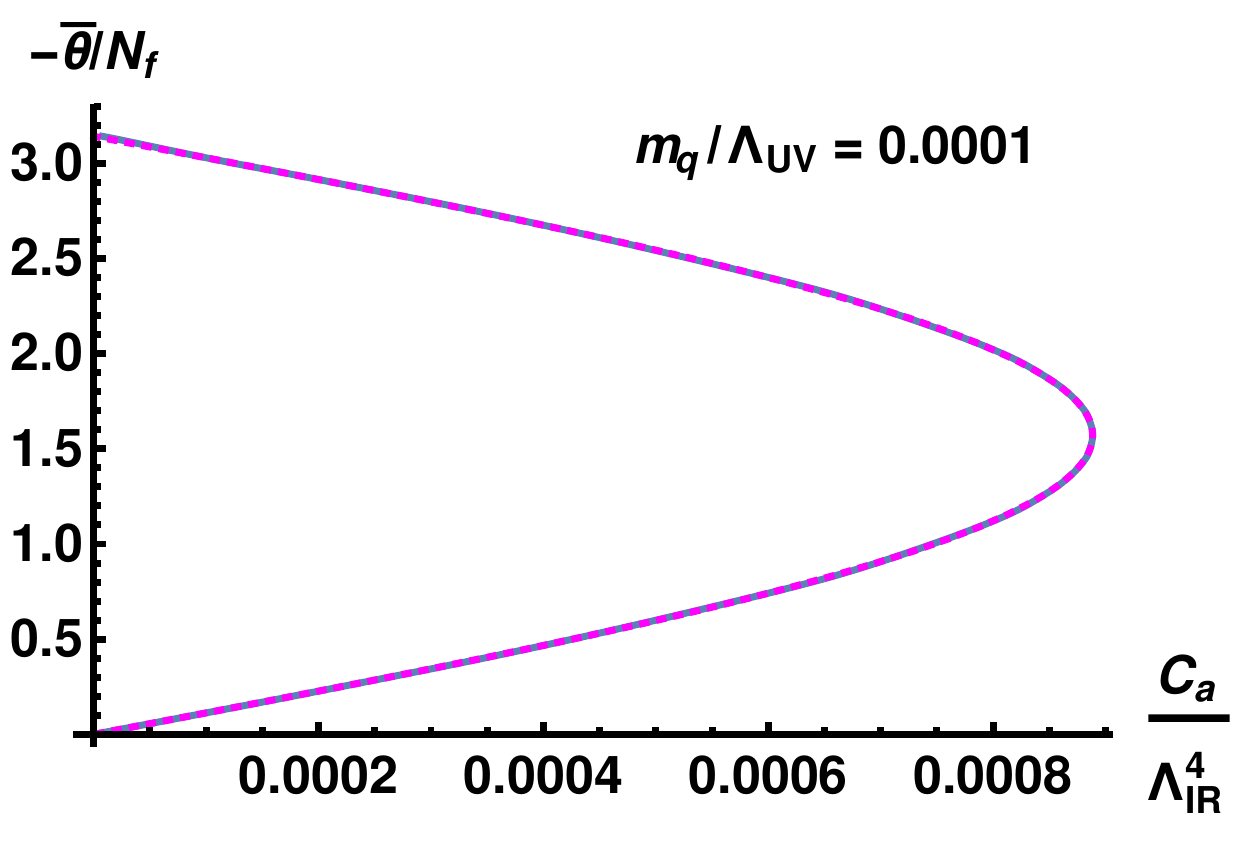}%
\hspace{2mm}\includegraphics[width=0.49\textwidth]{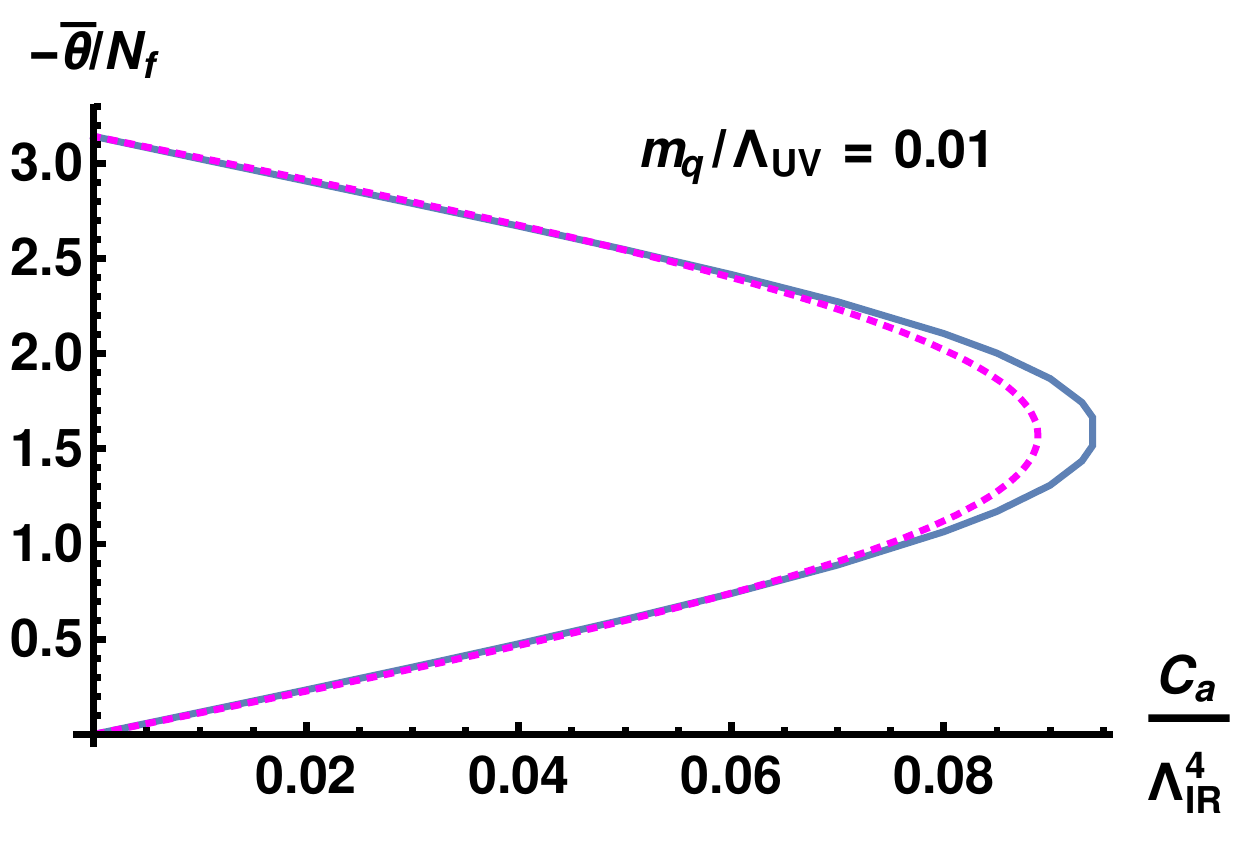}
\includegraphics[width=0.49\textwidth]{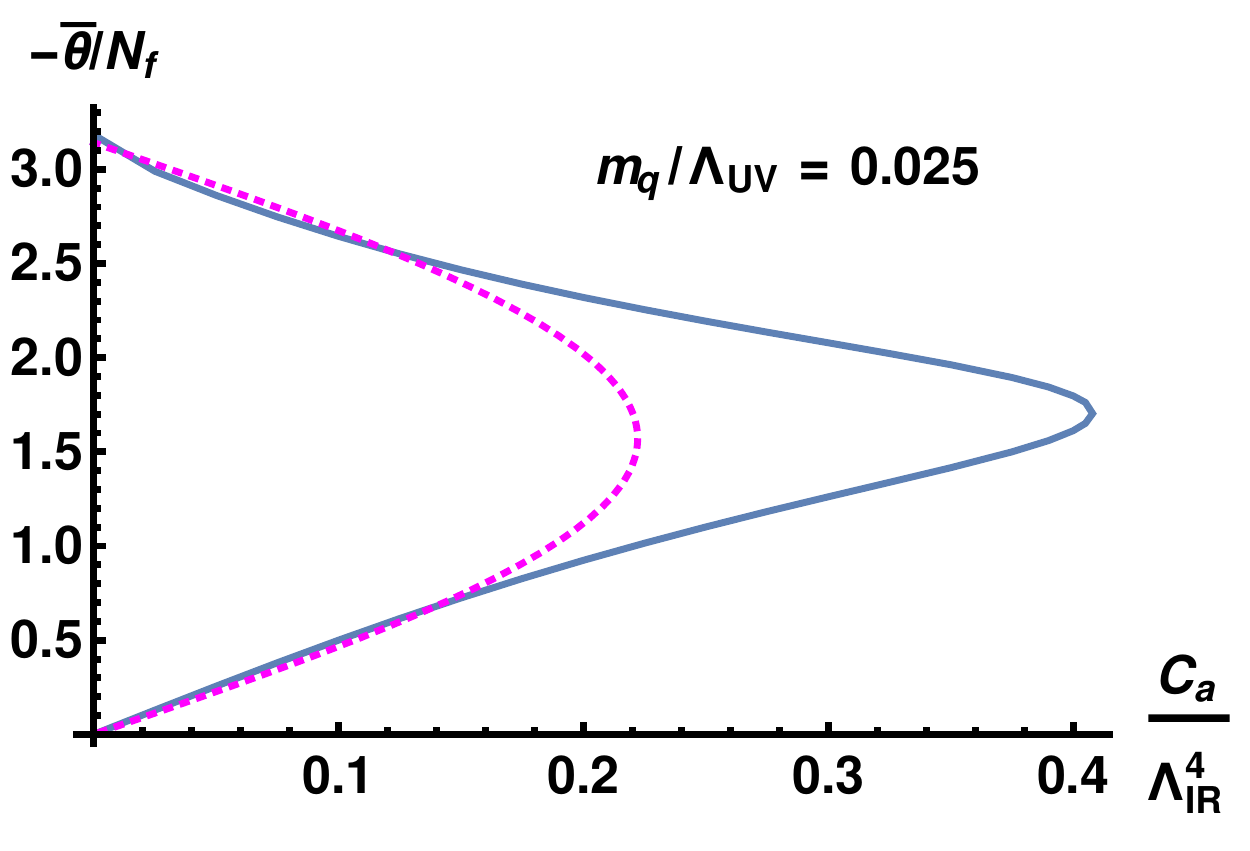}%
\hspace{2mm}\includegraphics[width=0.49\textwidth]{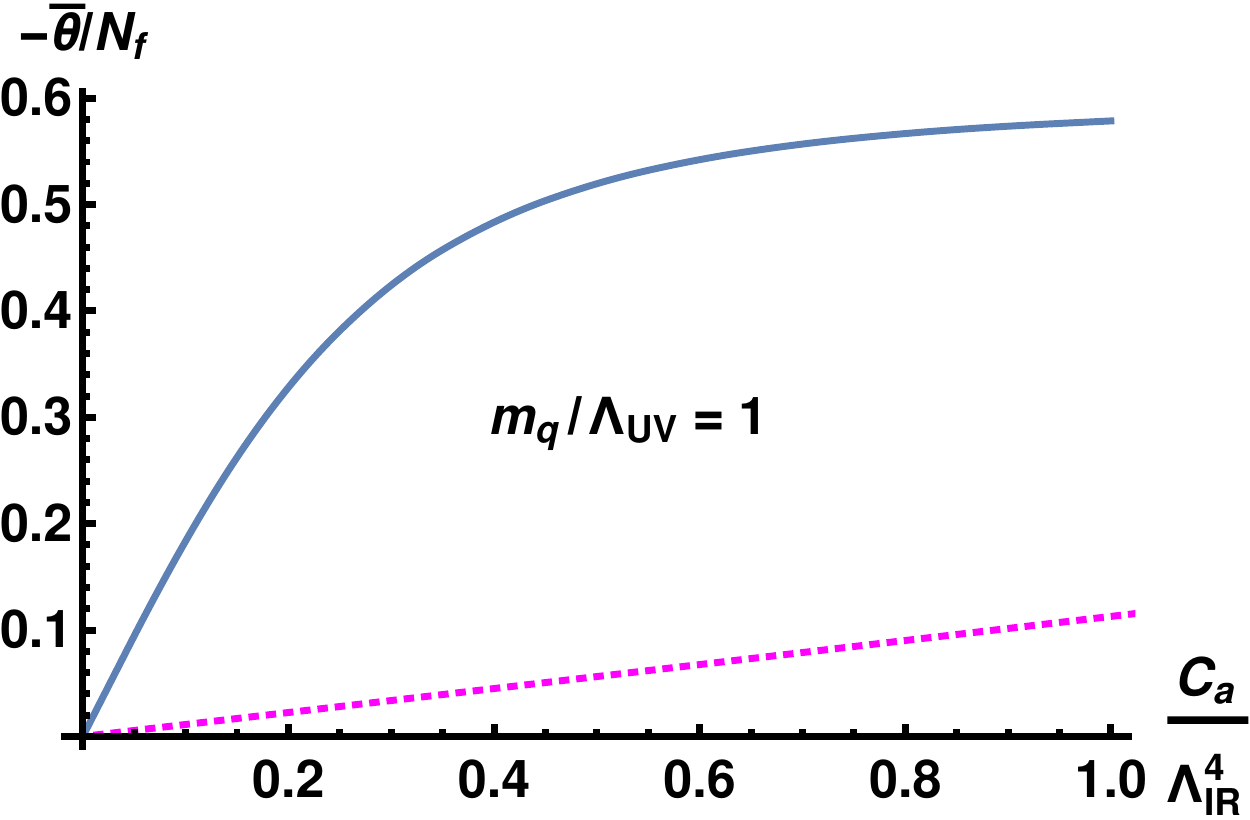}
\end{center}
\caption{The dependence of the (gauge invariant) $\bar \theta$-angle on $\Ca$ for the ``QCD-like'' potentials I at $x=2/3$ and for various fixed values of the quark mass as indicated in the plots. The blue curves are numerical data, and the dotted magenta curves are determined by the analytic approximation at small $m_q$ from~\protect\eqref{Cthetasmallmq}.}
\label{fig:thetavsC}\end{figure}

Implementation of the CP-odd sector removes the reflection symmetry: Because the phase of the tachyon is nontrivial, it is natural that $\tau$ is the absolute value of the complex tachyon. Therefore the quark mass is also defined as the absolute value of the source for the complex tachyon, and $T_0>0$.
At finite $\bar\theta$-angle we also have a second variable, the integration constant $\Ca$, which controls the value of the $\bar\theta$-angle. More precisely, the pair $(T_0,\Ca)$ can be mapped to ($m_q,\bar\theta$) after the background has been constructed.

We have studied the CP-odd backgrounds numerically, restricting our analysis to the region with positive $\Ca$ -- the solutions at negative $\Ca$ can be obtained by applying the CP transformation as pointed out at the end of Sec.~\ref{subsec:eoms}.
The procedure for creating the numerical solutions is essentially the same as discussed in~\cite{jk,Arean:2013tja,Jarvinen:2015ofa} -- the solutions are obtained by shooting from near the IR singularity, and the boundary conditions there are given by the known IR asymptotic expansions. As the axion and the phase of the tachyon could be integrated out of the equations of motion, there are essentially only two differences with respect to the equations at zero $\bar\theta$-angle: there is a new integration constant $\Ca$ and the tachyon equation of motion is now written in terms of the absolute value of the complex tachyon rather than its real part. As we have demonstrated above in Sec.~\ref{sec:thetaback} the IR asymptotics (and therefore also the IR boundary conditions) are unchanged, up to the possible appearance of some special solutions which will be discussed below. When presenting the numerical data, dimensionful quantities can be given either in UV units ($\LUV$) or in IR units ($\LIR$),
which are discussed in Sec.~\ref{sec:bg} and defined precisely in~\eqref{LUVdef} and~\eqref{LIRdef} in Appendix~\ref{app:Asymptotics}.

We discuss first details in the QCD-like phase ($0<x<x_c$ with $x_c-x = \morder{1}$), where a rich structure is found, and return to the dependence of the backgrounds on $x$ below. In this phase, the absolute value of the quark mass and the $\bar\theta$-angle depend on $T_0$ and $\Ca$ as depicted schematically in Fig.~\ref{fig:masscontours}.
Recall first what happens on the horizontal axis ($\Ca=0$) where the $\bar\theta$-angle vanishes and the tachyon is real. The real quark mass as a function of $T_0$ in this case is given in Fig.~\ref{fig:massdepzerotheta}, see~\cite{jk}. As we have already pointed out, at finite $\Ca$ we define $m_q$ as the absolute value of the source of the complex tachyon, whereas Fig.~\ref{fig:massdepzerotheta} shows the dependence of the real part of the source on $T_0$. Therefore, in order to compare to Fig.~\ref{fig:masscontours} (left), one needs first take the absolute value so that the negative values of $m_q$ in Fig.~\ref{fig:massdepzerotheta} are reflected to positive values.

Solutions are only found for $T_0>T_{0c}$, where the critical value $T_{0c}$ is the endpoint of the dashed curve in Fig.~\ref{fig:masscontours} and denoted by the vertical blue line in Fig.~\ref{fig:massdepzerotheta}.
The value of the quark mass oscillates as $T_0 \to T_{0c}$ from above, so that there are infinitely many zeroes (of which the three which occur at largest $T_0$'s are shown as red crosses in Fig.~\ref{fig:masscontours}). The first node (largest value of $T_0$) is the standard stable vacuum at zero quark mass, whereas the other nodes are unstable Efimov vacua. As one approaches the critical value $T_{0c}$, the background flows closer and closer to the IR fixed point, but misses it eventually due to the nonzero tachyon. It is also possible that there is only a finite amount of nodes on the horizontal axis. This can happen if the bulk mass of the tachyon satisfies the BF bound at the IR fixed point~\cite{jk,Arean:2013tja}, as is the case for potentials~I at low values of $x$~\cite{alho}.

Extending to the solutions with $\Ca \ne 0$ and therefore finite $\bar\theta$-angle,
the nodes are smoothed out, but the region at small $T_0$, where no regular solutions exist
(white in Fig.~\ref{fig:masscontours}), remains at least for small $\Ca$.
The structure of the sketch in Fig.~\ref{fig:masscontours} can be confirmed numerically for
the concrete choices of potentials~I that we have introduced.
As an example we show the dependence of the quark mass and the $\bar\theta$-angle on
$T_0$ and $\Ca$ for the ``QCD-like'' potentials~I in Fig.~\ref{fig:massthetaT0}.
The range of $T_0$ was chosen in the vicinity of the ``standard'' zero mass vacuum,
which is denoted by the rightmost cross in Fig.~\ref{fig:masscontours} and by the vertical
dashed red line in Fig.~\ref{fig:massdepzerotheta}. It can also be verified analytically
that $\bar\theta$ is quantized in units of $N_f\,\pi$ on the $C_a=0$ axis, as shown
in Fig.~\ref{fig:masscontours} (right), see Appendix~\ref{app:thetabackgrounds}.
The uniqueness and stability of the solutions is discussed in the same Appendix.

Notice that there are two types of curves of constant $m_q$ in Fig.~\ref{fig:masscontours} (left). First, some of the curves start from the horizontal axis, circle around some of the nodes, and return to the axis. Second, some curves start from the horizontal axis and exit the plot at its upper edge.
We plot in Fig.~\ref{fig:thetavsC} the value of the $\bar\theta$-angle at constant $m_q$, {\it i.e.}, along the curves,
for potentials~I. The plots for $m_q/\LUV=0.0001$,  $m_q/\LUV=0.01$, and $m_q/\LUV=0.025$ correspond to contours in Fig.~\ref{fig:masscontours} (left) which start from the standard $\bar\theta=0$ solutions, \ie on the interval marked with dark blue color, and return on the horizontal axis on the dark red interval, having $\bar \theta=-N_f \pi$. We will also show in Appendix~\ref{app:thetabackgrounds} why these curves end exactly at $\bar \theta=-N_f \pi$. The remaining plot at $m_q/\LUV=1$ in Fig.~\ref{fig:thetavsC} corresponds to a curve in Fig.~\ref{fig:masscontours} (left) which starts from the dark blue interval and exits the plot without returning to the horizontal axis, which leads to the solutions being found only for a finite\footnote{Notice that this is the case only for a single branch of solutions, which are connected by continuous deformations of the parameters. As argued in Sec.~\ref{sec:VQCDdef}, there are also disconnected branches which realize the $2\pi$ periodicity of $\bar\theta$, an taking them into account solutions are found for all values of $\bar\theta$.} range of $\bar\theta$. We also show the analytic small $m_q$ approximation (given below in~\eqref{Cthetasmallmq}) as dashed magenta curves. The value $m_q/\LUV=1$ is so large already that the small $m_q$ result does not work even as a rough approximation. If the value of $m_q$ is increased further the plot will remain essentially unchanged.

In Fig.~\ref{fig:RGflow} we study numerically the holographic RG flow of the field $\bar \ax = \ax + x\,\xi\, V_a$
which is invariant under the $U(1)_A$ transformation~\eqref{u1transf}. The field vanishes in the IR due to the
boundary condition $\bar\ax(\infty)=\ax(\infty)=0$, and its boundary value is $\bar\ax_0 = \bar\theta/N_c$. For the
left hand plot we have picked four points from the curve in the top right plot of Fig.~\ref{fig:thetavsC} at
$m_q/\LUV = 0.01$, which have pairwise the values $C_a/\LIR^4 =0.02$ and $C_a/\LIR^4 = 0.08$, but are on
different branches of the curve. The flow of $\bar \ax$ is determined by~\eqref{baraflow}. When $m_q/\LUV \ll 1$,
the two integrals in this equation affect the flow at different scales of $r$, which can also be seen from
Fig.~\ref{fig:RGflow}. The first integral is the only finite term in the probe limit $x \to 0$. It adds a contribution to the
flow at $r \sim 1/\LUV$, which is roughly proportional to $C_a$. Indeed the curves having the same $C_a$ overlap
in Fig.~\ref{fig:RGflow} (left) when $r \sim 1/\LUV$. The second integral is the flavor contribution which affects the
flow mostly at $r \sim \sqrt{m_q/\sigma}$. This term is dominant at small $r$ in the plot, and results in a different
flow for the curves which have the same $C_a$ but different branch. We will see in Sec.~\ref{sec:smallmq} that this
structure is analytically tractable in the limit $m_q \to 0$. Also, one can show that the flow on the upper branch of Fig.~\ref{fig:thetavsC} approaches
a step function as $C_a \to 0$ (see Appendix~\ref{app:thetabackgrounds}) and indeed the flow at $C_a/\LIR^4 = 0.02$ (dot-dashed green curve) in the left hand plot of Fig.~\ref{fig:RGflow} is already reminiscent of a step function.
In the right hand plot of Fig.~\ref{fig:RGflow} we plot
$\bar \ax(r)$ for $m_q/\LUV=1$ and for various values of $C_a$. In this case the RG flow is significant only for
$r\sim 1/\LUV$. One can check that the first integral in~\eqref{baraflow} dominates.

\begin{figure}[!tb]
\begin{center}
\includegraphics[width=0.49\textwidth]{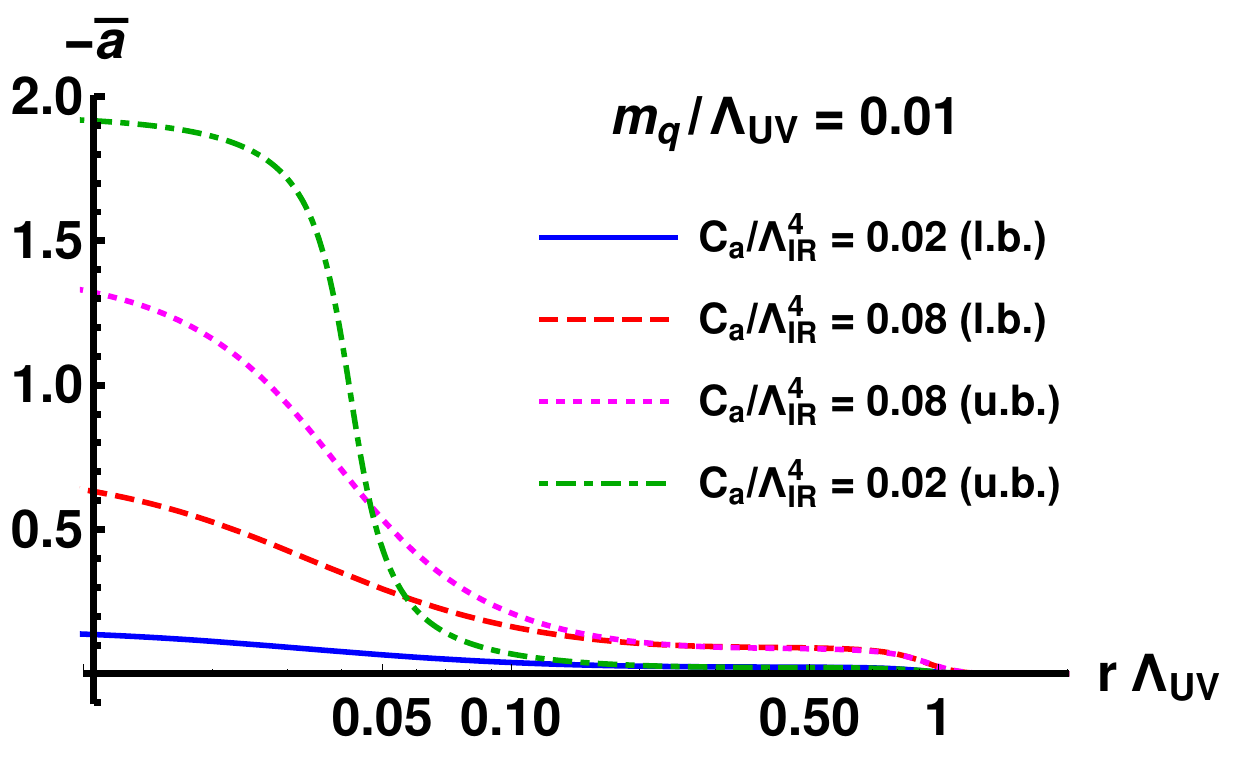}%
\hspace{2mm}\includegraphics[width=0.49\textwidth]{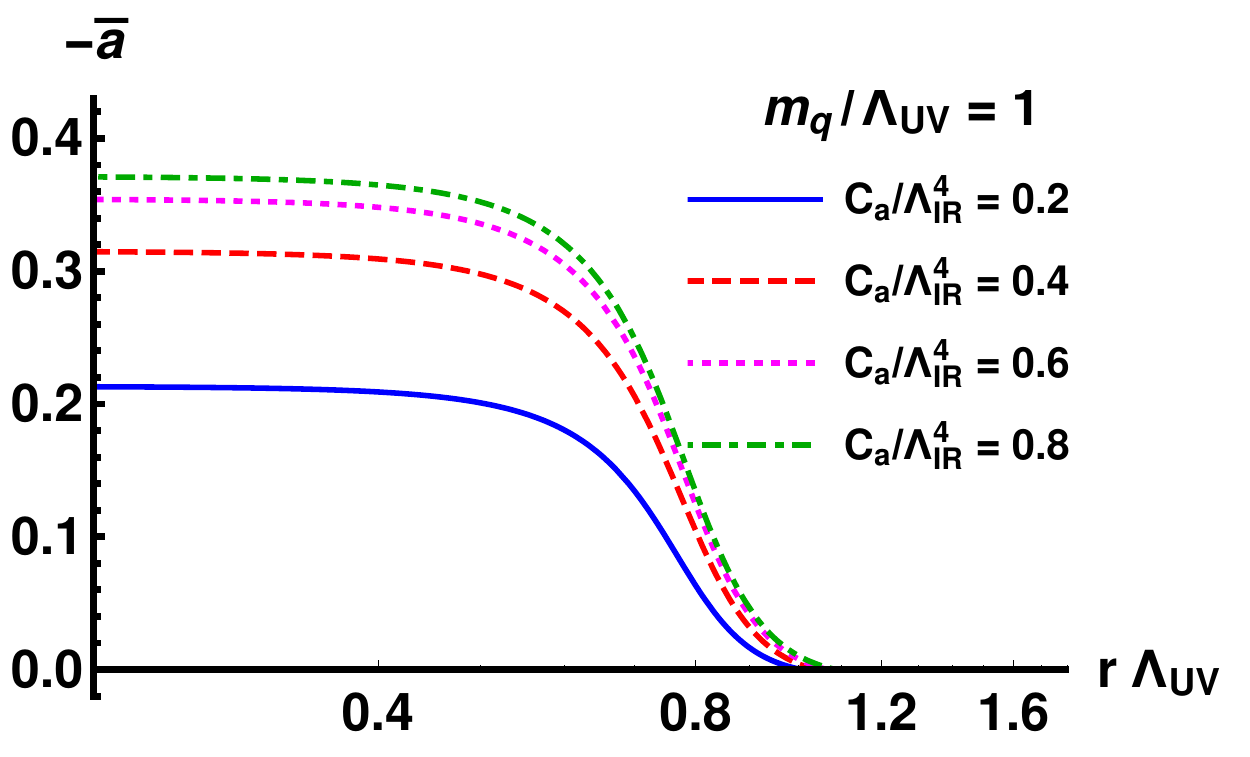}
\end{center}
\caption{The holographic RG flow of the gauge invariant field $\bar \ax$ for potentials~I at $x=2/3$. Left: $\bar \ax(r)$ at $m_q/\LUV = 0.01$. The solid blue and dot-dashed green curves have $C_a/\LIR^4 = 0.02$, while the dashed red and dotted magenta curves have $C_a/\LIR^4 = 0.08$. The solid blue and dashed red curves are for the lower branch (as denoted by ``l.b.'' in the legend) in the top right plot of Fig.~\protect\ref{fig:thetavsC}, while the dotted magenta and dot-dashed green curves are for the upper branch (denoted by ``u.b.'' in the legend). Right:  $\bar \ax(r)$ at $m_q/\LUV = 1$. The solid blue, dashed red, dotted magenta, and dot-dashed green curves have $C_a/\LIR^4 = 0.2$, $0.4$, $0.6$, and $0.8$, respectively.  }
\label{fig:RGflow}\end{figure}

\begin{figure}[!tb]
\begin{center}
\includegraphics[width=0.49\textwidth]{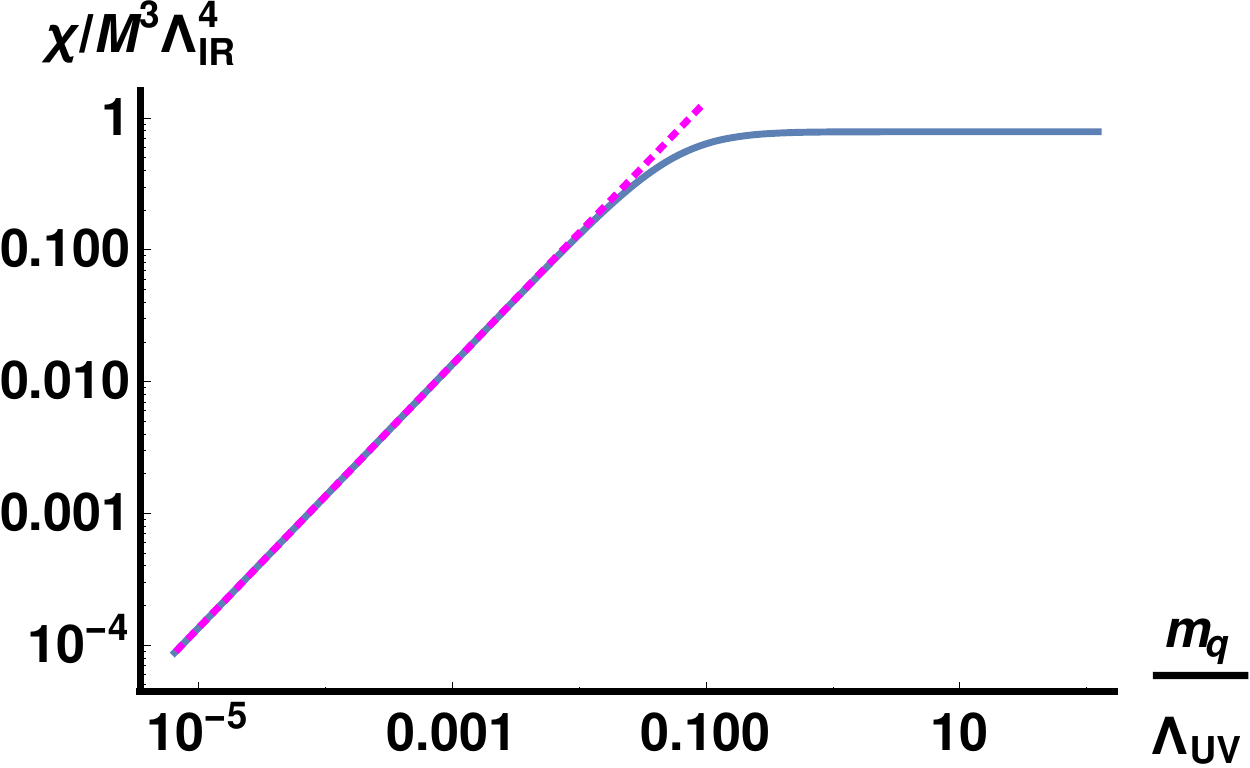}%
\hspace{2mm}\includegraphics[width=0.49\textwidth]{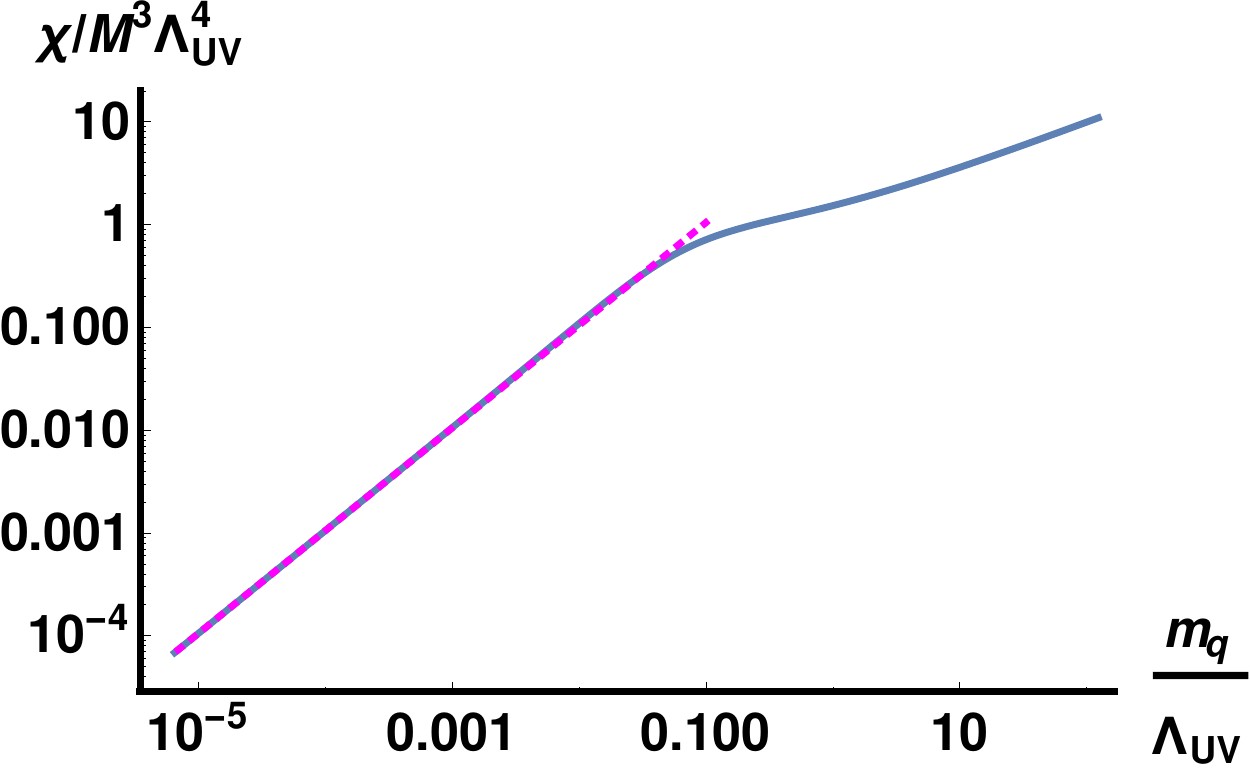}
\end{center}
\caption{The dependence of the topological susceptibility on the quark mass in IR units (left hand plot)
and in UV units (right hand plot) in the standard, dominant vacuum for potentials I with $x=2/3$.
The units were discussed in Sec.~\protect\ref{sec:bg} and they are defined in terms of the asymptotic expansions of
Appendix~\protect\ref{app:Asymptotics}. The blue solid curves are numerical data and the dashed magenta curves follow
the approximation at small $m_q$ given in Eq.~\protect\eqref{suskisrel}.}
\label{fig:suskis}\end{figure}

\begin{figure}[!tb]
\begin{center}
\includegraphics[width=0.49\textwidth]{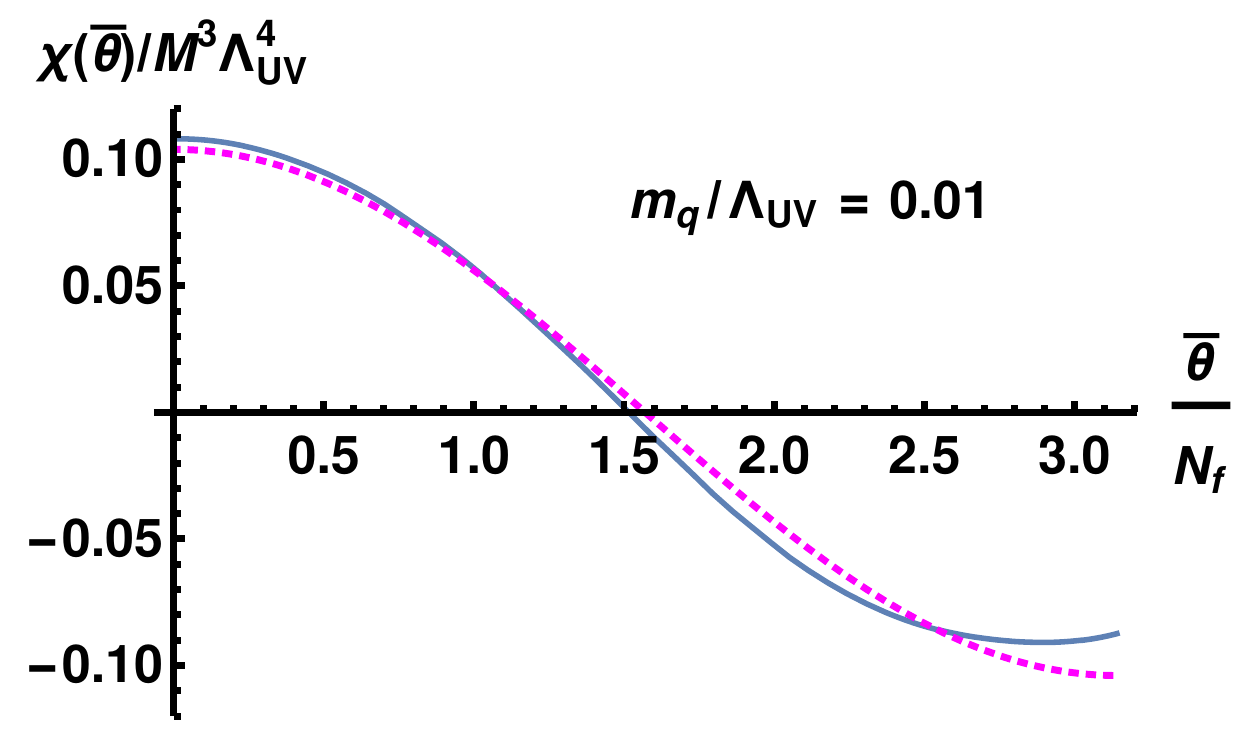}%
\hspace{2mm}\includegraphics[width=0.49\textwidth]{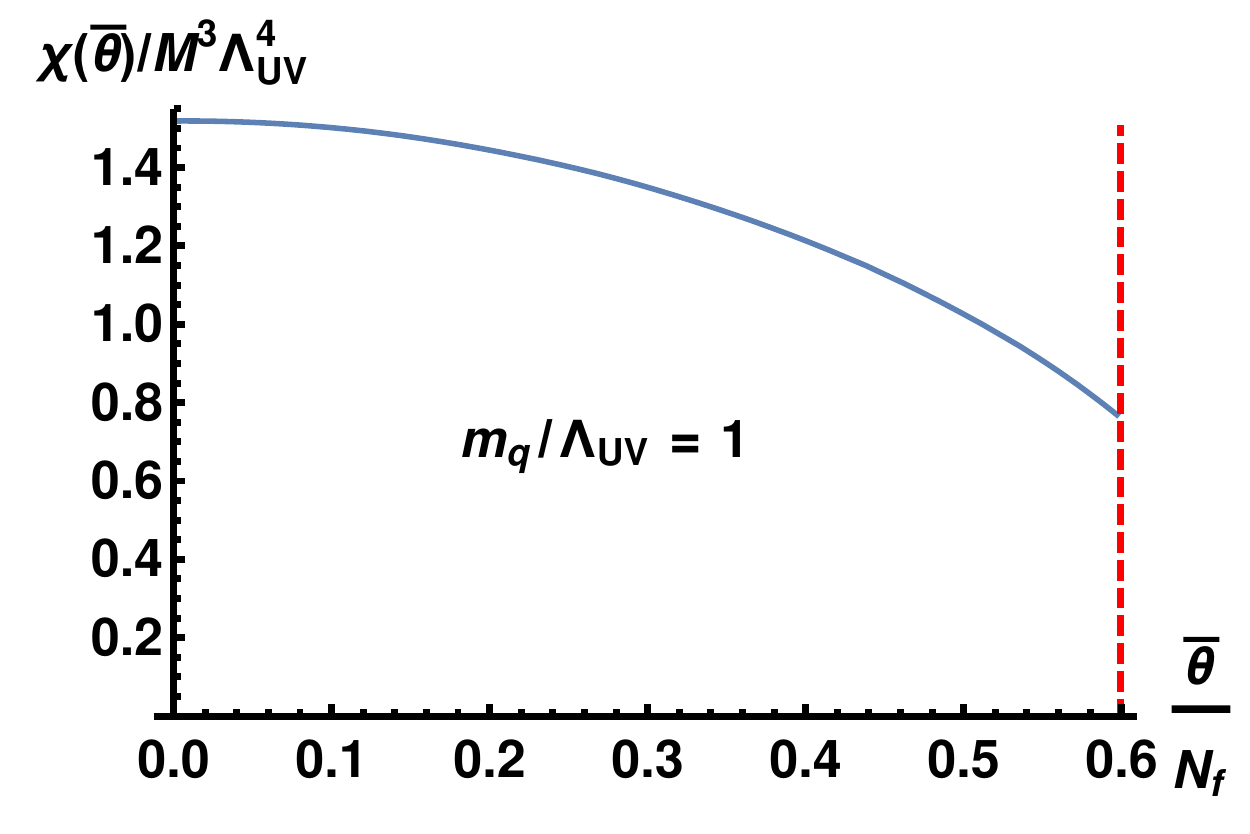}
\end{center}
\caption{The dependence of $\chi$ on the $\bar\theta$-angle for small (left hand plot) and relatively large
(right hand plot) quark mass for potentials I with $x=2/3$. On the left-hand plot, the magenta curve is
given by~\protect\eqref{suskisgen}.  The dashed red vertical line denotes the limiting value of the $\bar \theta$-angle as $C_a \to - \infty$ along the curve of the fixed mass value.}
\label{fig:suskistheta}\end{figure}

\begin{figure}[!tb]
\begin{center}
\includegraphics[width=0.49\textwidth]{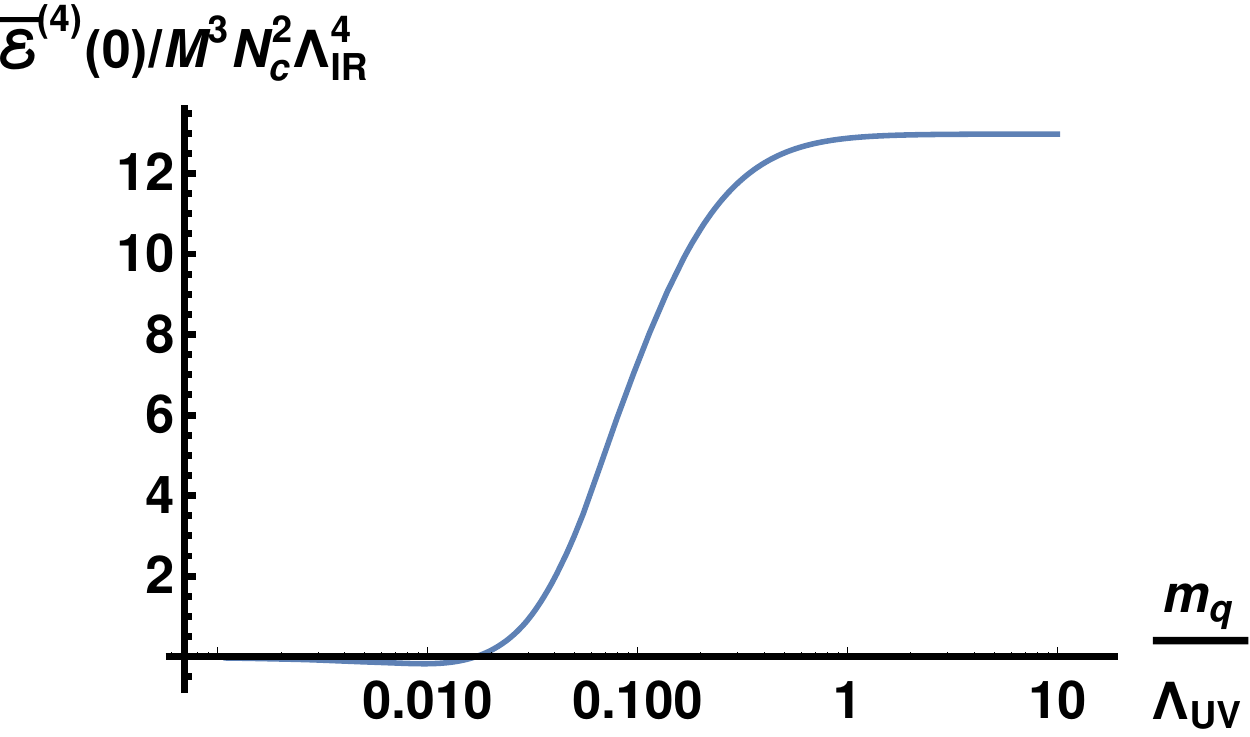}%
\hspace{2mm}\includegraphics[width=0.49\textwidth]{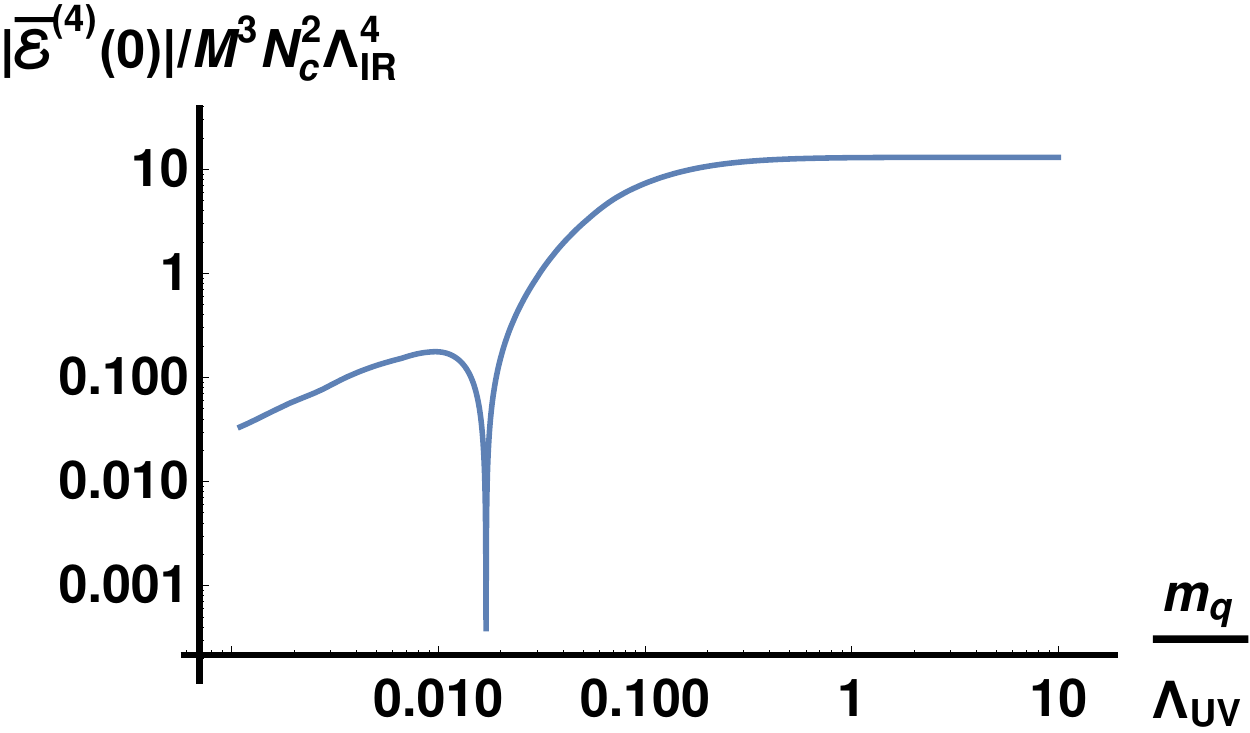}
\end{center}
\caption{The dependence of the fourth order derivative of the free energy at $\bar\theta =0 $ as a function of $m_q$ for potentials I with $x=2/3$. Left: linear scale. Right: logarithmic scale.
}
\label{fig:fourthder}\end{figure}

\begin{figure}[!tb]
\begin{center}
\includegraphics[width=0.49\textwidth]{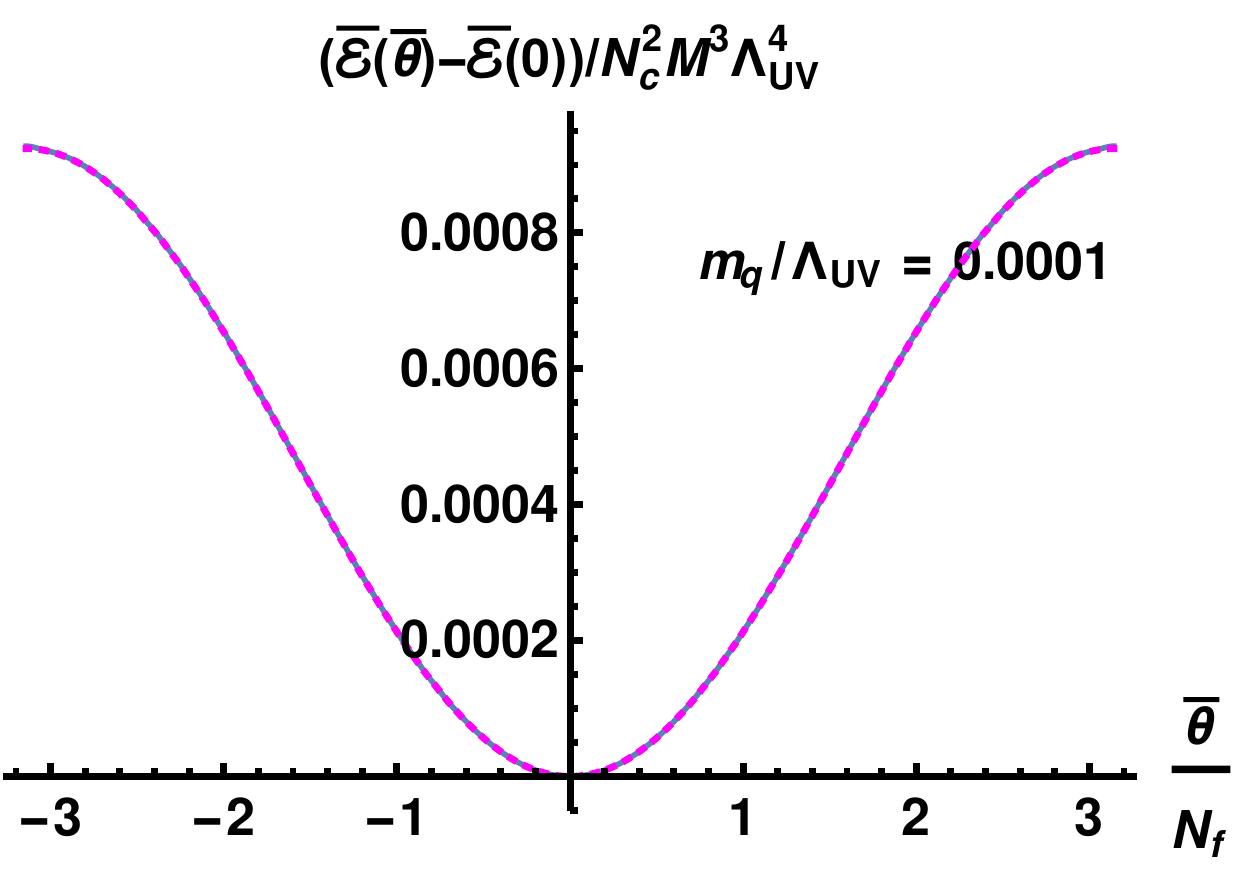}%
\hspace{2mm}\includegraphics[width=0.49\textwidth]{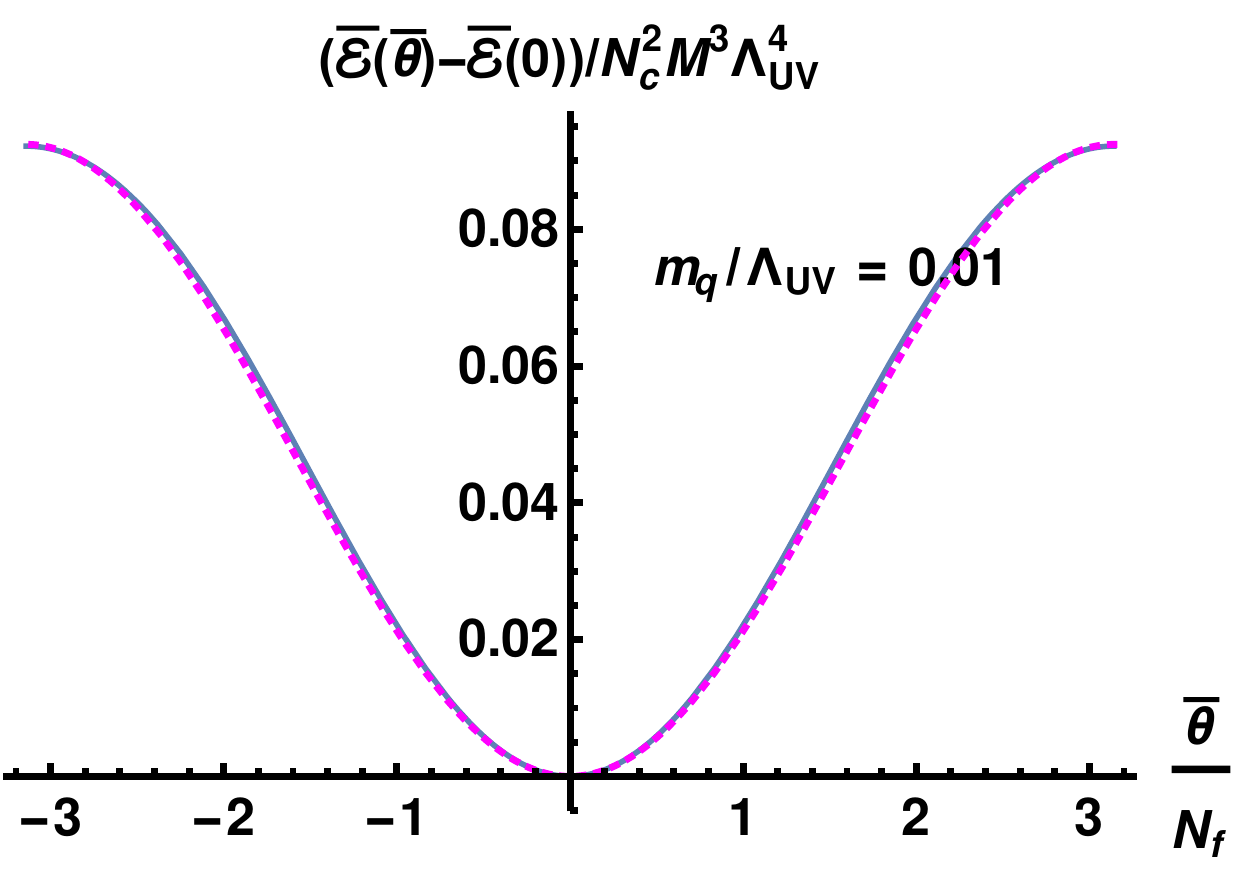}
\includegraphics[width=0.49\textwidth]{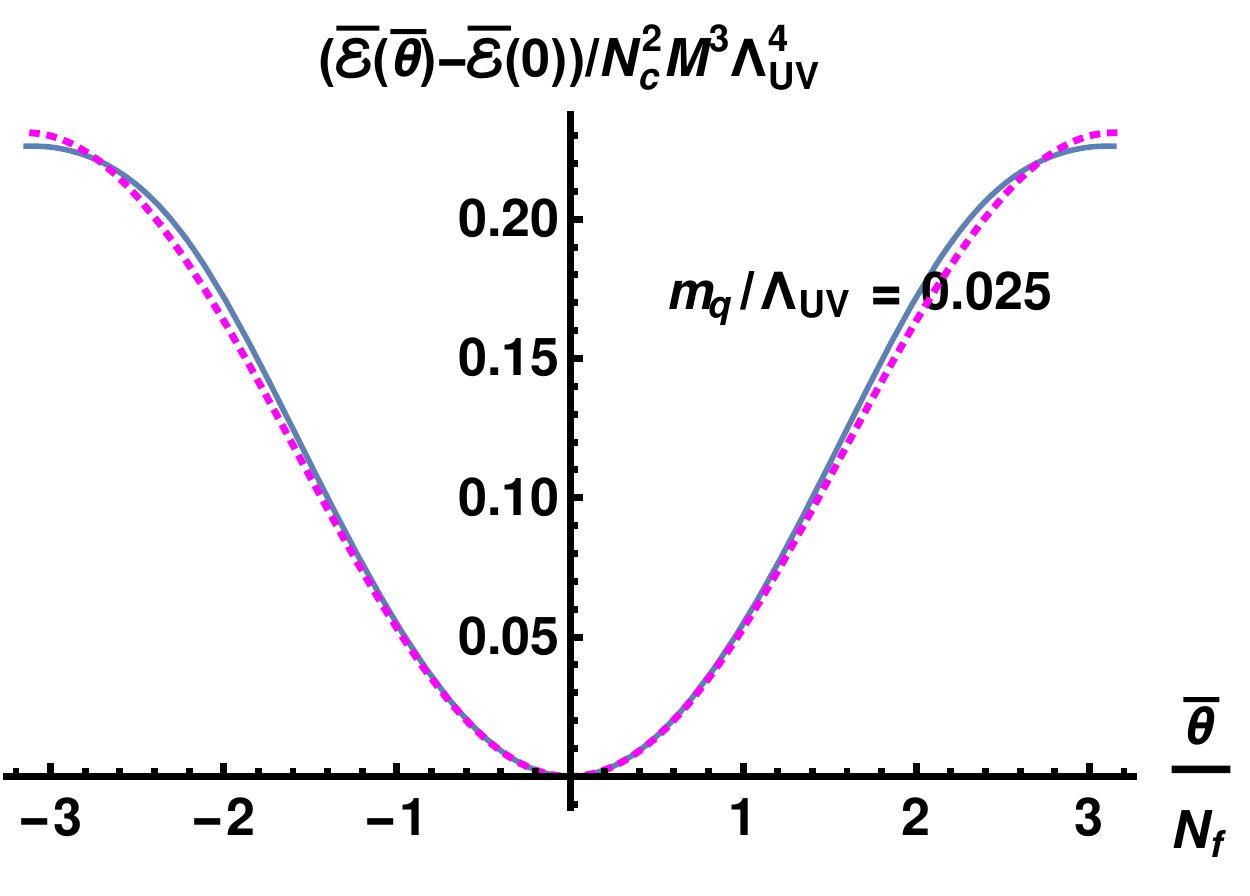}%
\hspace{2mm}\includegraphics[width=0.49\textwidth]{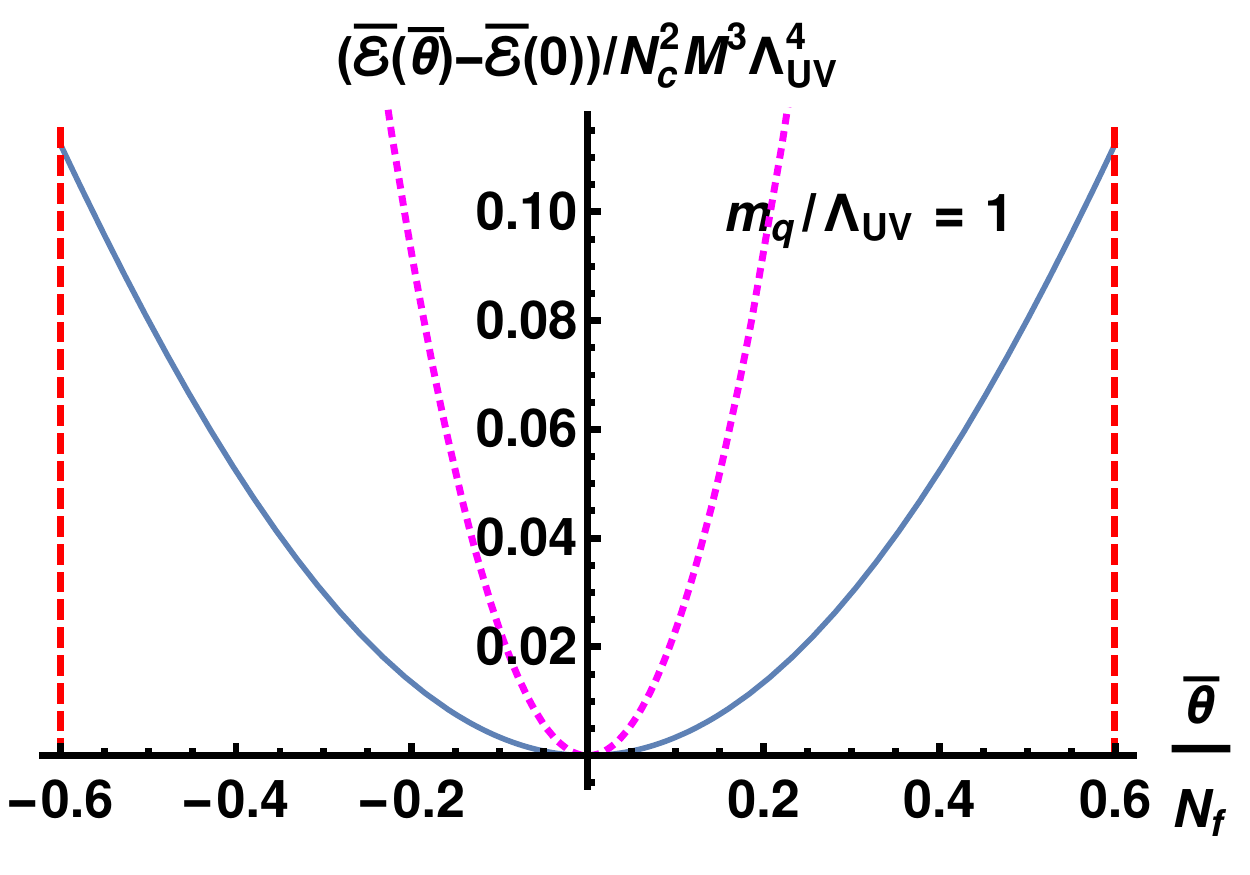}
\end{center}
\caption{The dependence of the free energy on the $\bar \theta$-angle for potentials I at $x=2/3$ and for various fixed values of the quark mass as indicated in the plots. (Notation as in Fig.~\protect\ref{fig:thetavsC}.) The dashed red vertical lines denote the limiting values of the $\bar \theta$-angle as $C_a \to \pm \infty$ along the curve of the fixed mass value.
}
\label{fig:Evstheta}\end{figure}

\subsection{Free energy and topological susceptibility\label{free}}

We analyzed above the contribution to the free energy from the CP-odd action $S_a$. However, the dependence of the free energy density on $\bar \ax_0$ is not fully captured by this contribution when $x$ is nonzero.
This is the case because $S_f$ depends on the derivative of the phase, $\xi'$, whose source varies as $\bar\ax_0$ is varied.
Therefore, we need to study the full energy density. This can be done quite simply since we only allow a variation of $\bar\ax_0$ while keeping the other sources (in particular $m_q$ and $\Lambda_\mathrm{UV}$) fixed. In this case we can read from~\eqref{dEres} that
\be \label{diffen}
 \delta \mathcal{E} = - M^3 N_c^2 \Ca\, \delta \bar \ax_0 = - M^3 N_c \Ca\, \delta \bar \theta \,.
\ee
Notice that this result is valid for any value of $\bar \ax_0$. Since the integral in~\eqref{axrelation} is positive, $\bar \ax_0 =0$ is the only minimum of the energy (for the branch of solutions continuously connected to $\bar \ax_0 =0$).

We may write the relation~\eqref{axrelation} as
\be \label{Gchirel}
  -M^3\Ca = G_\chi(\bar\theta)\bar\ax_0 = \frac{G_\chi(\bar\theta) \bar\theta}{N_c}
\ee
where
\be \label{Gchidef}
 G_\chi(\bar\theta) = M^3 \left[\int_0^\infty dr \left({1\over e^{3A} Z}+x\, f_a\, V_a\right)\right]^{-1} \ .
\ee
The  topological susceptibility (generalized to nonzero $\bar\theta$) therefore becomes, in terms of $G_\chi$,
\be \label{suskisdef}
 \chi(\bar\theta) \equiv \overline{\mathcal{E}}''(\bar\theta) = G_\chi(\bar\theta)+\bar\theta\, G_\chi'(\bar\theta)
\ee
where we used~\eqref{diffen} and~\eqref{Gchidef}.

For small $\ax_0$ integrating~\eqref{diffen} gives
\be \label{Esmallth}
 \overline{\mathcal{E}}(\bar \theta) -  \overline{\mathcal{E}}(0) = \frac{1}{2} N_c^2\, \chi\, \bar \ax_0^2 + \mathcal{O}\left(\bar \ax_0^4\right) = \frac{1}{2} \chi\, \bar \theta^2 + \mathcal{O}\left(\bar \theta^4\right)\,,
\ee
where $\chi = \chi(\bar\theta=0) = G_\chi(\bar\theta=0)$. We denote the energy in~\eqref{Esmallth} by $\overline{\mathcal{E}}$ rather than $\mathcal{E}$ in order to stress that it is the energy of the configuration obtained from the solution at $\bar\theta = 0 $ by continuously varying $\bar\theta$. In order to determine the final free energy in the dominant vacuum, $\mathcal{E}$, we will need to take into account the other branches of solutions.

We plot the topological susceptibility for potentials~I at $x=2/3$ as a function of the quark mass at $\bar\theta=0$ in Fig.~\ref{fig:suskis} and as a function of $\bar\theta$ at fixed $m_q$ in Fig.~\ref{fig:suskistheta}. The magenta curves are given by the small $m_q$ approximation which will be discussed in Sec.~\ref{sec:smallmq} and matches with effective field theory (this curve lies above the range of the plot in the right-hand plot of Fig.~\ref{fig:suskistheta}). We also notice that the susceptibility in IR units shown in Fig.~\ref{fig:suskis} (left) approaches a constant value at large $m_q$. This signals the decoupling of quarks, and the value is that of the YM limit (\ie IHQCD), which can be seen as follows. As the quark mass grows, the tachyon is sizeable except for a short interval up to $r \simeq 1/m_q$ in the UV. Outside this interval, the exponential behavior of the potential $V_a$ suppresses the second term in the integrand in~\eqref{Gchidef}. Therefore the second term is suppressed, and the leading contribution arises from the first term, which has the same functional form as the expression for $\chi$ in IHQCD~\cite{ihqcd}. Because this integral is dominated in the IR where the background approaches smoothly the YM (or IHQCD) background as the quark mass grows~\cite{Jarvinen:2015ofa}, the result for $\chi$ in this limit agrees with that of YM.

A comparison of the two plots in Fig.~\ref{fig:suskis} at large mass shows that $\chi$ only approaches a constant value when measured in IR units, which signals the fact that $\LUV$ and $\LIR$ are different at large quark mass as we now explain. The difference between these two scales might be surprising, since a large mass decouples the quarks so that the low energy dynamics is that of the YM theory, which only has a single energy scale.  The UV scale $\LUV$ differs from $\LIR$ because it is defined through the running of the 't Hooft coupling asymptotically in the UV where the quarks are not decoupled: the definition~\eqref{LUVdef} is not directly affected by the quark mass for any value of $m_q$. This can be seen explicitly in the UV expansions~\eqref{UVexpsapp}--\eqref{UVexpsapp2}: for $r\ll 1/m_q$
the backreaction of the tachyon is suppressed, no matter how large $m_q$ is. The relation between the energy scales can be found by requiring continuity  between the YM and full QCD behavior of the coupling at $r \sim 1/m_q$, which leads to~\cite{Jarvinen:2015ofa}
\be\label{ratiolargemq}
 \frac{\LUV}{\LIR} \sim \left(\frac{m_q}{\LUV}\right)^{b_0/b^\mathrm{YM}_0 -1} =  \left(\frac{m_q}{\LUV}\right)^{-2x/11}\,,
\ee
where $b_0$ and $b_0^\mathrm{YM}$ are the leading coefficients of the beta functions of QCD and YM theory, respectively. It was observed in~\cite{Jarvinen:2015ofa} that observables such as the glueball masses and thermodynamic variables similarly approach their YM values in IR units at large $m_q$, and $\LIR$ is therefore identified as the single energy scale of the YM theory.

Using $\chi \sim \LIR^4$ at large quark mass, together with the relation~\eqref{ratiolargemq},
gives the asymptotic large $m_q$ behavior of $\chi$:
\be \label{suskislargemq}
 \frac{\chi}{\LUV^4}  \sim \left(\frac{m_q}{\LUV}\right)^{4\left(1-b_0/b^\mathrm{YM}_0\right)} =  \left(\frac{m_q}{\LUV}\right)^{8x/11} \ , \qquad  \left(\frac{m_q}{\LUV} \gg 1\right) \ .
\ee

We also show the $m_q$-dependence of the fourth order coefficient in the expansion of the free energy around $\bar\theta = 0$ in Fig.~\ref{fig:fourthder}. For $m_q/\LUV\gg 1$, the coefficient approaches a constant in IR units, as was the case for topological susceptibility in Fig.~\ref{fig:suskis}. For small $m_q$ the coefficient vanishes in accordance with effective field theory~\cite{Witten:1980sp}.

The free energy for solutions at finite $\bar \ax_0$ can be obtained by integrating the differential~\eqref{diffen} numerically, using the dependence between $\Ca$ and $\bar \ax_0$ (or equivalently $\bar\theta/N_f$) given in Fig.~\ref{fig:thetavsC}. We present the results as a function of $\bar\theta/N_f$ at fixed quark mass in Fig.~\ref{fig:Evstheta}. The blue curves are numerical data and magenta dashed curves are given by the analytic result~\eqref{energysmallmq} at small $m_q$. We notice that the approximation works slightly better for the integrated energy
than for the relation between the $\bar\theta$-angle and $\Ca$ of Fig.~\ref{fig:thetavsC} where there is already significant deviation between the numerical data and the analytic approximation at $m_q/\LUV =0.025$.

The numerical result agrees with the generic argument presented above: the energy is minimized at $\bar\theta = 0$ for each value of the quark mass. For very small $m_q$ there are also additional Efimov vacua, which we will discuss more in Sec.~\ref{sec:spirals} and in Appendix~\ref{app:thetabackgrounds}, and argue that they are also subleading.
The final expression for the energy is then obtained by taking into account the periodicity of the theta
angle (see Sec.~\ref{sec:VQCDdef} and~\cite{Witten:1998uka}):
\be \label{Efinaldef}
 \mathcal{E}(\bar \theta) = \min_{k \in \mathbb{Z}}\ \overline{\mathcal{E}}\left(\frac{\bar \theta + 2\pi k}{N_f}\right) \,.
\ee
When writing down the last expression we recalled that for a single branch the free energy is naturally a function of $\bar \ax_0 = \bar\theta/N_f$.
Therefore the derivatives with respect to this variable are of the same order as the function, $\morder{N_c^2}$. We have found that the global minimum of $\overline{\mathcal{E}}$ is $\overline{\mathcal{E}}(0)$, so~\eqref{Efinaldef} is minimized for some $k$ which satisfies approximately $k\simeq -\bar\theta/2\pi$. Then the argument of $\overline{\mathcal{E}}$ in~\eqref{Efinaldef} is $\morder{1/N_f}$: there is always some value of $k$ such that $|\bar \theta + 2\pi k| \le \pi$. Therefore we may apply Taylor expansion for  $\overline{\mathcal{E}}$ around the origin, which gives the final result for the free energy
\be \label{freeenfinal}
 \mathcal{E}(\bar \theta) = \overline{\mathcal{E}}(0) + \frac{1}{2}\, \chi\, \min_{k \in \mathbb{Z}} \left(\bar \theta +2\pi k\right)^2
\ee
in the Veneziano limit. Here we recalled that the second derivative of $\overline{\mathcal{E}}$ is the topological susceptibility. The result is similar in form to that obtained in the 't Hooft limit~\cite{Witten:1998uka}.

\begin{figure}[!tb]
\begin{center}
\includegraphics[width=0.49\textwidth]{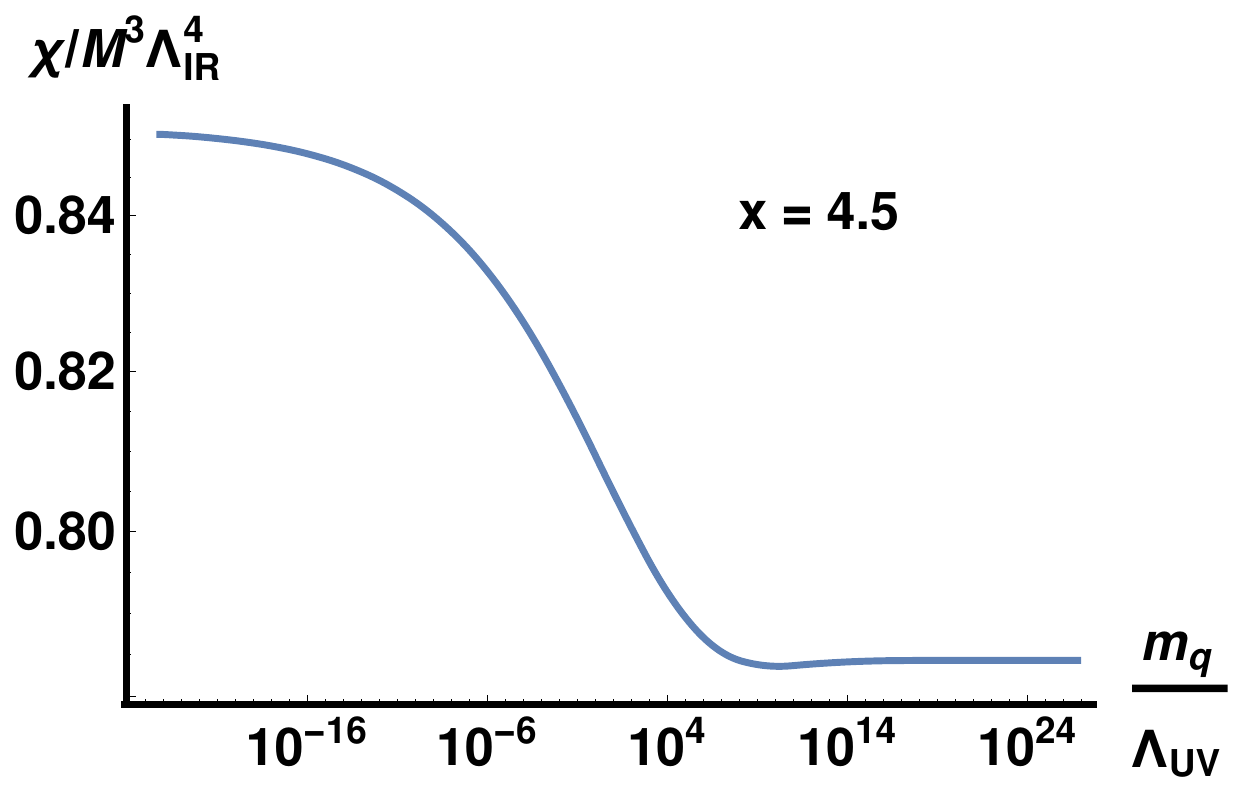}%
\hspace{2mm}\includegraphics[width=0.49\textwidth]{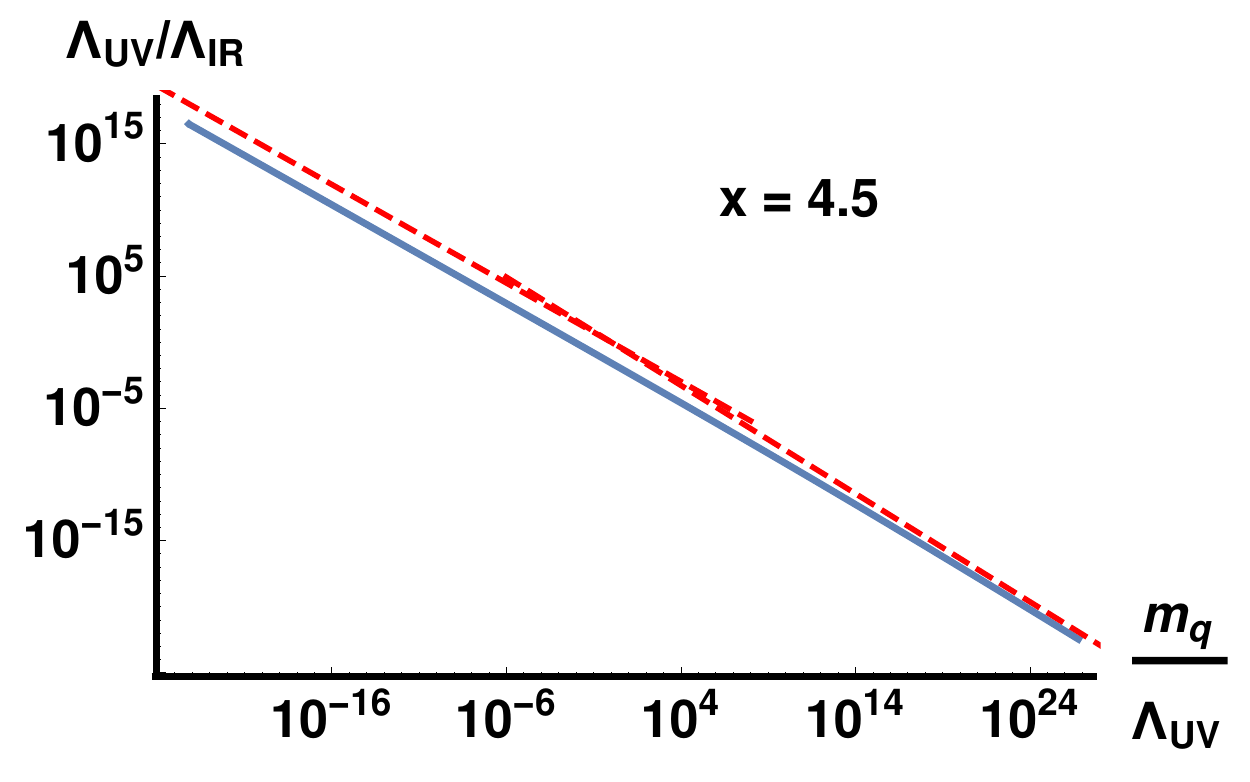}
\end{center}
\caption{Conformal window, $x=4.5$. Left: The dependence of the topological susceptibility on the quark mass in
IR units for potentials~I. Right: The ratio $\LUV/\LIR$ as a function of the quark mass (blue solid curve). The red dashed lines show the scaling relations~\protect\eqref{ratiosmallmq} and~\protect\eqref{ratiolargemq} at small and large quark masses, respectively.}
\label{fig:suskisCW}\end{figure}

\begin{figure}[!tb]
\begin{center}
\includegraphics[width=0.49\textwidth]{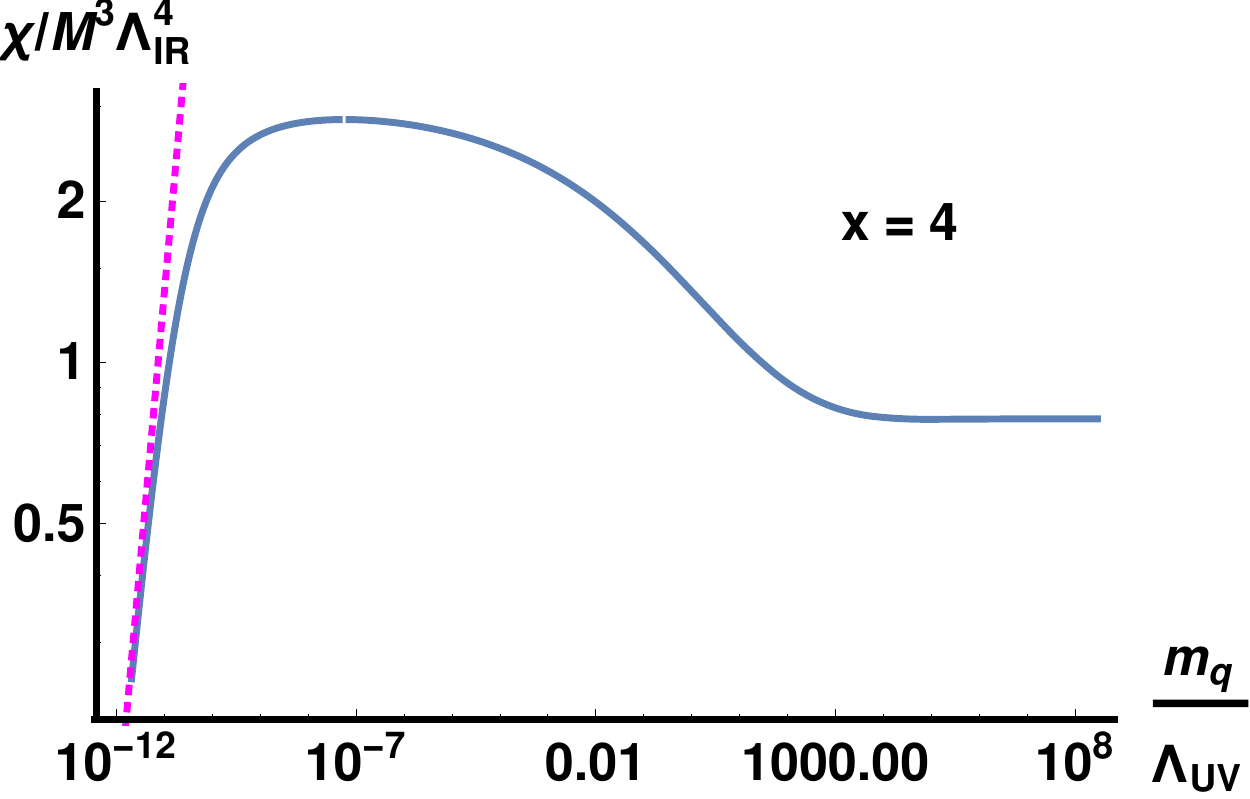}%
\hspace{2mm}\includegraphics[width=0.49\textwidth]{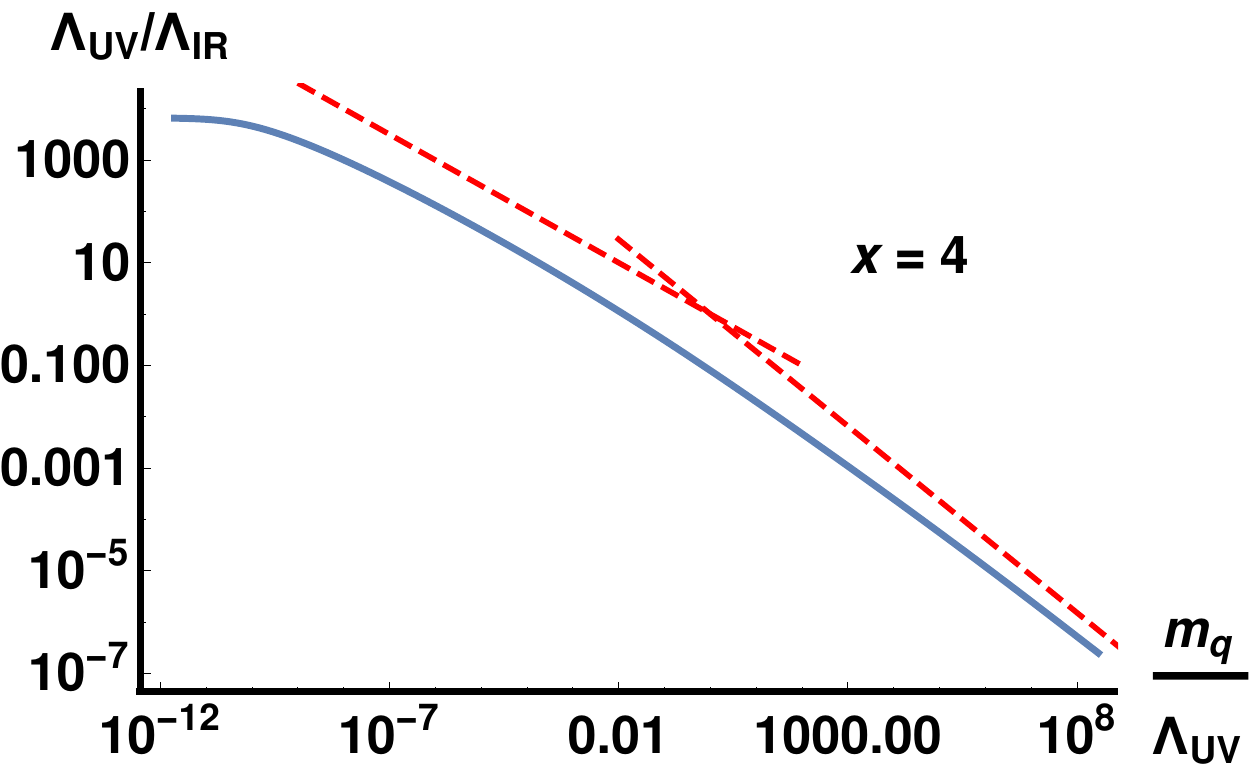}
\end{center}
\caption{Walking regime, $x=4$. Left: The dependence of the topological susceptibility on the quark mass in
IR units for potentials~I. The dashed magenta curve is given by
the approximation at small $m_q$ given in Eq.~\protect\eqref{suskisrel}. Right: The ratio $\LUV/\LIR$ as a function of the quark mass (blue solid curve). The red dashed lines show the scaling relations~\protect\eqref{ratiosmallmq} (with $\Delta_*=2$) and~\protect\eqref{ratiolargemq} at intermediate and large quark masses, respectively.}
\label{fig:suskiswalking}\end{figure}

\subsection{Dependence on $x$ and hyperscaling}

Above we restricted ourselves to the QCD-like regime with $x = \morder{1}$, but it is also interesting to study the vacua as a function of $x$.

 It was found, \cite{jk,Jarvinen:2015ofa}, that there is a BKT-type transition at some $x=x_c$ between the QCD-like phase and the conformal window as $x$ varies (at zero temperature). In the QCD phase,  the vacua with nontrivial tachyon and zero quark mass (corresponding to the crosses of Fig.~\ref{fig:masscontours}) only exist in the chirally broken phase where one of them is the energetically favored vacuum.

On the other hand, in the conformal window ($x_c<x<11/2= x_\mathrm{BZ}$), the picture is much simpler. At zero $\bar\theta$,
there is no spontaneous chiral symmetry breaking and the quark mass grows monotonically with $T_0$.
When $\bar\theta$ is nonzero, the situation is similar: in Fig.~\ref{fig:masscontours} the nodes on the horizontal axis are absent and the quark mass grows monotonically with $T_0$ for fixed $\Ca$.

The topological susceptibility in the conformal window ($x=4.5$) is shown for the ``QCD-like'' potentials~I in Fig.~\ref{fig:suskisCW} (left). For large quark mass, $m_q/\LUV \gg 1$, the susceptibility approaches the YM value as was the case in the QCD-like phase $0<x<x_c \simeq 4.083$. Consequently, the susceptibility in UV units obeys~\eqref{suskislargemq} in this limit. When $m_q/\LUV \to 0$ the susceptibility also approaches a finite, nonzero value in IR units. This is in agreement with earlier observations that mass gaps and decay constants are of  order  $\LIR$ for small $m_q$ in the conformal window, \cite{Jarvinen:2015ofa}. Note that the vertical axis of the plot does not start from the origin, and the UV and IR limiting values are actually rather close.

Overall, the dependence of CP-odd observables on $m_q$ is weak, when the observables are expressed in units of $\LIR$. For example, the relation between the source $\bar\theta$ and the VEV $C_a$ is similar to that of the bottom-right plot of Fig.~\ref{fig:thetavsC} for all values of $m_q$. The relation is determined by the IR behavior of the solutions of the various fields which are weakly dependent on $m_q$. While most observables are therefore proportional to the scale $\LIR$, the ratio $\LUV/\LIR$ depends strongly on the quark mass as shown in Fig.~\ref{fig:suskisCW} (right). The ratio obeys different power laws for $m_q/\LUV \ll 1$ and $m_q/\LUV \gg 1$, given in~\eqref{ratiosmallmq} and~\eqref{ratiolargemq}, respectively, and shown as red dashed lines in Fig.~\ref{fig:suskisCW} (right).

The hyperscaling relation (see~\cite{DelDebbio:2010ze}) for $\chi$ can be found by taking into account the dependence of the scales on $m_q$ in the limit of small $m_q$,
\be \label{ratiosmallmq}
\frac{\LIR}{\LUV} \sim \left(\frac{m_q}{\LUV}\right)^{\frac{1}{\Delta_*}}\,,
\ee
where $\Delta_*$ is the dimension of the quark mass at the IR fixed point~\cite{Jarvinen:2015ofa}. Because $\chi \sim \LIR^4$,
\be \label{chihyper}
 \frac{\chi}{\LUV^4} \sim \left(\frac{m_q}{\LUV}\right)^\frac{4}{\Delta_*} \sim \left(\frac{m_q}{\LUV}\right)^\frac{4}{1+\gamma_*} \ , \qquad \left(\frac{m_q}{\LUV} \ll 1\right)
\ee
where $\gamma_*$ is the anomalous dimension of the quark mass at the fixed point.

We then discuss the phase diagram near the conformal transition and, in particular, in the regime with walking behavior ($x_c-x \ll 1$). 
As we have already pointed out, the diagram of Fig.~\ref{fig:masscontours} has nodes for all $x$ within the interval $0<x<x_c$. However, they approach the dashed line where the solution ceases to exist as $x \to x_c$ from below.

We demonstrate  the approach to $x=x_c$ by studying numerically the topological susceptibility in the dominant vacuum.
It is shown at $\bar\theta=0$ as a function of the quark mass in Fig.~\ref{fig:suskiswalking} (left). We chose $x=4$, which is close to the critical value  $x_c \simeq 4.083$ for the potentials~I used here. Three separate regimes can be identified as the quark mass varies. For very small\footnote{As shown in~\cite{Jarvinen:2015ofa}, the boundaries between the three regimes are roughly at $m_q \sim \LUV\exp\left(-{2 \hat K}/{\sqrt{x_c-x}}\right)$ with $\hat K$ given in~\eqref{hatKres}, and at $m_q \sim \LUV$.} $m_q$, the topological susceptibility is proportional to $m_q$ and obeys the relation~\eqref{suskisrel} as in the QCD-like phase.
 In the intermediate regime, the topological susceptibility is close to the constant value associated to the IR CFT (that is approached in the walking region).
 Finally, for large $m_q$, $\chi/\LIR^4$ approaches the constant value associated with the QCD-like IR regime. In contrast, for lower values of $x$, far from the walking regime, the topological susceptibility (see the left plot in Fig.~\ref{fig:suskis}) contains only the first and third regimes above, while the intermediate regime is (not surprisingly) absent.

The dependence of the ratio $\LUV/\LIR$ on $m_q/\LUV$ is shown in Fig.~\ref{fig:suskiswalking}. We observe that the ratio takes a finite value in the limit $m_q \to 0$, as in the QCD-like phase, and shows similar behavior to the conformal window in the intermediate and large $m_q$ regimes. In particular, the scaling relation~\eqref{ratiosmallmq} with $\Delta_* =2$~\cite{Jarvinen:2015ofa} (shown as the dashed red line at intermediate quark masses), is consistent with the numerical results. Consequently, we obtain a hyperscaling-like relation
\be
\chi/\LUV^4 \propto (m_q/\LUV)^2
\label{I.0}\ee
 in the intermediate regime.

In summary, the dependence of the topological susceptibility (and the CP-odd physics in general) on $x$ and $m_q$, is qualitatively similar to that obtained for other observables (such as meson masses) at $\bar\theta=0$, see Fig.~2 in~\cite{Jarvinen:2015ofa}.

\subsection{The chiral limit and comparison to effective field theory} \label{sec:smallmq}

The solutions for the CP-odd fields can be studied analytically at small $m_q$, \ie in the vicinity of the nodes
of Fig.~\ref{fig:masscontours} (left).
As shown in Sec.~\ref{ssec:uvasymp}, in the UV, the complex tachyon satisfies the linearized Eq. \eqref{complextauUV},
where $\Ca$ does not appear explicitly.
In particular, that equation has the same form as the equation for the (real) tachyon at zero $\bar \theta$-angle
or equivalently at zero $\Ca$. The UV boundary data for the complex tachyon (but not necessarily
for its absolute value and its phase) is expected to behave smoothly as the IR boundary conditions are varied.
For small $m_q$ (that is for $m_q/\LUV \ll \sigma_0/\LUV^3$), we may therefore write the asymptotic
solution as
\be \label{smallmqtau}
 \frac{1}{\ell}\tau\,e^{i\xi}\simeq e^{i\xi_0}
m_q\,r\,(-\log(\Lambda r))^{-\rho} + \sigma_0 \,r^3\,(-\log(\Lambda r))^\rho\,,
\ee
where $\sigma_0$ is the real valued VEV for the standard solution at $m_q=0$,
and we neglected $\morder{m_q}$ corrections to the VEV term.\footnote{One can check that $\sigma_0$
is related to $\sigma$ (which was defined in terms of the UV expansion of the absolute value $\tau$) as
$\sigma \cos \xi_0 = \sigma_0 + \morder{m_q}$, and to the complex VEV defined in~\eqref{ctachuv} as
$e^{i\xi_0}\hat\sigma = \sigma_0 + \morder{m_q}$.}

Recall that~\eqref{smallmqtau} is not gauge invariant:
in particular $\xi_0$, or equivalently the phase of the quark mass, transforms under $U(1)_A$. So far we have not worried about gauge dependence, because we were mostly using gauge invariant variables, but it is convenient to fix the gauge now. We do this by requiring that the tachyon phase vanishes in the IR, $\xi \to 0$ as $r \to \infty$. This makes sense even when working with the UV expansions, because the phase $\xi$ only varies significantly in the UV region when the quark mass is small. That is, the tachyon is real up to $\morder{m_q}$ corrections in the IR region for this gauge choice. Continuity at $r\sim 1/\LIR$ implies the VEV term of the tachyon and $\sigma_0$ are real up to  $\morder{m_q}$ corrections.

Inserting the phase and absolute value from~\eqref{smallmqtau} in the phase equation of~\eqref{smalltaueqs} we find that
\be \label{Csmallmq}
 \Ca = -2 m_q\, \sigma_0\, \h_0\, W_0\, \ell^5\, \sin \xi_0 + \morder{m_q^2}\,.
\ee
Moreover, for the current gauge choice~\eqref{axrelation} implies
\be \label{thetaxi0exp}
 \bar \ax_0 = x\, \xi_0 - \Ca \int_0^\infty dr \left({1\over e^{3A} Z}+x\, f_a\, (V_a-1)\right)\,,
\ee
where the second term vanishes as $m_q \to 0$ (because $\Ca \to 0$ in this limit as seen from~\eqref{Csmallmq}) -- the possible singular contributions from $f_a$ near the tachyon nodes are regulated by the factor $1-V_a$ which also vanishes at the nodes. Therefore we find
\be \label{thetaxirel}
 \bar\theta = N_c\, \bar\ax_0 = N_f\, \xi_0 + \morder{m_q}\,.
\ee
Notice, however, that the first term in~\eqref{thetaxi0exp} vanishes as $x \to 0$ --
the integral which we dropped is actually much smaller than the term $x\xi_0$ if $x \gg m_q \sigma/\LIR^4$. That is, the limits $x \to 0$ and $m_q \to 0$ do not commute. We will take $x=\morder{1}$ first and return to the case of small $x$ below.

\subsubsection{Limit of small $m_q$ for $x=\morder{1}$}

The relation between the VEV $\Ca$ and the source $\bar\theta$ becomes
\be \label{Cthetasmallmq}
 \Ca = -2 m_q\, \sigma_0\, \h_0\, W_0\, \ell^5\, \sin \frac{\bar\theta}{N_f} + \morder{m_q^2}\,.
\ee
The free energy at small, constant $m_q$ can be then obtained\footnote{We are working around the standard solution, the rightmost cross of Fig.~\ref{fig:masscontours}, so that the integration starts from $\bar\theta =0$.} by integrating~\eqref{diffen}:
\begin{align}
 \overline{\mathcal{E}}(\bar \theta) -  \overline{\mathcal{E}}(0) &= 2 N_c\, N_f\, M^3\, m_q\, \sigma_0\, \h_0\,
W_0\, \ell^5 \left(1-\cos\frac{\bar\theta}{N_f}\right) + \morder{m_q^2}& \\
 &= -\langle \bar \psi\psi\rangle\big|_{m_q=0}  \left(1-\cos\frac{\bar\theta}{N_f}\right) m_q   + \morder{m_q^2}\,.&
 \label{energysmallmq}
\end{align}
Here we recall that the proportionality constant between the VEV $\sigma_0$ and the chiral condensate (for the standard solution at $m_q=0$) is $-2 N_c N_f M^3  \h_0 W_0 \ell^5$~\cite{ikp,Jarvinen:2015ofa}. The same result is also found by using chiral effective Lagrangians in the limit of small $m_q$~\cite{DiVecchia:1980yfw,Witten:1980sp} (see also the review~\cite{Vicari:2008jw}).

The (generalized) topological susceptibility reads
\begin{align}
 \chi(\bar\theta) &= - M^3 N_c\frac{d\Ca}{d\bar\theta} = \frac{2 M^3 m_q  \sigma_0 \h_0 W_0 \ell^5}{x}\,\cos \frac{\bar\theta}{N_f}+ \morder{m_q^2}&\\
 &= -\frac{ \langle \bar \psi\psi\rangle\big|_{m_q=0}}{N_f^2}\,  m_q\,\cos \frac{\bar\theta}{N_f}+ \morder{m_q^2}\,.
\label{suskisgen}
\end{align}
Taking here $\bar\theta \to 0$ we obtain
\be\label{suskisrel}
 \chi = \chi(\bar\theta=0) = -\frac{ \langle \bar \psi\psi\rangle\big|_{m_q=0}}{N_f^2}\, m_q+ \morder{m_q^2}\ \,,
\ee
which agrees with the well-known field theory result~\cite{Crewther:1977ce,Shifman:1979if}. The estimates~\eqref{Cthetasmallmq}--\eqref{energysmallmq} were compared to numerical data in Figs.~\ref{fig:thetavsC}--\ref{fig:Evstheta} where they give the dotted magenta curves.

\subsubsection{Limit of small $m_q$ for any $x$} \label{sec:smallmqanyx}

Notice that because the condensate is $\morder{N_fN_c}$, \eqref{suskisrel} diverges for $x \to 0$. This signals the breakdown of the small $m_q$ approximation. As we pointed out above, the limits $m_q \to 0$ and $x \to 0$ do not commute. This reflects properties of QCD: when $x \to 0$ the axial anomaly is effectively suppressed, the $\eta'$ meson becomes light, and must also be taken into account when analyzing the physics.

From~\eqref{suskisdef} and~\eqref{Gchidef} we see that the susceptibility approaches its YM value $\chi_\mathrm{YM}$ (defined in IR units as explained above) as $x \to 0$. Working directly with this equation we can obtain an improved estimate:
\be
 \chi^{-1} = \chi_\mathrm{YM}^{-1} -\frac{N_f^2}{\langle \bar \psi\psi\rangle\big|_{m_q=0}\ m_q}\left(1 +\morder{m_q}\right)\,.
\ee
This expression is valid at small $m_q$ but for all values of $x$ (within the QCD like regime, $0<x<x_c$), and agrees with chiral perturbation theory~\cite{Leutwyler:1992yt} (see also Sec.~\ref{sec:EFT}).

We can also derive formulae for the free energy which are valid for all values of $x$. Namely, as we pointed out above, the integral in~\eqref{thetaxi0exp} is only relevant when $x$ is small. But in this regime its second term is suppressed, and the first term is related to the YM topological susceptibility, so that we find\footnote{The precise scaling limit which determines which terms we keep here is that $x \LUV/m_q$ is fixed as $m_q \to 0$, but the expressions which we obtain will also remain valid for $x = \morder{1}$ and $m_q \to 0$. Notice that the condensate is $\morder{N_fN_c}$.}
\be
 \frac{\bar\theta}{N_f} = \xi_0 - M^3 C_a x^{-1} \chi_\mathrm{YM}^{-1} + \morder{m_q}
\ee
whereas the relation~\eqref{Csmallmq} is unchanged and therefore implies
\be \label{finalconst}
 M^3 N_f N_c C_a = \langle \bar \psi\psi\rangle\big|_{m_q=0} \, m_q\,\sin \xi_0 + \morder{m_q^2} = N_f^2\, \chi_\mathrm{YM}\left(\xi_0-\frac{\bar\theta}{N_f}\right) + \morder{m_q^2} \,.
\ee

We then compare to effective field theory results~\cite{Witten:1980sp} given explicitly in Sec.~\ref{sec:EFT}. Identifying the phase $\phi$ and the coupling $a$
(introduced in Eqs. \eqref{cc2} and \eqref{eq:mmatrix} respectively) as $\phi = \xi_0$ and
\be
 \frac{a}{m_\pi^2} = - \frac{N_fN_c\, \chi_\mathrm{YM}}{\langle \bar \psi\psi\rangle\big|_{m_q=0}\ m_q} \,, \qquad a = \frac{N_c \chi_\mathrm{YM}}{\hat f_\pi^2}\,,
\ee
where the latter form follows after the use of Gell-Mann-Oakes-Renner relation, the above conditions~\eqref{finalconst} match with~\eqref{c1} up to corrections suppressed by $m_q$. Moreover, imposing these conditions the differential~\eqref{diffen} integrates to
\be
 \overline{\mathcal{E}}(\bar\theta) - \overline{\mathcal{E}}(0) =  - \langle \bar \psi\psi\rangle\big|_{m_q=0}\ m_q \,\left(1-\cos\xi_0\right) + \frac{ \chi_\mathrm{YM}}{2}\left(N_f\xi_0-\bar\theta\right)^2 + \morder{m_q^2}
\ee
which agrees with the potential~\eqref{cc10} with the above identifications.

We also remark that the solutions to the last equality in~\eqref{finalconst}, i.e., $\xi_0(\bar\theta)$,
are unique only when $|\langle \bar \psi\psi\rangle\big|_{m_q=0} \, m_q| \le N_f^2\, \chi_\mathrm{YM}$.
When this condition is violated, $\xi_0(\bar\theta)$ has several branches, which is the case for
$x\ll m_q/\LUV$. Because the condensate is negative these branches first appear near
$\xi = (2n+1)\pi = \bar\theta/N_f$ as $x$ decreases, where $n$ is an arbitrary integer.

Finally, while we wrote the above formulae around the standard vacuum, \ie the rightmost cross of Fig.~\ref{fig:masscontours}, they hold also in the vicinity of other points with $m_q=0$ (the Efimov vacua) with minor changes. That is, we need to interpret $\sigma_0$ and $\langle \bar \psi\psi\rangle$ as the values of the corresponding $m_q=0$ solutions. In addition, the value of $\bar \theta$ should be chosen as depicted in Fig.~\ref{fig:masscontours} (right), so that the starting point of integration in~\eqref{energysmallmq} and~\eqref{suskisgen} has also changed. For the susceptibility this results in a factor $(-1)^n$ in~\eqref{suskisrel} near the $n$th Efimov vacuum:
\be\label{suskisrelEf}
\chi^{-1} = \chi_\mathrm{YM}^{-1} -\frac{(-1)^n N_f^2}{\langle \bar \psi\psi\rangle\big|_{n,\,m_q=0} \ m_q}\left(1 +\morder{m_q}\right),
\ee
where the chiral condensate is that of the $n$th Efimov vacuum at zero quark mass. The sign of the condensate is $-(-1)^n$~\cite{jk,Jarvinen:2015ofa}, so that both contributions to (the inverse of) $\chi$ are positive.

\subsection{Complex Efimov spirals} \label{sec:spirals}

It is possible to gain some analytic understanding of the structure of the solutions as the dashed line in Fig.~\ref{fig:masscontours} is approached where the theory flows closer and closer to the IR fixed point. That is, we can generalize the approach detailed in Sec.~5 of~\cite{Jarvinen:2015ofa} to the case of nonzero $\Ca$. First we briefly review the main points of the analysis at $\Ca=0$.

In the vicinity of the fixed point, when the BF bound is violated, the tachyon satisfies the linearized equation of motion, the solution of which can be written as a linear combination of
\begin{align}
\label{taumdef}
 \frac{\tau_m}{\ell} &\simeq \frac{m_q}{\LUV} K_m \left(r \LUV\right)^2\ \sin\left[\n\log\left(r\LUV\right) + \phi_m \right]\ ,& \quad &\left(\frac{1}{\LUV}\ll r \ll \frac{1}{\LIR}\right) & \\
\label{tausdef}
 \frac{\tau_\sigma}{\ell} &\simeq \frac{\sigma}{\LUV^3} K_\sigma \left(r \LUV\right)^2\ \sin\left[\n\log\left(r\LUV\right) + \phi_\sigma \right]\ ,& \quad &\left(\frac{1}{\LUV} \ll r \ll \frac{1}{\LIR}\right) &
\end{align}
which have zero VEV and quark mass, respectively (the value of the quark mass and the VEV are determined by the continuation of the solutions to the UV boundary where~\eqref{TUVres} holds). Here $\LUV \gg \LIR$ since the flow becomes close to the fixed point, and the parameters $\phi_i$ and $K_i$ are $\mathcal{O}(1)$ real numbers which can be determined by solving the tachyon equation of motion numerically. The parameter $\n$ is the imaginary part of the dimension of the quark mass at the fixed point: $\n = \mathrm{Im}\Delta_* = \sqrt{-m^2_{\t*} \ell_*^2-4}$, where $m_{\t*}$ and $\ell_*$ are the (imaginary) bulk tachyon mass and the AdS radius at the fixed point, respectively. The IR regular solution can be written in a similar form
\be \label{tauIR}
 \frac{\tau}{\ell} \simeq K_\mathrm{IR} \left(r \LIR\right)^2\ \sin\left[\n\log\left(r\LIR\right) + \phi_\mathrm{IR} \right] \ , \qquad \left(\frac{1}{\LUV} \ll r \ll \frac{1}{\LIR}\right) \ ,
\ee
but in this case it helps to write the dimensionful quantities in IR units -- then the $\mathcal{O}(1)$ parameters $K_\mathrm{IR}$ and $\phi_\mathrm{IR}$ take constant values as the fixed point is approached (\ie as $T_0 \to T_{0c}$ from above)~\cite{Jarvinen:2015ofa}. Expressing~\eqref{tauIR} as a linear combination of~\eqref{taumdef} and~\eqref{tausdef}, one finds the spiral equations
\be \label{spiraleqs}
\begin{split}
 \frac{m_q}{\LUV} &= \frac{K_\mathrm{IR} }{K_m}\,\frac{\sin\left(\phi_\mathrm{IR}-\phi_\sigma-\n \vs  \right)}{ \sin\left(\phi_m-\phi_\sigma\right)} \, e^{-2 \vs} \\
 \frac{\sigma}{\LUV^3} &= \frac{K_\mathrm{IR} }{K_\sigma }\,\frac{\sin\left(\phi_\mathrm{IR}-\phi_m-\n \vs  \right)}{\sin\left(\phi_\sigma-\phi_m\right)}\, e^{-2 \vs}
\end{split}
\ee
where the spiral is parametrized in terms of
\be
 \vs = \log\frac{\LUV}{\LIR}\ .
\ee
This kind of spiral structures are relatively common in holographic models, and have been studied in detail in a different context in~\cite{Iqbal:2011in}.

There is a simple asymptotic relation between $\vs$ and $T_0$.
The IR geometry has a well-defined limit as the fixed point is approached, and the leading perturbation to the geometry is driven by the minimal distance to the fixed point~\cite{Jarvinen:2015ofa}:
\be \label{massdepCW}
 T_0-T_{0c} \sim \l_*-\l_\mathrm{IR} \equiv \l_*-\l(r=1/\LIR) \sim \left( \frac{\LIR}{\LUV}\right)^{\delta} \ .
\ee
Here $r=1/\LIR$ is roughly the value of the coordinate where the growing tachyon field finally drives the flow away from the fixed point,
and $\delta $ is the derivative of the holographic beta function at the fixed point, given by~\cite{jk,Jarvinen:2015ofa}
\be
 \delta = \sqrt{4 -\frac{9 V_2 \lambda_*^2}{V_0}}-2\,.
\ee
The parameters $V_i$ are defined though the expansion of the effective potential at the fixed point~\cite{jk} as
\be
 V_\mathrm{eff}(\l) \equiv V_g(\l) - x V_{f0}(\l) = V_0 + V_2\left(\lambda-\lambda_*\right)^2 + \mathcal{O}\left(\left(\lambda-\lambda_*\right)^3\right) \ .
\ee
In terms of $u$, ~\eqref{massdepCW} becomes
\be \label{uT0rel}
  u\, \delta\simeq -\log(T_0-T_{0c}) + \mathrm{const.}\ , \qquad (T_0 \to T_{0c}) \,.
\ee

The spiral~\eqref{spiraleqs} admits a relatively simple generalization to nonzero $\bar \theta$-angle, as we will now show. It is natural to keep $\Ca$ in the IR units, \ie $\Ca/\LIR^4$, fixed as $T_0 \to T_{0c}$. Using the UV expansions of Sec.~\ref{sec:thetaback} in the tachyon equation of motion we immediately see that the effect of finite $\Ca$ is suppressed in the UV by $\mathcal{O}(r^4)$. But one can derive a more general result which holds for all $r\ll 1/\LIR$ and in particular near the fixed point. Namely, the equations~\eqref{smalltaueqs} and~\eqref{complextauUV} actually hold for all $r\ll 1/\LIR$ as can be verified by inserting the behavior of the tachyon $\tau \sim r^2$ and the metric $e^A \sim 1/r$ near the fixed point in the generic equations of motion for the tachyon~\eqref{taueqs}.

Since the complex tachyon therefore solves the same equation as the (real part of the) tachyon at $\Ca=0$, the solutions~\eqref{taumdef} and~\eqref{tausdef} are otherwise unchanged for $\Ca \ne 0$, but $m_q$ and $\sigma$ should be replaced by their complex counterparts in~\eqref{ctachuv}:
\be \label{UVcmap}
 m_q \mapsto m_q e^{i\xi_0} \ , \qquad \sigma \mapsto \hat \sigma e^{i\xi_0} \,.
\ee
Here  $\hat \sigma = \sigma + i\Ca/(2m_q\ell^5 \h_0 W_0)$, with $\sigma$ defined as the coefficient of the UV expansion of the absolute value of the tachyon in~\eqref{tauuv}.
This also means that the coefficients $K_i$ and $\phi_i$ in these equations are real and independent of $\Ca$ and the $\bar\theta$-angle.

The flow of the tachyon in the IR (for $r\sim 1/\LIR$), however, changes in a nontrivial manner. The tachyon is complex in the IR for generic $\Ca$, and therefore the coefficients of~\eqref{tauIR} must be allowed to take complex values:
\be
 K_\mathrm{IR} \mapsto K_\mathrm{IR} e^{i k_\mathrm{IR}} \ , \qquad \phi_\mathrm{IR} \mapsto \phi_\mathrm{IR} + i \varphi_\mathrm{IR} \ ,
\ee
and $T_{0c}$ also depends on $\Ca$. The result for the tachyon near the fixed point can be found by applying these maps to~\eqref{tauIR}:
\be \label{tauIRc}
\begin{split}
\frac{\tau e^{i\xi}}{\ell} &\simeq K_\mathrm{IR}e^{i k_\mathrm{IR}} \left(r \LIR\right)^2 \left\{ \cosh\varphi_\mathrm{IR}\sin\left[\n\log\left(r\LIR\right) +  \phi_\mathrm{IR} \right]\right. \\
&\qquad\left.+i \sinh\varphi_\mathrm{IR}\cos\left[\n\log\left(r\LIR\right) + \phi_\mathrm{IR} \right]\right\} \,.
\end{split}
\ee
Finally, the spiral equations~\eqref{spiraleqs} generalize to
\be \label{cspiraleqs}
\begin{split}
 \frac{m_q e^{i\xi_0}}{\LUV} &= \frac{K_\mathrm{IR}e^{i k_\mathrm{IR}} }{K_m}\,\frac{\sin\left(\phi_\mathrm{IR}-\phi_\sigma-\n \vs  \right)\cosh\varphi_\mathrm{IR} +i\cos\left(\phi_\mathrm{IR}-\phi_\sigma-\n \vs  \right)\sinh\varphi_\mathrm{IR}}{ \sin\left(\phi_m-\phi_\sigma\right)} \, e^{-2 \vs} \\
 \frac{\hat \sigma e^{i\xi_0}}{\LUV^3} &= \frac{K_\mathrm{IR} e^{i k_\mathrm{IR}}}{K_\sigma }\,\frac{\sin\left(\phi_\mathrm{IR}-\phi_m-\n \vs  \right)\cosh\varphi_\mathrm{IR} +i\cos\left(\phi_\mathrm{IR}-\phi_m-\n \vs  \right)\sinh\varphi_\mathrm{IR}}{\sin\left(\phi_\sigma-\phi_m\right)}\, e^{-2 \vs}\,.
\end{split}
\ee
These equations describe, among other things, the structure of the Efimov vacua near $\Ca=0$ as one approaches the dashed curved of Fig.~\ref{fig:masscontours} (left).

The equation for the phase in~\eqref{smalltaueqs}, $\xi' \simeq \Ca/(\h V_{f0} e^{3A}\t^2)$, leads to additional constraints. Inserting here the solution~\eqref{tauIRc} or the combination of~\eqref{taumdef} and~\eqref{tausdef}, recalling also the maps~\eqref{UVcmap}, gives the identities
\be
 \frac{\Ca}{\ell_*^3 V_{f0}(\l_*) \h(\l_*)} = -\ell^2 K_\mathrm{IR}^2 \LIR^4\n \cosh \varphi_\mathrm{IR} \sinh \varphi_\mathrm{IR} = \frac{\Ca K_m K_\s \LUV^4 \n \sin\left(\phi_m-\phi_\s\right)}{2 \h_0 W_0 \ell^3 } \,.
\ee
Here the first identity constrains the $\Ca$ dependence of $K_\mathrm{IR}$ and $\varphi_\mathrm{IR}$. Equating the first and third term proves directly that $\sin\left(\phi_m-\phi_\s\right)>0$, fixing the handedness of the spiral. It was pointed out in~\cite{Jarvinen:2015ofa} that this sign is also necessary for the chirally broken vacua to dominate over the chirally symmetric vacuum.

 \begin{figure}[!tb]
 \begin{center}
\includegraphics[width=0.49\textwidth]{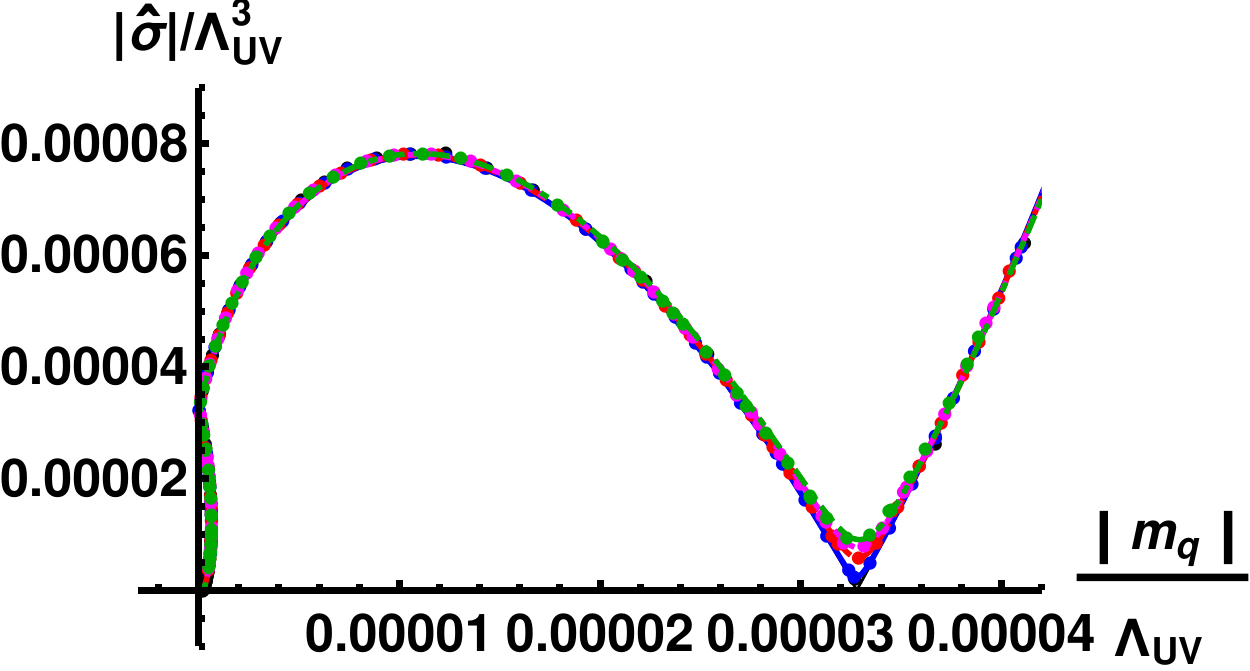}%
\hspace{2mm}\includegraphics[width=0.49\textwidth]{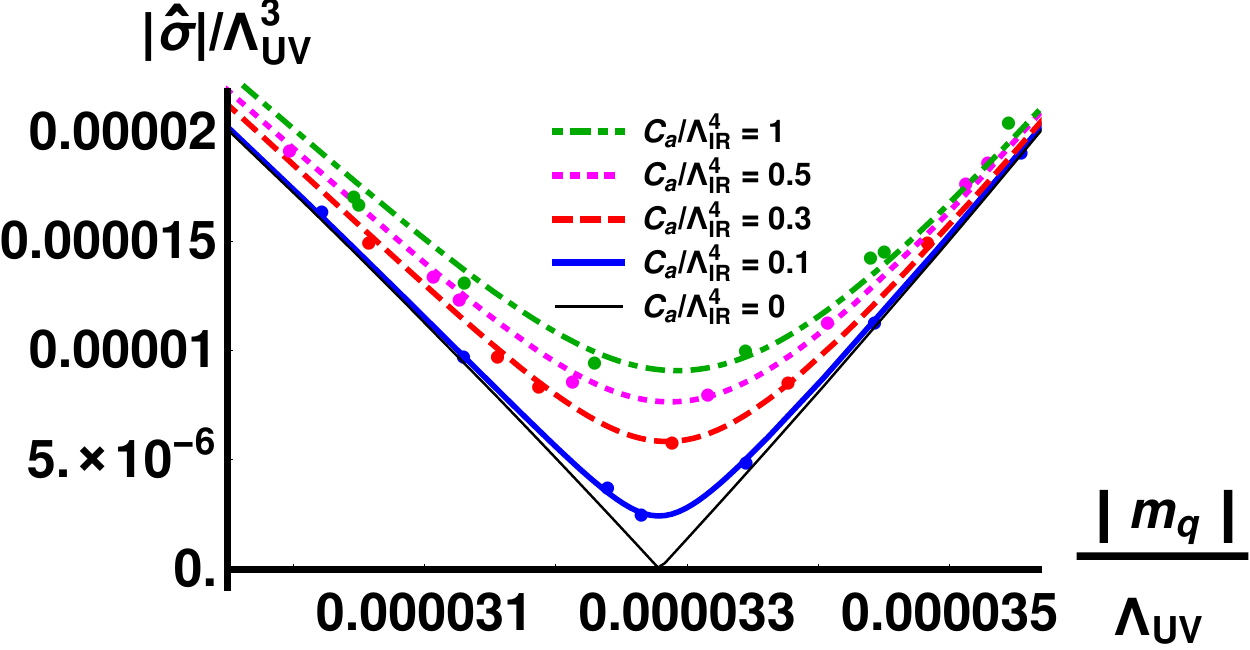}
\vspace{-3mm}

\includegraphics[width=0.49\textwidth]{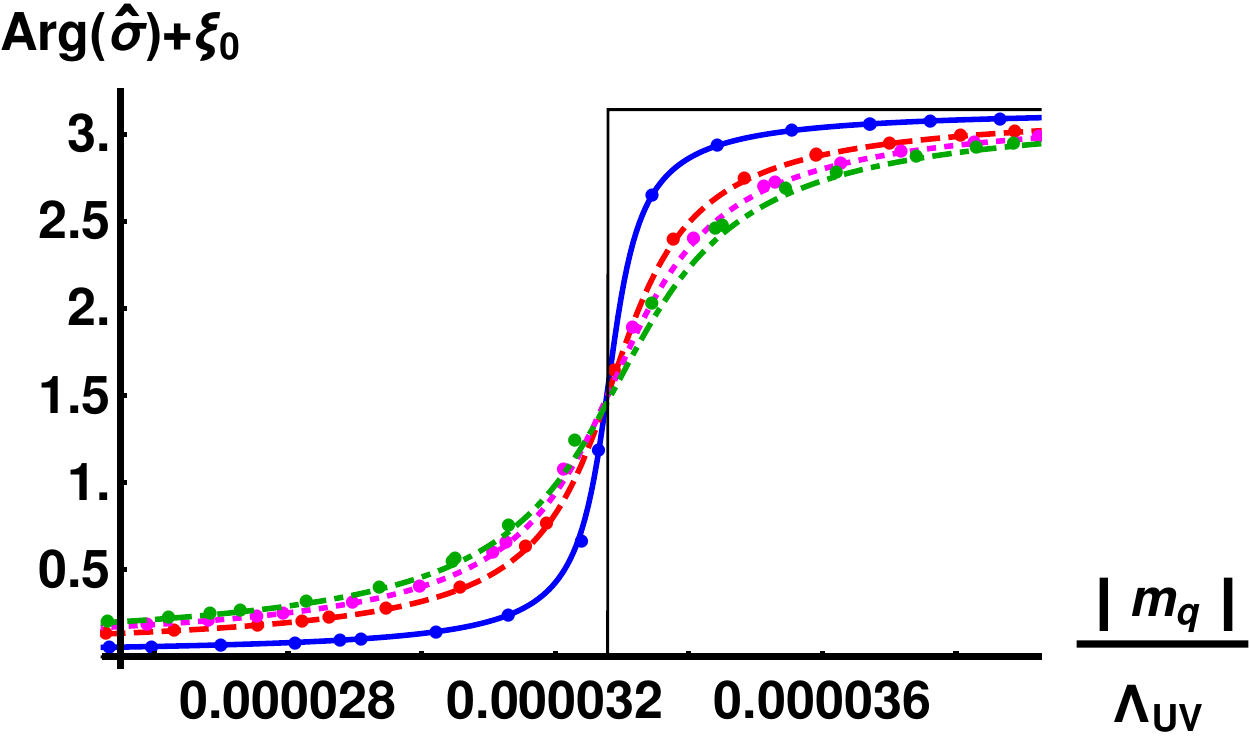}%
\hspace{2mm}\includegraphics[width=0.49\textwidth]{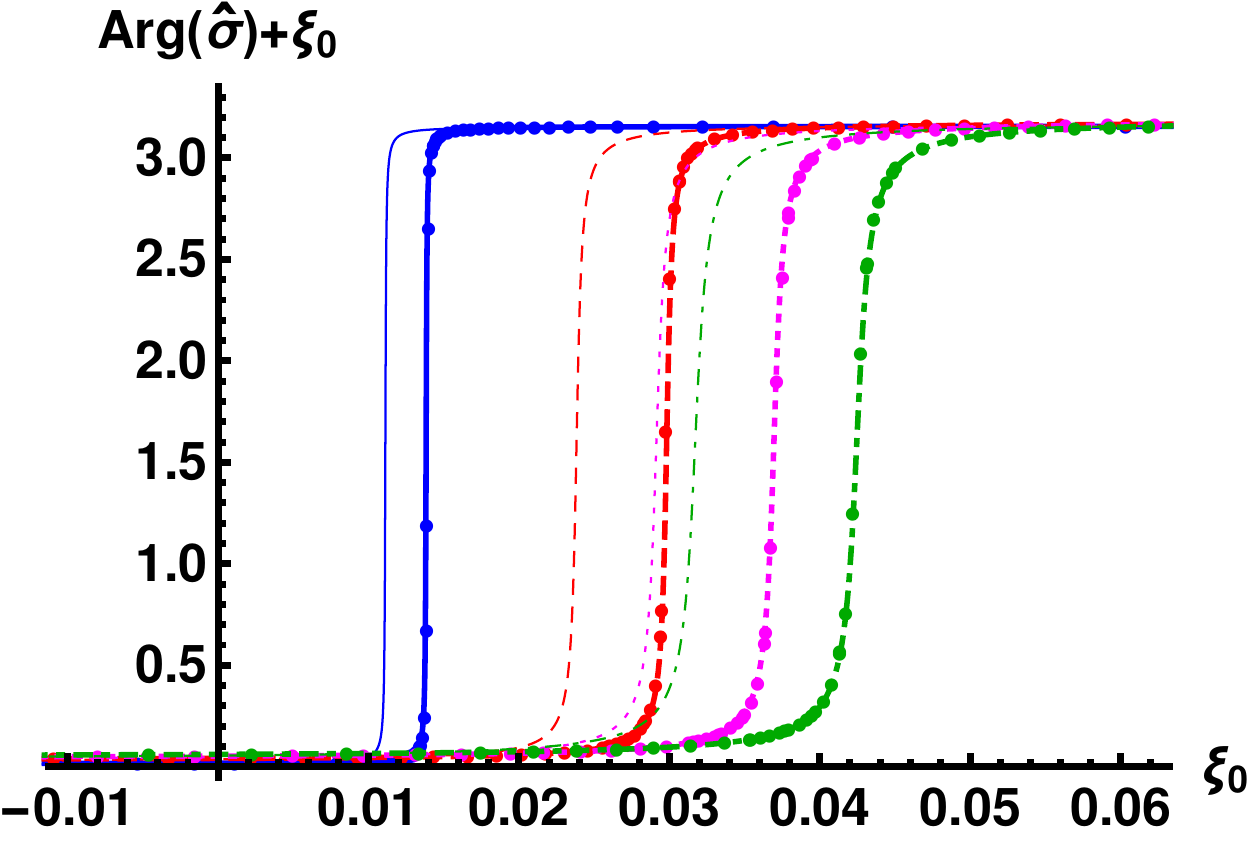}
\end{center}
 \caption{Sections of the complex spiral compared with data for potentials~I with $x=2.5$. The thin black, solid
 blue, dashed red, dotted magenta, and dot-dashed green curves have $C_a/\LIR^4=0$, $0.1$, $0.3$, $0.5$, and $1$,
 respectively.
 Top row: the absolute value of the quark mass and the condensate in two regions close to the origin
 of the spiral. Bottom row: the phase of the condensate in the same region as the
 absolute values in the top-right plot as a function of the absolute value of quark mass (left) and the phase of the quark mass (right).}
 \label{fig:spirals}\end{figure}

We compare the asymptotic formulae~\eqref{cspiraleqs} to numerical data for potentials~I at $x=2.5$
in Fig.~\ref{fig:spirals}. Since the formulae hold for small quark mass and the condensate, we plot
a section of the spiral (for the absolute value of both the source and the VEV) very close to the
origin, in the region where the tachyon solution has two nodes as $\Ca \to 0$, so the
solutions are identified as unstable (second) Efimov vacua.  As it turns out, the $\Ca$
dependence is relatively mild, in particular the complex phase factor $\varphi_\mathrm{IR}$
remains numerically small for all values of $\Ca$. Therefore our data and curves almost overlap
as seen in the top-left plot. In order to see the dependence on $\Ca$ we zoom in the region of
the top-left plot near the point where $|\hat\s|$ has a node for $\Ca=0$. The result is
shown in the top-right plot. The data (dots) follows the prediction from the formulae (curves)
well, even if the dependence on $\Ca$ is weak. Notice also that the various parameters of the curves in
these plots were not fitted to the data, but extracted directly from their definitions ({\em e.g.},
in~\eqref{tauIRc}).

The dependence on $\Ca$ can be seen more clearly in the plots on the bottom row of Fig.~\ref{fig:spirals} where we show the phases of the source and the VEV in the same region as for the top-right plot. In the bottom-right plot, the curve for $C_a=0$ is not shown because it has shrunk into a set of discrete points. For this plot, we fitted the value of the phase factor $k_\mathrm{IR}$ directly to the data.  We did this rather than using the definition in~\eqref{tauIRc} because the value of  $k_\mathrm{IR}$ obtained from the definition appeared to be clearly shifted with respect to data -- the spirals for these values of the phase are given by the thin curves in the plot. It is possible that this offset is a numerical effect -- the construction of an IR regular tachyon solution close enough to the fixed point for it to properly obey the asymptotic formula~\eqref{tauIRc} and the consequent four parameter fit to extract the numerical values of the parameters is demanding due to limited numerical precision.

\section{Spectra of singlet pseudoscalar bound states at $\bar\theta=0$} \label{sec:singletPSzeroth}

In order to compute the spectrum of mesons and glueballs one needs to study the fluctuations of all the
fields of V-QCD.
These fluctuations decouple into different sectors corresponding to glueballs and mesons with
$J^{PC}=0^{++},\; 0^{-+},\;1^{++},\; 1^{--},\; 2^{++}$, where $J$ stands for the spin and $P$ and
$C$ for the field properties under parity and charge conjugation respectively.
They can be further classified into two classes according to their transformation properties under the
flavor group: flavor non-singlet modes (expanded in terms of the generators of $SU(N_f)$) and flavor
singlet modes.

The general analysis of the fluctuations for the model at vanishing $\bar\theta$-angle was carried out
in \cite{Arean:2013tja}.
There the quadratic action for each sector was computed, and the spectra for all but one sector were
calculated numerically.
In this section we will analyze the one sector left out in \cite{Arean:2013tja}:
the flavor singlet pseudoscalar modes at vanishing $\bar\theta$-angle. We will restrict our study to the case $\bar\theta =0$ -- this is the case which is closest to ordinary QCD,
and at finite $\bar\theta$ solving the fluctuation equations would be technically very involved because the
singlet scalar and pseudoscalar mesons and glueballs would all mix.
This sector is made up of  the pseudoscalar $0^{-+}$ flavor
singlet meson and the the $0^{-+}$ glueball which mix due to the axial anomaly (realized by
the CP-odd sector).
Since we are in the Veneziano limit, the mixing takes place at leading order in $1/N_c$.
In the next section we will
study the spectra for backgrounds at finite $\bar\theta$-angle, restricting the analysis to
the flavor non-singlet sector.

The masses of the singlet pseudoscalar states are particularly important, because they contain the physics of
the $\eta^\prime$ meson, which is identified as the state with the lowest mass in this sector at small $x$.
We will demonstrate, both analytically and
numerically, that the mass of the $\eta^\prime$ meson obeys the Witten-Veneziano formula. Because of the
backreaction, the pseudoscalar glueballs and mesons mix nontrivially already at small $x$, which affects the
derivation of the Witten-Veneziano relation. We have not found a transformation ({\em e.g.}, a rotation in the
space of wave functions) which would remove this mixing, so we will need to study it carefully. Therefore our
derivation is more involved than the typical arguments in the
literature~\cite{Barbon:2004dq,Armoni:2004dc,Sakai:2004cn,ckp}.

\subsection{Pseudoscalar singlet fluctuations at $\bar\theta=0$}

We now write down the fluctuation equations for the pseudoscalar singlet sector.
First, the vector and axial vector combinations of the gauge fields are
\be
V_M = \frac{A_M^L + A_M^R}{2}\,,\qquad
A_M = \frac{A_M^L - A_M^R}{2}\,.
\label{VAdefsmain}
\ee
They contribute to both the singlet and non-singlet flavor sectors.
Next we write the complex tachyon field as
\be \label{Tfluctdef}
T(x^\mu,r)=\tau(r) \exp\left[i\thf(x^\mu,r)\right]\,,
\ee
where
$\tau$ is the background solution, and
$\thf$ is the
pseudoscalar flavor singlet fluctuation.

The flavor singlet pseudoscalar degrees of freedom
correspond to gauge invariant combinations of the longitudinal
part of the flavor singlet axial vector fluctuation $A^{\lVert S}_\m$, the pseudoscalar phase of the tachyon
$\thf$ and the axion field $a$.

We split these fields as
\begin{align}
A^{\lVert S}_\m(x^\mu, r) &= -\varphi_L(r)\,\partial_\m({\cal T}(x^\m))\,,\nonumber\\
\thf(x^\mu, r) &=2\varphi_\theta(r)\,{\cal T}(x^\m)\,,\nonumber\\
a(x^\mu, r) &=2\varphi_\mathrm{ax}(r)\,{\cal T}(x^\m)\,.
\label{defsamain}
\end{align}
The following combinations of the above fields:
\be
\begin{split}
P(r)&\equiv\varphi_\theta(r)-\varphi_L(r)\,,\\
Q(r)&\equiv\varphi_\mathrm{ax}(r)+x\,V_a(\l,\t)\,\varphi_L(r)\,,
\label{ginvpsrst}
\end{split}
\ee
are invariant under the residual gauge transformations (\ref{u1transf}). They correspond to the
pseudoscalar glueball ($0^{-+}$) and $\eta'$ meson towers.
They
satisfy the coupled differential equations (see Appendix~\ref{app:quadfluctdet} and~\cite{Arean:2013tja} for more details)
\begin{align}
&\partial_r\left[V_f\,e^{\Awf}\,\G^{-1}\,\gf^2\left(
-4e^{2\Awf}\,{V_f\,\h\,\t^2\over N_a+N_b}\,P'+{V_a'\over V_a}\,{N_b\over N_a+N_b}\,P
+{N_b\over x\,V_a\,(N_a+N_b)}Q'\right)\right]+\nonumber\\
&+4V_f\,e^{3\Awf}\,\G\,\h\,\t^2\,P-4e^{3\Awf}\,Z\,V_a\,Q=0\,,\label{cpsys1text}\\ \nonumber \\
&\partial_r\left[e^{3\Awf}\,Z\,\left(
4x\,e^{2\Awf}\,{V_a\,V_f\,\h\,\t^2\over N_a+N_b}\,P'+x\,{V_a'\,N_a\over N_a+N_b}\,P
+{N_a\over N_a+N_b}Q'\right)\right]+m^2\,e^{3\Awf}\,Z\,Q=0\,,\nonumber\\ \label{cpsys2text}
\end{align}
where  the primes denote derivatives with respect to $r$, and
$N_a$, $N_b$ and $G$ are given by the following expressions:
\be \label{NaNbdef}
N_a=V_f\left(4e^{2\Awf}\,\h\,\t^2-m^2\,\gf^2\right)\, ,\qquad
N_b=4x\,e^{2\Awf}\,Z\,V_a^2\,\G\, ,\qquad
G=\sqrt{1+e^{-2A}\kappa\,\tau'^2}\,.
\ee
Notice that $G$ is just the restriction of $\tG$ defined in \eqref{gdef} to the $\xi'=0$ case.

\subsection{Mass of the $\eta^\prime$ meson at small $x$}

We start by discussing the probe limit $x \to 0$ at nonzero but small quark mass. Because the terms depending on flavor are suppressed in the action, the fluctuation equations~\eqref{cpsys1text} and~\eqref{cpsys2text} admit solutions for which $P=\morder{1}$, $Q=\morder{x}$.\footnote{There is also another set of solutions which has a different $x$ dependence and will be identified with the glueballs in the limit $x \to 0$ as we shall see below.}
It is identified as the flavor mode at small $x$: to leading order in $x$, the $P$ component satisfies the same fluctuation equation as the wave function for the non-singlet pseudoscalar fluctuation described in~\cite{Arean:2013tja}. More precisely, $P$  is mapped to the difference $\psi_P-\psi_L$ of the radial wave function of the pion field and longitudinal gauge field  as suggested by the definition~\eqref{ginvpsrst}, and the variable $\hat P$ defined in~\eqref{hatteddefs} is mapped to $\hat \psi_P$.

Therefore one is led to expect that the flavor singlet and non-singlet
pseudoscalar mesons become degenerate as $x \to 0$. This is however not obvious since, as it turns out, the
convergence towards the $x=0$ solution is not uniform in $r$. In the IR there is no issue because the exponential
suppression of the potentials $V_f$ and $V_a$ decouples the glue from the flavor for all values of $x$. In the UV,
however, glue and flavor are nontrivially coupled at small $r$ for any positive $x$ (more precisely, when
$r \ll \sqrt{x}$), as seen from the UV expansions in Appendix~\ref{app:Asymptotics}.\footnote{Notice that the
situation is different from the scalar sector, where the glue and flavor were decoupled also asymptotically in the UV
and therefore the convergence toward $x = 0$ was uniform~\cite{Arean:2013tja}.} In principle this could lead to the
flavor singlet and non-singlet mesons having different UV boundary conditions in the limit $x \to 0$ and the masses
of the mesons being different. One can check by using the UV expansions
from Appendix~\ref{app:Asymptotics} that the boundary conditions are the same and therefore the singlet and
non-singlet states do become degenerate. In particular, $\eta^\prime$ becomes degenerate with the pions and its
mass obeys the Gell-Mann-Oakes-Renner (GOR) relation as $x \to 0$ as expected from the fact that the anomaly
vanishes when $x\to 0$. We will discuss this in more detail below.

\subsubsection{Limit of zero $x$} \label{sec:zeroxWV}

We wish to discuss what happens at small but finite $x$, but it is useful to recall first how the GOR relation arises from the fluctuation equations in the limit of zero $x$. We rewrite the fluctuation equations~\eqref{cpsys1text} and~\eqref{cpsys2text} as a first order system (for later convenience first at finite $x$)
\begin{align}
 \label{hatPdef}
 \hat P &= \frac{e^{3A}}{m_f^2-H_{PS}}\left[V_f\, \kappa\, \tau^2\, G^{-1} P' - V_a\, Z \left(Q'+x V_a' P\right) \right], \\
 \hat Q &= \frac{e^{3A}\, Z}{m_f^2\,(m_f^2-H_{PS})} \left[m_f^2 \left(Q'+x\, V_a'\, P\right) - \frac{4 e^{2 A}\, \kappa\, \tau^2 }{w^2}\frac{d}{dr}(Q+x\,V_a\,P)\right],\\
 \label{hatPeq}
 \hat P' &= e^{3 A} \left( V_a\, Z\, Q-G\, V_f\, \kappa\,  \tau ^2\, P\right), \\
 \label{hatQeq}
 \hat Q' &= -e^{3 A}\, Z\, Q\,,
\end{align}
where
\be \label{HPSdef}
 H_{PS} = \frac{4 e^{2 A}\, \kappa\,  \tau ^2}{w^2}+\frac{4 e^{2 A}\, \,x\, G\, V_a^2\, Z}{V_f\,w^2}\,,
\ee
and $m_f$ is the mass of the fluctuations.

As $x \to 0$ the fluctuations associated to the mesons satisfy~\eqref{hatPdef} and \eqref{hatPeq} with $Q$ and $x$ set to zero and including only the first term in~\eqref{HPSdef}.
These equations are the same as the fluctuation equations for the non-singlet pseudoscalar mesons~\cite{Arean:2013tja}, which signals the suppression of the axial anomaly as $x \to 0$.
The GOR relation is found by studying the fluctuation equations perturbatively at small $m_f^2$ and also taking $m_q \to 0$.  When $m_f^2=0$ there is a solution to the system which is normalizable in the IR but not in the UV. As seen from the expansions~\eqref{CP1UVsol} and~\eqref{CP2UVsol} in Appendix~\ref{app:Asymptotics} both $P$ and $\hat P$ approach finite values at the boundary. It is convenient to normalize the solution such that $P \to 1$ as $r \to 0$. Then the boundary value of $\hat P$,
\be
 C_P \equiv \lim_{r\to 0} \hat P(r) \ , \qquad (m_f^2 = 0)
\ee
can be related to the decay constant of the $\eta'$ meson (up to corrections $\morder{m_q}$). The relation is analogous to that for the pion decay constant, found in Appendix~E of~\cite{Jarvinen:2015ofa}. Since we have taken $x=0$ the decay constants of the pion and the $\eta'$ are actually equal. After a careful comparison to the analysis of~\cite{Jarvinen:2015ofa} we find
\be \label{fpiCPrel}
 f_\pi^2 = f_{\eta'}^2 =  M^3\, N_f\, N_c\, C_P + \morder{m_q}\, , \qquad (x=0)   \,.
\ee

To obtain the GOR relation we compute the leading order perturbation in $m_f^2$ and check when the solution becomes normalizable in the UV. As is always the case for the GOR relation (and as we will verify below) the relevant regime is close to the boundary ($r \sim \sqrt{m_q/\sigma}$), where the source and VEV terms of the tachyon are of the same order.
Therefore we can take $\hat P \simeq \mathrm{const.}$ and neglect the logarithmic corrections to the potentials in~\eqref{hatPdef}. We obtain
\be \label{PprimeGOR}
 P'(r) \simeq C_P\left(\frac{-4 r }{\ell\, W_0\, w_0^2} + \frac{ m_f^2\, r^3}{W_0\, \kappa_0\, \ell^3\, \tau^2}\right) \ .
\ee
Here the constants $w_0$, $\kappa_0$, and $W_0$ are the boundary values of $w$, $\kappa$, and $V_f$, respectively. We also approximated $e^A\simeq \ell/r$.
The first term in~\eqref{PprimeGOR} gives the weak $r$ dependence of the $m_f=0$ solution which can be neglected, but the second term is the important perturbation which becomes $\morder{1}$ when $m_q \sim m_f^2$. Integrating over $r$ and using the fact that $P \simeq 1$ when $m_f^2=0$ gives
\be
 P(r) \simeq 1 - \frac{ C_P\, m_f^2 }{W_0\, \kappa_0\, \ell^3 } \int_r^\infty \frac{ \hat r^3\,d \hat r}{\t(\hat r)^2} \,,
\ee
where the integral is dominated by the regime with $r\sim \sqrt{m_q/\sigma}$ as expected.
The solution is normalizable when $P$ vanishes at the boundary, which determines the mass of the $\eta'$. This leads to the GOR relation in the limit $x \to 0$:
\be
1 \simeq \frac{ C_P\, m_{\eta'}^2 }{W_0\, \kappa_0\, \ell^3 } \int_0^\infty \frac{ \hat r^3\,d \hat r}{\t(\hat r)^2} \simeq
\frac{ f_{\eta'}^2  \, m_{\eta'}^2 }{2 M^3\, N_f\, N_c\,  W_0\, \kappa_0\, \ell^5\, m_q\, \sigma } = -\frac{f_{\eta'}^2  m_{\eta'}^2 }{m_q\langle \bar \psi \psi \rangle} \,.
\ee

\subsubsection{Small but finite $x$} \label{sec:WVstrategy}

We then discuss the $\morder{x}$ contributions to the GOR relation for $\eta'$. At this order the coupling between the glue and the flavor can no longer be neglected and we need to study the complete system~\eqref{hatPdef}--\eqref{hatQeq}. But in the IR there is decoupling and we can unambiguously define the IR normalizable solutions for the glue and the flavor, denoted by $\psi^{(Q)}$ and $\psi^{(P)}$, respectively. The leading term of $\psi^{(P)}$ can be readily identified with the zero $x$ solution discussed above.  In order to determine the $\eta'$ mass we need to study the normalizability of both these solutions in the UV. More precisely, we would want to find the coefficients $C_-$ and $C_1$ of the non-normalizable terms of the UV asymptotic expressions in~\eqref{CmUVsol} and~\eqref{C2UVsol} for each of these two solutions. It is, however, easier to expand at $x=0$ before expanding at $r=0$ and therefore study the non-normalizable terms~\eqref{CQ2UVsol} and~\eqref{CP2UVsol}. Indeed by studying the
expansions one sees that the non-normalizable terms in the two sets of asymptotic expressions are mapped to (linear combinations of) each other (possibly up to highly suppressed terms) when the order of limits is changed. We present here a sketch on how the $\morder{x}$ corrections behave, and a systematic, precise treatment is done in Appendix~\ref{app:WV}.

We denote the solutions defined by the UV expansions~\eqref{CQ2UVsol} and~\eqref{CP2UVsol} by $\phi^{(Q)}$ and $\phi^{(P)}$, respectively. As the superscripts suggest, $\phi^{(P)}$ is the perturbative solution of the $P$ field with vanishing $Q$, and $\phi^{(Q)}$ is obtained by first solving $Q$ perturbatively, in analogy to the IR normalizable solutions. We further define
\be \label{coeffmats}
 \left(\begin{array}{c}
   \psi^{(P)}\\ \psi^{(Q)}
 \end{array}\right)
 =
 \left(\begin{array}{cc}
   C_{PP} & C_{PQ}\\ C_{QP} & C_{QQ}
 \end{array}\right)
 \left(\begin{array}{c}
   \phi^{(P)}\\ \phi^{(Q)}
 \end{array}\right) + \textrm{UV normalizable terms}\ .
\ee
A normalizable mode can be constructed as a linear combination of $\psi^{(P)}$ and $\psi^{(Q)}$ when the determinant of the coefficient matrix vanishes.

We will sketch here how the coefficient matrix is computed -- a detailed analysis will be given in Appendix~\ref{app:WV} by performing a systematic expansion in both $x$ and the (squared) bound state mass $m_f^2$.
The elements can be obtained essentially by computing the values of $P$ and $Q$ near the boundary for the solutions $\psi^{(P)}$ and $\psi^{(Q)}$: as seen by comparing to the expressions~\eqref{CQ2UVsol} and~\eqref{CP2UVsol}, $C_{IJ}$ is the value of $J$ for the solution $\psi^{(I)}$ (with $I,J$ taking the values $P,Q$).\footnote{Some care is needed because $\phi^{(Q)}$ also contains a logarithmically divergent term for $P$.}
The coupling between the glue and the flavor is irrelevant for the diagonal elements of the matrix. We have computed $\psi^{(P)}$ at leading order in $x$ above, from which we readily obtain that $C_{PP} \simeq 1 - m_f^2/m_\pi^2$. We may normalize $\psi^{(Q)}$ such that $C_{QQ} = 1$. The backreaction of flavor on glue is suppressed by $x$ but not vice versa, which leads to\footnote{This ensures that taking $x \to 0$ the determinant is $\propto C_{PP}$ and we will have $m_{\eta'} = m_\pi$.} $C_{PQ} = \morder{x}$ but  $C_{QP} = \morder{x^0}$. From the fluctuation equations we see\footnote{Solving~\eqref{hatPdef} for $P'$ gives $P' \sim Q'/\tau^2 + \cdots$ which leads to this enhancement.} that $C_{QP}$ is enhanced as $m_q \to 0$, which will also be proven in Appendix~\ref{app:WV}. Taking stock,
\be
 \left(\begin{array}{cc}
   C_{PP} & C_{PQ}\\ C_{QP} & C_{QQ}
 \end{array}\right)
 \simeq
 \left(\begin{array}{cc}
   1-\frac{m_f^2}{m_\pi^2} & \morder{x\ m_\pi^0}\\ \morder{m_\pi^{-2}\ x^0} & 1
 \end{array}\right) \ .
\ee
The determinant vanishes when $m_f$ equals the mass of the $\eta'$, which leads to the expected relation
\be \label{WVformulasketch}
 m_{\eta'}^2 = m_\pi^2 + \morder{x\ m_\pi^0}\ .
\ee

In order to compute the coefficient in the $\morder{x}$ term of~\eqref{WVformulasketch},
we solve the fluctuation equations in a systematic expansion of the wave functions at small
$x$ and $m_q$ in Appendix~\ref{app:WV}. This results in the Witten-Veneziano formula for the mass of the $\eta^\prime$ meson:
\be \label{etapmassfinal}
 m_{\eta'}^2 \simeq  m_\pi^2 + x \frac{N_f \,N_c\,\chi_\mathrm{YM}}{f_\pi^2} \,,
\ee
where $\chi_\mathrm{YM}$ is the topological susceptibility for Yang-Mills theory. For our conventions $f_\pi^2 =\morder{N_fN_c}$, so that the second term is indeed $\morder{x}$.

\begin{figure}[!tb]
\begin{center}
\includegraphics[width=0.49\textwidth]{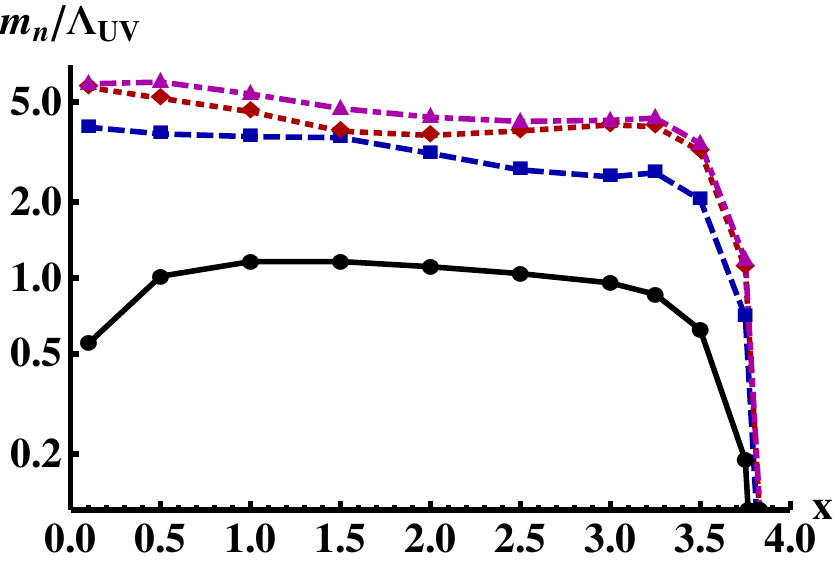}%
\hspace{2mm}\includegraphics[width=0.49\textwidth]{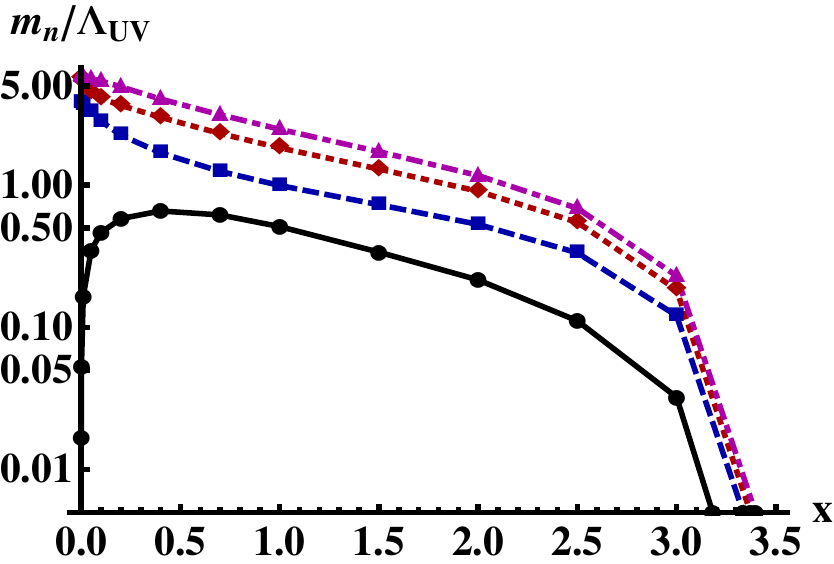}
\end{center}
\caption{Masses of the lowest four singlet pseudoscalar states in the logarithmic scale as a function of $x$. Left: potentials~I. Right: potentials~II.}
\label{fig:psmasses}\end{figure}

\subsection{Numerical results}
\label{ssec:numthzero}

We have computed the spectra of the singlet pseudoscalars numerically both for the potentials~I and
potentials~II defined above in Sec.~\ref{sec:potentials}. The numerical study was done by using the
fluctuation equations given in Appendix~\ref{app:quadfluctdet} as explained in~\cite{Arean:2013tja}
and in Appendix~G of~\cite{Jarvinen:2015ofa}.

The spectrum is shown in the logarithmic scale as a function of $x$ in Fig.~\ref{fig:psmasses}.
The light $\eta'$ state is best visible in the right hand plot for small $x$. The third lowest state is a glueball in the limit $x \to 0$ for both potentials\footnote{The mass of the glueball is actually the same for both potentials~I and~II because they only differ in the flavor sector.} whereas the other states are $\bar \psi \psi$ states. In the walking regime
all masses tend to zero obeying the Miransky scaling law. Apart from the light $\eta'$ meson at small $x$,
the dependence of the spectrum on $x$ is for both potentials very similar to that found for the singlet
scalars in~\cite{Arean:2013tja}. Notice that there is additional interesting level crossing structure for
potentials~I.

\begin{figure}[!tb]
\begin{center}
\includegraphics[width=0.49\textwidth]{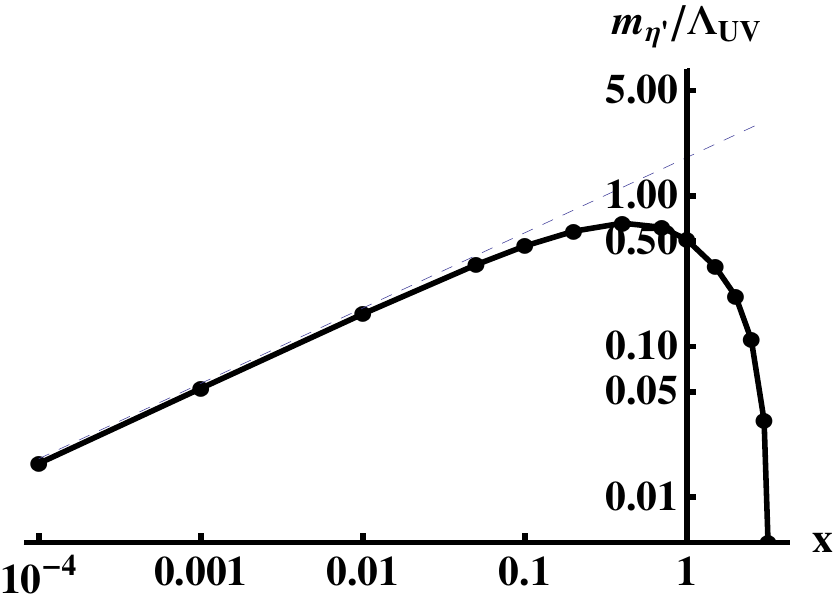}%
\hspace{2mm}\includegraphics[width=0.49\textwidth]{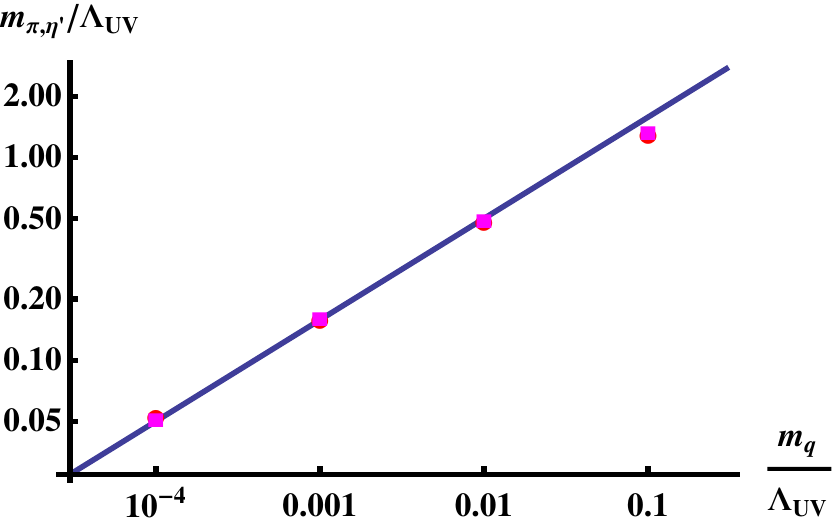}
\end{center}
\caption{The mass of the $\eta'$ meson for potentials~II. Left: The mass of $\eta'$ as a function of $x$ in log-log scale for $m_q=0$. The dashed blue line is a fit with the expected dependence $m_{\eta'}^2 \propto x$.  Right: the dependence of the mass of the $\eta'$ (red circles) and the pion mass (magenta squares) on $m_q$ at $x=0.0001$. The blue line is a fit to the GOR relation.}
\label{fig:etapmass}\end{figure}

We study the dependence of the mass of the $\eta'$  meson on $x$ and $m_q$ in more detail for potentials~II
(for which the numerical computations are much easier then for potentials~I) in Fig.~\ref{fig:etapmass}.
The left hand plot demonstrates that the dependence on $x$ at $m_q=0$ is that predicted by
Eq.~\eqref{etapmassfinal}. The right hand plot shows the data for the $\eta'$ and pion masses at very small
$x=0.0001$. The data points overlap perfectly, as predicted by Eq.~\eqref{etapmassfinal}. The dependence on
the quark mass matches with the GOR relation (blue line).

\section{Flavor non-singlet spectra at finite $\bar\theta$} \label{sec:nonsingletfl}
\label{sec:finitethspc}

\subsection{Fluctuations at finite $\bar\theta$ }
\label{ssec:thbfluc}

 We now study the quadratic fluctuations for the backgrounds with a nontrivial $\bar\theta$-angle studied in
Sec.~\ref{sec:thetaback}.
In those backgrounds both the tachyon phase $\xi$ and the QCD axion $\ag$ are nonvanishing,
which makes the analysis of the fluctuations, in particular the flavor singlet sector, more involved.
In the following we will restrict ourselves to the analysis of the flavor non-singlet sector.
This sector consists of the flavor non-singlet $1^{--}$ vector and
$1^{++}$ axial vector mesons, and the flavor non-singlet $0^{-+}$ pseudoscalar and $0^{++}$ scalar
mesons; and these last two get mixed in a parity breaking\footnote{Notice that charge conjugation remains
as a good quantum number, and therefore the vectors and axial vectors do not mix at finite $\bar\theta$.} finite $\bar\theta$ vacuum.

\subsubsection{Flavor non-singlet sector}
\label{flavornonsing}

This sector involves the $SU(N_f)$ part of the vector, axial vector, scalar, and pseudoscalar mesons.
The vector and axial vector fluctuations were defined in Eq. \eqref{VAdefsmain}. The scalar and pseudoscalar
mesons, which will mix in the presence of a nonzero phase of the tachyon, correspond to fluctuations of
the complex tachyon:
\begin{align}
&T=\left[\tau(r)+s(r,x)+\tilde s(r,x)\right]\exp\left[\xi(r)+\thf(r,x)+\tilde\pi(r,x)\right]\,,\\\nonumber
& {\rm with}\quad \tilde s(r,x) = \mathfrak{s}^a(r,x)\,t^a\,,\qquad \tilde \pi(r,x) = \pi^a(r,x)\,t^a\,.
\end{align}
Only the DBI piece of the action, \ie Eq. \eqref{generalact}, contributes to the non-singlet sector fluctuations.
In Appendix \ref{app:thbfluc} we write the resulting action up to quadratic order in
the fluctuations and derive the equations of motion. We now summarize them sector by sector.

\subsubsection*{Scalar-pseudoscalar mesons}
The fluctuations of the modulus and phase of the tachyon, and the longitudinal part of the axial vector
contribute to this sector.
We shall consider the following Ansatz for the three coupled fields:
\begin{align}
A_\mu^{||}=-\psi_L(r)\,\partial_\mu P^a(x)\,t^a\,,\quad
\tilde s = \psi_s(r)\,P^a(x)\,t^a\,,\quad
\tilde\pi=2\psi_p(r)\,P^a(x)\,t^a\,,
\end{align}
where $\partial_\mu\partial^\mu P^a(x) = m^2\, P^a(x)$.
As shown in Appendix \ref{app:thbfluc} the equations of motion for these fields can be recombined into
the two coupled equations (\ref{nsfpsmesons1}, \ref{nsfpsmesons2}) for the two fields
$\psi_s$, and $\hat\psi_l=e^{A}\,w^2\,V_f\,\tG^{-1}\,\psi_L'$.
The normalizable solutions of those equations will correspond to the scalar and pseudoscalar mesons, which
mix in a parity breaking finite $\bar\theta$-vacuum.


\subsubsection*{Vector mesons}

We consider  the Ansatz
\be
V_\mu=\psi_{\rm V}(r)\,{\cal V}_\mu^a(x)\,t^a\,,\quad
{\rm with}\quad \partial_\nu\partial^\nu {\cal V}_\mu^a(x)=m^2_{\rm V}\, {\cal V}_\mu^a(x)\,,
\label{vecans}
\ee
for the transverse part of the vector meson fluctuation (the longitudinal part can be set to zero).
The equation of motion for $V_\mu$ resulting from the Lagrangian \eqref{pscalu1s1tb} reduces to
\be
\frac{1}{V_f\, \gf^2\, e^{\Awf}\,\tilde\G}\,
\partial_r \left( V_f\, e^{\Awf}\,
{\gf^2 \over\tilde\G}\, \partial_r \psi_V \right)
+m_V^2\, \psi_V  = 0 \, .
\label{thnsvectoreom}
\ee

\subsubsection*{Axial vector mesons}

We shall take the following Ansatz for the transverse part of the axial vector mesons
\be
A_\mu^{\bot}=\psi_{\rm A}(r)\,{\cal A}_\mu^a(x)\,t^a\,,\quad
{\rm with}\quad \partial_\nu\partial^\nu {\cal A}_\mu^a(x)=m^2_{\rm A}\, {\cal A}_\mu^a(x)\,,
\label{axvecans}
\ee
The equation of motion for $A_\mu^{\bot}$ follows swiftly from the Lagrangian \eqref{pscalu1s1tb},
and in terms of this Ansatz takes the form
\be
\frac{\partial_r \left( V_f\, \gf^2\, e^{\Awf}\,
\tilde\G^{-1}\, \partial_r \psi_A \right)}{V_f\, \gf^2\, e^{\Awf}\,\tilde\G  }
-4 {\t^2\, \h\,e^{2 \Awf}\,G^2 \over \gf^2\,\tilde G^2}\,\psi_A  +m_A^2\, \psi_A = 0\,.
\label{thnsaxveom}
\ee

\subsection{The Gell-Mann-Oakes-Renner relation at finite $\bar\theta$}

It is possible to compute analytically the $\bar\theta$ dependence of the pion mass at small $m_q$, and to use this to write the generalization of the GOR relation at finite $\bar\theta$. We review here the key points of the computation and details are given in Appendix~\ref{app:gmor}.

The pion mass is found by analyzing the fluctuation equations for the pseudoscalar and scalar sectors in the UV and in the IR, and requiring match of the results in middle, where the regimes of applicability of the two results overlap when $m_q$ is small. In the UV analysis, it is essential to use the fluctuations of the real and imaginary parts of the tachyon, in terms of which the fluctuation equations decouple when $r \ll 1/\LUV$ and $m_q/\LUV \ll 1$. The most important difference with respect to the computation at $\bar\theta=0$ is that the background tachyon solution is replaced by the real part of the tachyon (see Appendix~\ref{app:gmor} for details). In the IR, or more precisely when $r \gg \sqrt{m_q/\sigma}$, it is enough to show that the mixing of the scalars and pseudoscalars is suppressed by $\morder{m_q^2}$, and consequently the IR solutions are the same as at $\bar\theta = 0$.

Matching the UV and IR approximations for $\sqrt{m_q/\sigma} \ll r \ll 1/\LUV$, where both of them are accurate, then fixes the pion mass. The only difference with respect to the result at $\bar\theta$ is that $m_q$is replaced by the source for the real part of the tachyon, {\it i.e.}, $m_q \cos \xi_0 \simeq m_q \cos(\bar\theta/N_f)$, where we also recalled the result~\eqref{thetaxirel}. The final result for the generalized GOR relation is therefore
\be \label{gmortext}
 f_{\pi,0}^2 m_\pi^2 =  -\langle \bar \psi\psi\rangle\big|_{m_q=0}\, m_q \cos \frac{\bar\theta}{N_f} +\morder{m_q^2} \,,
\ee
where $f_{\pi,0}$ is the pion decay constant at $\bar\theta=0$.  The result agrees with effective field theory (see, {\it e.g.},~\cite{diCortona:2015ldu}).

\begin{figure}[!tb]
\begin{center}
\includegraphics[width=0.49\textwidth]{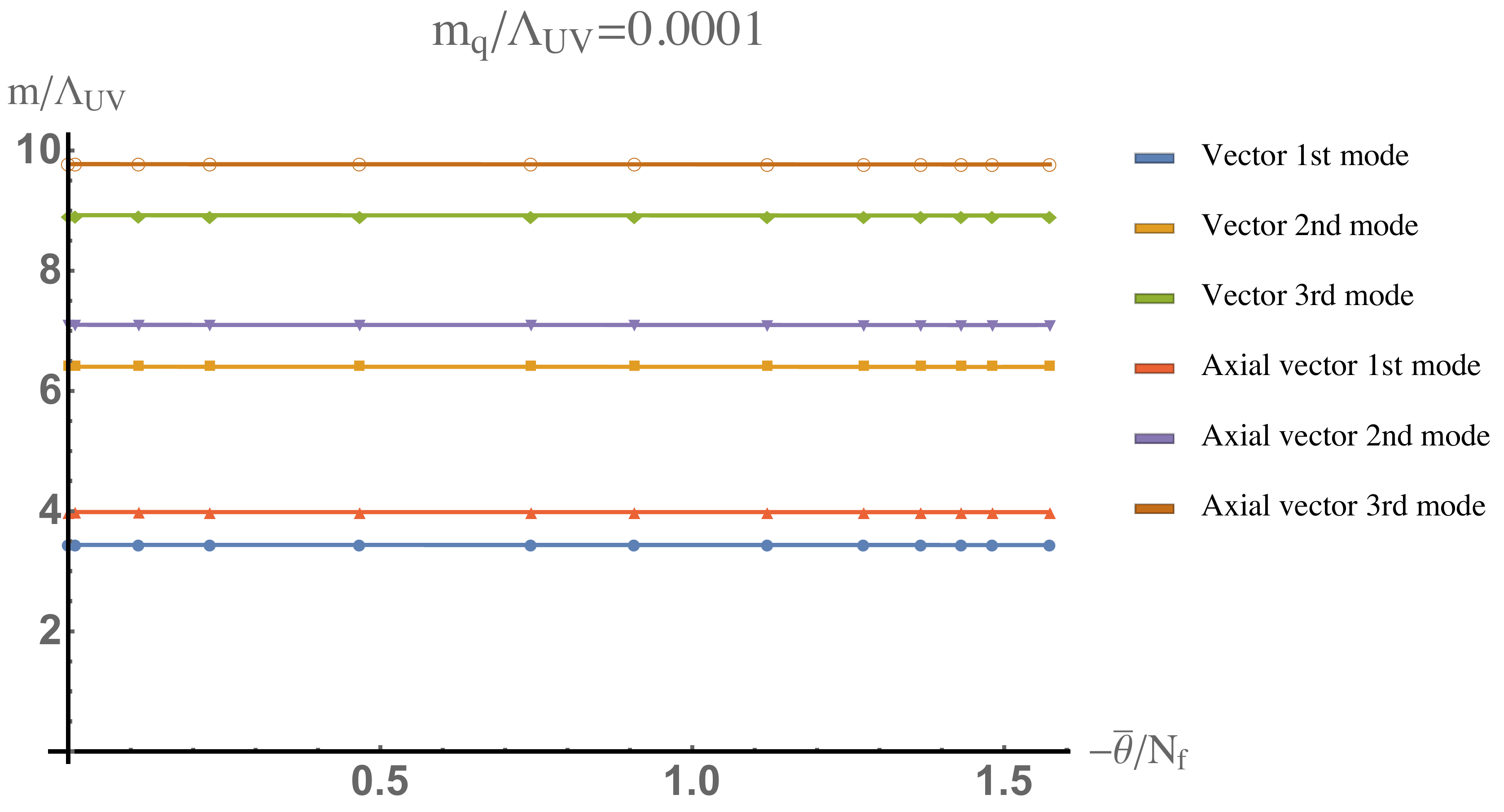}
\includegraphics[width=0.49\textwidth]{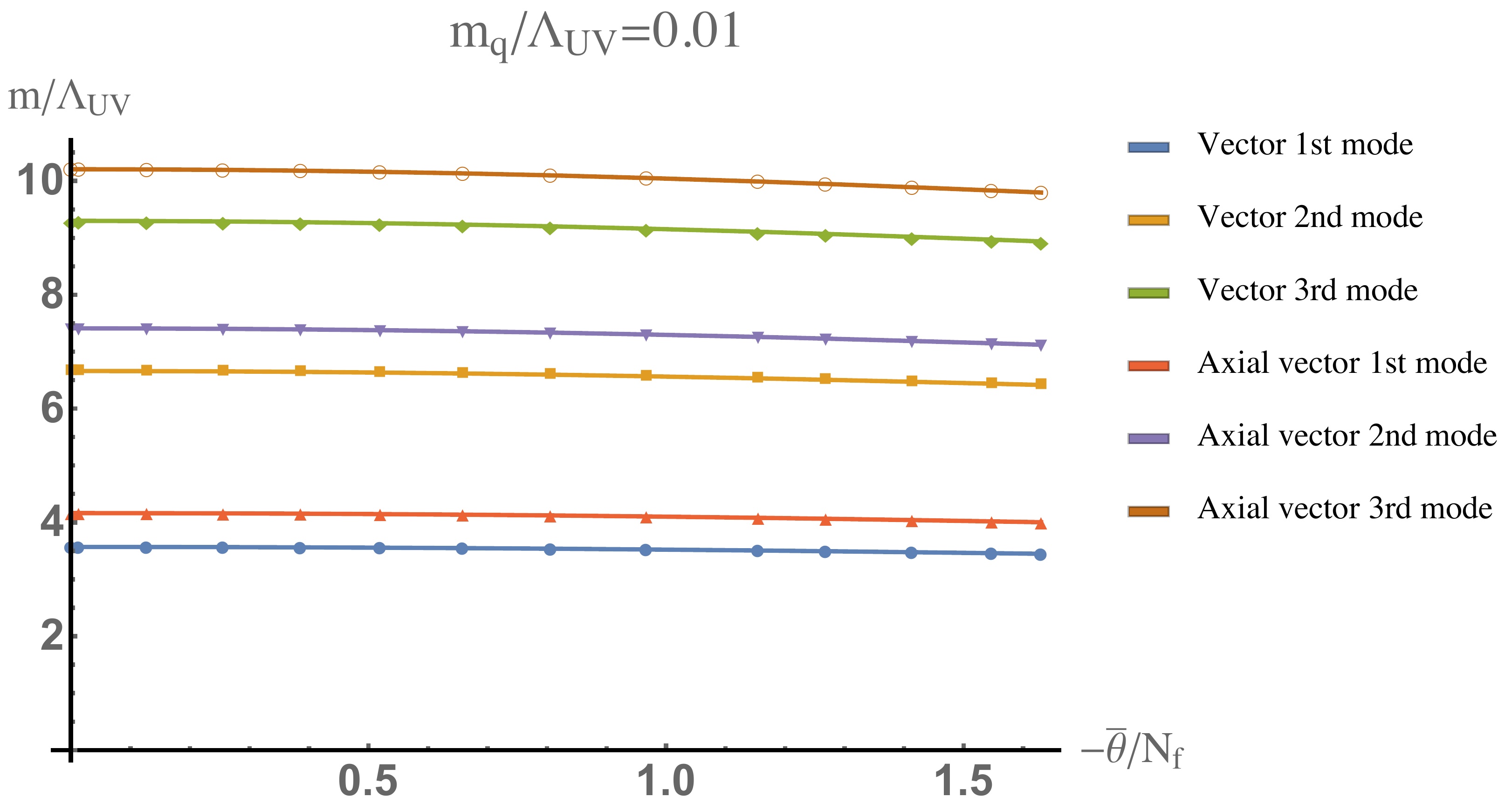}
\includegraphics[width=0.49\textwidth]{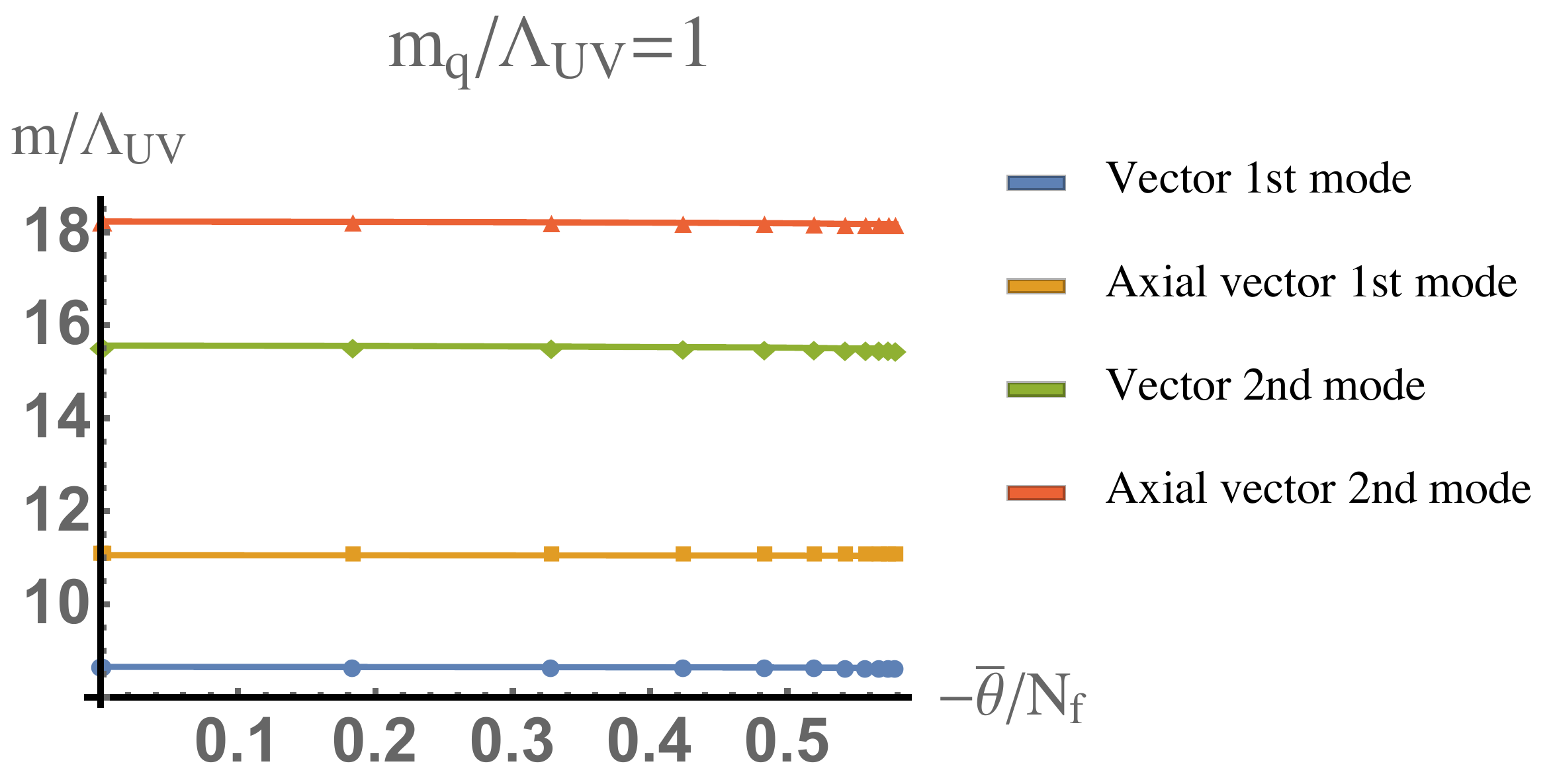}
\end{center}
\caption{The vector and axial vector masses for quark masses, $m_q/\Lambda_\mathrm{UV}=0.0001, 0.01, 1$, in terms of $\bar \theta /N_c$ for potentials I.}
\label{fig:vecmass}
\end{figure}

\subsection{Numerical analysis}

The flavor non-singlet spectra of vector, axial-vector, pseudoscalar and scalar mesons have been calculated for
different values of quark mass, as a function of the $\bar \theta$-angle. The full action of the model,
Eqs. (\ref{generalact}) and (\ref{samain}), is expanded to quadratic order in terms of the excitation fields defined
in section \ref{flavornonsing}.  The quadratic action of the flavored excitations in a background of nontrivial
$\bar \theta$ is presented in Eq.~(\ref{pscalu1s1tb}). The vector and axial-vector excitation equations are decoupled and are given by Eqs.~(\ref{thnsvectoreom}, \ref{thnsaxveom}), while the pseudoscalars and scalars are coupled because of the nonzero background $\bar \theta$-angle, Eqs.(\ref{eompsip}, \ref{eompsil}, \ref{eompsis}). The numerical procedure of determining the mass spectrum, both for coupled and decoupled excitations, is described in detail in~\cite{Arean:2013tja}. The computation consists basically of the solution of the excitation equations in the bulk spacetime with normalizable boundary conditions both at the boundary and the bottom of spacetime. The spectrum is calculated for potentials I, and for different values of quark masses $m_q/\Lambda_\mathrm{UV}=0.0001, 0.01, 1$.
As it is depicted in Fig.~\ref{fig:thetavsC}, in case of small quark mass, for any value of the integration parameter, $C_a$, there are two background solutions, corresponding to two different values of $\bar \theta$. For larger quark mass, only the lower branch survives, and it does not turn back to the horizontal axis ($m_q/\Lambda_\mathrm{UV}=1$ case). It has been found numerically that the spectrum is stable only in the lower branch of the solution. In the upper branch of solutions, it was found that one mode from the scalar channel has negative mass squared signaling an instability of the spectrum (see last plot in Fig. \ref{fig:psscmass}).

\begin{figure}[!tb]
\begin{center}
\includegraphics[width=0.51\textwidth]{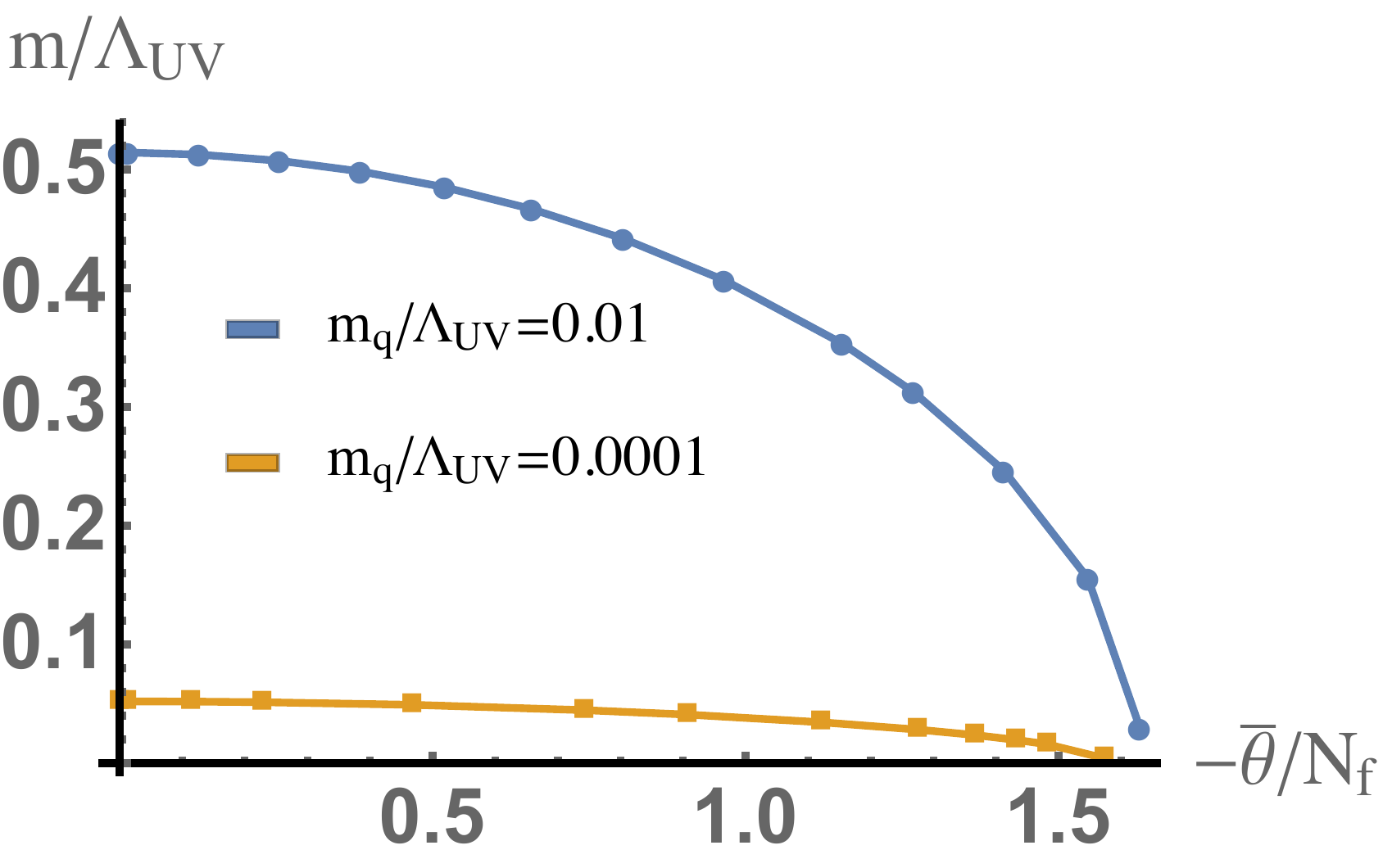}
\includegraphics[width=0.48\textwidth]{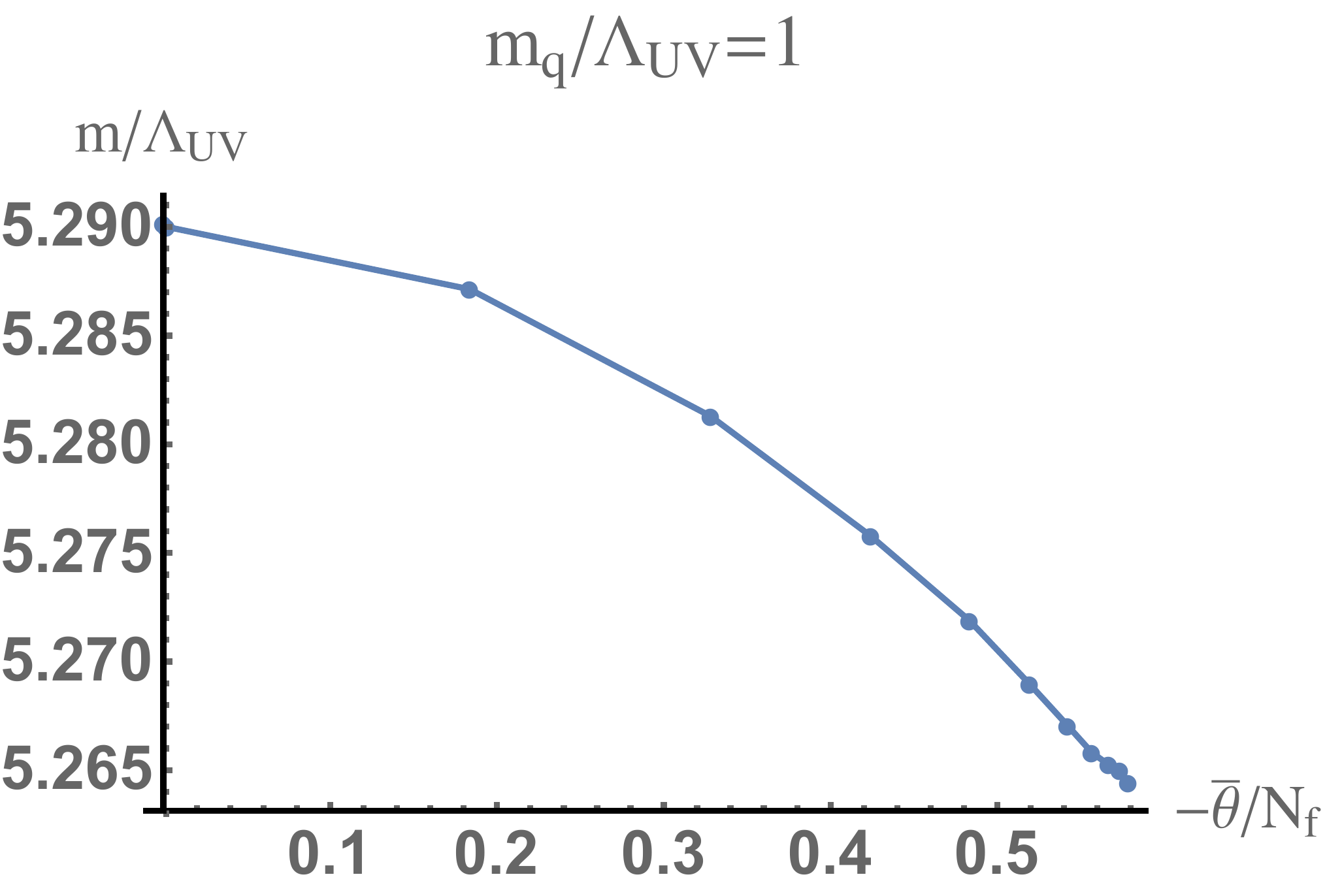}
\end{center}
\caption{The lowest mode masses of the pseudoscalar-scalar channel, for quark masses, $m_q/\Lambda_\mathrm{UV}=0.0001, 0.01$ (left plot) and $m_q/\Lambda_\mathrm{UV}= 1$ (right plot), in terms of $\bar \theta /N_c$ for potentials I. those modes correspond the the pion.}
\label{fig:pion}
\end{figure}

\begin{figure}[!tb]
\begin{center}
\includegraphics[width=0.49\textwidth]{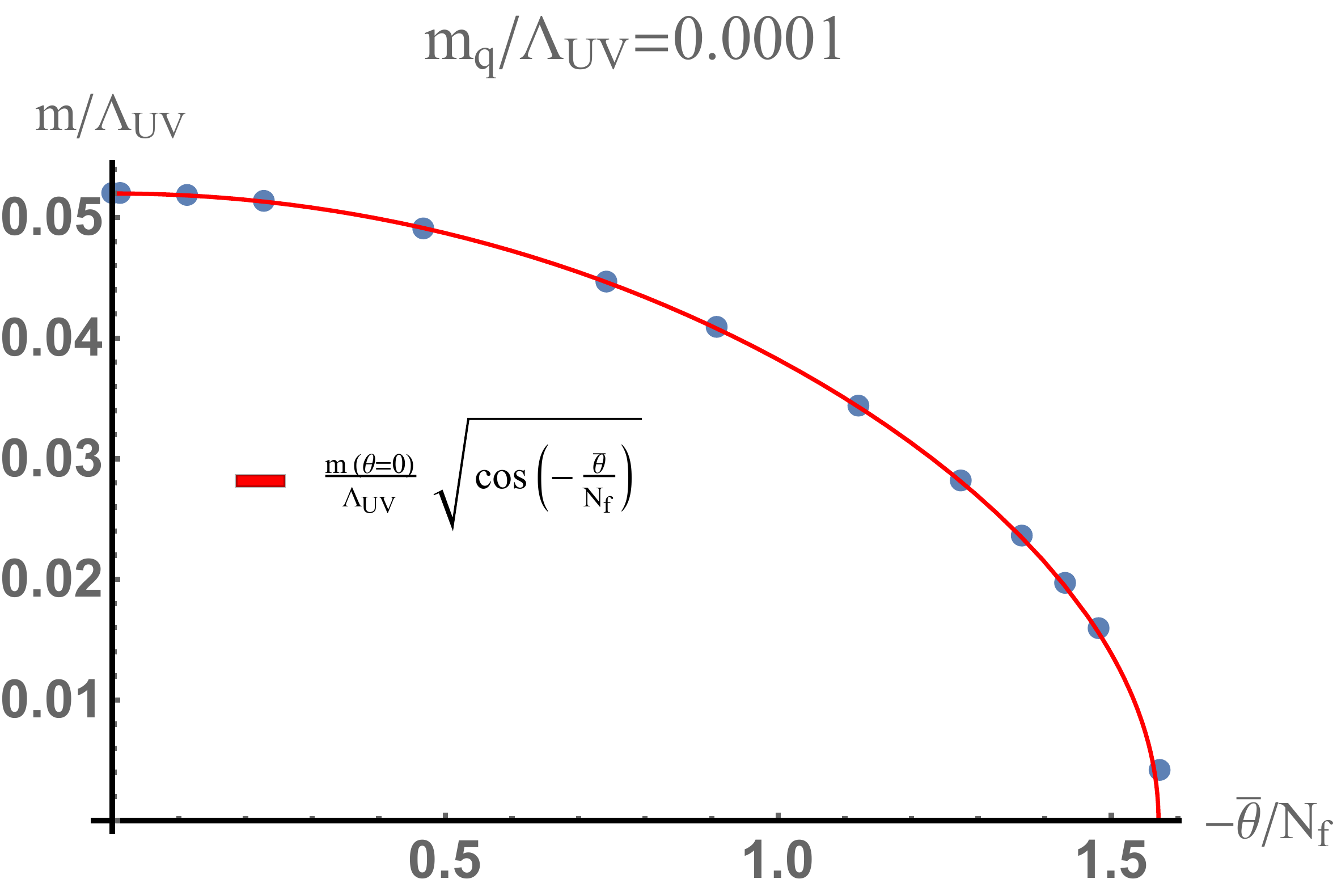}
\end{center}
\caption{The numerical result of the pion mass in terms of $\bar \theta$ in V-QCD is seen to be in perfect agreement with analytic formula (\protect\ref{pionthetaeq}).}
\label{piontheta}
\end{figure}

\begin{figure}[!tb]
\begin{center}
\includegraphics[width=0.49\textwidth]{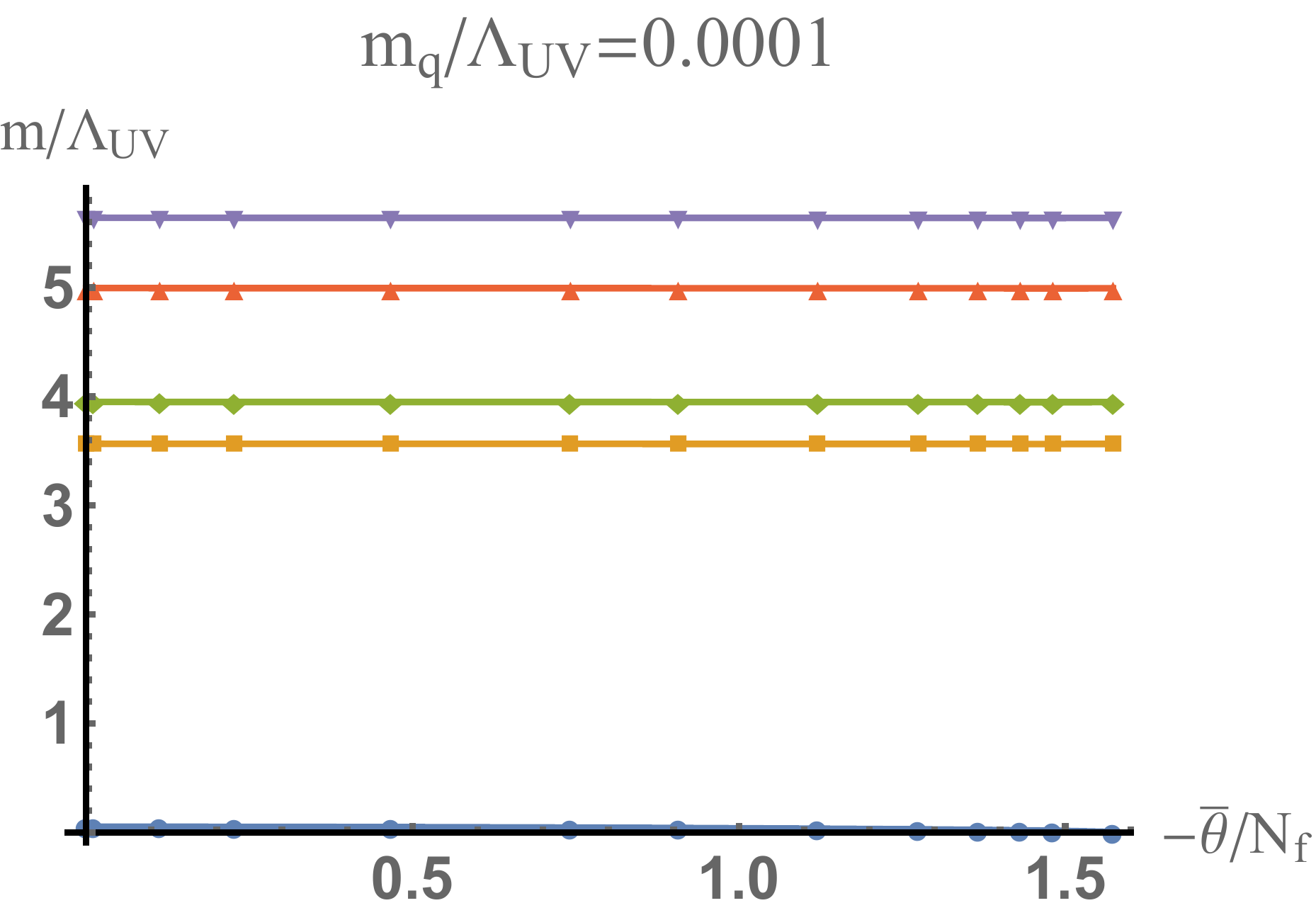}
\includegraphics[width=0.49\textwidth]{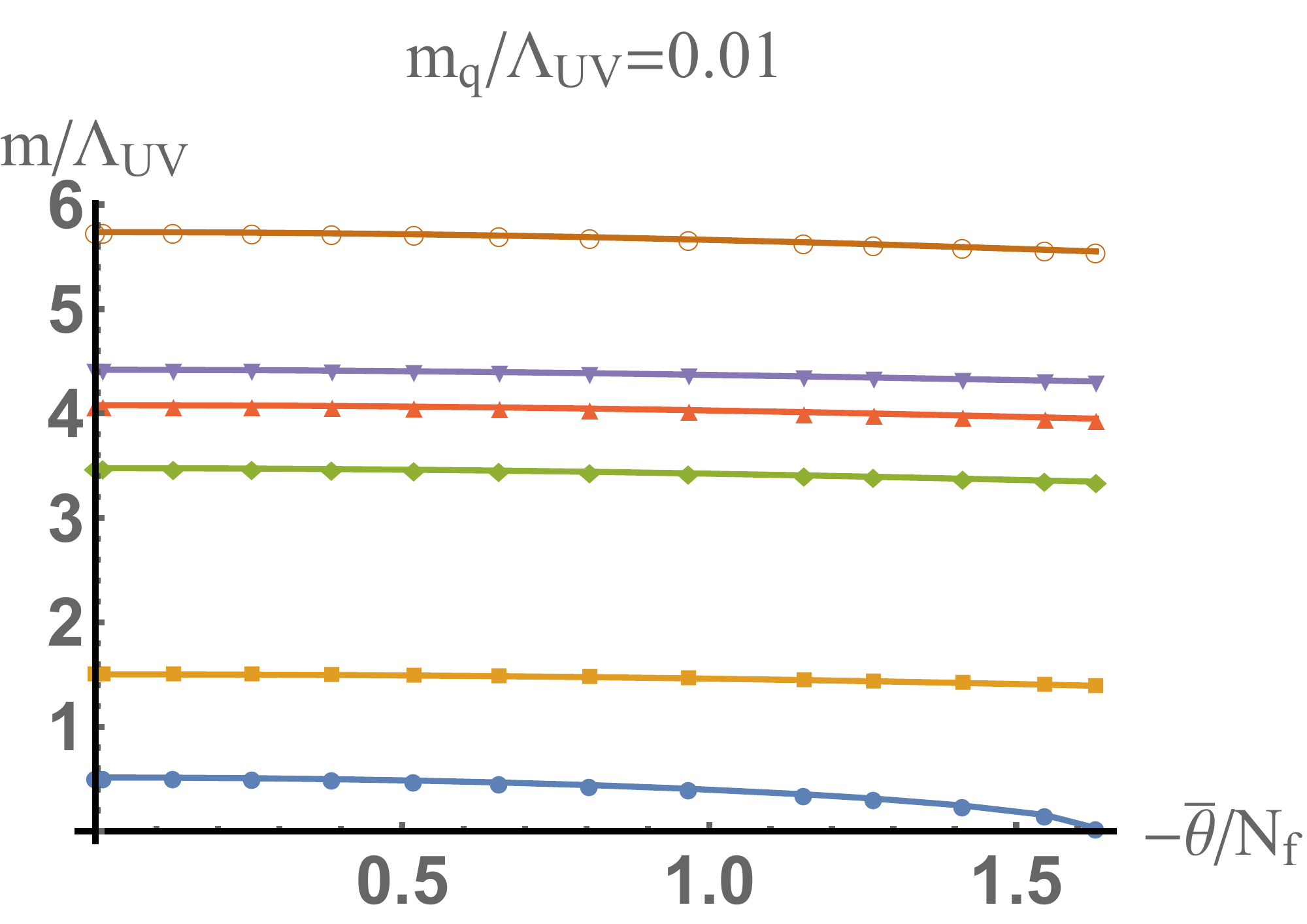}
\includegraphics[width=0.49\textwidth]{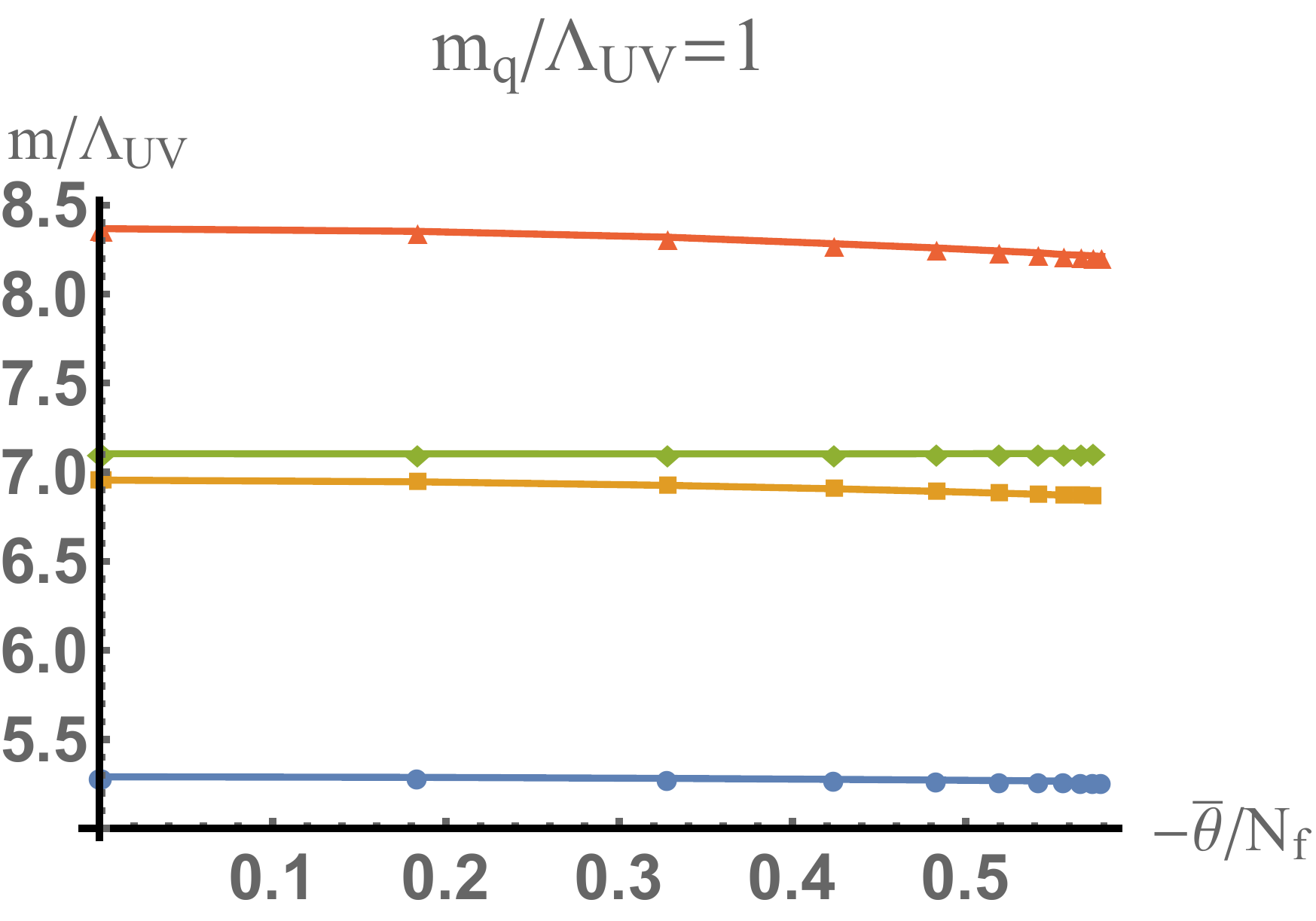}
\includegraphics[width=0.49\textwidth]{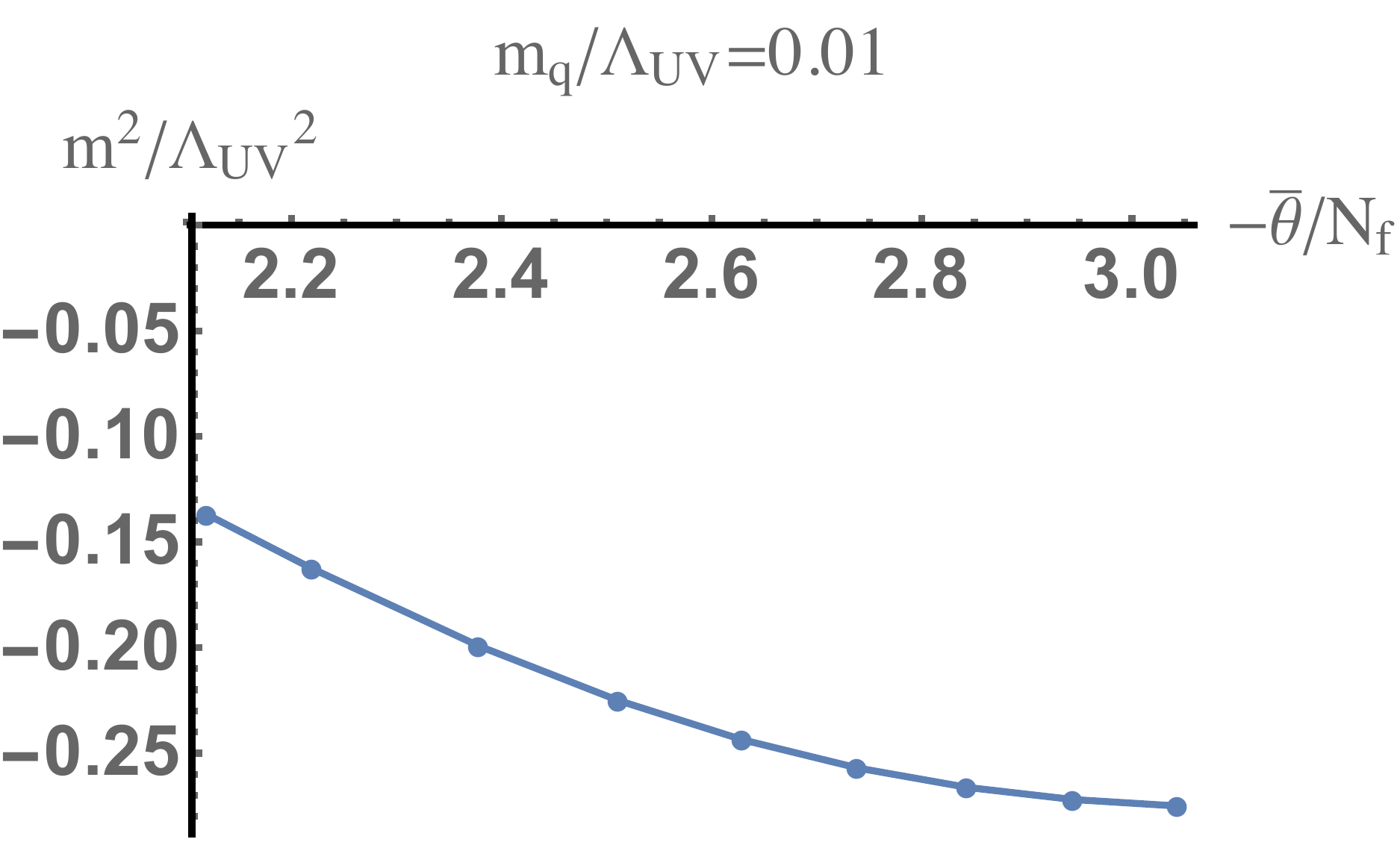}
\end{center}
\caption{The lowest mode masses of the pseudoscalar-scalar channel, for quark masses, $m_q/\Lambda_\mathrm{UV}=0.0001, 0.01, 1$, in terms of $\bar \theta /N_c$ for potentials I. The lightest mode correspond the the pion. In the last plot, it is shown that the pseudoscalar-scalar channel contains a mode of negative mass squared in the upper branch of solutions. The example which is shown here is for $m_q/ \Lambda_\mathrm{UV} = 0.01$.}
\label{fig:psscmass}
\end{figure}

In Fig. \ref{fig:vecmass}, we plot the three lowest masses of the vector and axial-vector mesons.  It observed that the vector and axial-vector masses have a mild dependence on $\bar \theta/N_c$. As it is expected, for larger quark masses the whole spectrum moves to higher meson masses, even though the difference in the spectrum for $m_q/\L_\mathrm{UV}=0.0001$ and $m_q/\L_\mathrm{UV}=0.01$ is small. The dependence of the pion mass on $\bar \theta/N_f$ is depicted in  Fig. \ref{fig:pion}. The pion mass decreases with increasing $\bar \theta$-angle. In case of small equal quark masses, the pion mass as a function of $\bar{\theta}$ is obtained from~\eqref{gmortext},
 \be
 m_{\pi}(\bar\theta)= m_{\pi}(0) \sqrt{\cos\left({\bar{\theta} \over N_f} \right)} \,.
 \label{pionthetaeq}
 \ee
 In case of $m_q/\L_\mathrm{UV}=0.0001$, we verified the above relationship numerically as it is seen in Fig.~\ref{piontheta}.  Finally, the pseudoscalar-scalar masses are presented in Fig. \ref{fig:psscmass}. It is noticed that the states do not mix at finite $\bar \theta$. The lowest state corresponds to the pion, the next two to scalar excitations and the highest to a pseudoscalar mode.

\addcontentsline{toc}{section}{Acknowledgments}

\section*{Acknowledgements}

We would like to thank F.~Bigazzi, A.~Cotrone, and K.~Rummukainen for discussions.
D.A. thanks the FRont Of
pro-Galician Scientists for unconditional support.
This work was supported in part by European Union's Seventh Framework Programme under grant agreements (FP7-REGPOT-2012-2013-1) no 316165 and the Advanced ERC grant SM-grav 669288.
The work of D.~Are\'an is
supported by the German-Israeli Foundation (GIF), grant 1156.
This work of I. Iatrakis is part of the D-ITP consortium, a program of the Netherlands Organisation for Scientific
Research (NWO) that is funded by the Dutch Ministry of Education, Culture and Science
(OCW).

\newpage

\appendix
\renewcommand{\theequation}{\thesection.\arabic{equation}}
\addcontentsline{toc}{section}{Appendices}
\section*{APPENDIX}

\section{UV and IR asymptotics of the background} \label{app:Asymptotics}

In this section we will present the asymptotic form of the background fields for the
choices of potentials relevant for the analysis in this work. Most of the expressions turn out to be independent of $\bar\theta$.
For a more general analysis
of the asymptotics at $\bar\theta=0$ we refer the reader to the appendix D of \cite{Arean:2013tja}.

\subsection{UV asymptotics} \label{subapp:UVback}
As explained in \cite{Arean:2013tja}, in the UV ($r\to0$) the tachyon decouples from
the glue fields $\l$ and $A$, whose asymptotic form is determined by the effective
potential
\be \label{Videfapp}
 V_{\rm eff}(\l)=V_g(\l)-x\, V_f(\l,0)=\frac{12}{\ell^2}\left[1 + V_1 \l +V_2
\l^2+\cdots \right] \,,
\ee
resulting in the following solutions
\begin{align} \label{UVexpsapp}
\Awf(r) \ =& -\log\frac{r}{\ell} + \frac{4}{9 \log(r \Lambda)}  \\
&+ \frac{
  \frac{1}{162} \left[95 \!-\! \frac{64 V_2}{V_1^2}\right] +
   \frac{1}{81} \log\left[-\log(r \Lambda)\right] \left[-23 \!+\! \frac{64
V_2}{V_1^2}\right]}{
  \log(r \Lambda)^2} +{\cal O}\left(\frac{1}{\log(r\Lambda)^3}\right) + \morder{m_q^2r^2} \nn  \\
  \label{UVexpsapp2}
  V_1 \l(r)=&-\frac{8}{9 \log(r \Lambda)} + \frac{
   \log\left[-\log(r \Lambda)\right] \left[\frac{46}{81} - \frac{128 V_2}{81
V_1^2}\right]}{\log(r \Lambda)^2}+{\cal
O}\left(\frac{1}{\log(r\Lambda)^3}\right) + \morder{m_q^2r^2}\, ,
\end{align}
where $\Lambda= \LUV$ defines the UV scale of the theory. We also wrote down the size of the leading corrections due to the tachyon. Notice, in particular, that these results are independent of $\bar\theta$.

The UV scale may be defined explicitly as
\be \label{LUVdef}
 \LUV = \lim_{r\to 0} \frac{1}{\ell} \exp\left[A -\frac{8}{9V_1 \l} + \left(\frac{23}{36}-\frac{16 V_2}{9 V_1^2}\right) \log \frac{9V_1\l}{8} \right] \ .
\ee

In order to solve for the tachyon one inserts the asymptotic solutions for $\l$, and $A$ into the equation
of motion for the tachyon. We also need the UV expansions of $a$ and $\kappa$, which read
\be
\kappa(\l)\sim\kappa_0(1+\kappa_1\,\l)\,,\qquad
a(\l) \sim a_0 \left( 1+a_1 \l \right);\qquad
{\rm with}\quad {\kappa_0\over a_0}={2\ell^2\over3}\,.
\ee
We discuss the asymptotics of the tachyon at finite $\bar\theta$ in Sec.~\ref{ssec:uvasymp}. We present here for reference the result at $\bar\theta = 0$, which reads
\begin{align}
\label{TUVres}
 \frac{1}{\ell}\tau(r) \ =\ &m_q\, r\,
(-\log(r\Lambda))^{-\rho} \left[1+ {\cal
O}\left(\frac{1}{\log(r\Lambda)}\right)\right]
\\ \nn
&
+\sigma\, r^3\,
(-\log(r\Lambda))^{\rho} \left[1+ {\cal
O}\left(\frac{1}{\log(r\Lambda)}\right)\right] \,,
\end{align}
with
\be
 \rho=  -\frac{4}{3}-\frac{4 \h_1}{3 V_1} \,.
\label{rhodef}
\ee

\subsection{IR asymptotics} \label{subapp:IRback}

We will only present here the discussion for the particular asymptotics of $V_g$ that matches
well with the IR properties of QCD \cite{ihqcd}.

\subsubsection{$\l$ and $A$}
\label{subsubapp:IRbackglue}

For regular potentials, the IR divergence of the tachyon decouples the tachyon and the axion from $\l$ and $A$ and therefore their asymptotics is independent of $\bar\theta$.
For a glue potential $V_g$ with the following IR asymptotic form\footnote{The factors of $8\pi^2$ were included because this leads to simple expressions for the coefficients $v_i$ for the potentials which we use.}
\be \label{Vgas}
 V_g(\l) = v_0 \left(\frac{\l}{8\pi^2}\right)^{4/3}\sqrt{\log \frac{\l}{8\pi^2}}\left[1 +
 {v_1\over\log\left(\frac{\l}{8\pi^2}\right)} +
 {v_2\over \log^2\left(\frac{\l}{8\pi^2}\right)} + \cdots \right] \,,
\ee
the asymptotic solution for the background glue fields reads
\begin{align}
&A = -\frac{r^2}{R^2}+\frac{1}{2}\log\frac{r}{R}+A_c + \mathcal{O}\left(r^{-2}\right)\,,
\label{IRresA}\\
&\log \l = \frac{3}{2}\frac{ r^2}{R^2}+\l_c+ \mathcal{O}\left(r^{-2}\right)\,,
\label{IRresl}
\end{align}
with
\be
A_c=-\log R-\frac{1}{2}\log v_0+
\frac{5}{4}\log 2+\frac{3}{4}\log 3 + \frac{23}{24}+\frac{4 v_1}{3}\,,\qquad
\l_c=-\frac{23}{16}-2 v_1 + \log (8\pi^2)\,.
\label{glueircons}
\ee
Here $R=1/\LIR$ sets the IR scale of the model. A possible explicit definition is
\be \label{LIRdef}
 \LIR = \lim_{r \to \infty} \frac{1}{r}\sqrt{\frac{2}{3} \log \l} \ .
\ee

\subsubsection{The tachyon}
We will consider the following asymptotics for the relevant potentials
\be \label{potIRas}
\h(\l) \sim \h_c\, \l^{-\kappa_p} (\log \l)^{-\kappa_\ell}\,, \qquad
a(\l) \sim a_c\, \l^{a_p} (\log \l)^{a_\ell} \,, \qquad
V_{f0}(\l) \sim v_c\, \l^{v_p} (\log \l)^{-v_\ell}\,,
\ee
where $\h_c$ and $a_c$ are assumed to be positive. In particular, we will focus on the special case $v_p=10/3$,
singled out by the requirement of having non-singular $\bar\theta\neq0$ backgrounds. A thorough analysis
of the acceptable asymptotics with $v_p<10/3$ was presented in the appendix D of \cite{Arean:2013tja}. At finite $\bar\theta$, the results for $\t$ there remain unchanged, and the asymptotics of $\xi$ can be found by substitution to~\eqref{xieqs2}. We consider two cases at $v_p=10/3$ in the following, both leading to acceptable IR asymptotics.

\paragraph{1.} $v_p=10/3$, $\kappa_p=4/3$, $\kappa_\ell>-3/2$, $a_p=a_\ell=0$, and $v_\ell>1-\kappa_\ell/2$.
This case results in the following asymptotic solution for the tachyon
\be
\tau(r)=\tau_0\exp\left[C_{\rm I}\left({r\over R}\right)^{3+2\kappa_\ell}\left(1+{\cal O}(r^{-2})
\right)
\right]\,,
\label{tauiras1}
\ee
with
\be
C_{\rm I}={2^{1-\kappa_\ell}\,3^{\kappa_\ell}\,a_c\,e^{2A_c+{4\over3}\l_c}\,R^2\over
(3+2\kappa_\ell)\,(2v_\ell+\kappa_\ell-2)}\,.
\label{tauirasc1}
\ee
Substituting these asymptotics in Eq.~(\ref{xieqs2}) we obtain for $\xi'$ in the IR
\begin{align}
\xi'\sim&-C_a\left({3\over2}\right)^{v_\ell+{\kappa_\ell\over2}}{e^{-4A_c-{8\over3}\l_c}\over
\sqrt{\kappa_c}\,v_c\,
}\,{\tau'\over r^{2-\kappa_\ell-2v_\ell}\,\tau^2}\nonumber\\
=&-C_a\,R^2\,e^{-2A_c}\,{2^{1-{3\over2}\kappa_\ell-v_\ell}\,3^{{3\over2}\kappa_\ell+v_\ell}\,e^{-{4\over3}\l_c}\,a_c
\over \tau_0\,(2v_\ell+\kappa_\ell-2)\,v_c\,\kappa_c^{3/2}}
\,\left({r\over R}\right)^{3\kappa_\ell+2v_\ell}
\,\exp\left[-C_{\rm I}\,\left(r\over R\right)^{3+2\kappa_\ell}\right]\,,
\label{eq:xipirs1}
\end{align}
where the prime stands now for the derivative with respect to the dimensionless variable\footnote{Notice
that, as can be seen from Eq.~\ref{axsol}, $C_a$ has units of $R^4$, and thus the combinations
$C_a\,R^2\,e^{-2A_c}$ in \eqref{eq:xipirs1}, and $C_a\,e^{-4A_c}$ in~\eqref{eq:xiprIRc2} are
 dimensionless.} $r/R$.
$\xi'$ vanishes exponentially, and therefore leads to a regular background in the IR.
Moreover, one
can easily check that for such an exponentially vanishing $\xi'$, all the new terms (proportional to $\xi'$) in the tachyon equation of motion (\ref{taueqs}) are exponentially suppressed, and thus the asymptotic solution (\ref{tauiras1})
is not modified by a nonzero $\xi'$.

\paragraph{2.} $v_p=10/3$, $\kappa_p=4/3$, $\kappa_\ell=-3/2$, $a_p=a_\ell=0$, and $v_\ell>7/4$.
This case results in a tachyon diverging power-like in the IR, namely
\be
\tau(r)=\tau_0\,\left({r\over R}\right)^{C_{\rm II}}\left(1+{\cal O}(r^{-2})\right)\,,
\label{tauiras2}
\ee
where
\be
C_{\rm II}={8\sqrt{2}\,a_c\,e^{2A_c+{4\over3}\l_c}\,R^2\over
3\sqrt{3}\,\kappa_c\,(4v_\ell-7)}\,.
\label{tauirasc2}
\ee
We now analyze the asymptotics of $\xi'$. First, the constraint (\ref{IRreq}),
namely $e^{4A}\,V_{f0}\,\sqrt{\kappa}\,\tau>0$, in terms of the IR asymptotics becomes
\be
C_{\rm II}>2v_\ell-7/2\,,
\label{IRc2const}
\ee
and therefore we have
\be
{7\over4}<v_\ell<{7\over4}+{C_{\rm II}\over2}\,,
\ee
while the equation (\ref{xieqs2}) for $\xi'$ reduces to
\be
\xi'\sim-{C_a\over \tau_0}\left({2\over3}\right)^{{3\over4}-v_\ell}
\,{e^{-4A_c-{8\over3}\l_c}\,C_{\rm II}\over
v_c\,\sqrt{\kappa_c}}\,\left({r\over R}\right)^{-C_{\rm II}-{9\over2}+2v_\ell}\,,
\label{eq:xiprIRc2}
\ee
which vanishes power-like (as $r\to\infty$) if the constraint (\ref{IRc2const}) is satisfied.
In addition, the new terms (proportional to $\xi'$) in the equation of motion
for $\tau$ (\ref{taueqs}) are suppressed by powers of $r$. Then the leading IR asymptotic form
\eqref{tauiras2} for $\tau$ is not modified by a nonzero $\xi'$.

\subsubsection{Special tachyon asymptotics}

We have also found special asymptotic solutions which are absent at vanishing $\bar\theta$-angle. For such special solutions, the two terms in~\eqref{irvineq} have the same asymptotic behavior:
\be \label{irspecbeh}
e^{4A}V_f\sqrt{\kappa} \t \sim |\Ca| V_a\,,\qquad (r \to \infty)\,.
\ee
Such solutions are linked to the regular tachyon solutions discussed above as follows. One may consider what happens as the single parameter $\tau_0$ (the same parameter as $T_0$ in Fig.~\ref{fig:masscontours})  which determines the normalization of the tachyon, decreases. Typically, assuming that the potentials admit a fixed point, there is a critical value $\t_{0c}$ such that the background flows closer and closer to the fixed point as $\t_0 \to \t_{0c}$ from above (or more precisely, the length of the interval of the bulk coordinate where the background is close to the fixed point, increases). Then when $\t_0\le \t_{0c}$, no solution with regular UV behavior exists. But it may also happen that there is no lower boundary for $\t_0$ in which case one can consider the limit $\t_0 \to 0$.

Taking $\t_0 \to 0$ at vanishing $\bar\theta$-angle, one expects that there is a region with small tachyon, $\t \ll 1$, already deep in the IR, where $\l \gg 1$. When there is a fixed point, for small enough $\t_0$ the tachyon is therefore much smaller than one when $\l>\l_*$, in the region corresponding to positive $\beta$-function. Consequently the tachyon decouples from the metric and the flow of the solution from the IR toward the UV stops without reaching the standard regular UV boundary. That is, a critical value $\t_{0c}$ exists, below which the UV regularity is lost. But when the $\bar\theta$-angle is nonzero, the situation is different, because the tachyon is complex. When $\t_0$ is decreased, and $\t$ becomes relatively small deep in the IR, the dynamics of the phase of the tachyon becomes important. As it turns out, the phase starts to evolve, backreacting on the behavior of the absolute value $\t$. Instead of approaching smoothly the origin, the value of the tachyon starts to rotate around it on the complex plane. The net effect is, as pointed out above, that~\eqref{irspecbeh} holds, and the tachyon does not decouple. In the limit $\t_0 \to 0$ one obtains a new asymptotic behavior, which is determined by~\eqref{irspecbeh}.

It is straightforward to compute this special tachyon asymptotics for any of the choices of potentials in the IR discussed above or in~\cite{Arean:2013tja}. As an example, we present the solution for a class of potentials which includes the potentials~I which were used in the numerical analysis in this article. That is, we take $\kappa_p=4/3$, $\kappa_\ell=0$, $a(\l) =\mathrm{const.}$, $V_a(\t) =\exp( -a \t^2 -a_l |\t|)$, and restrict to $0<v_p<10/3$, in which case the regular tachyon asymptotics is an exponential~\cite{Arean:2013tja}. The special asymptotics for the tachyon then reads
\be
 \tau = \frac{10 -3 v_p}{2 a_l} r^2+\frac{2 v_\ell-4}{a_l}\log (r) + \morder{\left(\frac{1}{r}\right)^0} \,.
\ee
The phase of the tachyon behaves as $\xi' \sim 1/\sqrt{r}$ as $r\to \infty$, and
\be
\frac{e^{8A}V_f^2 \kappa \t^2}{\Ca^2 V_a^2} -1 \sim \frac{1}{r}\,,\qquad (r \to \infty)\,,
\ee
so that~\eqref{irspecbeh} is indeed confirmed.

Notice that the special solution does not involve any additional integration constants (unlike the regular tachyon solutions which have one constant). Therefore it maps to a curve on the plane of physical parameters ($m_q/\LUV,\bar\theta$). It may happen though that the solution does not admit a regular UV boundary -- this is not guaranteed even if the dynamics of the complex tachyon prevents the tachyon from decoupling. Actually for the numerical values of parameters used in the numerical analysis of this article, it turns out that the special solution is always unphysical. However, if we decreased the value of $a_l$ in~\eqref{Valindef} from the chosen value, the solution would be physical.

\section{Equations of motion} \label{app:eoms}

In this appendix we present the full equations of motion arising from the action~\eqref{radialaction}, and discuss some of their consequences.
We first write down the equations of motion for the CP-odd fields $\ax $ and $A_M$. They read
\begin{align}
\label{axeom}
 0 &=\partial_M \left[g^{MN}\sqrt{-g}\,Z(\l)\, H_N\right] \,, \\\nonumber
\label{AMeom}
 0 &= 2 V_a(\l,\tau)\, g^{MN}\sqrt{-g}\,Z(\l)\,H_N  \\
  &\phantom{=}-\frac{1}{2}\,V_f(\l,\tau)\,\kappa(\l)\, \t^2\, \left(\partial_N \xi+2A_N\right)\sum_{k=+,-}\!\! \sqrt{-\det{\bf A}_{(k)}}\left(\left({\bf A}^{-1}_{(k)}\right)^{MN}\!+\left({\bf A}^{-1}_{(k)}\right)^{NM} \right)\nonumber \\
  &\phantom{=}+\frac{1}{4}\,\partial_N \!\left[V_f(\l,\tau)\,w(\l)\, \sum_{k=+,-}\!\! k\,\sqrt{-\det{\bf A}_{(k)}}\left(\left({\bf A}^{-1}_{(k)}\right)^{MN}\!-\left({\bf A}^{-1}_{(k)}\right)^{NM} \right)\right] \,,
\end{align}
where
\begin{align}
\label{HMdef}
 H_M &= \partial_M\ax+x\,\xi\,\partial_MV_a(\l,\tau)-2 x\,A_M\,V_a(\l,\tau)\nonumber\\
 &= \partial_M\bar \ax -x\,V_a(\l,\tau)\,\partial_M\xi-2 x\,A_M\,V_a(\l,\tau)\,,
\end{align}
${\bf A}_{(k)}$ were defined in~\eqref{radpmdef}, and $\bar \ax = \ax +x\, \xi\, V_a(\l,\t)$.
Due to invariance under~\eqref{u1transf}, we have\footnote{Considering $\bar\ax$ as the axion field (rather than $\ax$), as we will be doing below, we obtain otherwise the same equation as~\eqref{eomrel}, but without the $\ax$-term.}
\be \label{eomrel}
 \frac{\delta S}{\delta \xi(x)} = x V_a \frac{\delta S}{\delta \ax (x)} -\frac{1}{2} \partial_M \frac{\delta S}{\delta A_M (x)} \ .
\ee
Consequently the equation of motion for $\xi$ follows from~\eqref{axeom} and~\eqref{AMeom}.

The Einstein equations take the form
\be
 R_{MN} - \frac{1}{2}g_{MN} R = T_{MN}^g + T_{MN}^f + T_{MN}^a \,,
\ee
where
\begin{align}
 T_{MN}^g &= \frac{1}{2} g_{MN} \left[V_g-{4\over3}{(\partial_P\lambda)^2\over\lambda^2}\right] +\frac{4}{3}{\partial_M\lambda \partial_N\lambda \over\lambda^2} \,, \\
 T_{MN}^f &= -\frac{x V_f}{8 \sqrt{-g}} \sum_{k=+,-} \sqrt{-\det {\bf A}_{(k)}} \ g_{MP}
 \left[\left({\bf A}_{(k)}^{-1}\right)^{PQ}+\left({\bf A}_{(k)}^{-1}\right)^{QP}
 \right]
 g_{QN} \,,\\
 T_{MN}^a &= -\frac{1}{4} g_{MN}\, Z\, \left(H_P\right)^2 +\frac{Z}{2}\, H_M\, H_N \,.
\end{align}

The remaining equations of motions are those of the scalars $\l$ and $\t$. It is useful to keep $\bar \ax$, rather than $\ax$, fixed when varying the scalar fields. The equations can be written as
\begin{align}
 0&= \frac{8}{3\,\l} \partial_M \left[\frac{g^{MN}\sqrt{-g}\,\partial_N\l}{\l}\right] 
 + \sqrt{-g}\frac{dV_g}{d\l}- \frac{x}{2}\frac{\partial V_f}{\partial \l}\left(\sqrt{-\det{\bf A}_{(+)}}+\sqrt{-\det{\bf A}_{(-)}}\right)    \nonumber\\
 &\phantom{=} - \frac{\sqrt{-g}}{2}\frac{d Z}{d \l} \left(H_M\right)^2+\sqrt{-g}\,Z\, x\, \frac{\partial V_a}{\partial \l}\, \left(\partial_M \xi +2 A_M\right) g^{MN} H_N \nonumber\\
 &\phantom{=}-\frac{1}{4}x\,V_f\,\frac{\partial\kappa}{\partial \l}\, \left[
 (\partial_M \tau)\, (\partial_N \tau)
 +\tau^2(\partial_M \xi+2 A_M)\,(\partial_N \xi+2A_N)
 \right]\sum_{k=+,-}\!\! \sqrt{-\det{\bf A}_{(k)}}\left({\bf A}^{-1}_{(k)}\right)^{MN}\nonumber\\
 &\phantom{=} +\frac{1}{4}x\,V_f\,\frac{\partial w}{\partial \l}\, F_{MN}\sum_{k=+,-}\! k \sqrt{-\det{\bf A}_{(k)}}\left({\bf A}^{-1}_{(k)}\right)^{MN} \,
\end{align}
and
\begin{align}
 0&= \frac{1}{4}\, \partial_M\left[V_f\,\kappa\, \, \partial_N \tau\sum_{k=+,-}\!\! \sqrt{-\det{\bf A}_{(k)}}\left(\left({\bf A}^{-1}_{(k)}\right)^{MN}\!+\left({\bf A}^{-1}_{(k)}\right)^{NM}\right) \right]   \nonumber\\
 &\phantom{=} -\frac{1}{2}\,V_f\,\kappa\, \tau(\partial_M \xi+2 A_M)\,(\partial_N \xi+2A_N)
 \sum_{k=+,-}\!\! \sqrt{-\det{\bf A}_{(k)}}\left({\bf A}^{-1}_{(k)}\right)^{MN} \nonumber\\
 &\phantom{=} -\frac{1}{2}\frac{\partial V_f}{\partial \t}\left[\sqrt{-\det{\bf A}_{(+)}}+\sqrt{-\det{\bf A}_{(-)}}\right] + \sqrt{-g}\,Z\, \frac{\partial V_a}{\partial \t}\, \left(\partial_M \xi +2 A_M\right) g^{MN} H_N \,.
\end{align}

We will now argue that the gauge field vanishes for the background solution. To see this, we consider an Ansatz where all fields depend on $r$ only, and assume the Poincar\'e covariant form of the metric~\eqref{bame}. Since there are other four vectors than $A_\mu$ and no sources which break the Poincar\'e symmetry, we expect that $A_\mu =0$. In order to verify this, we notice that for the background Ansatz ${\bf A}_{(\pm)}$ are diagonal up to terms involving $A_\mu$. Therefore, as only radial derivatives are nonzero, the equations of motion for $A_\mu$ in~\eqref{AMeom} are indeed satisfied for $A_\mu=0$. Moreover, it is convenient to choose the gauge where $A_r=0$. This still leaves the freedom of transforming $\ax$ and $\xi$ by a constant $\varepsilon$ in~\eqref{u1transf}.

After setting the gauge fields to zero, we notice that ${\bf A}_{(+)}={\bf A}_{(-)}$ and both matrices are symmetric so that the last term in~\eqref{AMeom} vanishes. Inserting the $r$-dependent Ansatz, the only nontrivial equation is obtained for $M=r$, which simplifies to
\be \label{Areomfinal}
 V_a(\l,\t) e^{3A}\,Z(\l)\,H_r
  - \frac{V_f(\l,\tau)\,\kappa(\l)\, \t^2\, \xi'\,e^{3A}}{\tG}  = 0
\ee
with $\tG$ defined in~\eqref{gdef}.

\section{On the phase diagram at finite $\bar\theta$-angle} \label{app:thetabackgrounds}

We discuss here first the the branch structure and/or uniqueness of the background solutions at finite $\bar\theta$-angle.
In Fig.~\ref{fig:thetavsC} and at small quark mass, there are two or zero solutions at fixed $C_a$ and $m_q$, but as functions of the sources $m_q$ and $\bar\theta$ the solutions are typically  unique. One should notice, however, that these plots do not contain all possible solutions as we did not study the solutions near the Efimov vacua, \ie close to the leftmost crosses of~Fig.~\ref{fig:masscontours}. The sketch of Fig.~\ref{fig:masscontours} suggests that the mapping from $(T_0,\Ca)$ to ($m_q,\bar\theta$) is bijective also in the regime of Efimov vacua. This indeed turns out to be the case for generic values of $x$, but (as we demonstrate in Sec.~\ref{sec:smallmq}) for very small values of $x$ the mapping is not bijective: as $x \to 0$ a nontrivial branch structure as a function of $\bar\theta $ appears in the vicinity of the Efimov vacua.

Even at larger $x$ where the
nontrivial branch structure is absent, one should bear in mind that the $\bar\theta$-angle is periodic.
Near different crosses of Fig.~\ref{fig:masscontours} we encounter backgrounds for
which $\bar\theta$ differs by integer multiples of $2 \pi N_f$, but the quark mass is the same: this corresponds to
the change of phase of the tachyon by multiples of $2\pi$ and therefore the backgrounds cannot be distinguished
by using UV data. Consequently, solutions at high $m_q$ are unique, but the solutions which differ by $2\pi$
rotations of the tachyon in the UV appear at small $m_q$ and generalize the Efimov vacua to finite $\bar\theta$-
angle. The number of Efimov vacua grows with decreasing $m_q$ and becomes infinite for $m_q=0$.

We argue  now that the generalized Efimov vacua at finite $\Ca$ are unstable.
For fixed $m_q/\LUV \ll 1$, such Efimov vacua are found on half-rings that encircle the crosses of Fig.~\ref{fig:masscontours}. We have shown in~\cite{jk,Arean:2013tja,Jarvinen:2015ofa} that for $\Ca=0$ the Efimov vacua are perturbatively unstable. We remind the reader that in Fig.~\ref{fig:masscontours}, the Efimov vacua for $\Ca=0$ and with fixed (absolute value of) quark mass are found on the horizontal axis for various discrete values of $T_0$ and the stable standard vacuum is found on the horizontal axis to the right of the Efimov vacua. For these vacua, the free energy decreases with $T_0$.
This was proven analytically for vacua with high $n$ (\ie close to the dashed curve in Fig.~\ref{fig:masscontours}) and numerically for vacua with low $n$.
For the Efimov vacua at finite $\Ca$, the free energy on the half-rings around the crosses is typically monotonic, as is seen from the plots of Fig.~\ref{fig:Evstheta}. When this is the case, the question of ordering the saddle points at $\Ca \ne 0$ according to their energies simply boils down to the same question for the saddle points at $\Ca=0$. There are also cases (as one can see from the analysis of Sec.~\ref{sec:EFT}, and the discussion in Sec.~\ref{sec:smallmqanyx}), in particular at small $x$, where the energy is not monotonic on the half-rings. Even in this case, the variation of the free energy along the half-rings is of the order $\sim m_q |\langle\bar\psi\psi\rangle|_{n,m_q=0}$ where the chiral condensate is evaluated at the corresponding Efimov vacua at $m_q=0$. Because the condensate at the Efimov vacua are strongly suppressed with respect to the standard vacuum (see, e.g., Fig.~9  in~\cite{Jarvinen:2015ofa}, where the Efimov vacua are found near the origin), this variation is too small to overcome the energy difference between the Efimov vacua and the standard vacuum.

In Sec.~\ref{sec:finitethspc} we will also demonstrate numerically that even the solutions near the $m_q=0$ standard vacuum (the rightmost cross in Fig.~\ref{fig:masscontours}) are perturbatively unstable if $|\bar\theta| \gtrsim N_f \pi/2$. Therefore, all vacua in Fig.~\ref{fig:masscontours} left of the standard vacuum, \ie for $|\bar\theta| \gtrsim N_f \pi/2$, are perturbatively unstable. Moreover, if we take into account all branches of solutions, which are obtained by shifting $\bar\theta$ by multiples of $2\pi$, the dominant solutions are found in the immediate vicinity of the $\bar\theta=0$ standard solutions (marked with the blue line in Fig.~\ref{fig:masscontours} (right)), as we argue in Sec.~\ref{free}.

As shown on the horizontal axis of Fig.~\ref{fig:masscontours} (right), $\bar\theta$ takes values
quantized in units of $\pi N_f$ on this axis, corresponding to phase shifts in units of $\pi$ of the tachyon.
In order to prove this, we analyze the behavior of the tachyon solution near the horizontal axis of
Fig.~\ref{fig:masscontours}. As $\Ca \to 0$, the solution approaches smoothly the real valued solution
having $\Ca=0$ exactly. Between the rightmost cross on the real axis and the next cross to the left in
Fig.~\ref{fig:masscontours} (left), the (real part of the) tachyon at $\Ca=0$ has a single node at some $r = r_0$ -- notice that indeed such a
solution has\footnote{As we discussed above, in this plot the quark mass is defined as the real part of
the source for the tachyon (whereas at finite $\bar\theta$ we define $m_q$ as the absolute value of the source).
Therefore negative values are possible.} $m_q<0$ in the plot of Fig.~\ref{fig:massdepzerotheta}.
By definition $\tau$ is positive in our analysis at $\Ca \ne 0$, so a change of sign in the real part of
the tachyon must be realized through a shift in the tachyon phase $\xi$. From~\eqref{xieqs2} we see
that $\xi'$ is positive for positive $\Ca$, so the phase must jump by $+\pi$ at the node of the tachyon
as $\Ca \to 0$ from above: we have
\be
 \xi'(r) \to \pi \delta(r-r_0)\,,
\ee
where $r_0$ is the location of the node. By using~\eqref{axsol} for the gauge invariant contribution we obtain
\be
 \bar \ax ' = \ax ' + x\, (\xi\, V_a)' = x\, \xi'\, V_a + \morder{\Ca}  \to x\, \pi \delta(r-r_0)
\ee
as $\Ca \to 0$ from above, where we used the fact that $V_a =1$ at the tachyon node (as the tachyon vanishes by definition, $\t(r_0)=0$). Consequently,
\be
 \bar\theta  = N_c\, \bar \ax_0 \to - \pi N_f \,.
\ee
For solutions with more tachyon oscillations, which can be found on the horizontal axis of Fig.~\ref{fig:masscontours} closer to the dashed curve, one just needs to sum over the contributions from separate tachyon nodes. One finds that $\bar\theta \to -n N_f \pi$ as $\Ca \to 0$ from above, where $n$ is the number of nodes.

An interesting possibility is that the white region of Fig.~\ref{fig:masscontours} at small $T_0$ is absent at
large $\Ca$ so that the dashed curve ends on the vertical axis. Such behavior is observed for potentials~I for
some (small) choices of the coefficient $a_l$ in the function $V_a(\tau) = \exp(-a_q\,\t^2 -a_l\,\t)$
(but not for the choice of $a_l$ used in the numerical studies of this article).
This means that solutions exist at arbitrary small $T_0$, so it is natural to ask what happens in the
limit $T_0 \to 0$. As it turns out, the tachyon does not vanish in this limit, but assumes an asymptotic
behavior in the IR, which is different from the standard regular IR asymptotics
(see Appendix~\ref{app:Asymptotics}). For this special asymptotics, the two terms in~\eqref{irvineq} have the
same IR behavior:
\be
 e^{4A}V_{f}(\l,\t) \sqrt{\kappa(\l)} \t \simeq |\Ca| V_a(\t) \,, \qquad (r \to \infty)\,.
\ee
For potentials~I with the above choice of $V_a$ the tachyon diverges as $\t \sim r^2$, while the phase behaves as $\sim \sqrt{r}$ at large $r$ (see the end of Appendix~\ref{app:Asymptotics} for precise treatment). This behavior is enough to decouple the tachyon from the metric, which consequently follows the usual (Yang-Mills) asymptotics.
Unlike with the standard asymptotics, this tachyon asymptotics   involves no free parameters, and the only free parameter for this kind of solutions is $\Ca$. Therefore the solutions would define a curve on the $(m_q,\bar\theta)$-plane.

\section{Fluctuation equations for the singlet pseudoscalars at $\bar\theta=0$}
\label{app:quadfluctdet}
In \cite{Arean:2013tja} the whole set of fluctuations of V-QCD was studied at $\bar\theta=0$. Here we will focus on those contributing
to the singlet pseudoscalar sector. We need only consider the flavor singlet axial vector,
the phase of the tachyon, and the axion field.

The fluctuations of the left and right gauge fields can be written in terms of the vector and axial combinations
\be
V_M = \frac{A_M^L + A_M^R}{2}\, ,\qquad
A_M = \frac{A_M^L - A_M^R}{2}\,,
\label{VAdefstext}
\ee
with the associated field strengths being $V_{MN}$, $A_{MN}$.
For the axial vectors we first need to separate the transverse and longitudinal parts:
\be
A_\m (x^\mu, r)=A^\bot_\m (x^\mu, r)+ A^{\lVert}_\m(x^\mu, r) \,,
\ee
where $\partial^\n A^\bot_\n (x^\mu, r) = 0$, and the longitudinal term is the divergence of a scalar function.

For the axial vector modes we need to treat the flavor non-singlet and flavor singlet terms separately, and therefore we write
\begin{equation}
  A^{\bot}_\m(x^\mu, r) = A^{\bot F}_\m(x^\mu, r)+ A^{\bot S}_\m(x^\mu, r)\,,\quad
  A^{\lVert}_\m(x^\mu, r) = A^{\lVert F}_\m(x^\mu, r)+ A^{\lVert S}_\m(x^\mu, r) \, .
\label{ansaxlong}
\end{equation}
where the superscript $S$ ($F$) stands for the flavor singlet (non-singlet) part of the mode.
In the following we will deal only with the longitudinal flavor singlet part of the longitudinal axial vector mode
$A^{\lVert S}$, since
only this part contributes to the action of the singlet pseudoscalar sector.

On the other hand, the fluctuations of the tachyon are given by
\begin{equation}
 T=(\tau+s+\mathfrak{s}^at^a)e^{i\thf+i\,\pi^at^a}\,,
 \label{tfluctdef}
\end{equation}
where $t^a$ are the generators of $SU(N_f)$.
We are mostly interested in the standard vacuum for $0<x<x_c$ which gives rise to a nontrivial spectrum
\cite{Arean:2013tja}. Therefore, the background solution $\t(r)$ is nonzero and the phases
$\thf$, $\pi^a$ in~\eqref{tfluctdef} are well defined.

\subsection{Flavor singlet pseudoscalar mesons}

The quadratic action for all the fluctuations of V-QCD was computed in \cite{Arean:2013tja}.

Here we will write down the two pieces contributing to the flavor singlet pseudoscalar sector: $S_1$ coming from the DBI
piece (\ref{generalact}), and $S_2$ from the CP-odd sector (\ref{samain}). We write each separately:
\begin{align}
\label{pscalu1s1}
S_1\ =&-M^3 N_c^2\,{x\over2} \int d^4 x\, dr\,V_f(\l,\t)\, e^{\Awf}\, G^{-1}
\\\nonumber
&\times\bigg[
\gf(\l,\t)^2\,\left(\partial_r A^{\lVert S}_\m\right)^2 + e^{2\Awf}\h(\l,\t)\,\t^2\,(\partial_r\thf)^2
+e^{2\Awf}\,\G^2\,\h(\l,\t)\,\t^2\left(\partial_\m\thf+2 A^{\lVert S}_\m\right)^2
\bigg]\,,
\end{align}
and
\be
S_2=-{M^3\, N_c^2\over2} \int d^4 x\, dr\,Z(\l)\,e^{3\Awf}\left[
\left(\partial_\m a-2x\,V_a(\l,\t)\, A^{\lVert S}_\m\right)^2
+\left(\partial_r a+x\,\partial_r V_a(\l,\t)\,\thf\right)^2
\right]\,,\label{pscalu1s2}
\ee
where
\be
 G = \sqrt{1+e^{-2A}\kappa(\l) \t'^2}
\ee
and we have set $A_r=0\,$.

We split the fields in the action as
\begin{align}
A^{\lVert S}_\m(x^\mu, r) &= -\varphi_L(r)\,\partial_\m({\cal T}(x^\m))\,,\nonumber\\
\thf(x^\mu, r) &=2\varphi_\theta(r)\,{\cal T}(x^\m)\,,\nonumber\\
a(x^\mu, r) &=2\varphi_\mathrm{ax}(r)\,{\cal T}(x^\m)\,.
\label{defsamainapp}
\end{align}
The following combinations of the above fields
\begin{align}
P(r)&\equiv\varphi_\theta(r)-\varphi_L(r)\,,\nonumber\\
Q(r)&\equiv\varphi_\mathrm{ax}(r)+x\,V_a(\l,\t)\,\varphi_L(r)\,,\nonumber\\
R(r)&\equiv \varphi_\mathrm{ax}'(r)+x\,\partial_r V_a(\l,\t)\,\varphi_\theta(r)\,,
\label{ginvpsmain}
\end{align}
are invariant under the residual gauge transformations (\ref{u1transf}).
Only two of them are
independent, and they realize the pseudoscalar glueball ($0^{-+}$) and $\eta'$ meson towers.

Indeed, following~\cite{Arean:2013tja} $R$ can be eliminated from the fluctuation equations, which are found by varying Eqs.~\eqref{pscalu1s1} and~\eqref{pscalu1s2}. The result may be written as
\begin{align}
&\partial_r\left[V_f\,e^{\Awf}\,\G^{-1}\,\gf^2\left(
-4e^{2\Awf}\,{V_f\,\h\,\t^2\over N_a+N_b}\,P'+{V_a'\over V_a}\,{N_b\over N_a+N_b}\,P
+{N_b\over x\,V_a\,(N_a+N_b)}Q'\right)\right]+\nonumber\\
&+4V_f\,e^{3\Awf}\,\G\,\h\,\t^2\,P-4e^{3\Awf}\,Z\,V_a\,Q=0\,,\label{cpsys1}\\ \nonumber \\
&\partial_r\left[e^{3\Awf}\,Z\,\left(
4x\,e^{2\Awf}\,{V_a\,V_f\,\h\,\t^2\over N_a+N_b}\,P'+x\,{V_a'\,N_a\over N_a+N_b}\,P
+{N_a\over N_a+N_b}Q'\right)\right]+m^2\,e^{3\Awf}\,Z\,Q=0\,,\nonumber\\ \label{cpsys2}
\end{align}
where  $N_a$ and $N_b$ are given by the following expressions:
\be \label{NaNbdefapp}
N_a=V_f\left(4e^{2\Awf}\,\h\,\t^2-m^2\,\gf^2\right)\, ,\qquad
N_b=4x\,e^{2\Awf}\,Z\,V_a^2\,\G\,.
\ee

{\bf Change of variables}:
For the analysis of the mass of the $\eta'$meson, it is useful to define the conjugate variables of $P$ and $Q$:
\begin{align} \label{hatteddefs}
\Psi = \left(\begin{array}{c}
           \hat P\\\hat Q
          \end{array}\right)= \nonumber\\
        = \frac{1}{N_a+N_b}&\left(
\begin{array}{cc}
\frac{V_f e^A\gf^2}{4 G}\left(-N_b-4 e^{2 A}V_f \kappa  \tau ^2\right) & \frac{V_f e^A\gf^2}{4 G}\frac{N_b}{x V_a} \\
e^{3 A}Z\left(4 x e^{2 A}V_a V_f \kappa  \tau ^2-N_a x V_a\right)m^{-2} & e^{3 A}Z N_a m^{-2}
\end{array}
\right)
\left(\begin{array}{c}
            P'\\ (Q + x\, V_a\, P)'
          \end{array}\right).
          \end{align}
Then the terms in the square brackets in~\eqref{cpsys1} and~\eqref{cpsys2} can be expressed in terms of $\hat P$ and $\hat Q$. Taking a suitable linear combination of these equations, differentiating once, and after some simplifications, the fluctuation equations can be written as
\be
\Psi'' + C_1^{-1}\, \partial_r C_1\, \Psi' + M\, \Psi = 0\,,
\ee
where
\begin{align}
C_1 &= e^{-3 A}\left(
\begin{array}{cc}
\frac{1}{G V_f \kappa  \tau ^2} & \frac{V_a}{G V_f \kappa  \tau ^2} \\
\frac{V_a x}{G V_f \kappa  \tau ^2} & \frac{1}{Z}+\frac{V_a^2 x}{G V_f \kappa  \tau ^2}
\end{array}
\right)\,,\nonumber \\ \nonumber \\
M &= \left(
\begin{array}{cc}
G^2 m^2-\frac{4 e^{2 A} \left( GV_a^2 x Z+G^2 V_f \kappa  \tau ^2 \right)}{V_f \gf^2} & e^{-2A}\h \t'^2 m^2 V_a \\
\frac{4 e^{2 A} G V_a x Z}{V_f \gf^2} & m^2
\end{array}
\right)\,.
 \end{align}

\subsection{Flavor singlet axial vector mesons}
It was shown in \cite{Arean:2013tja} that the action for the singlet sector of the (transverse) axial
vector modes has an extra term coming from the CP-odd sector. The action is given by
\begin{align}
S_A\ =& -{M^3 N_c^2\over2}\!\int\! d^4x\, dr
\bigg\{ x\,V_f(\l,\t)\, e^\Awf\, \G^{-1}\!\bigg[\,
{1 \over 2}\,\G^2 \, \gf(\l,\t)^2 A_{\mu\nu}\,A^{\mu\nu} +\nonumber\\
&+
\,\gf(\l,\t)^2\, \partial_r A^{\bot S}_\mu\, \partial_r A^{\bot S\,\mu}+4\h(\l,\t)\,\tau^2\, e^{2 \Awf}\,\G^{2}\,A^{\bot S}_\mu A^{\bot S\,\mu}\bigg]+ \nonumber\\
&+4x^2\,Z(\l)\,e^{3\Awf}\,V_a(\l,\t)^2\,A^{\bot S}_\mu A^{\bot S\,\mu}
\bigg\} \,,
\label{axacti2}
\end{align}
where indeed the last contribution originates from the CP-odd action (\ref{samain}) while all the other terms come from the
flavor sector piece (\ref{generalact}). Taking the following Ansatz:
\begin{equation}
A^{\bot S}_\m(x^\mu, r) =\varphi_A(r)\, {\cal X}_\m(x^\m)\,,
\end{equation}
the resulting fluctuation equation takes the form
\begin{align}
&\frac{1}{V_f(\l,\t)\,\gf(\l,\t)^2\,e^\Awf \G  }\,
\partial_r \left( V_f(\l,\t)\, \gf(\l,\t)^2\, e^{\Awf}
\G^{-1} \partial_r \varphi_A \right)
+m_V^2 \varphi_A+\nonumber\\
&-4\left[x\,e^{2\Awf}{Z(\l)\,V_a(\l,\t)^2\over V_f(\l,\t)\,\G\,\gf(\l,\t)^2}+ {\t^2 e^{2 \Awf}
\over \gf(\l,\t)^2}\h(\l,\t)\right]\varphi_A= 0\,.
\label{axvectoreom2}
\end{align}

\section{Fluctuations of the $\bar\theta$-backgrounds}
\label{app:thbfluc}

In this section we will derive the equations of motion for the
fluctuations in backgrounds corresponding to nonzero $\bar\theta$-vacua of the dual theory.
As mentioned in Sec. \ref{ssec:thbfluc} we will only consider the flavor non-singlet sector,
which consists of the vector and axial vector mesons, together with the scalar and pseudoscalar
fluctuations of the complex tachyon which we write as
\begin{align} \label{tachyonpaf}
&T=\left[\tau(r)+s(r,x)+\tilde s(r,x)\right]\exp\left[\xi(r)+\thf(r,x)+\tilde\pi(r,x)\right]\,,\\\nonumber
& {\rm with}\quad \tilde s(r,x) = \mathfrak{s}^a(r,x)\,t^a\,,\qquad \tilde \pi(r,x) = \pi^a(r,x)\,t^a\,.
\end{align}
Only the DBI piece of the action, given by Eq. \eqref{generalact}, contributes to the non-singlet sector
fluctuations. Up to quadratic order in the fluctuations it reads
\begin{align}
\label{pscalu1s1tb}
S_1\ =&-{1\over2}M^3 N_c\,{\mathbb Tr} \int d^4x\, dr\,V_f(\l,\t)\, e^{3\Awf}\,\tG^{-1}
\\\nonumber
&\times\bigg\{
\left[2\h\, {\partial_\t V_f  \over V_f }\,\t'+(\partial_\t\h)\,\t'\left(1+\tG^{-2}\right)
-2e^{-2A}\,\tG^{-2}\h^2\,\t\,\t'\,\xi'^2\right]
\tilde s\,\tilde s'
\\\nonumber
&+\bigg[
{\partial_\t^2 V_f  \over V_f }\,e^{2A}\,\tG^2+{\partial_\t V_f  \over V_f }\,(\partial_\t \h)\,(\t'^2+\t^2\,\xi'^2)
+2{\partial_\t V_f  \over V_f }\,\h\,\t\,\xi'^2+{\partial_\t^2 \h  \over 2}\,(\t'^2+\t^2\,\xi'^2)\\\nonumber
&+(\partial_\t\h)(1+\tG^{-2})\,\t\,\xi'^2-{e^{-2A}  \over 4\tG^2}(\partial_\t\h)^2(\t'^2+\t^2\,\xi'^2)^2
+\h\,\tG^{-2}\,(1+\h\,e^{-2A}\,\t'^2)\,\xi'^2\bigg]\tilde s^2\\\nonumber
&+\left[\h\,\tG^{-2}(1+\h\,e^{-2A}\,\t^2\,\xi'^2)\right]\tilde s'^2
+\left[\h\,(1+\h\,e^{-2A}\,\t^2\,\xi'^2)\right](\partial_\mu\tilde s)^2\\\nonumber
&+\left[\h\,\tG^{-2}\,\t^2(1+\h\,e^{-2A}\,\tau'^2)\right]\tilde\pi'^2
+\left[\h\,\t^2(1+\h\,e^{-2A}\,\t'^2)\right](\partial_\mu\tilde \pi+2A_\mu)^2\\\nonumber
&+2\xi'\left[{\partial_\t V_f  \over V_f }\,\h\,\t^2+\h\,\t\left(2-\tG^{-2}\,e^{-2A}\,\h\,\t^2\,\xi'^2\right)
+{\t^2\over2}\,(\partial_\t\h)(1+\tG^{-2})\right]\tilde s\,\tilde\pi'\\\nonumber
&-\left[2e^{-2A}\,\tG^{-2}\,\h^2\,\t^2\,\t'\,\xi'\right]\tilde s'\,\tilde\pi'
-\left[2e^{-2A}\h^2\,\t^2\,\t'\,\xi'\right](\partial_\mu\tilde\pi +2A_\mu)\,\partial^\mu\tilde s\\\nonumber
&+w^2\,e^{-2A}\left[{\tG^2\over2}(A_{\mu\nu}\,A^{\mu\nu}+V_{\mu\nu}\,V^{\mu\nu})+A_\mu'^2+V_\mu'^2\right]
\bigg\}\,.
\end{align}
The vector and axial vector mesons correspond to the transverse part of the fields $V_\m$ and $A_\m$
respectively, and their equations of motion can be readily obtained from the above Lagrangian.
They are Eqs. (\ref{thnsvectoreom}) and (\ref{thnsaxveom}) in the main text.

\subsection{Scalar-pseudoscalar mesons}

This sector consists of the fluctuations of the modulus and phase of the tachyon, and the longitudinal
part of the axial vector, which we decompose as
\begin{align} \label{spsdecomp}
A^{\lVert F}_\mu=-\psi_L(r)\,\partial_\mu P^a(x)\,t^a\,,\qquad
\tilde s = \psi_s(r)\,P^a(x)\,t^a\,,\qquad
\tilde\pi=2\psi_p(r)\,P^a(x)\,t^a\,,
\end{align}
where $\partial_\mu\partial^\mu P^a(x) = m^2\, P^a(x)$.
The corresponding equations of motion read:
\begin{align}
&{\tG\over V_f\,e^{3A}}\,\partial_r\left[
V_f{e^{3A}\over \tG}\left(4P_2\,\psi_p'+M_1\,\psi_s+M_2\,\psi_s'\right)\right]+
m^2\left[4P_1\,(\psi_p-\psi_L)+M_3\,\psi_s\right]=0\,,\label{eompsip}\\
&{\tG\over V_f\,e^{3A}}\,\partial_r\left(V_f\,{e^A\over \tG}\,w^2\,\psi_L'\right)+4P_1\,(\psi_p-\psi_L)
+M_3\,\psi_s=0\,,\label{eompsil}\\
&\label{eompsis}
{\tG\over V_f\,e^{3A}}\,\partial_r\left[V_f{e^{3A}\over \tG}\left(2S_3\,\psi_s'+S_1\,\psi_s+2M_2\,\psi_p'\right)\right]\\\nonumber
&\quad+2m^2\,M_3\,(\psi_p-\psi_L)+2m^2\,S_4\,\psi_s
-S_1\,\psi_s'-2S_2\,\psi_s-2M_1\,\psi_p'=0\,,
\end{align}
where the different coefficients are given by
\begin{align} \label{coefstheq}
& P_1=\h\,\t^2(1+\h\,e^{-2A}\,\t'^2)\,,\qquad P_2=\h\,\tG^{-2}\,\t^2(1+\h\,e^{-2A}\,\t'^2)
\,,\\\nonumber
& S_1=2\h\, {\partial_\t V_f  \over V_f }\,\t'+(\partial_\t\h)\,\t'\left(1+\tG^{-2}\right)
-2e^{-2A}\,\tG^{-2}\h^2\,\t\,\t'\,\xi'^2\,,\\\nonumber
&S_2=\bigg[
{\partial_\t^2 V_f  \over V_f }\,e^{2A}\,\tG^2+{\partial_\t V_f  \over V_f }\,(\partial_\t \h)\,(\t'^2+\t^2\,\xi'^2)
+2{\partial_\t V_f  \over V_f }\,\h\,\t\,\xi'^2+{\partial_\t^2 \h  \over 2}\,(\t'^2+\t^2\,\xi'^2)\\\nonumber
&\qquad+(\partial_\t\h)(1+\tG^{-2})\,\t\,\xi'^2-{e^{-2A}  \over 4\tG^2}(\partial_\t\h)^2(\t'^2+\t^2\,\xi'^2)^2
+\h\,\tG^{-2}\,(1+\h\,e^{-2A}\,\t'^2)\,\xi'^2\bigg]\,,\\\nonumber
&S_3=\h\,\tG^{-2}(1+\h\,e^{-2A}\,\t^2\,\xi'^2)\,,
\qquad S_4=\h\,(1+\h\,e^{-2A}\,\t^2\,\xi'^2)\,,\\\nonumber
&M_1=2\xi'\left[{\partial_\t V_f  \over V_f }\,\h\,\t^2+\h\,\t\left(2-\tG^{-2}\,e^{-2A}\,\h\,\t^2\,\xi'^2\right)
+{\t^2\over2}\,(\partial_\t\h)(1+\tG^{-2})\right]\,,\\\nonumber
&M_2=-2e^{-2A}\,\tG^{-2}\,\h^2\,\t^2\,\t'\,\xi'\,,\quad M_3=-2e^{-2A}\,\h^2\,\t^2\,\t'\,\xi'\,.
\end{align}
The equations (\ref{eompsip} - \ref{eompsis}) can be recombined into two equations for two variables.
They read
\begin{align}
&\psi_s''+\partial_r(\log C_1)\psi_s'+{1\over C_1}\left(M+\partial_r C_3+m^2\,C_2\right)\psi_s+{m^2\over2 C_1}\left[
\partial_r\left({M_2\over P_2}\right)-{M_1\over P_2}\right]\hat\psi_l=0\,,\label{nsfpsmesons1}\\
&K\,P_2\,\partial_r\left({\hat\psi_l'\over K\,P_1}\right)-4e^{2A}\,{P_2\over w^2}\,\hat\psi_l+m^2\hat\psi_l
+K\left[P_2\,\partial_r\left({M_3\over P_1}\right)-M_1\right]\psi_s
=0\,,\label{nsfpsmesons2}
\end{align}
in terms of
\begin{align}
&\hat\psi_l=e^{A}\,w^2\,V_f\,\tG^{-1}\,\psi_L'\,,\\
&K=e^{3A}\,V_f\,\tG^{-1}\,,\qquad
M=K\left({M_1^2\over 2P_2}-2S_2\right)\,,\qquad
C_1=2e^{3A}\,V_f\,\h\,\tG^{-3}
\,,\\
&C_2=2\h\,e^{3A}\,V_f\,{\tG\over G^2}
\,,\qquad
C_3=K\left(S_1-2{M_1\,M_2\over4P_2}\right)\,.
\end{align}
In addition, one can solve for $\psi_p$ from the equation
\be
\psi_p'={m^2\over4}\,{1\over K\,P_2}\,\hat\psi_l-{M_2\over4P_2}\,\psi_s'-{M_1\over 4P_2}\,\psi_s\,.
\ee

\section{UV asymptotics of the meson wave functions} \label{app:Asymptoticswf} \label{app:UVAsymptotics}
In this section we shall study the UV ($r\to0$) asymptotic solutions of the system of coupled differential equations
(\ref{cpsys1})-(\ref{cpsys2}) satisfied by the wave functions of the flavor singlet pseudoscalar modes.

We will first study the system (\ref{cpsys1})-(\ref{cpsys2}) in the UV ($r\to0$) region.
The UV asymptotics of the background were presented in Appendix \ref{subapp:UVback}.
We will consider the following UV asymptotics for the remaining potentials determining the action
of the fluctuations:
\begin{align}
\label{potsmiscUV}
V_{f0}(\l) &\sim W_0\left(1+W_1\,\l\right)\,,\qquad
V_a(\l,\t)\sim  1 - b_0 \tau^2 \,,\nonumber \\
Z(\l)&\sim Z_0\left(1+c_1 \lambda
\right)\,,\qquad\quad
w(\l) \sim w_0 \left(1+w_1 \lambda \right)\,.
\end{align}

\subsection{Zero quark mass}

The UV expansions at $m_q = 0$ are given by
\begin{align}
\label{CpUVsolmq0}
\left(
\begin{array}{c}
 P \\
 Q \\
 \hat{P} \\
 \hat{Q}
\end{array}
\right)&=C_+ \left(
\begin{array}{c}
 -r^{\alpha _+} (-\log (r \Lambda ))^{p_+} \left(1+\mathcal{O}\left(\log (r \Lambda )^{-1}\right)\right) \\
 x r^{\alpha _+}  (-\log (r \Lambda ))^{p_+} \left(1+\mathcal{O}\left(\log (r \Lambda )^{-1}\right)\right)\\
 \frac{x r^{-2+\alpha _+} \ell ^3  (-\log (r \Lambda ))^{p_+} Z_0}{-2+\alpha _+} \left(1+\mathcal{O}\left(\log (r \Lambda )^{-1}\right)\right)\\
 \frac{x r^{-2+\alpha _+} \ell ^3(-\log (r \Lambda ))^{p_+} Z_0}{2-\alpha _+} \left(1+\mathcal{O}\left(\log (r \Lambda )^{-1}\right)\right)
\end{array}
\right)
\\
\label{CmUVsolmq0}
&+C_- \left(
\begin{array}{c}
 -r^{\alpha _-}  (-\log (r \Lambda ))^{p_-} \left(1+\mathcal{O}\left(\log (r \Lambda )^{-1}\right)\right)\\
 x r^{\alpha _-} (-\log (r \Lambda ))^{p_-}\left(1+\mathcal{O}\left(\log (r \Lambda )^{-1}\right)\right)  \\
 \frac{x r^{-2+\alpha _-} \ell ^3 (-\log (r \Lambda ))^{p_-} Z_0}{-2+\alpha _-} \left(1+\mathcal{O}\left(\log (r \Lambda )^{-1}\right)\right) \\
 \frac{x r^{-2+\alpha _-} \ell ^3 (-\log (r \Lambda ))^{p_-} Z_0}{2-\alpha _-}\left(1+\mathcal{O}\left(\log (r \Lambda )^{-1}\right)\right)
\end{array}
\right)
\\
\label{C1UVsolmq0}
&+C_1 \left(
\begin{array}{c}
 -\frac{m^2 r^{-2} (-\log (r \Lambda ))^{-2 \rho }}{2 \ell ^5 \s^2 W_0 \kappa _0} \left(1+\mathcal{O}\left(\log (r \Lambda )^{-1}\right)\right) \\
 -\frac{m^2 r^4  \left(-3 x b_0 w_0^2 Z_0+w_0^2 W_0 \kappa _0+x \ell ^2 Z_0 \kappa _0\right)}{2 \ell ^3 Z_0 \left(-2 w_0^2 W_0+x \ell ^2 Z_0\right) \kappa _0} \left(1+\mathcal{O}\left(\log (r \Lambda )^{-1}\right)\right)\\
 \frac{3 m^2 r^2  w_0^2 \left(-x b_0 Z_0+W_0 \kappa _0\right)}{8 w_0^2 W_0 \kappa _0-4 x \ell ^2 Z_0 \kappa _0} \left(1+\mathcal{O}\left(\log (r \Lambda )^{-1}\right)\right)\\
 1-\frac{m^2 r^2  \left(-3 x b_0 w_0^2 Z_0+w_0^2 W_0 \kappa _0+x \ell ^2 Z_0 \kappa _0\right)}{8 w_0^2 W_0 \kappa _0-4 x \ell ^2 Z_0 \kappa _0}\left(1+\mathcal{O}\left(\log (r \Lambda )^{-1}\right)\right)
\end{array}
\right)
\\
\label{C2UVsolmq0	}
&+C_2 \left(
\begin{array}{c}
 1-\frac{m^2 r^2  \log (r \Lambda )}{ 8 } \left(1+\mathcal{O}\left(\log (r \Lambda )^{-1}\right)\right)\\
 \frac{r^6 \ell ^2  (-\log (r \Lambda ))^{2 \rho } \s^2 x W_0 \left( \ell^2 \kappa _0 - 6 b_0 w_0^2\right)}{x \ell^2 Z_0 - 6 W_0 w_0^2} \left(1+\mathcal{O}\left(\log (r \Lambda )^{-1}\right)\right)\\
 \frac{r^4 \ell ^5  (-\log (r \Lambda ))^{2 \rho } \s^2 w_0^2 W_0 \left(x b_0 Z_0-W_0 \kappa _0\right)}{12 w_0^2 W_0 - 2\ell^2 x Z_0} \left(1+\mathcal{O}\left(\log (r \Lambda )^{-1}\right)\right)\\
 \frac{r^4 \ell ^5  (-\log (r \Lambda ))^{2 \rho } \s^2 x W_0 \kappa _0\left( \ell^2 \kappa _0 - 6 b_0 w_0^2\right)}{-4x \ell^2 Z_0 +24  W_0 w_0^2}\left(1+\mathcal{O}\left(\log (r \Lambda )^{-1}\right)\right)
\end{array}
\right) \,,
\end{align}
where we have defined
\be
 \alpha_\pm = 1\pm \sqrt {1+{4 \,x \, \ell^2 \, Z_0 \over W_0 \, w_0^2 }}
\ee
and
\begin{align}
p_\pm & =\pm{16 \,x \, \ell^2  \, Z_0 \left(V_1-c_1-W_1-2w_1\right) \over 9 V_1 W_0 \, w_0^2 \sqrt{1+{4 \,x \, \ell^2  \, Z_0 \over W_0 \, w_0^2 }} } \,.
\end{align}
The parameter $\rho$ was defined in~\eqref{rhodef} and $m$ is the mass of the fluctuation.

\subsection{Finite quark mass}

The UV expansions at finite $m_q$ are given by
\begin{align}
\label{CpUVsol}
\left(
\begin{array}{c}
 P \\
 Q \\
 \hat{P} \\
 \hat{Q}
\end{array}
\right)&=C_+ \left(
\begin{array}{c}
 -r^{\alpha _+} (-\log (r \Lambda ))^{p_+} \left(1+\mathcal{O}\left(\log (r \Lambda )^{-1}\right)\right) \\
 x r^{\alpha _+}  (-\log (r \Lambda ))^{p_+} \left(1+\mathcal{O}\left(\log (r \Lambda )^{-1}\right)\right)\\
 \frac{x r^{-2+\alpha _+} \ell ^3  (-\log (r \Lambda ))^{p_+} Z_0}{-2+\alpha _+} \left(1+\mathcal{O}\left(\log (r \Lambda )^{-1}\right)\right)\\
 \frac{x r^{-2+\alpha _+} \ell ^3(-\log (r \Lambda ))^{p_+} Z_0}{2-\alpha _+} \left(1+\mathcal{O}\left(\log (r \Lambda )^{-1}\right)\right)
\end{array}
\right)
\\
\label{CmUVsol}
&+C_- \left(
\begin{array}{c}
 -r^{\alpha _-}  (-\log (r \Lambda ))^{p_-} \left(1+\mathcal{O}\left(\log (r \Lambda )^{-1}\right)\right)\\
 x r^{\alpha _-} (-\log (r \Lambda ))^{p_-}\left(1+\mathcal{O}\left(\log (r \Lambda )^{-1}\right)\right)  \\
 \frac{x r^{-2+\alpha _-} \ell ^3 (-\log (r \Lambda ))^{p_-} Z_0}{-2+\alpha _-} \left(1+\mathcal{O}\left(\log (r \Lambda )^{-1}\right)\right) \\
 \frac{x r^{-2+\alpha _-} \ell ^3 (-\log (r \Lambda ))^{p_-} Z_0}{2-\alpha _-}\left(1+\mathcal{O}\left(\log (r \Lambda )^{-1}\right)\right)
\end{array}
\right)
\\
\label{C1UVsol}
&+C_1 \left(
\begin{array}{c}
 \frac{m^2 r^2 (-\log (r \Lambda ))^{2 \rho }}{2 \ell ^5 m_q^2 W_0 \kappa _0} \left(1+\mathcal{O}\left(\log (r \Lambda )^{-1}\right)\right) \\
 -\frac{m^2 r^4  \left(x b_0 w_0^2 Z_0+w_0^2 W_0 \kappa _0-x \ell ^2 Z_0 \kappa _0\right)}{2 \ell ^3 Z_0 \left(-2 w_0^2 W_0+x \ell ^2 Z_0\right) \kappa _0} \left(1+\mathcal{O}\left(\log (r \Lambda )^{-1}\right)\right)\\
 \frac{m^2 r^2  w_0^2 \left(x b_0 Z_0-W_0 \kappa _0\right)}{8 w_0^2 W_0 \kappa _0-4 x \ell ^2 Z_0 \kappa _0} \left(1+\mathcal{O}\left(\log (r \Lambda )^{-1}\right)\right)\\
 1-\frac{m^2 r^2  \left(x b_0 w_0^2 Z_0+w_0^2 W_0 \kappa _0-x \ell ^2 Z_0 \kappa _0\right)}{8 w_0^2 W_0 \kappa _0-4 x \ell ^2 Z_0 \kappa _0}\left(1+\mathcal{O}\left(\log (r \Lambda )^{-1}\right)\right)
\end{array}
\right)
\\
\label{C2UVsol}
&+C_2 \left(
\begin{array}{c}
 1+\frac{m^2 r^2  \log (r \Lambda )}{-2+4 \rho } \left(1+\mathcal{O}\left(\log (r \Lambda )^{-1}\right)\right)\\
 \frac{r^2 \ell ^2  (-\log (r \Lambda ))^{-2 \rho } m_q^2 W_0 \kappa _0}{Z_0} \left(1+\mathcal{O}\left(\log (r \Lambda )^{-1}\right)\right)\\
 \frac{\ell ^3  (-\log (r \Lambda ))^{-2 \rho } m_q^2 w_0^2 W_0 \left(-x b_0 Z_0+W_0 \kappa _0\right)}{2 x Z_0} \left(1+\mathcal{O}\left(\log (r \Lambda )^{-1}\right)\right)\\
 \frac{\ell ^5  (-\log (r \Lambda ))^{1-2 \rho } m_q^2 W_0 \kappa _0}{1-2 \rho }\left(1+\mathcal{O}\left(\log (r \Lambda )^{-1}\right)\right)
\end{array}
\right)\,.
\end{align}

If we set $x \to 0$ first the expansions become
\begin{align}
\label{CQ1UVsol}
 \left(
\begin{array}{c}
 P \\
 Q \\
 \hat{P} \\
 \hat{Q}
\end{array}
\right)&=C_1^{(Q)} \left(
\begin{array}{c}
 \frac{2 r^2}{\ell  w_0^2 W_0}\left(1+\mathcal{O}\left(\log (r \Lambda )^{-1}\right)\right) \\
 \frac{m^2 r^4}{4 \ell ^3 Z_0}\left(1+\mathcal{O}\left(\log (r \Lambda )^{-1}\right)\right) \\
 -1+\frac{m^2 r^2}{8}\left(1+\mathcal{O}\left(\log (r \Lambda )^{-2}\right)\right) \\
 1-\frac{m^2 r^2}{8}\left(1+\mathcal{O}\left(\log (r \Lambda )^{-2}\right)\right)
\end{array}
\right)
\\
\label{CQ2UVsol}
&+C_2^{(Q)} \left(
\begin{array}{c}
 \frac{2 \ell ^2 \log (r \Lambda ) Z_0}{w_0^2 W_0}\left(1+\mathcal{O}\left(\log (r \Lambda )^{-1}\right)\right) \\
 1+\frac{m^2 r^2}{4}\left(1+\mathcal{O}\left(\log (r \Lambda )^{-1}\right)\right) \\
 -\frac{\ell ^3 Z_0}{2 r^2}\left(1+\mathcal{O}\left(\log (r \Lambda )^{-1}\right)\right) \\
 \frac{\ell ^3 Z_0}{2 r^2}\left(1+\mathcal{O}\left(\log (r \Lambda )^{-1}\right)\right)
\end{array}
\right)\,,
\\
\left(
\begin{array}{c}
 P \\
\label{CP1UVsol}
 \hat{P}
\end{array}
\right)&=C_1^{(P)}\left(
\begin{array}{c}
 \frac{m^2 r^2 (-\log (r \Lambda ))^{2 \rho }}{2 \ell ^5 m_q^2 W_0 \kappa _0} \left(1+\mathcal{O}\left(\log (r \Lambda )^{-1}\right)\right)\\
 1-\frac{1}{4} m^2 r^2 \left(1-\frac{\rho }{\log (r \Lambda )}+\mathcal{O}\left(\log (r \Lambda )^{-2}\right)\right)
\end{array}
\right)
\\
\label{CP2UVsol}
&+C_2^{(P)} \left(
\begin{array}{c}
 1+\frac{m^2 r^2 \log (r \Lambda )}{2 (-1+2 \rho )}\left(1+\mathcal{O}\left(\log (r \Lambda )^{-1}\right)\right) \\
 \frac{\ell ^5 (-\log (r \Lambda ))^{1-2 \rho } m_q^2 W_0 \kappa _0}{1-2 \rho }\left(1+\mathcal{O}\left(\log (r \Lambda )^{-1}\right)\right)
\end{array}
\right)
\end{align}
where $Q=0=\hat Q$ for the last two functions.

\section{Proof of the Witten-Veneziano formula} \label{app:WV}

In order to prove the Witten-Veneziano formula we follow the strategy outlined in Sec.~\ref{sec:WVstrategy}. We need to compute the coefficient matrix in~\eqref{coeffmats} at leading nontrivial order by studying the behavior of the IR normalizable solutions in the UV. To do this precisely we expand the IR normalizable solutions $\psi^{(P)}$ and $\psi^{(Q)}$ systematically at small $x$ and $m_f^2$ with $m_f$ being the mass of the fluctuation mode.
We write the expansions of the various fields as
\be
 F = \sum_{M,N=0}^\infty F_{MN}\,  m_f^{2M}\, x^N\ , \qquad (F = P, Q, \hat P, m_f^2\hat Q) \ .
\ee
Notice that in addition to $x$ and $m_f^2$ we have another small parameter $m_q$. Therefore we should study how the components $F_{MN}$ behave at small $m_q$. In particular some of the components are proportional to $m_q^{-1} \sim m_\pi^{-2}$ (as we have already seen in Sec.~\ref{sec:WVstrategy}) and therefore subleading terms in the expansion may contribute at leading order in the $\eta'$ mass. We need to identify such terms.

For the  $\psi^{(P)}$ solution we expect that $Q=\morder{x}=\hat Q$ and therefore we can set $Q_{0M}=0=\hat Q_{0M}$ for all $M$. Substituting the expansions in the fluctuation equations~\eqref{hatPdef}--\eqref{hatQeq} we find
\be
P_{00}' = -\frac{4 e^{-A} G}{V_f w^2} \hat P_{00}\ , \qquad \hat P_{00}' = -e^{3 A} G  V_f \kappa  \tau^2P_{00} \ .
\ee
We choose the IR normalizable solution which is recognized as the $m_f=0=x$ solution discussed in Sec.~\ref{sec:zeroxWV}, with the normalization $P_{00} \to 1$ and $\hat P_{00} \to C_P$ as $r \to 0$. Recall that $C_P$ was related to the pion decay constant in~\eqref{fpiCPrel}, up to corrections suppressed by $m_q$ and $x$. The other components $F_{MN}$ can be solved iteratively from the fluctuation equations with the boundary condition that they vanish in the IR, which uniquely defines $\psi^{(P)}$.

Going higher order in $m_f^2$ we find that
\be
 P_{10}' = \frac{e^{-3 A} G \left( w^2\hat P_{00}-4 e^{2 A}  \kappa  \tau ^2 \hat P_{10}\right)}{V_f w^2 \kappa  \tau ^2}\ , \qquad \hat P_{10}'=  -e^{3 A} G V_f \kappa  \tau ^2P_{10} \ .
\ee
Here $\hat P_{10}$ is not enhanced as $m_q \to 0$, but $P_{10}$ is.
The same arguments as when discussing~\eqref{PprimeGOR} give\footnote{To be precise, the corrections to the integral in~\eqref{P10lim} are larger than $\morder{m_q^0}$ because we did not include the logarithmic corrections to the potentials, but in the final expression the corrections are indeed $\morder{m_q^0}$ as shown in Appendix~D of \cite{Jarvinen:2015ofa}.}
\be \label{P10lim}
 \lim_{r \to 0}P_{10} =  -\frac{ C_P }{W_0 \kappa_0 \ell^3 } \int_0^\infty \frac{  r^3d  r}{\t( r)^2} + \morder{m_q^0} = -\frac{1}{m_\pi^2} +\morder{m_q^0} \ .
\ee
The higher order corrections in $x$ in the equations lead to
\be
P_{01}' = -\frac{4 e^{-A} G}{V_f w^2} \hat P_{01}\ ,\qquad \hat P_{01}' = -e^{3 A} \left( V_a^2 ZP_{00}+G  V_f \kappa  \tau^2 P_{01}\right) \ .
\ee
We notice that these terms are not enhanced as $m_q \to 0$ and may be neglected as their contributions are suppressed by $\morder{x}$.
Summarizing, we obtain the anticipated result
\be
 C_{PP} \simeq \lim_{r \to 0} (P_{00}+m_f^2 P_{10}) = 1- \frac{m_f^2}{m_\pi^2} \ .
\ee
We also obtain the exact solutions
\be
 \hat Q_{01} = 0 \ , \qquad Q_{01} = -V_a P_{00} \ .
\ee
This is remarkable since it allows us to compute the leading contribution to the mixing coefficient $C_{PQ}$. It is given by
\be
 C_{PQ} \simeq  x \lim_{r \to 0} Q_{01} = - x \ ,
\ee
where we used the fact that $V_a \to 1$ at the boundary as required by the implementation of the axial anomaly. Again it may be checked that higher order terms in the series expansion contribute negligibly to $C_{PQ}$.

We then go on discussing the solution $\psi^{(Q)}$. The solution is defined by setting
\be
 \hat Q_{00} = \mathrm{const.} = -C_Q\,,
\ee
which leads to
\be
 Q_{00}(r) = C_Q \int_r^\infty \frac{d\hat r}{e^{3A} Z} \ .
\ee
Normalizing $Q_{00}$ to unity at the boundary (so that $C_{QQ}\simeq 1$)
\be \label{chiCQrel}
 C_Q = \left(\int_0^\infty \frac{d r}{e^{3A} Z}\right)^{-1} = \frac{\chi_\mathrm{YM}}{M^3} \,,
\ee
where $\chi_\mathrm{YM}$ is the (Yang-Mills) topological susceptibility.
Again solving the fluctuation equations perturbatively gives
\be
 P_{00}' = -\frac{e^{-3 A} G \left(C_Q V_a w^2+4 e^{2 A}  \kappa  \tau ^2\hat P_{00}\right)}
 {V_f w^2 \kappa  \tau ^2}\ ,\qquad
 \hat P_{00}' = e^{3 A} \left( V_a Z Q_{00}-G  V_f \kappa  \tau ^2 P_{00}\right) \ .
\ee
As $m_q \to 0$ the most important contribution is the term $\propto C_Q$, and after integration
\be
 \lim_{r\to 0}P_{00}(r) =  \frac{ C_Q }{W_0 \kappa_0 \ell^3 } \int_0^\infty \frac{  r^3d  r}{\t( r)^2} + \morder{m_q^0} = \frac{C_Q}{C_P m_\pi^2} +\morder{m_q^0} \ ,
\ee
where we again could set $V_a$ to unity near the boundary. Therefore $C_{QP} \simeq C_Q/C_Pm_\pi^2$.

Collecting the results,
\be
 \left(\begin{array}{cc}
   C_{PP} & C_{PQ}\\ C_{QP} & C_{QQ}
 \end{array}\right)
 \simeq
 \left(\begin{array}{cc}
   1-\frac{m_f^2}{m_\pi^2} & -x\\ \frac{C_Q}{C_P m_\pi^2} & 1
 \end{array}\right) \ .
\ee
Requiring the determinant of this matrix to vanish
marks the point $m_f=m_{\eta'}$, and results in the Witten-Veneziano formula
\be \label{etapmassfinalapp}
 m_{\eta'}^2 \simeq m_\pi^2 + x \frac{C_Q}{C_P} \simeq  m_\pi^2 + x \frac{N_f N_c\chi_\mathrm{YM}}{f_\pi^2} \ ,
\ee
where we used~\eqref{fpiCPrel} and~\eqref{chiCQrel} in the last step.
Recall that for our normalization conventions $f_\pi^2 = \morder{N_fN_c}$.

\section{The Gell-Mann-Oakes-Renner relation at finite $\bar\theta$}\label{app:gmor}

It is possible to compute the dependence of the mass of the pion on $\bar\theta$ in the QCD-like regime ($0<x<x_c$ and $x_c-x = \morder{1}$) when the quark mass $m_q$ is small. This can be done by solving the fluctuation equations in different approximations near the boundary and in the IR and by requiring that the two results agree.

The general fluctuation equations for the scalar and pseudoscalar mesons are given in Eqs.~\eqref{nsfpsmesons2} in Appendix~\ref{app:thbfluc}. However, the (appropriately normalized) pion wave function is localized near the boundary, for $r \sim \sqrt{m_q/\sigma}$ (see computation for V-QCD at $\bar\theta = 0$
in Appendices~E and~F of~\cite{Jarvinen:2015ofa}). In order to see how the pion wave function behaves in this neighborhood at finite $\bar\theta$, it is useful to rewrite the fluctuations of the tachyon using a decomposition into real and imaginary parts rather then the absolute value and phase.

To make the argument precise, we write
\be \label{Treimdef}
 T = \tau e^{i \xi} + s_r^a t^a + i s_i^a t^a
\ee
for the flavor nonsinglet fluctuations instead of~\eqref{tachyonpaf}. Moreover we decompose the wave functions of the fluctuations to radial and spatial parts as in~\eqref{spsdecomp} and denote the radial wave functions as $\psi_r$ and $\psi_i$.
When $r \ll 1/\LUV$ and when the squared mass of the fluctuations is $\morder{m_q}$, as we expect for the lowest mode which will be identified as the pion below, the tachyon field of~\eqref{Treimdef}, and therefore also the functions $\psi_r$ and $\psi_i$, satisfy the same equation~\eqref{complextauUV} as the background:
\be \label{complexTeq}
 \psi_{r/i}'' +  \pa_r\log\left(  e^{3A} \h V_{f0}\right)\, \psi_{r/i}'  - e^{2A} m_{\t}^2 \psi_{r/i} = \morder{m_q\,\psi_{r/i}}\,, \qquad \left(r \ll \frac{1}{\LUV}\right) \,,
\ee
where the leading correction arises from the terms involving the (squared) mass of the fluctuations.

We denote the linearly independent non-normalizable and normalizable solutions of~\eqref{complexTeq} as $\tau_m$ and $\tau_\sigma$, respectively, so that the background solution reads
\be \label{bgUV}
 \tau(r) e^{i\xi(r)} = m_q e^{i\xi_0}\, \tau_m(r) + \left(\sigma_0+ \morder{m_q}\right)\, \tau_\sigma (r) \,, \qquad \left(r \ll \frac{1}{\LUV}\right) \,.
\ee
Here we
denoted the value of the VEV at $m_q=0$ by $\sigma_0$ and dropped corrections $\morder{m_q}$ to the VEV term as in the analysis of Sec.~\ref{sec:smallmq}. We have chosen the gauge where $\xi \to 0$ as $r \to \infty$, so that $\xi \simeq \bar\theta/N_f$ (see~\eqref{thetaxirel}). The imaginary part of the pion mode is UV normalizable and therefore given as
\be
 \tau \cos(\xi) \, \psi_p = \tau_r\,\psi_p = \psi_i = C_p \tau_\sigma \,,
\ee
where
\be
 \tau_r = \tau \cos (\xi) \simeq m_q \cos (\xi_0) \, \tau_m + \sigma_0\, \tau_\sigma
\ee
is the real part of the background solution. Now we find that
\be
  \psi_p' = C_p \frac{\tau_\sigma' \tau_r-\tau_\sigma \tau_r'}{\tau_r^2}
\ee
and therefore
\be
 V_{f0} \kappa e^{3A} \tau_r^2  \psi_p' = C_p  V_{f0} \kappa e^{3A}\,(\tau_\sigma' \tau_r-\tau_\sigma \tau_r') \,.
\ee
This expression is constant for $r \ll 1/\LUV$, up to correction suppressed by $\morder{m_q}$, as can be seen analyzing~\eqref{complexTeq} -- the latter form is proportional to the Wronskian (see also Appendix~D in~\cite{Jarvinen:2015ofa}). The value of the constant can be computed by taking $r \to 0$ on the right hand side:
\be \label{UVresult}
 V_{f0} \kappa e^{3A} \tau_r^2  \psi_p' = C_p \lim_{r \to 0}(\tau_\sigma' \tau_r-\tau_\sigma \tau_r') = 2 C_p W_0 \kappa_0 \ell^5 m_q \cos \xi_0 \,.
\ee

To complete the computation of the pion mass we need to make contact with the fluctuations in the IR. We note that the tachyon background depends on $m_q$ smoothly when written as a complex field:
\be
 \tau e^{i \xi} = \tau_0 + \morder{m_q} \,,
\ee
where $\tau_0$ is the (real) background at $m_q=0$.
As $\tau_0$ does not have nodes in the IR, the phase $\xi$ is $\morder{m_q}$. The phase is the source of parity violation, and therefore the mixing between the scalar and pseudoscalar fluctuations is controlled by it. That is, all coefficients of the mixing terms in the fluctuation equations~\eqref{eompsip}--\eqref{eompsis} in Appendix~\ref{app:thbfluc} are $\morder{m_q}$ in the IR. Therefore we can choose a basis with an IR normalizable mode, relevant for the pion, which has $\psi_p =\morder{1}$ and $\psi_s =\morder{m_q}$. Taking stock, the fluctuation equation~\eqref{eompsip} for the pion mode becomes
\be \label{piflucteq}
  {1\over G\,V_f\,\kappa\,\tau^2\, e^{3A}}\,\partial_r\left(
{V_f\,\kappa\,\tau^2\,e^{3A}\over G}\,\psi_p'\right)+
m^2\,(\psi_p-\psi_L)= \morder{m_q^2} \,,\qquad \left(r \gg \sqrt{\frac{m_q}{\sigma_0}}\right) \,,
\ee
where the precise range of validity can be seen by inserting the UV expression~\eqref{bgUV} into the fluctuation equations. This implies that in the IR the fluctuation equation takes the same form as for $\bar\theta=0$. In particular, there is not dependence on $\bar\theta$, apart from possible dependence through the mass of the fluctuation $m$.

When $m=0$ exactly, the solutions to~\eqref{piflucteq} are given by
\be \label{pionsols}
{ V_f\,\kappa\,\tau^2\,e^{3A}\over G }\,\psi_p' = \mathrm{const.} \,,\qquad \psi_p = \mathrm{const.}
\ee
The first solution is non-normalizable in the IR, but the second solution is normalizable (when $m_q=0$ also it is identified as the pion mode). For $m$ small but nonzero the terms mix so that the normalizable mode is constant to leading order in $m$ but also includes a component $\propto m^2$ corresponding to the first term which can be computed by integrating~\eqref{piflucteq}.

When $\sqrt{m_q/\sigma_0} \ll r \ll 1/\LUV$ we can match the UV and IR behavior of the fluctuations, construct the pion mode, and determine its mass. From the IR analysis we learned that the pion is dominated by the second solution in~\eqref{pionsols}, but there is also a small component corresponding to the first solution (which can be obtained in principle by integrating~\eqref{piflucteq}), which satisfies
\be \label{IRresult}
 V_f\,\kappa\,\tau^2\,e^{3A}\, \psi_p' = k_p\, m_\pi^2\, \psi_p + \morder{m_q^2} \,,\qquad \left(\sqrt{\frac{m_q}{\sigma_0}}\ll r \ll \frac{1}{\LUV}\right) \,,
\ee
where the proportionality coefficient $k_p$ is independent of $m_q$ and $\bar\theta$. Notice that for $r\ll 1/\LUV$ we were able to approximate $G=1$ and that~\eqref{piflucteq} implies that the left hand side of~\eqref{IRresult} is indeed constant in this regime. Comparing to the result of the UV analysis in~\eqref{UVresult} (noticing that $C_p \simeq \psi_p$ and $\tau_r\simeq \tau$) we see that
\be
 k_p m_\pi^2 =  2 W_0 \kappa_0 \ell^5 m_q \cos \xi_0 + \morder{m_q^2}
\ee
As $k_p$ was independent of $m_q$ and $\bar\theta$, we find that $m_\pi^2 \propto m_q \cos \xi_0$. Comparing to the GOR relation at $\bar\theta=0$~\cite{Jarvinen:2015ofa}, the proportionality constant is found to be $k_p = f_{\pi,0}^2/M^3 N_fN_c\sigma_0$, where $f_{\pi,0}$ is the pion decay constant at $\bar\theta=0$. The final result therefore reads
\be \label{gmorfin}
 f_{\pi,0}^2 m_\pi^2 =  -\langle \bar \psi\psi\rangle\big|_{m_q=0}\, m_q \cos \frac{\bar\theta}{N_f} +\morder{m_q^2} \,,
\ee
where we inserted~\eqref{thetaxirel} and the relation between the condensate and $\sigma_0$.

\end{document}